\documentclass[12pt,a4paper,twoside]{report}
\pdfoutput=1
\usepackage{hyperref}
\usepackage{mathrsfs}
\usepackage{array}
\usepackage{slashed}
\usepackage{subcaption}
\usepackage[utf8]{inputenc}
\usepackage[russian,greek,english]{babel}
\usepackage{amsmath}
\usepackage{amssymb}
\usepackage{graphicx}
\usepackage{color}
\usepackage[table]{xcolor}
\usepackage{rotating}
\usepackage{float}
\usepackage{appendix}
\usepackage[top=3cm,bottom=5cm,left=2.8cm,right=2.8cm]{geometry}
\usepackage{cite}
\usepackage{lmodern}
\usepackage{pgf}
\usepackage{multirow}
\usepackage[absolute,overlay]{textpos} 
\usepackage{multicol} 
\usepackage{calc} 
\usepackage{tikz} 
\usepackage{pifont}
\usepackage{yfonts}
\usepackage{caption}
\hypersetup{linktoc=all,bookmarksopen=true,bookmarksopenlevel=0,bookmarksnumbered,
    colorlinks,
    citecolor=blue,
    filecolor=black,
    linkcolor=blue,
    urlcolor=blue,
		pdfstartview=FitH
}
\usepackage{fancyhdr}
\fancypagestyle{plain}{
\fancyhf{} 
\fancyhead[OR,EL]{\sffamily
\bfseries \nouppercase \thepage} 
}
\definecolor{bordeau}{rgb}{0.3515625,0,0.234375}

\label{The thesis}
\newcommand{\PhDTitleFR}{Objets astrophysiques compacts en gravité modifiée} 
\newcommand{\PhDkeywordsFR}{Gravité modifiée, théories scalaire-tenseur, trous noirs} 
\newcommand{\PhDsumFR}{Vingt années se sont écoulées depuis la découverte de l'expansion accélérée de l'Univers, ravivant l'intérêt pour les théories alternatives de la gravité. Ajouter un champ scalaire à la métrique habituelle de la relativité générale est l'une des manières les plus simples de modifier notre théorie de la gravité. En parallèle, nos connaissances sur les trous noirs et les étoiles à neutrons sont en plein essor, grâce notamment au développement de l'astronomie par ondes gravitationnelles. Cette thèse se situe au carrefour entre les deux domaines : elle étudie les propriétés des objets compacts dans les théories tenseur-scalaire généralisées. Nous commençons par rappeler les théorèmes d'unicité essentiels établis depuis les années soixante-dix. Après avoir présenté le théorème d'unicité pour les trous noirs en théorie de Horndeski, nous l'étendons aux étoiles. La deuxième partie de cette thèse détaille les différentes manières de contourner ce théorème. Parmi elles, nous présentons des solutions où la dépendance temporelle du champ scalaire permet de le raccorder à une solution cosmologique, mais aussi des trous noirs statiques et asymptotiquement plats. Dans la troisième partie, nous établissons un critère important pour la stabilité de ces solutions, qui s'appuie sur leur structure causale. C'est aussi l'occasion d'étudier la propagation des ondes gravitationnelles au voisinage de trous noirs, et de sélectionner les théories dans lesquelles les ondes gravitationnelles se propagent à la même vitesse que la lumière.} 

\newcommand{\PhDTitleEN}{Compact astrophysical objects in modified gravity} 
\newcommand{\PhDkeywordsEN}{Modified gravity, scalar-tensor models, black holes} 
\newcommand{\PhDsumEN}{Twenty years have passed since the discovery of the accelerated expansion of the Universe, reviving the interest for alternative theories of gravity. Adding a scalar degree of freedom to the usual metric of general relativity is one of the simplest ways to modify our gravitational theory. In parallel, our knowledge about black holes and neutron stars is booming, notably thanks to the advent of gravitational wave astronomy. This thesis is at the crossroads between the two fields, investigating the properties of compact objects in extended scalar-tensor theories. We start by reviewing essential no-hair results established since the seventies. After discussing the no-hair theorem proposed for black holes in Horndeski theory, we present its extension to stars. The second part of the thesis investigates in detail the various ways to circumvent this theorem. These notably include solutions with a time-dependent scalar field in order to match cosmological evolution, but also static and asymptotically flat configurations. In a third part, we establish an important stability criterion for these solutions, based on their causal structure. It is also the occasion to study the propagation of gravitational waves in black hole environments, and to select the theories where gravitational waves travel at the same speed as light.} 

\newcommand{\be}{\begin{equation}}
\newcommand{\ee}{\end{equation}}
\newcommand{\pd}{\partial}
\newcommand{\nd}{\nabla}
\newcommand{\nn}{\nonumber\\}
\newcommand{\mc}{\mathcal}
\newcommand{\tx}{\mathrm}

\newcommand{\Lb}{\Lambda_\tx{b}}
\newcommand{\Leff}{\Lambda_\tx{eff}}
\newcommand{\tr}{\tilde{r}}
\newcommand{\Mp}{M_\mathrm{Pl}}
\newcommand{\cmark}{\ding{51}}
\newcommand{\xmark}{\ding{55}}
\newcommand{\hpos}{12.5}																		

\label{Personal}
\newcommand{\PhDname}{M. Antoine Leh\'ebel} 
\newcommand{\NNT}{2018SACLS204} 
\newcommand{\ecodocnum}{564} 
\newcommand{\ecodoctitle}{\'Ecole Doctorale de Physique en \^Ile de France} 
\newcommand{\ecodocacro}{EDPIF}	
\newcommand{\PhDspeciality}{Physique} 
\newcommand{\PhDworkingplace}{l'Universit\'{e} Paris-Sud} 
\newcommand{\defenseplace}{Orsay} 
\newcommand{\defensedate}{2 juillet 2018} 
\newcommand{\logoEt}{\includegraphics[scale=1]{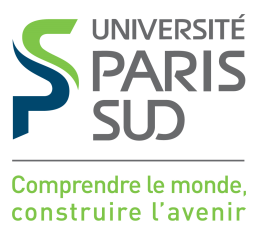}} 
\newcommand{\logoED}{\includegraphics[scale=.2]{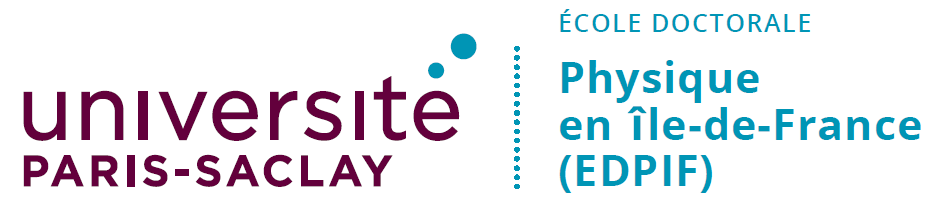}} 
\newcommand{\vpos}{0.1} 

\label{Jury}
\newcommand{\jurynameA}{Marios Petropoulos}
\newcommand{\juryadressA}{\it{Centre de Physique Théorique}, \'Ecole Polytechnique}
\newcommand{\jurygradeA}{Directeur de Recherche}
\newcommand{\juryroleA}{Pr\'{e}sident du jury} %
\newcommand{\jurynameB}{Karim Noui}
\newcommand{\juryadressB}{\it{Institut Denis Poisson}, Tours}
\newcommand{\jurygradeB}{Maître de Conférences}
\newcommand{\juryroleB}{Rapporteur}
\newcommand{\jurynameC}{Antonio Padilla}
\newcommand{\juryadressC}{\it{Center for Astronomy and Particle Physics}, Université de Nottingham}
\newcommand{\jurygradeC}{Professeur}
\newcommand{\juryroleC}{Rapporteur}
\newcommand{\jurynameD}{Eugeny Babichev}
\newcommand{\juryadressD}{\it{Laboratoire de Physique Théorique}, Université Paris-Sud}
\newcommand{\jurygradeD}{Chargé de Recherche}
\newcommand{\juryroleD}{Examinateur}
\newcommand{\jurynameE}{David Langlois}
\newcommand{\juryadressE}{\it{AstroParticule et Cosmologie}, Université Paris-Diderot}
\newcommand{\jurygradeE}{Directeur de Recherche}
\newcommand{\juryroleE}{Examinateur}
\newcommand{\jurynameF}{Christos Charmousis}
\newcommand{\juryadressF}{\it{Laboratoire de Physique Théorique}, Université Paris-Sud}
\newcommand{\jurygradeF}{Directeur de recherche}
\newcommand{\juryroleF}{Directeur de th\`ese}

\label{Document}

\begin{document}
\renewcommand{\arraystretch}{1.3}
\setcounter{tocdepth}{1}

\thispagestyle{empty}
\hypersetup{pageanchor=false}
\begingroup
\pagenumbering{gobble}
\fontsize{12pt}{14pt}\selectfont
\newgeometry{textheight=150ex,top=30pt,headheight=30pt,headsep=30pt,inner=80pt,left=2.8cm,right=2.8cm}
\begin{textblock}{5}(0,0)
	\textblockcolour{bordeau}
	\includegraphics [scale=1]{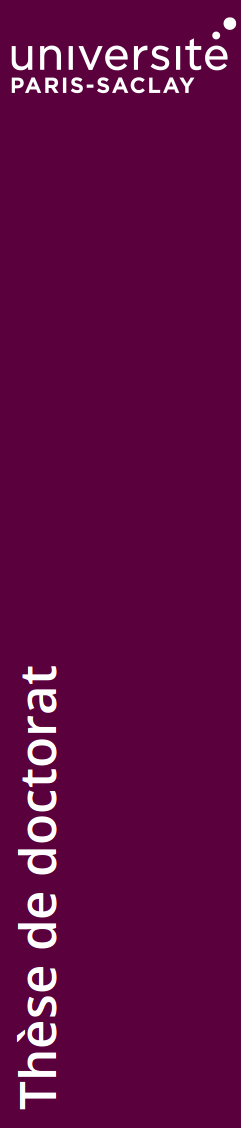}
	\vspace{300mm}
\end{textblock}

\begin{textblock}{1}(0.3,3)
	\Large{\rotatebox{90}{\color{white}{NNT : \NNT}}}
\end{textblock}

\begin{textblock}{1}(\hpos,\vpos)
	\textblockcolour{white}
\logoEt
\end{textblock}

\begin{textblock}{10.3}(5.4,3)
	\textblockcolour{white}
	
	\color{bordeau}
	\begin{flushright}
		\huge{\PhDTitleFR} \bigskip 
		\vfill
		\color{black} 
		\normalsize {Thèse de doctorat de l'Université Paris-Saclay} \\
		préparée à \PhDworkingplace \\ \bigskip
		\vfill
		\'Ecole doctorale n$^{\circ}$\ecodocnum ~\ecodoctitle ~(\ecodocacro)  \\
		
		\small{Spécialité de doctorat: \PhDspeciality} \bigskip 
		\vfill  
		\footnotesize{Thèse présentée et soutenue à \defenseplace, le \defensedate, par} \bigskip
		\vfill
		\Large{\textbf{\textsc{\PhDname}}} 
		\vfill
	\end{flushright}
	
	\color{black}
	\vspace{2cm}
	\begin{flushleft}
		
		\small Composition du Jury :
	\end{flushleft}

	\small
	\newcolumntype{L}[1]{>{\raggedright\let\newline\\\arraybackslash\hspace{0pt}}m{#1}}
	\newcolumntype{R}[1]{>{\raggedleft\let\newline\\\arraybackslash\hspace{0pt}}lm{#1}}
	
	\label{jury} 																				
	\begin{flushleft}
	\begin{tabular}{@{} L{10cm} R{4cm}}
		\jurynameA  & \juryroleA \\ \jurygradeA test, \juryadressA \\
		\jurynameB  & \juryroleB \\ \jurygradeB, \juryadressB \\
		\jurynameC  & \juryroleC \\ \jurygradeC, \juryadressC \\
		\jurynameD  & \juryroleD \\ \jurygradeD, \juryadressD \\
		\jurynameE  & \juryroleE \\ \jurygradeE, \juryadressE \\
		\jurynameF  & \juryroleF \\ \jurygradeF, \juryadressF \\
	\end{tabular} 
	\end{flushleft}   
\end{textblock} 
\restoregeometry
\endgroup

\newpage\null\thispagestyle{empty}\newpage
\newpage\null\thispagestyle{empty}\newpage
\hypersetup{pageanchor=true}

\pagenumbering{roman}

\chapter*{Remerciements}
\addcontentsline{toc}{chapter}{Remerciements}
\markboth{Remerciements}{Remerciements}

\foreignlanguage{greek}{Πρώτα απ 'όλα, σας ευχαριστώ θερμά για την εποπτεία της διατριβής μου, Χρήστου. Δεν θα ήθελα να ήθελα να βρω έναν καλύτερο διευθυντή διατριβών. Είτε πρόκειται για την εντυπωσιακή σας επιστημονική διαίσθηση είτε για την προσοχή σας σε μένα, σας χρωστάω πολλά. Σας ευχαριστώ επίσης για την εμπιστοσύνη μου και μου δίνει τόσο μεγάλη ελευθερία για αυτή τη διατριβή.}
\foreignlanguage{russian}{Евгений, сердечно благодарю вас за ваше постоянное присутствие в течение этих трех лет диссертации. Я многому научился у вас в самых разных областях физики, и всегда будет приятно продолжать сотрудничество с таким талантливым и заботливым исследователем.}

Merci beaucoup à Karim d'avoir relu avec soin ce manuscrit, et d'avoir joué le rôle de tuteur scientifique au cours de ces trois années passées au LPT. Many thanks to Tony as well for having accepted to referee this thesis, and for the interesting questions and comments.
Je remercie également Marios d'avoir présidé le jury durant la soutenance, avec toute l'autorité et l'efficacité qui sont siennes ; merci à David d'avoir mis à contribution son expertise pour évaluer mon travail.

Cette thèse doit beaucoup à mes échanges très enrichissants avec certains experts des théories tenseur-scalaire. J'exprime notamment ma profonde reconnaissance à Gilles Esposito-Farèse, pour sa bienveillance, son investissement ainsi que pour toutes les connaissances qu'il m'a permis de partager. 
Voglio ringraziare Marco Crisostomi per tutte le nostre discussioni scientifiche, ma anche per i bei momenti trascorsi insieme negli ultimi anni.
\foreignlanguage{russian}{Спасибо также Алексу Викману за интересные семинары и дискуссии, которые мы провели}. 
And I would like to thank Matt Tiscareno for having given me a first great experience in the world of research, and for the many recommendation letters I have bothered him for since!

Ces trois années passées au LPT furent très agréables, en particulier grâce à la générosité et la gentillesse de Marie-Agnès Poulet. Merci à Sébastien Descotes-Genon pour son travail de directeur, attentif à chacun dans l'avalanche de courriers que je l'imagine recevoir. Je remercie les autres membres du groupe de cosmologie : Bartjan Van Tent, Sandro Fabbri, Robin Zegers, Scott Robertson et Renaud Parentani, avec qui nous avons si souvent discuté autour d'un repas ou d'un café. 
Hvala, Damir, zbog vašeg stalnog interesa za mnoge od nas. 
Merci beaucoup aussi à Gatien, parrain toujours attentif.
Ce fut aussi un plaisir de passer tout ou partie de ces années avec les étudiants du LPT : Hermès Bélusca, Luiz Vale-Silva, Thibault Delepouve, Luca Lionni, Olcyr Sumensari, Mat\'ias Rodr\'iguez-V\'azquez, Andreï Angelescu, Gabriel Jung, Hadrien Vroylandt, Timothé Poulain, Ma\'ira Dutra, Florian Nortier et Nicolas Delporte. 
Luiz, você ainda me deve um restaurante para a copa do mundo. Olcyr, pretendo ir a Pádua um dia desses.
Estoy esperando impaciente que regreses a París, Mat\'ias.
Andrey, nu îndrăznesc să spun singurul lucru pe care mi l-ai învățat în limba română. Ar putea fi mai bine.

Pour tous les bons souvenirs de ces années cachanaises, je remercie affectueusement mes colocataires (au moins à temps partiel) Jérémy, Mathias, Tim, \'Elo et Max. Un grand merci pour la chanson de thèse et tant d'autres choses à Brigitte, Géraldine, Antoine et Romain. Merci à Adrien et Steven pour leurs analyses footballistiques pointues. Et merci aussi à Séverin, Amaudric, Armand, Camille, Hélèna, Delphine et Grégory pour leur présence ou leurs encouragements.

Enfin, merci de tout coeur à ma famille ; à mes parents, pour m'avoir donné depuis tout petit la curiosité scientifique et l'envie d'apprendre toujours davantage ; à ma grand-mère, à mes frères Philippe et Guillaume, à Alice, Karen et Patrick pour leur présence affectueuse ; à mes oncles, tantes et cousins, tout particulièrement Claire et Hélène.

\newpage
\thispagestyle{empty}
\cleardoublepage
\chapter*{R\'esum\'e}
\addcontentsline{toc}{chapter}{R\'esum\'e}
\markboth{R\'esum\'e}{R\'esum\'e}

En 1915, Einstein proposait sa nouvelle théorie de l'interaction gravitationnelle, la relativité générale. Celle-ci a drastiquement changé notre compréhension de l'espace et du temps. Bien que la relativité générale soit maintenant une théorie centenaire,elle a passé une impressionnante liste de tests et reste le point de départ de toute discussion à propos de la gravité.

\section*{Succès et défis de la physique moderne}

Quand la relativité générale est née, elle fournissait une explication cohérente de l'avance du périhélie de Mercure. Cependant, cette théorie manquait cruellement d'autres tests car, à cette époque, les mesures pouvaient difficilement atteindre la précision nécessaire. La première confirmation expérimentale de la déviation de la lumière est dûe à Eddington en 1919 \cite{Dyson:1920cwa}, au cours d'une éclipse solaire (la fiabilité de son expérience fut cependant remise en question). Quarante ans plus tard, Pound et Rebka utilisèrent la précision sans précédent offerte par l'effet M\"ossbauer pour mesurer le décalage vers le rouge gravitationnel de la lumière tombant d'une tour de vingt-deux mètres de haut \cite{Pound:1960zz}. Cette dernière expérience acheva ainsi la série des trois tests proposés par Einstein pour sa théorie en 1916 \cite{Einstein:1916vd}. En outre, la découverte des pulsars dans les années soixante et soixante-dix fournit une vérification indirecte de l'émission des ondes gravitationnelles, en particulier via l'étude du pulsar binaire de Hulse et Taylor \cite{Hulse:1974eb}. Depuis 1980, une quantité importante d'expériences a été mise en place pour vérifier les prédictions de la relativité générale, avec une précision croissante (et un succès croissant). Enfin, des ondes gravitationnelles ont été détectées directement grâce à des interféromètres gravitationnels pour la première fois en 2015 \cite{Abbott:2016blz}. Cette dernière observation est aussi totalement cohérente avec l'existence des trous noirs, dont nous parlerons plus en détail dans quelques paragraphes.

Cette liste de tests locaux est déjà très impressionnante, mais la relativité générale offre bien davantage dans le cadre de la cosmologie, c'est-à-dire de la physique aux échelles largement supérieures à la taille d'une galaxie. Conformément aux prédictions de la relativité générale, Hubble observa en 1929 que l'Univers est en expansion \cite{Hubble:1929ig}. \`A cause de cette expansion, on s'attend théoriquement à ce que l'Univers soit de plus en plus chaud lorsque l'on remonte le temps. C'est donc seulement après un instant donné de l'histoire de l'Univers que les particules chargées se sont combinées pour former des atomes, autorisant ensuite la lumière à voyager librement. En remontant encore plus loin dans le temps, on peut aussi prédire ce que devrait être de nos jours l'abondance relative des éléments les plus légers, comme l'hydrogène ou l'hélium. Ces deux prédictions ont été confirmées expérimentalement ; la première par l'observation du fond diffus cosmologique \cite{Penzias:1965wn}, la seconde par l'observation du spectre lumineux des quasars notamment \cite{Pettini:2012ph}.

Penchons-nous plus en détail sur la découverte de l'accélération de l'expansion de l'Univers en 1998 \cite{Perlmutter:1998np}, grâce au suivi des supernovas de type Ia. La composante d'énergie qui génère cette accélération est dénommée énergie noire. \`A ce stade, soulignons qu'il ne s'agit pas d'une faille dans la théorie de la relativité générale. L'action d'Einstein-Hilbert pour la relativité générale s'écrit
\be
S_\tx{EH}=\dfrac{\Mp^2}{2}\displaystyle\int{\tx{d}^4x\sqrt{-g}\,R},
\label{eq:EHactionfr}
\ee
où $R$ est le scalaire de Ricci 
et $\Mp=1/\sqrt{8\pi G}$, avec $G$ est la constante de Newton. Une constante cosmologique $\Lambda$ peut légitimement être ajoutée à l'action ci-dessus. Cela conduit à une phase d'expansion accélérée tardive, et la valeur de $\Lambda$ doit être spécifiée à partir des observations. Nous expliquerons plus loin pourquoi la valeur particulière prise par $\Lambda$ pose problème. 

Malgré tous 
ses succès, 
la relativité générale ne peut pas être la théorie ultime de la 
gravité. On peut formuler deux sortes d'objections. Premièrement, il existe des phénomènes qui ne trouvent aucune explication dans le cadre 
de la relativité générale. On peut citer 
la nécessité d'une phase d'expansion exponentielle dans l'Univers jeune (dénommée inflation), ou encore la présence dans l'Univers d'un type de matière qui n'interagit que gravitationnellement (la matière noire). Deuxièmement, il y a des problèmes de nature purement théorique. Ceux-ci ne doivent pas être pris à la légère. 

Tout d'abord, la relativité générale est une théorie non-renormalisable, contrairement au modèle standard de la physique des particules. En d'autres termes, la relativité générale perd son caractère prédictif au-delà de l'échelle de Planck, c'est-à-dire vers $10^{19}$~GeV. Trouver une théorie quantique de la gravité, qui serait prédictive à toutes les échelles d'énergie, est toujours un défi majeur de la physique moderne. Cependant, en principe, nous n'avons pas besoin de connaître la totalité de la structure de la théorie à haute énergie pour décrire ce qui se passe à basse énergie. Dans cette thèse, nous considérerons toujours la gravité comme un processus classique.

\`A côté de cette question, on trouve les problèmes de naturel. En théorie quantique des champs, tout paramètre est la somme d'une valeur « nue » fixée et de corrections quantiques, sous la forme d'un développement en série des constantes de couplage. Cette série est générée par la création et l'annihilation de particules virtuelles. Un problème de naturel apparaît lorsque la valeur mesurée d'un paramètre est largement inférieure à celle des corrections quantiques associées. Cette situation manque de naturel parce qu'elle requiert un ajustement fin entre la valeur nue et les corrections quantiques, afin qu'il ne reste qu'une très faible contribution finale. La principale grandeur 
qui présente un problème de naturel est la constante cosmologique $\Lambda$.
D'après les observations, elle est estimée à $\Lambda\simeq10^{-65}$~GeV$^2$. Avec une coupure ultraviolette $E_\tx{max}$ à l'échelle de Planck, on s'attend à des corrections d'ordre $\delta\Lambda\simeq E_\tx{max}^4/\Mp^2=\Mp^2\simeq 10^{37}$~GeV$^2$. En d'autres termes, si la valeur nue de la constante cosmologique déviait d'un pour $10^{123}$ des corrections quantiques, l'Univers se comporterait de façon totalement différente. En réalité, un calcul plus sérieux, utilisant la régularisation dimensionnelle plutôt qu'une coupure ultraviolette, donne $\delta\Lambda\simeq 10^{-29}$~GeV$^2$ \cite{Martin:2012bt}. Il reste tout de même un ajustement de cinquante-cinq ordres de grandeur. 
Le problème de la constante cosmologique est encore renforcé si l'on considère les transitions de phase dans l'Univers jeune. Celles-ci font varier la constante cosmologique du même ordre de grandeur que les corrections quantiques $\delta\Lambda$. Le problème semble inextricable dans le cadre de la relativité générale.

\section*{Modifications infrarouges de la gravité}

La valeur peu naturelle de la constante cosmologique est probablement la raison la plus frappante pour modifier notre théorie de la gravité aux grandes échelles. Dans cette thèse, nous nous concentrons sur les théories scalaire-tenseur, caractérisées par l'ajout d'un degré de liberté scalaire, couplé de manière non minimale à la gravité. On peut se demander quelle est la théorie scalaire-tenseur la plus générale que l'on puisse écrire. Pour avoir affaire à une théorie saine, il faut éviter la présence d'un type d'instabilité dénommé fantôme d'Ostrogradski. Il y a deux possibilités pour y arriver : que les équations du champ soient du deuxième ordre, ou que le Lagrangien décrivant la théorie possède une propriété nommé dégénérescence \cite{Woodard:2006nt}. La première possibilité a été explorée par Horndeski en 1974 \cite{Horndeski:1974wa}. Il a déterminé le Lagrangien scalaire-tenseur le plus général avec des équations du champ du deuxième ordre. Cette théorie peut être paramétrée par quatre fonctions arbitraires $G_2$, $G_3$, $G_4$ et $G_5$ du champ scalaire $\varphi$ et de la densité cinétique $X=-\pd_\mu\varphi\,\pd^\mu\varphi/2$ :
\be
\label{eq:Hfr}
S_\tx{H} = \displaystyle\int \tx{d}^4x \sqrt{-g} \,\left(\mc{L}_2+\mc{L}_3+\mc{L}_4+\mc{L}_5\right),
\ee
avec
\begin{align}
\mc{L}_2 &=G_2(\varphi,X) ,
\label{eq:L2fr}
\\
\mc{L}_3 &=-G_3(\varphi,X) \,\Box \varphi ,
\label{eq:L3fr}
\\
\mc{L}_4 &= G_4(\varphi,X) R + G_{4X} \left[ (\Box \varphi)^2 -\nabla_\mu\pd_\nu\varphi \,\nabla^\mu\pd^\nu\varphi\right] ,
\label{eq:L4fr}
\\
\begin{split}
\mc{L}_5 &= G_5(\varphi,X) G_{\mu\nu}\nabla^\mu \pd^\nu \varphi - \frac{1}{6}\, G_{5X} \big[ (\Box \varphi)^3 - 3\,\Box \varphi\, \nabla_\mu\pd_\nu\varphi\,\nabla^\mu\pd^\nu\varphi
\\
&\quad + 2\,\nabla_\mu\pd_\nu\varphi\, \nabla^\nu\pd^\rho\varphi\, \nabla_\rho\pd^\mu\varphi \big],
\label{eq:L5fr}
\end{split}
\end{align}
où un $X$ en indice signifie une dérivée par rapport à $X$. Comme mentionné ci-dessus, la théorie de Horndeski n'est pas la plus générale que l'on puisse écrire tout en évitant un fantôme d'Ostrogradski. On peut considérer des théories avec des équations du champ d'ordre supérieur, tant que le Lagrangien est dégénéré. Les premiers termes sains qui furent obtenus dans cette classe \cite{Gleyzes:2014dya} s'écrivent :
\begin{align}
\mc{L}_4^{\tx{bH}} &= F_4(\varphi,X)\, \epsilon^{\mu \nu \rho \sigma} {\epsilon^{\alpha \beta \gamma}}_\sigma\, \pd_\mu\varphi \,\pd_\alpha\varphi\, \nd_\nu \pd_\beta\varphi \,\nd_\rho \pd_\gamma\varphi ,
\label{eq:bH4fr}
\\
\mc{L}_5^{\tx{bH}} &=F_5(\varphi,X)\, \epsilon^{\mu \nu \rho \sigma} \epsilon^{\alpha \beta \gamma \delta}\, \pd_\mu\varphi \,\pd_\alpha\varphi \,\nd_\nu \pd_\beta\varphi \,\nd_\rho \pd_\gamma\varphi \,\nd_\sigma \pd_\delta\varphi.
\label{eq:bH5fr}
\end{align}
avec des fonctions $F_4$ et $F_5$ libres. N'importe quelle combinaison de $G_4$ et $F_4$, ou $G_5$ et $F_5$, ou encore $F_4$ et $F_5$ (la présence des termes $G_2$ et $G_3$ ne présentant pas d'importance) donne une théorie saine. Cependant, un mélange arbitraire de $G_4$, $G_5$, $F_4$ et $F_5$ donne une théorie instable. Nous appelons théorie de \textit{Horndeski et au-delà} le sous-ensemble obtenu en combinant les Lagrangiens (\ref{eq:L2fr})--(\ref{eq:L5fr}) et (\ref{eq:bH4fr})-(\ref{eq:bH5fr}) qui reste sain. Notons qu'il existe un ensemble plus large de théories saines, nommées théories scalaire-tenseur dégénérées d'ordre supérieur \cite{Langlois:2015cwa,Langlois:2015skt,Achour:2016rkg,BenAchour:2016fzp,Crisostomi:2016tcp,Crisostomi:2016czh}. Nous nous restreindrons cependant dans cette thèse à la théorie de Horndeski et au-delà, telle que définie ci-dessus, ce qui permet déjà d'appréhender les caractéristiques essentielles de ces modèles.

\section*{Les trous noirs}

La manière la plus simple de définir un trou noir est probablement la suivante : une région de l'espace temps où l'attraction gravitationnelle est si intense que ni la matière ni la lumière ne peuvent s'en échapper. Le concept n'est pas entièrement spécifique à la relativité générale, et fut évoqué par des scientifiques du XVIII$^\tx{\grave{e}me}$ et du XIX$^\tx{\grave{e}me}$ siècle. En 1916, Schwarzschild proposa sa célèbre solution, bien qu'elle ne fût pas interprétée comme un trou noir à l'époque. Ce n'est qu'à la fin des années cinquante qu'on lui donna ce sens. Kerr trouva la solution exacte pour un trou noir en rotation en 1963 \cite{Kerr:1963ud}. On conjectura à l'époque qu'il existait très peu de solutions pour un trou noir à l'équilibre. Des théorèmes furent établis, prouvant qu'en relativité générale, un trou noir au repos est entièrement caractérisé par sa masse, son moment cinétique et sa charge électrique \cite{Israel:1967wq,Carter:1971zc,Robinson:1975bv}. Ces résultats et les suivants sont généralement appelés théorèmes de calvitie, parce qu'ils imposent que les champs de matière (autres que le champ électromagnétique) sont dans un état trivial. Lorsque ce n'est pas le cas, par exemple si un champ scalaire est non trivial, on dit que la solution possède une chevelure (scalaire dans ce cas).  

En relativité générale, on pense que les trous noirs sont principalement de deux types : trous noirs stellaires et trous noirs supermassifs. Les trous noirs de masse stellaire se forment par l'effondrement gravitationnel d'étoiles suffisamment lourdes. Les trous noirs supermassifs ont des masses de l'ordre du million de masses solaires, et leur processus de formation et plus incertain (ils se forment probablement par absorption d'étoiles ou fusion avec d'autres trous noirs). Les trous noirs sont intrinsèquement difficiles à observer. Cependant, l'accrétion de matière par les trous noirs s'accompagne de radiation électromagnétique. L'émission de rayons X durant l'accrétion est ce qui amena à proposer Cygnus X-1 comme le premier trou noir jamais détecté en 1972 \cite{Webster:1972bsw}. Par ailleurs, la trajectoire des étoiles à proximité de Sagittarius A$^*$, au centre de la Voie lactée, montre que 4,3 millions de masses solaires sont compactées dans une sphère d'un rayon inférieur à $2\cdot10^{-3}$ années lumières \cite{Gillessen:2008qv}. Ceci ne prouve pas l'existence des trous noirs supermassifs, mais constitue un indice très fort. Le but de l'Event Horizon Telescope est d'observer le disque d'accrétion qui entoure Sagittarius A$^*$. Les résultats sont attendus pour fin 2018. Enfin, la preuve la plus directe de l'existence des trous noirs nous vient de la détection d'ondes gravitationnelles en 2015 \cite{Abbott:2016blz} (et de nombreuses fois depuis). Tout comme nous détectons la lumière, qui est le secteur dynamique de l'électromagnétisme, nous détectons à présent la partie dynamique de la gravité. En principe, n'importe quel objet se déplaçant émet des ondes gravitationnelles, mais seule la fusion de deux trous noirs, ou de deux étoiles à neutrons, relâche assez d'énergie pour être détectable par nos interféromètres gravitationnels. Jusqu'ici, les signaux détectés concordent avec les attentes théoriques. Dans les années à venir, une avalanche de données sera disponible grâce aux détecteurs basés au sol, nous permettant de tester la relativité générale dans le plus fort régime d'énergie disponible. Pour finir, le lancement de l'interféromètre spatial Laser Interferometer Space Antenna est prévu pour 2034. Cette expérience essaiera notamment d'observer la fusion de trous noirs supermassifs.

\section*{Principaux résultats de la thèse}

Les théories scalaire-tenseur possèdent des propriétés cosmologiques intéressantes. En parallèle, il est important de savoir si des objets compacts (trous noirs et étoiles) peuvent exister dans ces théories, et si oui, à quel point ces objets sont similaires à ceux rencontrés en relativité générale. Ces interrogations sont le point de départ de ma thèse. Les premiers outils indispensables pour cette analyse sont les théorèmes de calvitie. Dans la première partie, nous discutons un théorème de calvitie préalablement établi en théorie de Horndeski, conçu pour les trous noirs statiques, à symétrie sphérique et asymptotiquement plats. Nous étendons ce théorème aux étoiles sous des hypothèses très similaires.

Cependant, la théorie de Horndeski et au-delà est complexe, et de nombreuses hypothèses sont nécessaires pour établir les théorèmes de calvitie. En conséquence, il existe aussi de nombreuses manières d'arriver à des solutions qui possèdent une chevelure. Nous explorons ces voies dans la deuxième partie de la thèse. L'un des résultats essentiels est que, lorsque le champ scalaire joue le rôle de l'énergie noire, les trous noirs (ou les étoiles) possèdent une chevelure de manière générique. Nous montrons ceci en analysant l'effet des termes cubiques et quartiques les plus simples en théorie de Horndeski. Lorsque l'on impose une symétrie $\mathbb{Z}_2$ sur le secteur scalaire (c'est-à-dire dans le cas quadratique et quartique), il est facile de trouver des solutions exactes.  Certaines reproduisent exactement les solutions de la relativité générale, et notamment un espace-temps de Schwarzschild-de Sitter avec des propriétés d'auto-ajustement simples. Dans le cas cubique, il faut recourir à l'intégration numérique pour trouver des solutions de type trou noir, mais il en existe, qui possèdent des propriétés d'auto-ajustement similaires aux précédentes.

Cette partie est également l'occasion d'étudier quels modèles possèdent des solutions asymptotiquement plates avec un champ scalaire non trivial et statique (par opposition au cas où son évolution temporelle est dictée par la cosmologie). Il est plus difficile de trouver de telles solutions, et seuls quelques modèles permettent en fait de contourner le théorème dans ce cas. C'est la présence (ou l'absence) de termes spécifiques dans le Lagrangien qui autorise les solutions à s'écarter de la relativité générale. Nous examinons en détail ces termes dans le cas des secteurs quartiques et quintiques de la théorie de Horndeski et au-delà. Il est toujours légitime d'étudier des modèles quintiques dans ce cadre, où le champ scalaire ne joue aucun rôle à l'échelle cosmologique. Dans le cas où la théorie est invariante par translation, le Lagrangien quintique qui impose un champ scalaire non trivial est équivalent à un couplage linéaire entre le champ scalaire et l'invariant de Gauss-Bonnet. Ce modèle n'admet pas de solutions de type trou noir régulières, sauf si l'on autorise la norme du courant de Noether (associé à l'invariance par translation) à diverger sur l'horizon. 

Une fois que des solutions avec chevelure sont connues, l'étape suivante est d'étudier leur stabilité (et pour finir leur formation par effondrement). Le but de la troisième partie est d'étudier la stabilité de certaines solutions avec chevelure présentées dans le reste de la thèse. Nous nous concentrons sur les solutions où le champ scalaire correspond à l'énergie noire. Dans ce cas, l'imbrication entre dépendance spatiale et temporelle rend caduque l'utilisation de critères de stabilité usuels. En particulier, nous prouvons que lorsque le Hamiltonien est non borné inférieurement (ce qui est habituellement interprété comme un fantôme) dans certains systèmes de coordonnées, la solution peut malgré tout être stable. \`A la place, nous établissons le critère de stabilité correct : les cônes causaux associés à chaque degré de liberté (scalaire, spin 2 et matière) doivent avoir en commun une direction de genre temps et une surface de Cauchy de genre espace. Ce résultat n'est pas limité à une théorie scalaire-tenseur spécifique, et pourrait se révéler intéressant dans divers modèles de gravité modifiée. Nous appliquons ce critère à la solution de Schwarzschild-de Sitter auto-ajustée présentée plus tôt dans la thèse. Il existe une fenêtre de stabilité pour les paramètres de la théorie (qui semble ne pas dépendre de la présence d'un trou noir). Les conditions de stabilité empêchent cependant de passer d'une grande constante cosmologique nue à une faible valeur effective compatible avec les observations. 

Comme conséquence directe de l'analyse de stabilité linéaire, nous sommes aussi capables de calculer la vitesse des ondes gravitationnelles dans une solution à symétrie sphérique et fortement courbée (par opposition aux solutions cosmologiques faiblement courbées, où le résultat était déjà connu). Lorsque la vitesse des ondes lumineuses et gravitationnelles est la même aux échelles cosmologiques, elle reste identique dans l'environnement fortement courbé d'un trou noir. Nous présentons finalement une classe de modèles de Horndeski et au-delà qui passent les tests des ondes gravitationnelles ainsi que les tests locaux, et qui fournissent un vrai mécanisme d'auto-ajustement de la constante cosmologique.

Cette thèse a donné lieu aux publications scientifiques listées ci-dessous :
\begin{itemize}
\item E. Babichev, C. Charmousis, G. Esposito-Farèse et A. Lehébel, \textit{Hamiltonian vs stability and application to Horndeski theory} 
\item E. Babichev, C. Charmousis, G. Esposito-Farèse et A. Lehébel, \textit{Stability of Black Holes and the Speed of Gravitational Waves within Self-Tuning Cosmological Models, Phys. Rev. Lett.} \textbf{120} (2018) 241101
\item A. Lehébel, E. Babichev et C. Charmousis, \textit{A no-hair theorem for stars in Horndeski theories, JCAP} \textbf{1707} (2017) 037
\item E. Babichev, C. Charmousis et A. Lehébel, \textit{Asymptotically flat black holes in Horndeski theory and beyond, JCAP} \textbf{1707} (2017) 037
\item E. Babichev, C. Charmousis, A. Lehébel et T. Moskalets, \textit{Black holes in a cubic Galileon universe, JCAP} \textbf{1609} (2016) 011
\item E. Babichev, C. Charmousis et A. Lehébel, \textit{Black holes and stars in Horndeski theory, Class. Quant. Grav.} \textbf{33} (2016) 154002
\end{itemize}

\newpage
\thispagestyle{empty}
\cleardoublepage
{\hypersetup{linkcolor=black}
\pdfbookmark[0]{\contentsname}{toc}
\tableofcontents}

\newpage

\pagenumbering{arabic}

\chapter*{Introduction}
\addcontentsline{toc}{chapter}{Introduction}
\markboth{Introduction}{Introduction}

In 1915, Einstein proposed his new theory of gravitational interaction, general relativity. It drastically changed our understanding of space and time. Although general relativity is now a centenarian theory, it has passed an impressive list of tests and remains the starting point of any discussion about gravity.

\section*{Successes and challenges of modern physics}
\addcontentsline{toc}{section}{Successes and challenges of modern physics}

When general relativity was formulated, it provided a consistent explanation of the perihelion advance of Mercury. However, the theory was sorely lacking in other tests, because at that time, measurements could hardly reach the necessary precision. The first experimental confirmation of light deflection was made by Eddington in 1919 \cite{Dyson:1920cwa}, during a solar eclipse (although the reliability of his experiment was later questioned). Forty years later, Pound and Rebka used the unprecedented precision offered by M\"ossbauer effect to measure the gravitational redshift of light falling from a twenty-two meter high tower \cite{Pound:1960zz}. This completed the three classical tests proposed by Einstein for his theory in 1916 \cite{Einstein:1916vd}. On top of this, the discovery of pulsars during the sixties and seventies provided an indirect check of gravitational wave emission, notably through the study of the Hulse-Taylor binary pulsar \cite{Hulse:1974eb}. Since 1980, a number of experiments have been set up to check the local predictions of general relativity, with increasing precision (and increasing success). Finally, gravitational waves were detected directly thanks to gravitational interferometers for the first time in 2015 \cite{Abbott:2016blz}. This last observation was also fully consistent with the existence of black holes, about which we will say more in a couple of paragraphs.

This list of local tests is already impressive, but general relativity has far more to offer in the framework of cosmology, that is physics at lengthscales well above the size of a galaxy. In accordance with general relativity predictions, the Universe was found to be in expansion by Hubble in 1929 \cite{Hubble:1929ig}. Because of this expansion, one theoretically expects that the Universe was warmer and warmer when going back in time. Thus, it is only after a given time in the History of Universe that charged particles combined together to form atoms, thence allowing light to travel freely. Going back even further in time, one is also able to predict what the relative abundance of light elements, like hydrogen or helium, should be nowadays. These two predictions were verified experimentally; the first one through the observation of the cosmic microwave background \cite{Penzias:1965wn}, the second one through spectral observation of quasar light notably \cite{Pettini:2012ph}.

Let us discuss separately the discovery of the accelerated expansion of the Universe in 1998 \cite{Perlmutter:1998np}, through the survey of type Ia supernovae. The energy component that drives this acceleration is designated under the name of dark energy. At this stage, it should be emphasized that it is not a flaw in the theory. The usual Einstein-Hilbert action for general relativity reads
\be
S_\tx{EH}=\dfrac{\Mp^2}{2}\displaystyle\int{\tx{d}^4x\sqrt{-g}\,R},
\label{eq:EHaction}
\ee
where $R$ is the Ricci scalar\footnote{Throughout this thesis, we will use the notations of Wald \cite{Wald:1984rg}, except for the distinction between Latin and Greek indices; the signature of the metric is $(-,+,+,+)$. Unless specified otherwise, we work in units where the speed of light $c$ and the reduced Planck constant $\hbar$ are equal to unity.} and $\Mp=1/\sqrt{8\pi G}$, $G$ being Newton's constant. A cosmological constant term $\Lambda$ may legitimately be added in the above action. Doing so leads to a late phase of accelerated expansion, and the value of $\Lambda$ should be specified according to observations. We will explain later why the specific value of $\Lambda$ appears problematic. Note that, when refering to general relativity, we generically mean the action (\ref{eq:EHaction}) supplemented with a cosmological constant term $\Lambda$, that is:
\be
S_\tx{GR}=\dfrac{\Mp^2}{2}\displaystyle\int{\tx{d}^4x\sqrt{-g}\,(R-2\Lambda)},
\label{eq:GRaction}
\ee

As just evoked, general relativity accounts for most observations in the Solar System and over cosmological distances. On the other hand, in a laboratory, it becomes harder and harder to test gravity as lengthscale decreases. Most accurate checks of the inverse square law, for instance, cannot probe distances below a few tenths of micrometers (see e.g., \cite{Chiaverini:2002cb}). Below these scales, physics is described by the standard model of particle physics. The framework of standard model is quantum field theory, which combines quantum mechanics and special relativity in a coherent way. From quantum mechanics, quantum field theory keeps the Hilbert space structure. From special relativity, it keeps the invariance under Poincaré group transformations. Then, each species of particles is described by a fermionic or bosonic field, which corresponds to an irreducible and unitary representation of the Poincaré group. The dynamical structure and interactions between the different species are then specified by a Lagrangian density. In a minimalist fashion, this Lagrangian density may be written
\be
\mc{L}_\tx{SM}=-\dfrac14 F_{\rho\sigma}F^{\rho\sigma}+i\bar\psi \slashed D \psi-(D_\rho H)^\dagger D^\rho H + \bar\psi\, Y \psi H - \mu^2 |H|^2 - \lambda |H|^4.
\label{eq:SM} 
\ee
In the above expression, $\psi$ denotes collectively the fermionic fields (that is, quarks, leptons and their associated neutrinos). The term $F_{\rho\sigma}F^{\rho\sigma}$ stands for the sum of bosonic field strengths squared; these include the electroweak bosons and the gluons. Additionally, one scalar field is present in the model: the Higgs boson $H$, with its potential parametrized by $\mu$ and $\lambda$. Bosons interact with fermions through the gauge covariant derivative $D_\rho$. Finally, the non-zero vacuum expectation value of the Higgs boson generates a mass for fermions (other than neutrinos) through the Yukawa couplings $Y$, and some bosons acquire a mass through the $(D_\rho H)^\dagger D^\rho H$ term.

The above compact form hides the fact that nineteen parameters must be provided to fully specify the Lagrangian density. The standard model does not make any prediction about their precise value, and they must all be measured experimentally. This is however not a shortcoming. The standard model as such is consistent, and it provides incredibly precise predictions, notably concerning the value of the fine-structure constant \cite{Mohr:2015ccw}. 

Despite all these successes, the standard model together with general relativity cannot be the ultimate theory of nature. One can formulate two types of objections. First, there exist phenomena that do not find any explanation in the framework of the standard model, nor general relativity. Such examples are the oscillation of neutrinos (implying that they are massive), the apparent necessity for an exponential expansion phase in the early Universe (known as inflation), as well as the presence in the Universe of a matter species that interacts only through gravitation (dark matter). Secondly, there are problems of purely theoretical nature. These are not to be underestimated. Let us draw a parallel with the situation of Newtonian gravity around 1910. The observed advance of the perihelion of Mercury was deviating from the predictions of Newton's theory. This argument falls in the first category, the unexplained phenomena. However, rather than this problem, what lead Einstein to his theory of general relativity was mostly the theoretical inconsistency of the instantaneous propagation of gravity, i.e., a purely theoretical argument against Newton's theory. Standard model and general relativity face several of such problems nowadays.

First of all, general relativity is a non-renormalizable theory, as opposed to the standard model. In other words, general relativity loses its predictive power above the Planck energy scale, that is around $10^{19}$~GeV. It is still a major challenge of modern physics to find a quantum theory of the gravitational interaction, that would be predictive at all energy scales. However, in principle, we do not need to know the whole high-energy structure of the theory to treat lower energy scales. In this thesis, we will always consider gravitational interaction as a classical process. 

Aside of this question are the so-called naturalness problems. In quantum field theory, all parameters are the sum of a fixed ``bare'' value and of quantum corrections, under the form of a series expansion in powers of the coupling constants. This series is generated by creation and annihilation of virtual particles. Naturalness problems occur when the measured value of the parameter is much smaller than the quantum corrections. This situation is unnatural because it requires a fine adjustment between the bare value and the quantum corrections, so that they leave a very small overall contribution. The two main quantities\footnote{Another fine-tuning issue is the so-called strong CP problem. It corresponds to the absence in the standard model of a specific gluon-gluon interaction allowed by the symmetries. The dimensionless quantity that parametrizes this interaction is constrained by experiment to be smaller than $10^{-10}$.} that exhibit this fine-tuning problem are the cosmological constant $\Lambda$ and the Higgs boson mass $M_\tx{H}=\sqrt{-2\mu^2}$. Assuming an ultraviolet cutoff $E_\tx{max}$ for the standard model, the leading order\footnote{From a slightly different viewpoint, one can allow for a fine tuning between the bare value and the leading-order correction. Indeed, these two quantities (bare value and first correction) are actually divergent. One cancels out these two infinities, so why not canceling out two (finite) large numbers? However, such a tuning is completely spoiled by the next-to-leading-order correction, and further ones. Thus, the quantum corrections one would need to cancel depend heavily on the ultraviolet completion of the theory, which we do not know. This problem is known as radiative instability \cite{Kaloper:2014dqa}.} corrections to the Higgs boson mass $\delta M_\tx{H}$ are of order $Y E_\tx{max}$ \cite{tHooft:1979rat}. The Higgs boson mass itself is $M_\tx{H}=125$~GeV \cite{Aad:2012tfa,Chatrchyan:2012xdj}. The Yukawa couplings $Y$ being of order unity, the situation may be considered natural if $\delta M_\tx{H}\simeq E_\tx{max} \simeq 10^3$~GeV (not too large with respect to $M_\tx{H}$). If $E_\tx{max} \simeq 10^{16}$~GeV, characteristic scale of grand unified theories, the bare mass must be tuned with the quantum corrections at a level of one part in $10^{14}$. It is even worse if the cutoff of the standard model is assumed to be at Planck scale, around $10^{19}$~GeV. This fine tuning was one of the reasons to introduce supersymmetry, and to expect it to show up around $10^3$~GeV. However, the Large Hadron Collider now probes these energy scales, and shows no sign of new physics. Concerning the cosmological constant, the situation is much worse. Observationally, it is estimated to be $\Lambda\simeq10^{-65}
$~GeV$^2$. Assuming a sharp cutoff $E_\tx{max}$ at Planck scale, one expects quantum corrections of order $\delta\Lambda\simeq E_\tx{max}^4/\Mp^2=\Mp^2\simeq 10^{37}$~GeV$^2$. In other words, if the bare value of the cosmological constant was deviating of more than one part in $10^{123}$ of the quantum corrections, the Universe would behave entirely differently. Actually, a more involved computation, using dimensional regularization rather than a sharp cutoff, gives $\delta\Lambda\simeq 10^{-29}$~GeV$^2$ \cite{Martin:2012bt}. This still leaves a fine tuning of fifty-five orders of magnitude. Contrary to the Higgs boson mass, one cannot hope to cure the problem by a breakdown of the theory at a relatively low energy scale. Indeed, through the naive cutoff approach, the theory should already fail at $E_\tx{max}\simeq 1$~meV for the value of $\Lambda$ to be natural. The cosmological constant problem is further reinforced by considering phase transitions in the early Universe, like electroweak or quantum chromodynamics phase transitions. These will de-tune the value of the cosmological constant by amounts of similar magnitude as the quantum corrections $\delta\Lambda$. The problem seems inextricable in the framework of general relativity together with the standard model.

\section*{Infrared modifications of gravity}
\addcontentsline{toc}{section}{Infrared modifications of gravity}

The unnatural magnitude of the cosmological constant is probably the most vivid reason for trying to modify our theory of gravity over large distances. However, scientists did not wait for this discovery to explore alternative theories of gravity, pushed sometimes only by theoretical curiosity. It is certainly beyond the scope of this thesis to review extensively all modifications of gravity that were proposed. We will focus on scalar-tensor theories, characterized by a scalar degree of freedom that is non-minimally coupled to the metric, and say a few words about higher-dimensional models as well as massive gravity (insofar as they are related to scalar-tensor theories).

In 1961, Brans and Dicke proposed that the Newton constant $G$ is actually not a constant and may vary with spacetime location, thus behaving as a scalar field \cite{Brans:1961sx}. One may define a scalar field $\varphi=G_\tx{b}/G^{-1}$, where $G_\tx{b}$ is a bare Newton constant, a priori different from the one measured in Cavendish experiments (defined this way, $\varphi$ has no mass dimension). This scalar field is given some dynamics through a kinetic term $g^{\mu\nu}\pd_\mu\varphi\,\pd_\nu\varphi=(\pd\varphi)^2$:
\be
S_\tx{BD}= \dfrac{1}{16\pi G_\tx{b}}\displaystyle\int{\tx{d}^4x\sqrt{-g}\left[\varphi R -\dfrac{\omega}{\varphi}(\pd\varphi)^2\right]},
\label{eq:BDaction}
\ee
where $\omega$ is assumed to be a (dimensionless) constant. This action describes the gravitational sector, and is supplemented by a matter action, with matter fields minimally coupled to the metric. It can accommodate all general relativistic solutions, by simply assuming that $\varphi$ is a constant. Therefore, from experiment, one can only put bounds on the $\omega$ parameter; the most constraining bound on $\omega$ comes from the Shapiro delay measured by the Cassini spacecraft \cite{Alsing:2011er}, and yields $\omega\gtrsim40000$ (general relativity being restored in the large $\omega$ limit).

A quite straightforward extension of Brans-Dicke theory, often designated as scalar-tensor theories ---~though all other theories we will encounter in this thesis may also be called scalar-tensor theories~--- was proposed by Wagoner in 1970 \cite{Wagoner:1970vr}. In comparison with Brans-Dicke theory, Eq.~(\ref{eq:BDaction}), the coupling $\omega$ is now allowed to depend on the scalar field $\varphi$, and a potential term $V(\varphi)$ is added:
\be
S_\tx{ST}= \dfrac{1}{16\pi G_\tx{b}}\displaystyle\int{\tx{d}^4x\sqrt{-g}\left[\varphi R -\dfrac{\omega(\varphi)}{\varphi}(\pd\varphi)^2-V(\varphi)\right]},
\label{eq:STaction}
\ee
An arbitrary function of $\varphi$ in front of the Ricci scalar would not make the action more generic, because of possible field redefinitions. At this point, one can wonder what is the most general scalar-tensor theory that can legitimately be written. To answer this question, let us make a little detour through purely metric theories. In this framework, Lovelock proved an essential result in 1971 \cite{Lovelock:1971yv}. He established that, in four dimensions, the action of general relativity, Eq.~(\ref{eq:GRaction}), is the only one that generates divergence-free second-order field equations\footnote{Lovelock actually worked directly with the field equations, but he also proved that the rank-2 tensors he obtained correspond to Lagrangian densities; in four dimensions, the only relevant densities are a cosmological constant, the Ricci scalar and the Gauss-Bonnet invariant, which is a mere boundary term. Similar results were obtained by Cartan, Weyl and Vermeil long ago, but under slightly more restrictive assumptions (see the bibliography of Ref.~\cite{Lovelock:1971yv} for detailed references).}. The point of requiring second-order field equations is to avoid a type of instability called Ostrogradski ghost. Generically, higher than second-order field equations imply that the canonical momentum of some degree of freedom contributes linearly to the Hamiltonian density of the theory. This is the case whenever the Lagrangian satisfies a condition called non-degeneracy \cite{Woodard:2006nt}. As soon as the Ostrogradski degree of freedom is coupled to any other, the vacuum of the theory becomes unstable. Indeed, higher and higher energy modes of the coupled degree of freedom will be populated, while the Ostrogradski degree of freedom will compensate with an increasingly lower negative energy. 

To avoid the presence of this unphysical degree of freedom (called a ghost), one can either impose second-order field equations, or degeneracy of the Lagrangian. In the framework of scalar-tensor theories, the first option was analyzed by Horndeski in 1974. Following the work of Lovelock for purely metric theories, he determined the most general theory involving a metric and a scalar field, requiring divergence-free second-order field equations \cite{Horndeski:1974wa}. His results remained unsung for thirty years. They were rediscovered independently in the past decade \cite{Nicolis:2008in,Deffayet:2009wt,Deffayet:2009mn,Deffayet:2011gz}, motivated this time by the large cosmological constant problem\footnote{Generalized vector theories were also developed, already by Horndeski \cite{Horndeski:1976gi}, and more recently (see for example \cite{Deffayet:2010zh,Heisenberg:2014rta,Tasinato:2014eka}).}. This is the reason why Horndeski theory is also known by the name of Generalized Galileons. The two theories were proven to be equivalent by Kobayashi et al. in \cite{Kobayashi:2011nu}. Explicitly, Horndeski theory may be labeled in terms of four arbitrary functions $G_2$, $G_3$, $G_4$ and $G_5$ of the scalar $\varphi$ and the kinetic density $X=-\pd_\mu\varphi\,\pd^\mu\varphi/2$:
\be
\label{eq:H}
S_\tx{H} = \displaystyle\int \tx{d}^4x \sqrt{-g} \,\left(\mc{L}_2+\mc{L}_3+\mc{L}_4+\mc{L}_5\right),
\ee
with
\begin{align}
\mc{L}_2 &=G_2(\varphi,X) ,
\label{eq:L2}
\\
\mc{L}_3 &=-G_3(\varphi,X) \,\Box \varphi ,
\label{eq:L3}
\\
\mc{L}_4 &= G_4(\varphi,X) R + G_{4X} \left[ (\Box \varphi)^2 -\nabla_\mu\pd_\nu\varphi \,\nabla^\mu\pd^\nu\varphi\right] ,
\label{eq:L4}
\\
\begin{split}
\mc{L}_5 &= G_5(\varphi,X) G_{\mu\nu}\nabla^\mu \pd^\nu \varphi - \frac{1}{6}\, G_{5X} \big[ (\Box \varphi)^3 - 3\,\Box \varphi\, \nabla_\mu\pd_\nu\varphi\,\nabla^\mu\pd^\nu\varphi
\\
&\quad + 2\,\nabla_\mu\pd_\nu\varphi\, \nabla^\nu\pd^\rho\varphi\, \nabla_\rho\pd^\mu\varphi \big],
\label{eq:L5}
\end{split}
\end{align}
where a subscript $X$ stands for the derivative with respect to $X$, $G_{\mu\nu}$ is the Einstein tensor, and $\Box\varphi=\nd_\mu\pd^\mu\varphi$. Usual scalar-tensor theories, in the fashion of Eq.~(\ref{eq:STaction}), are of course a subclass of the Horndeski action. They correspond to vanishing $G_3$ and $G_5$, and to $G_4=\varphi/(16\pi)$, $G_2=[2X\omega(\varphi)/\varphi-V(\varphi)]/(16\pi)$. 

As mentioned in the previous paragraph, Horndeski theory is still not the most generic theory one may write in order to avoid the presence of an Ostrogradski ghost. It is sometimes possible to allow for higher than second-order field equations, as long as the dynamical structure of the Lagrangian exhibits a degeneracy. A natural path that leads to consider healthy degenerate theories is to consider the so-called \textit{disformal} transformations (first introduced by Bekenstein \cite{Bekenstein:1992pj}). They consist in a field redefinition of the metric:
\be
\tilde g_{\mu\nu}= C(X,\varphi)g_{\mu\nu}+D(X,\varphi)\,\pd_\mu\varphi\,\pd_\nu\varphi,
\label{eq:disformal}
\ee 
with some arbitrary functions $C$ and $D$ (when $D=0$, it reduces to a conformal transformation). Reference \cite{Bettoni:2013diz} first introduced these transformations in the framework of Horndeski theories, but still trying to avoid higher-order field equations. Indeed, in general, writing the Horndeski action $S_\tx{H}[g_{\mu\nu},\varphi]$ in terms of $\tilde g_{\mu\nu}$ generates higher-order derivatives (since $\tilde g_{\mu\nu}$ already contains derivatives of $\varphi$). Reference \cite{Zumalacarregui:2013pma} was the first to show that, although the field equations associated with $\tilde g_{\mu\nu}$ are of higher order, the resulting theory might be healthy\footnote{As long as the disformal transformation (\ref{eq:disformal}) is invertible, pure scalar-tensor theories formulated in terms of $(g_{\mu\nu},\,\varphi)$ and $(\tilde g_{\mu\nu},\,\varphi)$ are equivalent. However, the presence of matter makes the two formulations different according to whether it is minimally coupled to $g_{\mu\nu}$ or $\tilde g_{\mu\nu}$.}. This reference proved that, applying an arbitrary disformal transformation on the Einstein-Hilbert action, Eq.~(\ref{eq:EHaction}), the field equations can still be recast in a form that involves no higher than second-order time derivatives. 

Applying the disformal transformation (\ref{eq:disformal}) to the Lagrangians $\mc{L}_2$ and $\mc{L}_3$ leaves them in the same class. However, a purely disformal transformation depending on $X$ only ---~i.e., with $C=1$ and $D(X)$~--- applied on $\mc{L}_4$ and $\mc{L}_5$ separately yields the following terms \cite{Gleyzes:2014dya,Gleyzes:2014qga}:
\begin{align}
\mc{L}_4^{\tx{bH}} &= F_4(\varphi,X)\, \epsilon^{\mu \nu \rho \sigma} {\epsilon^{\alpha \beta \gamma}}_\sigma\, \pd_\mu\varphi \,\pd_\alpha\varphi\, \nd_\nu \pd_\beta\varphi \,\nd_\rho \pd_\gamma\varphi ,
\label{eq:bH4}
\\
\mc{L}_5^{\tx{bH}} &=F_5(\varphi,X)\, \epsilon^{\mu \nu \rho \sigma} \epsilon^{\alpha \beta \gamma \delta}\, \pd_\mu\varphi \,\pd_\alpha\varphi \,\nd_\nu \pd_\beta\varphi \,\nd_\rho \pd_\gamma\varphi \,\nd_\sigma \pd_\delta\varphi.
\label{eq:bH5}
\end{align}
Again, we use the notation of Wald \cite{Wald:1984rg} for $\epsilon_{\mu \nu \rho \sigma}$ (in particular, $\epsilon_{0123}=\sqrt{-g}$). One may consider free $F_4$ and $F_5$ functions, and combine the above terms with the Horndeski ones. These new terms were first proposed in \cite{Gleyzes:2014dya}, and are known as beyond Horndeski (or Gleyzes-Langlois-Piazza-Vernizzi) terms. Generically, a couple of quartic functions $F_4$ and $G_4$ can be mapped back to a pure Horndeski model $\mc{L}_4$. This is also true for a couple of quintic functions $F_5$ and $G_5$: they can in general be mapped to a pure quintic model $\mc{L}_5$ \cite{Gleyzes:2014dya,Gleyzes:2014qga}. However, if both $F_4$ and $F_5$ are present, they cannot be mapped to a pure Horndeski model. 

An important work \cite{Gleyzes:2014dya,Gleyzes:2014qga,Lin:2014jga,Deffayet:2015qwa,Langlois:2015cwa,Langlois:2015skt,Achour:2016rkg,BenAchour:2016fzp,Crisostomi:2016tcp,Crisostomi:2016czh} was carried out to explore the Hamiltonian structure of the beyond Horndeski terms, Eqs.~(\ref{eq:bH4})-(\ref{eq:bH5}), together with standard Horndeski ones, Eq.~(\ref{eq:H}). The presence or absence of the quadratic and cubic terms $\mc{L}_2$ and $\mc{L}_3$, does not matter for this analysis. Then, any combination of $G_4$ and $F_4$ (purely quartic model), or $G_5$ and $F_5$ (purely quintic model), or else $F_4$ and $F_5$ (purely beyond Horndeski model) leads to a degenerate ---~and thus healthy~--- model. Coherently, Ref.~\cite{Gleyzes:2014qga} also showed that, if a beyond Horndeski model can be disformally related to a Horndeski one, then the field equations may be written in a way that contains no more than second order time derivatives. On the contrary, the simultaneous presence of $G_4$, $G_5$, $F_4$ and $F_5$ generically leads to a non-degenerate Lagrangian with a deadly Ostrogradski degree of freedom. In the rest of this thesis, we will refer to the healthy subset of the sum of Lagrangians (\ref{eq:L2})--(\ref{eq:L5}) and (\ref{eq:bH4})-(\ref{eq:bH5}) as \textit{Horndeksi and beyond} theory. The corresponding action will be noted $S_\tx{bH}$.

There actually exists an even larger framework of degenerate scalar-tensor theories, which encompasses Horndeski and beyond theory. These models can be sorted by powers of the second derivatives of the scalar field, $\nd_\mu\pd_\nu\varphi$. They go by the name of Degenerate Higher-Order Scalar-Tensor (DHOST) theories, or extended scalar-tensor theories \cite{Langlois:2015cwa,Langlois:2015skt,Achour:2016rkg,BenAchour:2016fzp,Crisostomi:2016tcp,Crisostomi:2016czh}. They were fully investigated up to cubic order. A priori nothing forbids an arbitrary high order \cite{Crisostomi:2016czh}. On the other hand, Ref.~\cite{Langlois:2017mxy} has found that only the subclass of degenerate higher-order theories that are in relation with Horndeski theory through a disformal transformation, Eq.~(\ref{eq:disformal}), exhibits a healthy Newtonian limit. For concreteness and simplicity, in this thesis, we will focus on the subclass of Horndeski and beyond theory as defined above. In other words, we will not consider degenerate higher-order theories, in particular models that can be obtained from Horndeski and beyond theory through a conformal transformation $C(\varphi,X)$. Horndeski and beyond theory should capture the essential features of the healthy degenerate higher-order theories (some of these features being specific to beyond Horndeski models). Besides, in the case where the scalar field is a dark energy candidate, recent experiments ruled out the degenerate higher-order theories that are of higher order than quadratic in $\nd_\mu\pd_\nu\varphi$, as we will see in detail in Part \ref{part:3}.

Let us also emphasize that the above parametrization for Horndeski and beyond theory is only a choice among others. Appendix \ref{ap:otherparam} references other existing parametrizations, with the corresponding dictionary to switch from one to the other. Most of the terms (scalar or metric) in the above Lagrangian densities are not renormalizable. This is however not a problem since the motivation here is not to propose a renormalizable theory of gravity. Another objection comes from the effective field theory point of view. In such an approach, \textit{all} terms that are not forbidden by the symmetries of the theory must be present; the non-renormalizable ones are sorted by inverse powers of the ultraviolet cutoff scale of the theory; they come with dimensionless coefficients that are assumed to be natural (in the sense explained above). Therefore, from this point of view, it is hard to justify considering specific terms only among the Horndeski and beyond class. One should keep this caveat in mind, especially in Part \ref{part:3}, where a special relation between $F_4$ and $G_4$ is assumed in order to pass gravitational wave tests.

As mentioned above, there exist many other modified theories of gravity. Among these, let us cite higher-dimensional models and massive gravity because they often exhibit a scalar-tensor limit. This is the case for instance of the Dvali-Gabadadze-Porrati model \cite{Dvali:2000hr}. This model assumes that matter is located on a four-dimensional brane, inside a five-dimensional bulk spacetime. Gravity is allowed to ``leak'' in the fifth dimension. Expanding the five-dimensional metric and integrating out the fifth dimension, one can rearrange the perturbations into a tensorial, a vector and a scalar degrees of freedom (all in four dimensions). In a certain limit, the degrees of freedom decouple from each other; in this decoupling limit, a cubic interaction term is left over in the scalar sector. The associated action corresponds to $G_2\propto X$ and $G_3\propto X$ in Horndeski notation, Eq.~(\ref{eq:H}). More details are given in Sec.~\ref{sec:cubictdep}. Massive gravity constitutes another example. Indeed, a massive spin-2 field generically has six degrees of freedom, one of them being a ghost. It is only very recently that (non-linear) theories without ghosts were constructed \cite{deRham:2010ik,deRham:2010kj,Hassan:2011hr,Hassan:2011ea}. These healthy theories again exhibit a structure with a tensorial degree of freedom, a vector and a scalar one. Again, one can define a limit in which the scalar mode decouples from the others, leaving effectively a scalar-tensor theory.

\section*{Black holes}
\addcontentsline{toc}{section}{Black holes}

Generalized Galileon theory, the Dvali-Gabadadze-Porrati model or massive bi-gravity were all studied intensively for cosmological reasons, and their potential ability to generate a more natural accelerated expansion of the Universe. This corresponds to an extremely weak gravity regime. On the other hand, these theories were much less studied on local scales, in the intermediate and strong gravity regimes. Before exposing the purposes of this thesis, let us recall briefly some concepts and observations from the strongest gravity regime we have access to, that is black holes.

The simplest way to characterize a black hole is probably to define it as a region of spacetime where the pull of gravity is so strong that neither matter nor radiation can escape\footnote{In general relativity, a more mathematically accurate definition can be given, namely a region that is not in the causal past of the future null infinity \cite{Wald:1984rg}.}. The concept is not entirely specific to general relativity, and had already been evoked by 18th and 19th century scientists. In 1916, Schwarzschild proposed his famous solution, though it was not interpreted as a black hole at the time. It was not before the late fifties that it was given this interpretation. Kerr found the exact solution for a black hole in rotation in 1963 \cite{Kerr:1963ud}. It was then conjectured that very few solutions exist for steady-state black holes. Some theorems were established, proving that in general relativity, a stationary black hole is entirely defined upon specification of its mass, electric charge and angular momentum \cite{Israel:1967wq,Carter:1971zc,Robinson:1975bv}. These results and subsequent are generically designated by the name of no-hair theorem, or more exactly no-hair theorems, since there exist many with specific hypotheses. We will review these theorems in more detail in Chapter \ref{ch:nohairGR}. Around the same time, other important results were established for black holes. Among these are singularity theorems, which prove that the central singularity at the center of a black hole is not a mere symmetry artifact, but is actually always present \cite{Penrose:1964wq,Hawking:1969sw}. Black hole thermodynamics also established thermodynamical laws for black holes and lead to the notion of thermal Hawking radiation \cite{Bardeen:1973gs,Hawking:1974rv}. There remain nowadays open questions about black holes, such as the information loss paradox. Without a full theory of quantum gravity, it will probably be difficult to bring a definitive answer to these questions.
 
In general relativity, black holes are believed to be mostly of two types: stellar mass black holes and supermassive black holes. Stellar mass ones are formed by the collapse of sufficiently heavy stars. Supermassive ones have masses of the order of millions of solar masses, and their formation process is more uncertain (they likely form through absorbing stars or merging with other black holes). Black holes are essentially difficult to observe. However, the accretion of matter by black holes is accompanied by electromagnetic radiation. The emission of X-rays during an accretion process is what lead to propose Cygnus X-1 as the first black hole ever detected in 1972 \cite{Webster:1972bsw}. On the other hand, the motion of stars near Sagittarius A$^*$, at the galactic center of Milky Way, shows that 4.3 million solar masses are compacted in a sphere of radius smaller than $2\cdot10^{-3}$ light years \cite{Gillessen:2008qv}. This does not prove the existence of supermassive black holes, but is a strong hint. The Event Horizon Telescope goal is to observe the hot accretion disk that should surround Sagittarius A$^*$. The results are expected by the end of the year 2018. Finally, the most direct proof of the existence of black holes was offered by the detection of gravitational waves in 2015 \cite{Abbott:2016blz} (and several times since). Just as we detect light, which is the dynamical sector of electromagnetism, we can now detect the dynamical part of gravity. In principle, any massive body creates gravitational waves while moving; but only the fusion of two black holes, or neutron stars, releases enough energy to be detectable by actual gravitational interferometers. So far, the signals that were detected fit the theoretical expectations. In the forthcoming years, an avalanche of data is going to be available through earth-based interferometers, allowing us to test general relativity in the strongest energy regime available (where effective field theory invites us to expect deviations if there are some). Finally, the launch of the Laser Interferometer Space Antenna (LISA) is expected in 2034. This experiment will notably try to observe the fusion of supermassive black holes.

\section*{Structure of the thesis}
\addcontentsline{toc}{section}{Structure of the thesis}

This thesis gathers the knowledge and results collected during the time of my PhD. For the reader's convenience, I summarize here the structure of the thesis. It is divided in three parts, each of them subdivided in several chapters. The overall aim of the thesis is the study of compact objects in Horndeski and beyond theory.

The first part is devoted to no-hair theorems, both in metric and scalar-tensor theories. Chapter \ref{ch:nohairGR} reviews the powerful no-hair theorems that exist in general relativity, each time with their precise assumptions. It also discusses their extension to scalar-tensor theories in the fashion of the action (\ref{eq:STaction}), and clarifies the notion of hair for compact objects. Then, Chapter \ref{ch:nohair} is devoted to a black hole no-hair theorem that was proposed a few years ago in the framework of Horndeski theory. We further detail its assumptions and potential extensions. In the same spirit, we show in Chapter \ref{ch:nohairstars} that an analogue theorem can be proven to hold in the case of stars rather than black holes. In other words, under very similar assumptions, the only solutions allowed for stars are the general relativistic configurations.

Of course, these no-hair results hold only under certain hypotheses, that might prove physically relevant or not. It is the aim of Part \ref{part:2} to investigate all possible ways to circumvent the theorem. We show that there actually exist many successful ways to build black hole or star solutions with non-trivial scalar hair. In Chapter \ref{ch:tdep}, we justify through cosmology the introduction of time dependence for the scalar field while keeping a static metric. This way, we detail the construction of black holes with scalar hair in cubic and quartic Horndeski sectors; the former sector does not have reflection symmetry for the scalar field, while the latter does. Chapter \ref{ch:static} is devoted to isolated objects, with a static scalar field and an asymptotically flat geometry (thus closer to the target of the initial no-hair theorem). Still, we exhibit some new classes of black hole solutions in this context. Some of them stem from non-analytic Lagrangians, but we also discuss the cause of a linear coupling between the scalar field and the Gauss-Bonnet invariant. Chapters \ref{ch:tdep} and \ref{ch:static} additionally summarize what has been done to extend these non-trivial black hole solutions to the case of stars, and the potential deviations one can expect from general relativity in the case of neutron stars.

Part \ref{part:3} is further built on the non-trivial solutions presented in Part \ref{part:2}. It regroups results about stability of these solutions and propagation of gravitational waves. Although these two concepts may appear disjoint, they rely on the same calculations, namely linear perturbations of the background solution. In Chapter \ref{ch:pert}, we expose the formalism of linear perturbations, and use the effective metrics in which matter, gravitational and scalar perturbations propagate to establish stability criteria. We discuss the conclusions that can be drawn from the bounded or not character of the quadratic Hamiltonian, notably showing that an unbounded from below Hamiltonian \textit{does not} necessarily imply an instability. Last, Chapter \ref{ch:wave} is also based on the effective metrics mentioned above, but with in mind the speed of propagation of gravitational waves. We show that, starting from the solutions presented in Part \ref{part:2}, it is easy to construct solutions where gravitational waves propagate at the same speed as light; this remains true even in highly curved backgrounds, like the neighborhood of a black hole.

The thesis ends with a summary of the main results, and an outlook for further investigations in the field. This thesis gave rise to the publications listed herinbelow:
\begin{itemize}
\item E. Babichev, C. Charmousis, G. Esposito-Farèse and A. Lehébel, \textit{Hamiltonian vs stability and application to Horndeski theory} 
\item E. Babichev, C. Charmousis, G. Esposito-Farèse and A. Lehébel, \textit{Stability of Black Holes and the Speed of Gravitational Waves within Self-Tuning Cosmological Models, Phys. Rev. Lett.} \textbf{120} (2018) 241101
\item A. Lehébel, E. Babichev and C. Charmousis, \textit{A no-hair theorem for stars in Horndeski theories, JCAP} \textbf{1707} (2017) 037
\item E. Babichev, C. Charmousis and A. Lehébel, \textit{Asymptotically flat black holes in Horndeski theory and beyond, JCAP} \textbf{1707} (2017) 037
\item E. Babichev, C. Charmousis, A. Lehébel and T. Moskalets, \textit{Black holes in a cubic Galileon universe, JCAP} \textbf{1609} (2016) 011
\item E. Babichev, C. Charmousis and A. Lehébel, \textit{Black holes and stars in Horndeski theory, Class. Quant. Grav.} \textbf{33} (2016) 154002
\end{itemize}

\newpage
\thispagestyle{empty}
\cleardoublepage
\part{No-hair theorems}

\newpage
\thispagestyle{empty}
\cleardoublepage
\chapter{The hair of black holes}\label{ch:nohairGR}

The aim of this first chapter is to lay the foundations of the discussion we will pursue in this thesis. As mentioned in the introduction, the idea that ``black holes have no hair'' dates back from the late sixties. Many and various results have been established in this direction, as well as counter-examples. Instead of starting straight away with Horndeski theory and its extensions, it will be useful to review the history of what was achieved earlier, with a particular accent on scalar-tensor theories. We start this chapter with a reminder of some useful definitions. Then, we go through the variety of no-hair theorems, underlining their assumptions and briefly indicating possible ways out.

\section{Some definitions}

The statements we will make on black holes require the definition of some geometrical concepts. We will follow the definitions given in Wald \cite{Wald:1984rg}. First of all, the no-hair theorems often refer to the end point of gravitational collapse, when the black hole is quiescent. This may be encoded in the notion of a \textit{stationary} spacetime. Consider a spacetime manifold $M$ equipped with a  metric $g_{\mu\nu}$. Suppose there exists a one-parameter group of isometries $\phi_t$ ---~that is, for all $t$, $\phi_t:M\to M$ is a diffeomorphism such that the pullback of $g_{\mu\nu}$, $\phi_t^*g_{\mu\nu}$, is equal to $g_{\mu\nu}$. At any point of spacetime, a vector $\xi^\mu$ might be associated to $\phi_t$, that locally generates the orbits of this group. Suppose additionally that this vector is everywhere timelike, $g_{\mu\nu}\,\xi^\mu\xi^\nu<0$. Then, $(M,g_{\mu\nu})$ is called stationary. Alternatively, one can define a stationary spacetime as possessing a timelike Killing vector field. A \textit{static} spacetime is a stationary spacetime which possesses a hypersurface $\Sigma$ that is orthogonal to the orbits of the group of isometries. $\Sigma$ can be carried over through $\phi_t$, and as a consequence spacetime is foliated by hypersurfaces orthogonal to the orbits. Staticity is thus a stronger requirement than stationarity. 

In addition, compact astrophysical objects are usually more or less spherical. Therefore, it will be useful, as an approximation, to define \textit{spherically symmetric} spacetimes. A spacetime is said to be spherically symmetric if the group of its isometries contains SO(3) as a subgroup. The orbits of this subgroup define 2-spheres that can be sorted according to their area $A$, and thus to their areal radius $r$ defined through $A=4\pi r^2$. For a spacetime that is both static and spherically symmetric, one can always construct coordinates in which the metric takes the form
\be
\tx{d}s^2=-h(r)\,\tx{d}t^2 + \dfrac{\tx{d}r^2}{f(r)}+r^2(\tx{d}\theta^2+\sin^2\theta\,\tx{d}\phi^2).
\label{eq:statmetric}
\ee
We will use the above metric often throughout this thesis. More realistic physical systems, in particular rotating ones, can often be considered \textit{axisymmetric}. An axisymmetric spacetime is one which possesses a one-parameter group of isometries $\chi_\phi$ with closed spacelike curves. An axisymmetric and stationary spacetime is called so only if it is both axisymmetric and stationary, and if the action of the two isometries commute.

Another geometrical notion that is very useful in the context of no-hair theorems is the notion of asymptotic flatness. In usual general relativity, the concept of asymptotically flat spacetime has been given a more and more elaborate definition \cite{Penrose:1962ij,Penrose:1965am,Geroch:1972up,Ashtekar:1978zz,1980grg2.conf...37A}. The price to pay for a coordinate-independent definition is a refinement of technicalities. Furthermore, these definitions, beyond their complexity, were designed for general relativity and do not obviously extend to scalar-tensor theories. For our purposes, it will be enough to stick with a simple definition of asymptotic flatness, in the preferred system of coordinates associated with spherical symmetry. We will call a spacetime with line-element (\ref{eq:statmetric}) asymptotically flat if\footnote{In our notations, $\mc{F}(x)\underset{x \rightarrow a}{=}\mc{O}[\mc{G}(x)]$ means that there exists a bounded function $M$ defined on a neighborhood of $a$ such that $\mc{F}(x)=M(x)\,\mc{G}(x)$. Therefore, for instance, $f(r)\underset{r \rightarrow \infty}{=}\mc{O}(1/r)$ means that $f$ decays at most like $1/r$, but can decay faster.}
\begin{align}
h(r) &\underset{r \rightarrow \infty}{=}1+\mc{O}\left(\dfrac1r\right),~~~~h'(r)\underset{r \rightarrow \infty}{=}\mc{O}\left(\dfrac{1}{r^2}\right),
\label{eq:hflat}
\\
f(r) &\underset{r \rightarrow \infty}{=}1+\mc{O}\left(\dfrac1r\right),~~~~f'(r)\underset{r \rightarrow \infty}{=}\mc{O}\left(\dfrac{1}{r^2}\right).
\label{eq:fflat}
\end{align}
A slightly more generic definition, for a non-spherically symmetric spacetime, is to require that there exists a coordinate system $(t,x,y,z)$ such that
\be
g_{\mu\nu}\underset{r \rightarrow \infty}{=} \eta_{\mu\nu}+\mc{O}\left(\dfrac1r\right),
\ee
where $r=(x^2+y^2+z^2)^{1/2}$. We will call a spacetime pseudostationary when it is stationary ``far enough'', i.e., if there exists a Killing vector field that is timelike when $r\to\infty$ (or more precisely, close to the future and past null infinites, $\mathscr{I}^+$ and $\mathscr{I}^-$). This is the case of the usual vector $\pd_t$ of Kerr solution ---~in the form given in Eq.~(\ref{eq:Kerr}) for instance~--- which is a timelike Killing vector only outside of an oblate region. Finally, one can define the concept of a \textit{black hole} in an asymptotically flat spacetime (again, as a region that does not lie in the past of the future null infinity, $\mathscr{I}^+$). The topological boundary of this region is called the \textit{event horizon}. These notions may be extended to spacetimes with suitable asymptotic properties (such as open Friedmann-Lemaître-Robertson-Walker spacetimes).

Additional physical input is in general required in the proof of no-hair theorems. It often comes under the form of so-called energy conditions. There exist several of them; we will present the weak and strong energy conditions. It should be noted that the weak energy condition does not imply the strong one. The \textit{weak energy condition} is the assumption that the energy density associated to any field will be seen as positive by any observer. If the field in question is described by an energy-momentum tensor $T_{\mu\nu}$, it means that, for any 4-velocity $\xi^\mu$ (therefore timelike),
\be
T_{\mu\nu}\,\xi^\mu\xi^\nu \geq0.
\label{eq:wec}
\ee
The \textit{strong energy condition} is the assumption that, for any 4-velocity $\xi^\mu$,
\be
\left(T_{\mu\nu}-\dfrac12\,Tg_{\mu\nu}\right)\,\xi^\mu\xi^\nu \geq0.
\label{eq:sec}
\ee
In general relativity, this last condition may be given the meaning that timelike geodesic congruences are convergent, or that ``matter must gravitate towards matter'' \cite{Curiel:2014zba}. These energy conditions are hypotheses, assumed to hold for physically relevant types of matter. We will now see how they articulate with the various concepts defined in this section. Section \ref{sec:nohairGR} reviews a family of theorems established about (electro-)vacuum black holes in general relativity. Then, in Sec.~\ref{sec:scalarhair}, no-hair theorems for scalar-tensor theories are presented.

\section{The hair of vacuum black holes in general relativity}
\label{sec:nohairGR}

Israel first proved a theorem for static vacuum solutions in general relativity \cite{Israel:1967wq} (the vacuum assumption means that the only non-trivial field is the metric, all matter fields are assumed to vanish). He additionally assumed asymptotic flatness, regularity of the Riemann tensor squared $R_{\mu\nu\rho\sigma}R^{\mu\nu\rho\sigma}$, a spherical topology for the constant redshift surfaces (other plausible geometries include the toroidal one) and a regularity property called non-degeneracy for the horizon. Under these assumptions, the only solution is the Schwarzschild geometry. The important point is that spherical symmetry is not assumed. Carter proved a similar theorem, assuming pseudostationarity and axisymmetry instead of staticity \cite{Carter:1971zc}. In this case, he found only discrete families of solutions, with either one or two free parameters. Robinson further showed that only the Kerr family is suitable \cite{Robinson:1975bv}. Kerr black holes are fully characterized by a mass parameter $m$ and a rotation parameter $a=J/m$, $J$ being the angular momentum of the black hole. These results were strengthened by the findings of Hawking \cite{Hawking:1971vc}. The weak energy condition, Eq.~(\ref{eq:wec}), allows to get rid of the assumption of axisymmetry, and implies that the spatial sections of the event horizon must have the topology of a 2-sphere (excluding for instance toroidal geometries). These last results go by the name of rigidity theorem.

If, instead of assuming vacuum solutions, one allows the electromagnetic field to be non-trivial, one can prove similar results. The only physically viable solutions of the Einstein-Maxwell theory are inside the Kerr-Newman family (see \cite{Chrusciel:2012jk} for a review). In this case, the solutions are characterized by a third free parameter, the electric charge of the black hole $Q$.

This sequence of theorems lead to a conjecture: it is likely that, during a gravitational collapse that forms a black hole, all matter will either fall into the black hole or be expelled away. If this is true, the above theorems will apply. It is thus reasonable to assume that any quiescent black hole is in the Kerr-Newman family. However, this intuition might prove wrong: what if another field, which is not the metric nor the electromagnetic field, can develop a non-trivial structure during gravitational collapse, which survives in the equilibrium state? Such a structure would be what we call \textit{hair}. We will distinguish between primary and secondary hair. We call \textit{primary} hair non-trivial profiles described by an additional free parameter (on top of mass, angular momentum and possibly electric charge). This new parameter may take values either in a discrete or a continuous set. The first example of black holes with primary hair was found in the context of Einstein-Yang-Mills theory \cite{Volkov:1989fi}. When a solution differs from the Kerr-Newman ones because an additional field has a non-trivial structure, but at the same time it is still described in terms of mass, angular momentum and electric charge, we say this solution exhibits \textit{secondary} hair. Note that there may exist very soft hair: solutions for which the geometry is identical to the Kerr-Newman metric, while some field does not vanish. This is the case of stealth solutions, that we will describe in Chapter \ref{ch:tdep}. To know whether realistic solutions with hair exist in some given theory, a first step is to find an explicit solution of the field equations. Then, if these solutions are meant to represent the endpoint of gravitational collapse, it is important to check their stability, and ---~often numerically~--- their formation through the collapse of a star.

\section{No scalar hair theorems}
\label{sec:scalarhair}

As we just saw, the electromagnetic field can acquire a non-trivial structure in a stationary black hole spacetime. One may legitimately wonder whether this is true also for other fields; if some scalar field is allowed not to vanish, do we obtain a more general class of black holes, parametrized by a scalar charge in addition to Kerr-Newman parameters? There actually exist a family of theorems proving that scalar fields, as opposed to the electromagnetic field, must be in a trivial configuration for quiescent black holes. We review these theorems in this section, together with some paths to circumvent them. For more details on the hair of black holes in scalar-tensor theories, the reader may report to \cite{Herdeiro:2015waa,Sotiriou:2015pka}. Reference \cite{Volkov:2016ehx} also reviews solutions with Yang-Mills fields and in massive gravity.

\subsection{Bekenstein's theorem and its extensions}

A first no-hair theorem was obtained for canonical scalar fields without potential in \cite{Penney:1968zz,Chase}. Then, Bekenstein proposed a theorem that forbids scalar hair \cite{Bekenstein:1972ny}, using a procedure that was declined in many variants, and lead to several extensions \cite{Bekenstein:1971hc,Bekenstein:1972ky}. Let us give this theorem and go through its relatively short proof, to understand where the various assumptions play their role. Bekenstein's theorem deals with the following canonical theory:
\be
S_\tx{can} = \dfrac{\Mp^2}{2}\displaystyle\int{\tx{d}^4x\sqrt{-g} \left[R-(\pd\varphi)^2-V(\varphi)\right]}.
\label{eq:canscal}
\ee
In this action, $\varphi$ is chosen to have no mass dimension. It is assumed that:
\begin{enumerate}
\item spacetime is pseudostationary and the scalar field inherits the symmetries of spacetime,
\item spacetime is asymptotically flat, and the scalar field decays at least as fast as $1/r$ at spatial infinity (in adapted coordinates),
\item the scalar field is regular ($C^1$) on the horizon,
\item the potential obeys $\varphi \,V_\varphi \geq 0$, and the weak energy condition holds for the scalar energy-momentum tensor\footnote{To avoid confusion with the radial derivative that we will usually note with a prime, we write $V_\varphi$ for the derivative of $V$ with respect to $\varphi$.}.
\end{enumerate}
Under these conditions, the scalar field is necessarily in a trivial configuration, and the only solutions belong to the Kerr family. Before giving the proof, let us note that a no-hair theorem exists for Horndeski and beyond theory; as we will see in Chapter \ref{ch:nohair}, the structure of the assumptions is similar (although they are stronger, which makes the latter theorem weaker). In passing, the condition $\varphi \,V_\varphi \geq 0$ is satisfied for instance in the case of a massive potential $V(\varphi)=M^2\varphi^2$.

The proof goes as follows. Thanks to the weak energy condition, Hawking's rigidity theorem applies, and spacetime is axisymmetric. Thus, there exist a Killing vector $\xi^\mu$ associated to stationarity, and another one $\chi^\mu$ associated to axisymmetry. The scalar field equation for the action (\ref{eq:canscal}) reads:
\be
2\,\Box\varphi-V_\varphi(\varphi)=0.
\label{eq:bek0}
\ee
Let us integrate this equation, multiplied by $\varphi$, over a 4-volume $\mc{V}$ (to better picture $\mc{V}$, the reader may refer to Fig.~\ref{fig:tubes} of Chapter \ref{ch:nohairstars}, that will be used in a similar context). This 4-volume is delimited by a first hypersurface $\mc{S}_1$, that will hug part of the black hole horizon $\mc{H}$, a timelike hypersurface $\mc{S}_2$ far enough from the black hole, and two hypersurfaces $\mc{T}_1$ and $\mc{T}_2$. $\mc{T}_1$ can be chosen as a $\tau=\tx{constant}$ hypersurface, where $\tau$ is the affine parameter of the geodesics generated by $\xi^\mu$. $\mc{T}_2$ is then the hypersurface generated when ``shifting'' $\mc{T}_1$ by a constant amount of $\tau$, along the geodesics generated by $\xi^\mu$. Taking $\mc{S}_1$ far enough in space, and $\mc{T}_1$, $\mc{T}_2$ far enough in time (or at least in affine parameter $\tau$), one may cover the whole exterior region of the black hole. Note that the procedure of integrating the scalar field equation is common to many no-hair theorems, notably the one we will prove in Chapter \ref{ch:nohairstars}. In the present case,
\begin{align}
0&=\displaystyle\int_\mc{V}{\tx{d}^4x\sqrt{-g}\, (2\,\Box\varphi-V_\varphi)\,\varphi}
\\
&=\displaystyle\int_\mc{V}{\tx{d}^4x\sqrt{-g} \,(-\pd_\mu\varphi\,\pd^\mu\varphi-V_\varphi\,\varphi)}+\displaystyle\int_{\pd\mc{V}}{\tx{d}^3x\sqrt{|\gamma|} \,n^\mu\,\pd_\mu\varphi},
\label{eq:bek1}
\end{align}
where Eq.~(\ref{eq:bek1}) is obtained through integration by parts, $\pd\mc{V}=\mc{S}_1\cup\mc{S}_2\cup\mc{T}_1\cup\mc{T}_2$ is the boundary of $\mc{V}$, $\gamma_{\mu\nu}$ is the metric induced over the boundary, and $n^\mu$ is the normal to this boundary. By construction, the contributions of $\mc{T}_1$ and $\mc{T}_2$ in the last term of Eq.~(\ref{eq:bek1}) exactly cancel. The contribution of the faraway hypersurface $\mc{S}_2$ vanishes when it is taken to spatial infinity, because of the second assumption of the theorem. Only the contribution of $\mc{S}_1$ remains. To eliminate it, let us note that the event horizon of a stationary and asymptotically flat spacetime is a Killing horizon \cite{Hawking:1973uf}. Thus, when $\mc{S}_1\to\mc{H}$, $n^\mu$ tends towards a Killing vector, and it can only be a linear combination of $\xi^\mu$ and $\chi^\mu$. At the same time, $\varphi$ respects the isometries (assumption 1), which can be translated as $\xi^\mu\,\pd_\mu\varphi=0$, $\chi^\mu\,\pd_\mu\varphi=0$. As a consequence of this and the regularity of $\varphi$ on the horizon, $n^\mu\,\pd_\mu\varphi$ tends towards zero. Equation (\ref{eq:bek1}) thus tells us that:
\be
\displaystyle\int_\mc{V}{\tx{d}^4x\sqrt{-g}\,\pd_\mu\varphi\,\pd^\mu\varphi}=-\displaystyle\int_\mc{V}{\tx{d}^4x\sqrt{-g}\, V_\varphi\,\varphi}.
\ee
Under the last assumption of the theorem, the right-hand side is negative. Since $\xi^\mu\,\pd_\mu\varphi=0$, it is clear that the gradient of $\varphi$ cannot be timelike in the region where $\xi^\mu$ is timelike (far from the black hole). It is further argued in \cite{Hawking:1972qk} that $\pd_\mu\varphi$ can only be spacelike anywhere outside the black hole. Thus, if $\pd_\mu\varphi\neq0$ at any point of spacetime, the left-hand side will be positive. This is a contradiction, and the only way out is to set $\varphi=\tx{constant}$. This proves the theorem.

One can wonder what becomes of this theorem for more generic scalar-tensor theories. It was notably extended to Brans-Dicke theory, Eq.~(\ref{eq:BDaction}), by Hawking \cite{Hawking:1972qk}. Faraoni and Sotiriou extended it later \cite{Sotiriou:2011dz} to usual scalar-tensor theories, Eq.~(\ref{eq:STaction}). The argument is based on a field redefinition. Indeed, let us rewrite here the scalar-tensor action (\ref{eq:STaction}) with a slight change of notations:
\be
S_\tx{ST}= \dfrac{1}{16\pi G_\tx{b}}\displaystyle\int{\tx{d}^4x\sqrt{-\tilde{g}}\left[\Phi \tilde R -\dfrac{\omega(\Phi)}{\Phi}\,\tilde{g}^{\mu\nu}\pd_\mu\Phi\,\pd_\nu\Phi-U(\Phi)\right]}.
\label{eq:STaction2}
\ee
$S_\tx{ST}$ is thus defined in terms of the dynamical variables $\tilde g_{\mu\nu}$ and $\Phi$. The metric $\tilde g_{\mu\nu}$, often called \textit{Jordan frame} metric, is the physical metric to which matter (other than the scalar field) is assumed to be minimally coupled. Here, such matter fields are taken to be zero. Now, writing the action (\ref{eq:STaction2}) in terms of
\begin{align}
g_{\mu\nu}&=\Phi \,\tilde g_{\mu\nu},
\label{eq:gEinstein}
\\
\varphi&=\displaystyle\int{\dfrac{\tx{d}\Phi}{\Phi}\,\sqrt{\dfrac{2\,\omega(\Phi)+3}{2}}},
\label{eq:phiEinstein}
\end{align}
one recovers a canonical scalar field action:
\be
S_\tx{ST}= \dfrac{1}{16\pi G_\tx{b}}\displaystyle\int{\tx{d}^4x\sqrt{-g}\,\left[R -g^{\mu\nu}\pd_\mu\varphi\,\pd_\nu\varphi-V(\varphi)\right]},
\ee
where we defined $V(\varphi)=U(\Phi)/\Phi^2$. The metric $g_{\mu\nu}$ is called the \textit{Einstein frame} metric, because the dynamical part of the spin-2 degree of freedom takes the form of a standard Einstein-Hilbert action. Provided the above field redefinition is not singular, Bekenstein's theorem applies under the same assumptions. Note that in both cases (canonical scalar field and scalar-tensor theories), the assumption that $\varphi \,V_\varphi\geq0$ might be replaced by the assumption that $V_{\varphi\varphi}\geq0$. The proof is exactly similar, except that one has to multiply the scalar field equation, Eq.~(\ref{eq:bek0}), by $V_\varphi$ instead of $\varphi$.

No scalar hair theorems have also been slightly extended in at least two other directions. First, some potentials clearly do not respect the condition $\varphi V_\varphi\, \geq0$, or $V_{\varphi\varphi}\geq0$; this is the case for instance of the Higgs potential, Eq.~(\ref{eq:SM}). Efforts to improve the theorem in this direction have lead to extensions in the simplified case of spherically symmetric and static solutions. In this case, if a canonical scalar field respects the strong energy condition, it is necessarily trivial \cite{0264-9381-9-9-016,Heusler:1996ft}. Bekenstein, also in the case of spherical symmetry and staticity, was able to extend his no-hair results to an arbitrary number of scalar fields, and to scalar-tensor models with non-canonical kinetic terms \cite{Bekenstein:1995un}. Second, one may allow the scalar field not to respect the symmetries of spacetime ---~only the associated energy-momentum tensor cannot break these symmetries. For canonical scalar fields, as well as for arbitrary Lagrangian functionals of $\varphi$ and the kinetic density $X$, it was proven that the scalar field cannot depend on time in a stationary spacetime \cite{Pena:1997cy,Graham:2014ina}. We will however see that this idea is successful in the case of Horndeski and beyond theory.

\subsection{Some solutions with scalar hair}

Outside of the range of no-hair theorems, there actually exist solutions with scalar hair, either exact or numerical. We will now briefly describe some of them, insisting on why they are not ruled out by the above theorems. Chronologically, the first of these solutions was derived by Bocharova et al. \cite{Bocharova:1970skc} in USSR, and slightly later by Bekenstein in the USA \cite{Bekenstein:1974sf,Bekenstein:1975ts}. It is a solution of the following theory:
\be
S_\tx{BBMB}=\dfrac{1}{4\pi G}\displaystyle\int{\tx{d}^4x\sqrt{-g}\left[\dfrac{R}{4}-\dfrac12(\pd\varphi)^2-\dfrac{1}{12}R\varphi^2\right]},
\label{eq:BBMB}
\ee
which possesses a conformally invariant scalar field equation (i.e., this equation is invariant under $g_{\mu\nu}\to\Omega^2\,g_{\mu\nu}$ and $\varphi\to\varphi/\Omega$ with an arbitrary function $\Omega$). Note however that the whole action is not conformally invariant, due to the presence of the Einstein-Hilbert term. The theory (\ref{eq:BBMB}) possesses the following solution with non-trivial $\varphi$:
\begin{align}
\tx{d}s^2&=-\left(1-\dfrac{m}{r}\right)^2\tx{d}t^2+\dfrac{\tx{d}r^2}{(1-m/r)^2}+r^2(\tx{d}\theta^2+\sin^2\theta \,\tx{d}\phi^2),
\label{eq:BBMB1}
\\
\varphi&=\dfrac{\sqrt{3}m}{r-m},
\label{eq:BBMB2}
\end{align}
where $m$ is free and represents the gravitational mass of the black hole in Planck units (if one does not set Newton's consant $G$ to a unit value, the gravitational mass is $m/G$). The geometry is identical to the one of a Reissner-Nordstr\"om extremal black hole (that is, a spherically symmetric black hole with maximal electric charge). This is a typical example of secondary hair, as defined in Sec.~\ref{sec:nohairGR}: the scalar field is not trivial, but it is entirely fixed in terms of the mass parameter $m$. There are two reasons why this solution is not excluded by existing no-hair theorems. First, the scalar field is not regular at the horizon. Whether this is a physical issue is arguable, since the scalar field singularity does not render the metric itself singular. We will encounter the same problem for Horndeski and beyond theory in Sec.~\ref{sec:GB}. Note however that the solution (\ref{eq:BBMB1})-(\ref{eq:BBMB2}) was claimed to be unstable against perturbations \cite{Bronnikov:1978mx}. Another reason is that the field redefinition that is required to bring the action into canonical form is ill-defined at $r=2m$. To conclude with this model, it was shown that the action (\ref{eq:BBMB}) does not possess other spherically symmetric, static and asymptotically flat solutions than Schwarzschild's when $\varphi$ is finite on the horizon \cite{Zannias:1994jf}.

Another remarkable way to bypass the no-hair theorem in general relativity was found by Herdeiro and Radu \cite{Herdeiro:2014goa,Herdeiro:2015gia}. They simply used a canonical complex scalar field (that can be seen as a pair of real fields):
\be
S_\tx{HR}=\dfrac{1}{4\pi}\displaystyle\int{\tx{d}^4x\sqrt{-g}\left(\dfrac{R}{4}-\pd_\mu\varphi\,\pd^\mu\varphi^*-M^2\varphi\,\varphi^*\right)},
\label{eq:HR}
\ee
where $M$ is the mass of the scalar field. The solutions they found were built numerically, assuming a stationary and axisymmetric ansatz for the metric. Crucially, the scalar field does not respect these symmetries. It explicitly depends on both $t$ and $\phi$ (the coordinates respectively associated with the timelike and spacelike Killing vectors):
\be
\varphi=\psi(r,\theta)\, \tx{e}^{i(m\phi-\omega t)},
\ee
where $\psi$ is a real function, $\omega$ is a frequency and $m$ an integer (for periodicity). Clearly, the energy-momentum tensor associated with $\varphi$ respects the symmetries of the spacetime, and Einstein's equations are consistent. Numerical solutions can be found when $\omega$ belongs to a continuous interval. It thus constitutes an example of primary hair.

Non-trivial solutions were also found when quadratic curvature terms are included, such as the Gauss-Bonnet density $\hat G$, defined as:
\be
\hat G=R^{\mu\nu \alpha \beta} R_{\mu\nu \alpha \beta}-4R^{\mu\nu}R_{\mu\nu}+R^2.
\label{eq:GB}
\ee
A term proportional to $\tx{e}^{\varphi}\, \hat G$ naturally arises in the effective action of heterotic string theory, where $\varphi$ is called the dilaton field. Perturbative \cite{Campbell:1990fu,Mignemi:1992nt,Mignemi:1993ce} and numerical \cite{Kanti:1995vq,Alexeev:1996vs,Torii:1996yi} solutions to this theory were shown to exhibit secondary hair. In Sec.~\ref{sec:GB}, we will come back in detail on a model that can be viewed as the linear expansion of the $\tx{e}^{\varphi}\, \hat G$ term.

Other possibilities include scalar multiplets (known as Skyrmions), coupling of the scalar field to gauge fields or specific potentials that do not respect the positivity conditions (see \cite{Herdeiro:2015waa} for detailed references). There clearly exist many ways to build black holes with scalar hair. Part \ref{part:2} of this thesis will be devoted to the exploration of similar tracks in the framework of Horndeski and beyond theory. But first, let us present the no-hair results that have been obtained in this theory so far.
\newpage
\thispagestyle{empty}
\cleardoublepage
\chapter{A black hole no-hair theorem in Horndeski theory}\label{ch:nohair}

The scalar-tensor theories studied in Chapter \ref{ch:nohairGR} were the most generic one may construct using only the first derivatives of the scalar field. However, as we saw in the introduction, it is known since the work of Horndeski \cite{Horndeski:1974wa} that sensible actions can be built using also the second derivatives of the scalar field. The dynamical structure of the scalar and tensor degrees of freedom become intricated, and Bekenstein's procedure to prove the absence of scalar hair breaks down. However, a no-hair theorem was proposed by Hui and Nicolis \cite{Hui:2012qt} in the context of Horndeski theory. It affects the shift-symmetric version of Horndeski theory, that is the subclass of Horndeski theory such that the action is invariant under the transformation $\varphi\rightarrow\varphi+C$, where $C$ is an arbitrary constant. This subclass is obtained by imposing that all Horndeski functions depend on $X$ only:
\be
\forall \; i, \quad G_i(\varphi,X)=G_i(X).
\label{eq:SSH}
\ee
It is in fact equivalent to require that only the field equations possess this shift-symmetry \cite{Sotiriou:2014pfa}. To this global invariance is associated a conserved Noether current $J^\mu$:
\be
\label{eq:currentdef}
J^\mu=\frac{1}{\sqrt{-g}}\, \frac{\delta S[\varphi]}{\delta (\partial_\mu\varphi)}.
\ee
This current obeys a conservation equation, $\nd_\mu J^\mu =0$. It is a crucial tool in the proof of the theorem, and an ubiquitous quantity in all this thesis. Explicitly, for the shift-symmetric Horndeski action (\ref{eq:H})-(\ref{eq:SSH}), it reads, with the same notations as used in the introduction:
\be
\begin{split}
J^\mu &= -\pd^\mu\varphi \left\{\vphantom{\dfrac12} G_{2X} - G_{3X}\, \Box \varphi +G_{4X} R + G_{4XX} \left[ (\Box \varphi)^2 -\nabla_\rho\pd_\sigma\varphi \,\nabla^\rho\pd^\sigma\varphi \right] \right.
\\
 &\quad+G_{5X} G^{\rho\sigma}\nabla_{\rho}\pd_{\sigma}\varphi -\dfrac{G_{5XX}}{6} \left[ (\Box \varphi)^3 - 3\,\Box \varphi\,\nabla_\rho\pd_\sigma\varphi \,\nabla^\rho\pd^\sigma\varphi \right.
\\
 &\quad+ \left.\left. 2\,\nabla_\rho\pd_\sigma\varphi \,\nabla^\sigma\pd^\lambda\varphi\, \nabla_\lambda\pd^\rho\varphi \right] \vphantom{\dfrac12}\right\} -\pd^\nu X \left\{\vphantom{\dfrac12} - \delta^\mu_\nu\, G_{3X} + 2\, G_{4XX} (\Box \varphi \,\delta^\mu_\nu \right.
\\
 &\quad- \nabla^\mu\pd_\nu \varphi)+  G_{5X} G^\mu{}_\nu -\dfrac12 \,G_{5XX} \left[ \delta^{\mu}_{\nu}(\Box\varphi)^2 - \delta^{\mu}_{\nu}\,\nd_\rho\pd_\sigma\varphi\,\nd^\rho\pd^\sigma\varphi\right.
\\
 &\quad-\left.\left.\vphantom{(\Box\varphi)^2} 2\,\Box\varphi\, \nd^\mu\pd_\nu\varphi +2\, \nd^\mu \pd_\rho \varphi\, \nd^\rho \pd_\nu \varphi \right] \vphantom{\dfrac12}\right\} +2\,G_{4X} R^{\mu}{}_{\rho}\, \pd^\rho \varphi 
\\
 &\quad+ G_{5X} \left( -\Box \varphi\, R^\mu{}_\rho \,\pd^\rho\varphi + R_{\rho\nu}{}^{\sigma\mu}\, \nd^\rho\pd_\sigma\varphi\, \pd^\nu\varphi + R_\rho{}^\sigma \pd^\rho\varphi \, \nd^\mu\pd_\sigma\varphi \right).\vphantom{\dfrac12}
\label{eq:covcur}
\end{split}
\ee
In the next section, we give a detailed version of the no-hair theorem first formulated by Hui and Nicolis. Then, we discuss its extensions in several directions and its limits.

\section{The theorem}
\label{sec:theorem}

We will start  with the no-hair theorem, and then go through its proof. In addition to asymptotic flatness, it relies on several classes of assumptions. First, it assumes some specific symmetries for the background solution. Then, it requires the norm of the Noether current introduced above to be bounded everywhere. Whether this assumption is physical or not is arguable, as we will see notably in Sec.~\ref{sec:GB}. Finally, it assumes that the action is regular enough. The theorem reads as follows:

Consider a shift-symmetric Horndeski model as in (\ref{eq:H})-(\ref{eq:SSH}) where $G_2$, $G_3$, $G_4$ and $G_5$ are arbitrary functions of $X$. Let us now suppose that:
\begin{enumerate}
\item spacetime is spherically symmetric and static and the scalar field respects these symmetries,
\item spacetime is asymptotically flat, the gradient of $\varphi$ vanishes at spatial infinity,
\item the norm of the current $J^2$ is finite on the horizon,
\item there is a canonical kinetic term $X\subseteq G_2$ in the action and the $G_i$ functions are analytic at the point $X=0$.
\end{enumerate} 
Under these hypotheses, one can conclude that $\varphi$ is constant and thus the only black hole solution is locally isometric to Schwarzschild spacetime.  

The proof of the theorem is divided in two steps; first, one can prove that the radial component of the current vanishes. Secondly, one can use this result to establish that the scalar field is trivial. Let us start by making use of spherical symmetry and staticity. The metric may thus be written as in Eq.~(\ref{eq:statmetric}), while the scalar can depend on the radial coordinate only:
\begin{align}
\tx{d} s^2 &= -h(r) \,\tx{d} t^2 + \dfrac{\tx{d} r^2}{f(r)} + r^2 (\tx{d}\theta^2 + \sin^2 \theta\, \tx{d}\phi^2),
\\
\varphi &= \varphi(r).
\label{eq:statphi}
\end{align}
In this system of coordinates, only the radial component of the current is non-trivial. Thus, the scalar field equation reads
\be
\nd_\mu J^\mu = \dfrac{\tx{d}}{\tx{d}r}\left(\sqrt{\dfrac{h}{f}}\, r^2 J^r\right) = 0,
\label{eq:covEphi}
\ee
the generic solution of which is:
\be
J^r=\dfrac{Q}{r^2} \,\sqrt{\dfrac{f}{h}},
\label{eq:scalarcharge}
\ee
where $Q$ is a free integration constant. It follows that the norm of the current is
\be
J^2=\dfrac{Q^2}{r^4h}.
\label{eq:Jsquared}
\ee
The event horizon of the black hole is located at the point where $h$ first vanishes, starting from $r\to\infty$. Thus, we see that if $Q\neq0$, the norm of the current diverges when approaching the horizon of the black hole, which is excluded by the third assumption of the theorem. The only way out is to set $Q=0$, i.e., to impose that the Noether current vanishes identically. It seems quite reasonable that, in a static geometry, the black hole cannot accrete a continuous flux of scalar current. We have thus completed the first step of the proof. To go further, we need the explicit form of $J^r$:
\be
\begin{split}
\label{eq:JrHorni}
J^r=&-f \varphi' G_{2X} - f\,\dfrac{r h' + 4 h}{r h} X G_{3X} + 2 f \varphi' \dfrac{f h - h + r f h'}{r^2 h}G_{4X} 
\\
&+ 4 f^2 \varphi' \dfrac{h +rh'}{r^2h} X G_{4XX} -f h' \dfrac{1 - 3 f}{r^2h} X G_{5X} +2\, \dfrac{h' f^2}{r^2h} X^2 G_{5XX}
,\end{split}
\ee
where a prime stands for a derivative with respect to $r$. Let us now use the last assumption of the theorem, about the regularity of the action, together with the second one about asymptotic flatness. Under these assumptions, Eq.~(\ref{eq:JrHorni}) can be put under the form
\be
J^r  = \varphi' \mathcal{J}(\varphi',f,h'/h,r),
\label{eq:branches}
\ee
where
\be
\begin{split}
\label{eq:curlyJ}
\mc{J}=&-f G_{2X} +f^2\,\dfrac{r h' + 4 h}{2r h}\varphi' G_{3X} + 2 f\,\dfrac{f h - h + r f h'}{r^2 h}G_{4X} 
\\
&+ 4 f^2 \,\dfrac{h +rh'}{r^2h} X G_{4XX} +f^2h' \dfrac{1 - 3 f}{2\,r^2h} \varphi'G_{5X} -\dfrac{h' f^3}{r^2h} \varphi'X G_{5XX}
\end{split}
\ee
depends analytically on $\varphi'$ since $X=-\varphi'^{\,2} f/2$ and the $G_i$ functions are analytic. The vanishing of $J^r$ locally splits the solutions into two branches: at any point, either $\varphi'$ or $\mc{J}$ vanishes. Let us demonstrate that, under the assumptions of the theorem, $\varphi'=0$ is the solution. Indeed, a careful examination of the various terms in (\ref{eq:JrHorni}) shows that $\mc{J}(0,1,0,r) = -G_{2X}(0)=-1$. Because of asymptotic flatness and because the scalar field derivative is assumed to vanish at infinity, $(\varphi',f,h'/h)\to(0,1,0)$ when $r\to\infty$. Therefore, $\mathcal{J}$ tends towards $-1$ at infinity. Let us consider a radius $r_0$ large enough for $\mathcal{J}$ to remain negative when $r\geq r_0$. Since $J^r$ cancels exactly everywhere, so does $\varphi'$ in the whole outer region $r\geq r_0$. As a consequence, the contribution of the scalar to the field equations vanishes in this region, and the solution is uniquely given by general relativity: it is Schwarzschild geometry. We must finally establish that $\varphi'$ also cancels in the inner region\footnote{One could think that the system of field equations together with the general relativistic initial conditions at $r=r_0$ constitute a well posed Cauchy problem, thus forcing the solution to be general relativity in the inner region. However, because of the branch structure, the Cauchy-Kowalevski theorem does not apply.}. For $r\geq r_0$, since $\varphi'$ vanishes, the expression of $\mc{J}$ is easily derived from (\ref{eq:curlyJ}):
\be
\mc{J}=-f+2\,G_{4X}(0)G^{rr}
\label{eq:curlyJright}
,\ee
where $G^{rr}$ is the $(rr)$ component of Einstein tensor. Since in the region under consideration, the geometry is uniquely given by Schwarzschild solution, $G^{rr}$ vanishes identically. Note that it will not be true any more in the case of a star, where even the general relativity solution has $G^{rr}\neq0$ in the presence of matter. Still, in the vacuum, $\mc{J}$ is trivially equal to $-f$. When entering the region $r<r_0$, $\mc{J}$ remains non-zero by continuity. Accordingly, $\varphi'$ does not deviate from 0. This in turns implies that $\mc{J}$ remains equal to $-f$. Thus, down to the horizon where $f$ first vanishes, one has no other choice than to stick to the $\varphi'=0$ branch. One is left with Schwarzschild geometry as the unique solution over $r_\tx{S}< r<+\infty$, where $r_\tx{S}$ is the Schwarzschild radius of the black hole. This proves the theorem. Note that the theorem can be straightforwardly extended to the shift-symmetric version of Horndeski and beyond theory $S_\tx{bH}$, that is when imposing
\be
\forall \;i, \quad G_i(\varphi,X)=G_i(X),\qquad F_i(\varphi,X)=F_i(X),
\label{eq:SS}
\ee
by requiring also that the function $F_4$ and $F_5$, introduced in Eqs.~(\ref{eq:bH4})-(\ref{eq:bH5}), are analytic around the point $X=0$ \cite{Babichev:2017guv}.

\section{Generalization to stationary solutions?}
\label{sec:slowrot}

In comparison with the theorems presented in Chapter \ref{ch:nohairGR}, the no-hair theorem of Sec.~\ref{sec:theorem} seems relatively weak. In particular, the assumption of spherical symmetry and staticity is very restrictive. Realistic physical solutions rotate, even if the rotation might be slow. A natural question to ask is then whether the theorem can be extended to include stationary spacetimes, thus covering realistic cases. So far, no generic proof was found in the case of rotating solutions. However, no solution with scalar hair was found either that violates the assumption of the theorem only because it is stationary, rather than static and spherically symmetric. Since the stationary case appears very complex, one can start by assuming that the rotation is very slow, and can be treated as a perturbation with respect to spherical symmetry. This is the principle of the Hartle-Thorne formalism \cite{Hartle:1967he,Hartle:1968si}. In Horndeski theory, the slow rotation limit was first discussed in \cite{Sotiriou:2013qea}. Generically, a stationary and axisymmetric solution can be written:
\be
\tx{d} s^2 \underset{\epsilon\to0}{=} -h(r)\, \tx{d}t^2 + \dfrac{\tx{d} r^2}{f(r)} + r^2 (\tx{d}\theta^2 + \sin^2 \theta\, \tx{d}\phi^2)-2\,\epsilon \sin^2\theta\, \omega(r,\theta)\, \tx{d}t\,\tx{d}\phi+ \mc{O}(\epsilon^2),
\label{eq:HartleThorne}
\ee
where $\epsilon$ parametrizes the rotation speed, $h$ and $f$ are given by the spherically symmetric solution, and $\omega$ is determined as a function of the spherically symmetric solution for the scalar and the metric. In general relativity, $\omega$ depends on $r$ only and obeys the following equation:
\be
\omega''+\dfrac{\omega'}{2} \left(\dfrac{f'}{f}+\dfrac{8}{r}-\dfrac{h'}{h}\right)=0.
\label{eq:omegaslowrot}
\ee
The metric (\ref{eq:HartleThorne}) is the most general that is invariant under a simultaneous change of the sign of the rotation parameter $\epsilon$ and of the sign of either $\phi$ or $t$. The scalar field solution can be expanded as well in terms of $\epsilon$. Because of the background symmetry, at any order in perturbation theory, it can depend only on $\theta$ and $r$:
\be
\varphi(r,\theta)\underset{\epsilon\to0}{=}\varphi^{(0)}(r)+\epsilon \varphi^{(1)}(r,\theta)+ \mc{O}(\epsilon^2),
\label{eq:phislowrot}
\ee
where $\varphi^{(0)}$ is the spherically symmetric solution, and $\varphi^{(1)}$ the first order correction due to rotation. Reference \cite{Sotiriou:2013qea} noticed that reversing the sense of the rotation, i.e., changing the sign of $\epsilon$, would clearly change the solution (\ref{eq:phislowrot}) even when $t$ or $\phi$ is reversed, because these two variables do not enter the linear correction. As a consequence, if the no-hair theorem applies at the level of spherical symmetry, it also applies at first order in $\epsilon$. Indeed, when the theorem applies, $f$ and $h$ are given by the Schwarzschild solution, and $\varphi^{(0)}$ vanishes. Then, $\omega$ reduces to its general relativity counterpart and $\varphi^{(1)}$ vanishes identically as explained above.

This result goes in a favorable direction for the extension of the theorem to stationary and axisymmetric solutions. Another result that points towards the same direction was given in \cite{Barausse:2015wia,Barausse:2017gip}. It is explained in more detail in Chapter \ref{ch:nohairstars}. It basically states that, under assumptions 2 and 4 of the no-hair theorem of Sec.~\ref{sec:theorem}, binary stars cannot develop a non-trivial scalar profile. The gravitational wave emission is then expected to be identical to general relativity. For the precise conditions of validity of this statement, see Chapter \ref{ch:nohairstars}.

Finally, one can also look at the slow rotation limit from a different perspective. Reference \cite{Maselli:2015yva} asked the following question: if the background solution already has scalar hair, how does the slowly rotating solution behave? Does it follow Eq.~(\ref{eq:omegaslowrot}), or another master equation? The answer clearly depends on the theory under consideration. For further discussion, one might report to Paragraph \ref{sec:slowrotJohn} and Sec.~\ref{sec:GB}.
\newpage
\thispagestyle{empty}
\cleardoublepage
\chapter{Extending the result to stars}
\label{ch:nohairstars}

Black holes are not the only type of objects where huge curvatures effects are present. In the sequence that leads to the formation of a black hole, stars go through different steady states, such as white dwarfs and neutron stars. For instance, neutron stars of mass $m\simeq2\,M_\odot$ have an estimated radius of $R\simeq10$~km \cite{Guillot:2013wu}. The associated Schwarzschild radius is $R_\tx{S}\simeq3$~km. The gravitational field is therefore already very intense close to and inside such objects. The investigation of neutron stars is really a timely subject. Indeed, in 2017, gravitational wave detectors observed for the first time the inspiral and fusion of a binary neutron star system \cite{TheLIGOScientific:2017qsa,Monitor:2017mdv,GBM:2017lvd}. This observation and upcoming ones open an interesting and enlightening window on gravity.

In the context of alternatives theories of gravity, different aspects of compact stars must be studied. The first step is to investigate the properties of solitary stars. How different from general relativity can scalar-tensor strong field solutions be? When considering star solutions, the uniqueness theorems break down, already in general relativity\footnote{This is actually a good feature of general relativity. Newton's gravity predicts a whole variety of gravitational fields around planets or stars. We chart these fields thanks to artificial satellites in the Solar System. A theory that would allow for a unique gravitational field in the vacuum would be ruled out at once.}. Would the introduction of a fundamental scalar field enrich the spectrum of general relativistic star solutions, and in what way? Second, one can study the properties of binary neutron star systems. Since the seventies, we have information on such systems through the study of pulsars. This information has been used to put strong bounds on scalar-tensor theories, Eq.~(\ref{eq:STaction}), see e.g., \cite{Freire:2012mg}. As already mentioned, gravitational wave interferometers have also started detecting binary neutron star mergers. So far, only one event was detected, with very little information on the post-merger phase. However, many observations will likely follow. The emitted gravitational waves carry information on binary systems, but also on the details of the equation of state of neutron stars. The latter is poorly known up to now, and the post-merger phase will provide a lot of constraints on hypothetical models \cite{Takami:2014zpa}.

In this chapter, based on \cite{Lehebel:2017fag}, we provide the first step, namely a no-hair theorem, investigating the spherically symmetric and static star configurations in scalar-tensor theories with minimal matter coupling. This result is in very close relation with the black hole no-hair theorem of Chapter \ref{ch:nohair}. We show that, generically, the scalar profile is trivial, still under the key requirement that only derivatives of the scalar field are present in the action (with some extra technical assumptions). We first prove in Sec.~\ref{sec:Jr} that in a regular, spherically symmetric and static spacetime, there cannot be a non-vanishing and time-independent scalar flow. In Sec.~\ref{sec:nohairstar}, we use this fact in the framework of Horndeski and beyond theories, and examine the various assumptions that lead us to the no-hair result. We enumerate the possible ways out, that will be investigated in more detail in Part \ref{part:2}. At every step, we make the link with black hole configurations, underlining the similarities and differences with stars. Finally, we also discuss the results of Ref.~\cite{Barausse:2017gip}.

\section{No influx on stars}
\label{sec:Jr}

We will study the shift-symmetric Horndeski and beyond theory, but instead of considering vacuum solutions for standard matter, we allow its presence with minimal coupling to the metric. Namely, we analyze the following action:
\be
S= S_\tx{bH;SS}\left[g_{\mu\nu}; \varphi\right]+S_\tx{m}\left[g_{\mu\nu}; \Psi\right],
\label{eq:bHSSstar}
\ee
where the shift-symmetric version of the beyond Horndeski action $S_\tx{bH;SS}$ can be read from Eqs.~(\ref{eq:H}), (\ref{eq:bH4})-(\ref{eq:bH5}) and (\ref{eq:SS}). It involves the scalar field $\varphi$ and the metric $g_{\mu\nu}$; the matter action $S_\tx{m}$ contains matter fields collectively denoted as $\Psi$. The latter obey the weak equivalence principle: they couple minimally to the metric. Let us restrict our attention to a spherically symmetric and static geometry which is regular everywhere (that is, we assume the metric and scalar field are $C^1$ everywhere). Let us also assume that no horizon for the metric is present across the spacetime, thus excluding black holes form our analysis; rather, we have in mind star configurations. The scalar field is assumed to respect the symmetries of spacetime. Again, the theorem will be based on the Noether current associated with shift symmetry, $J^\mu$, Eq.~(\ref{eq:currentdef}). Under the above assumptions, the most general ansatz for the metric and scalar field are the ones given in Eqs.~(\ref{eq:statmetric})-(\ref{eq:statphi}). The only potentially non-vanishing component of $J^\mu$ is still the radial one. In a static configuration, it seems problematic that a non-vanishing flux can indefinitely flow towards the origin of coordinates, especially if this point cannot be a singularity hidden behind some horizon. Indeed, we will now prove that such star configurations are forbidden. In terms of the scalar charge introduced in Eq.~(\ref{eq:scalarcharge}), this means that $Q$ will have to vanish. To draw a parallel, Maxwell's equation in vacuum $\tx{div}(\textbf{E})=0$ also locally allows a radial electric field $\textbf{E} = C \textbf{r}/r^3$. However, integration over an extended domain imposes that $\textbf{E}$ actually vanishes in the absence of a charge distribution. Similarly here, $J^\mu$ will vanish because it is not sourced. 

\begin{figure}[ht]
\begin{center}
\includegraphics[width=.5\textwidth]{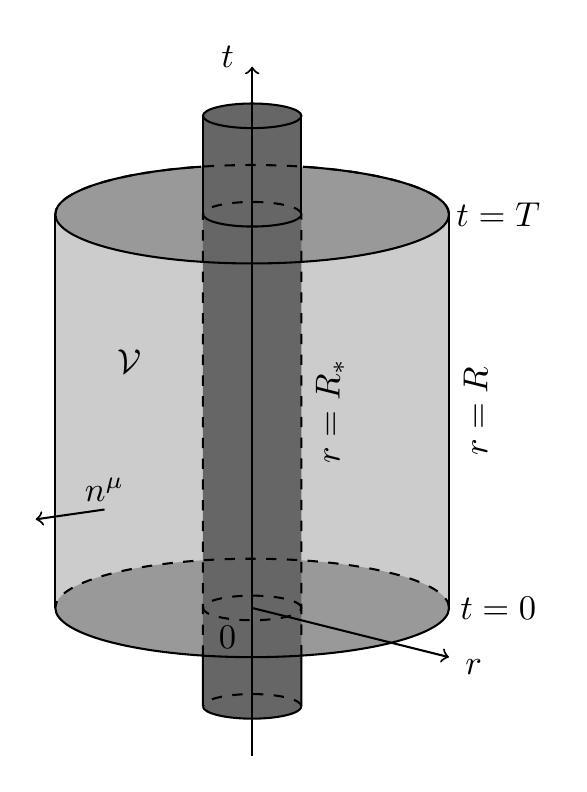}
\caption{Schematic representation of the spacetime. The vertical axis corresponds to time, and only two spatial dimensions are represented transversally. The light gray cylinder represents the 4-volume $\mathcal{V}$ over which the scalar field equation is integrated. The dark gray cylinder of radius $R_\ast$ is the worldtube of some star, located around the origin. Intermediate gray surfaces are constant time slices. $n^\mu$ is the outward-pointing unit vector normal to the boundary of $\mathcal{V}$.}
\label{fig:tubes}
\end{center}
\end{figure}

To prove that $J^\mu$ is indeed zero, let us integrate the conservation equation (\ref{eq:covEphi}) over a particular 4-volume $\mathcal{V}$. This volume is defined as the interior of a 2-sphere of radius $R$ between time $t = 0$ and $T$, as displayed in Fig.~\ref{fig:tubes}. Matter fields are not required to be located in a compact region. To set the ideas though, one can think of a star located at the origin of coordinates, with matter fields present below the surface $r = R_\ast$.
We have explicitly required that the geometry is regular everywhere, in particular at the origin of coordinates. Therefore, $\mathcal{V}$ is a compact manifold with boundary $\pd \mathcal{V}$. From Fig.~\ref{fig:tubes}, it is easy to understand that $\pd \mathcal{V}$ consists of the top, bottom and side of the light gray cylinder. More precisely, the top and bottom are the interior of 2-spheres of radius $R$, at time $T$ and 0 respectively; the side $\mathcal{S}$ is the Cartesian product of the 2-sphere of radius $R$ with the segment of time $[0; T]$. The Gauss-law version of Stokes theorem\footnote{Stokes theorem normally applies to $C^1$ functions. However, it might be extended to fields that have a $1/r^2$ pole at some point, as is usually done in electrostatics. Here, Eq.~(\ref{eq:scalarcharge}) tells us that $J^r$ precisely has a $1/r^2$ pole, or identically vanishes.} for $J^\mu$ then reads:
\be
\displaystyle\int_\mathcal{V}{\nd_\mu J^\mu} = \displaystyle\oint_{\pd\mathcal{V}}{n_\mu J^\mu},
\ee
where $n^\mu$ is the outward-pointing unit vector normal to $\pd \mathcal{V}$. The left-hand side vanishes because of current conservation. The integral over the top and bottom of the cylinder vanishes because $n^\mu$ and $J^\mu$ are orthogonal on these surfaces. Thus,
\begin{align}
0 &= \displaystyle\int_\mathcal{S}{n_\mu J^\mu}
\\
&=\displaystyle\int_\mathcal{S}{\tx{d}t \,\tx{d}^2\Omega \,\dfrac{R^2}{\sqrt{f(R)}}\sqrt{h(R)}\, J^r(R)}
\\
&= 4 \pi\, T R^2 \sqrt{\dfrac{h(R)}{f(R)}} J^r(R).
\end{align}
This is valid for arbitrary $R$. The only solution is to set $J^r=0$, or equivalently $Q=0$. Therefore, a permanent influx of scalar current is forbidden for a star configuration. 

A calculation similar to the one presented in this section was first carried out in \cite{Barausse:2017gip}. Note also that, so far, the proof follows a path similar to the one developed in \cite{Yagi:2015oca}. Reference \cite{Yagi:2015oca} deals in particular with a special type of Lagrangian, where a linear coupling between the scalar field and the Gauss-Bonnet invariant is present (see Sec.~\ref{sec:GB} for more details about this model). Such a theory is included in the class (\ref{eq:H})-(\ref{eq:SS}) up to boundary terms, because the Gauss-Bonnet invariant is a total divergence. Due to this fact precisely, the associated scalar field equation can be put under the form of a conservation equation. Using this fact and under some assumptions on the faraway behavior of the metric, the authors of \cite{Yagi:2015oca} prove that the scalar field cannot exhibit a $1/r$ decay at infinity. In the next section, we show that shift symmetry together with certain hypotheses ---~notably excluding the case of linear coupling to Gauss-Bonnet invariant~--- allows to establish a considerably stronger result for the very general Horndeski and beyond theory $S_\tx{bH}$.

The result established in this section is similar to the first step of the proof of Sec.~\ref{sec:theorem} for black holes. Of course, the latter does not assume that spacetime is regular everywhere. Instead, it requires that the norm of the current $J^2$ is finite at any point, especially when approaching the horizon. Interestingly, this extra assumption imposes that $J^r=0$ everywhere. In the situation examined in this section, the regularity of spacetime unequivocally imposes that $J^r$ vanishes. When studying a black hole, one thus has to impose an additional regularity assumption on the current which is not a trivial requirement. The question in this case amounts to know whether a divergence of $J^2$ has any observable effect. Black hole solutions with such behavior were exhibited in \cite{Sotiriou:2014pfa}. Although not stated in \cite{Sotiriou:2014pfa}, the norm of the current diverges at the horizon for the exhibited solutions \cite{Babichev:2016rlq}. However, the metric is regular everywhere. This issue is discussed with more detail in Sec.~\ref{sec:GB}.

\section{No scalar field around stars}
\label{sec:nohairstar}

The previous section allowed us to prove that the radial component of the current has to vanish in regular geometries. Similarly to the case of black holes, one now needs to use this in order to conclude that the scalar field is in a trivial configuration. For a Horndeski and beyond Lagrangian, the radial component of the current is given in Appendix \ref{ap:fieldeqs}, Eq.~(\ref{eq:staticJrbH}). On top of the assumptions about the symmetry of the background solution, let us assume hypotheses 2 and 4 of the black hole no-hair theorem; that is, roughly speaking, asymptotic flatness and regularity of the action. Then, one can follow the same procedure as in Sec.~\ref{sec:theorem}. The solutions split into two branches. Far away, the scalar field is necessarily in the trivial branch. This implies that $\mc{J}$, introduced in Eq.~(\ref{eq:branches}), is given according to Eq.~(\ref{eq:curlyJright}) in the outer region. We rewrite this equation here for convenience:
\be
\mc{J}=-f+2\,G_{4X}(0)G^{rr}
.\ee
The situation is now slightly different with respect to the black hole case. Indeed, when matter fields are present, the $(rr)$ component of Einstein tensor does not vanish any more. We thus see that the presence of a linear term in $G_4(X)$ is important. If $G_{4X}(0)=0$, the end of the proof is exactly similar to the one for the black hole case: the scalar field is forced to follow the trivial branch because $\mc{J}$ never vanishes. Indeed, we assumed that the spacetime under study contains no black hole; therefore, $f$ does not vanish at any point, and $\varphi$ remains trivial down to $r=0$.

If $G_{4X}(0)\neq0$, the end of the proof requires a bit more work. Let us start by splitting the $(rr)$ metric field equation in the following way:
\be
G^{rr}=\dfrac{1}{\Mp^2}(T^{(\varphi)\,rr}+T^{(\tx{m})\,rr})
\label{eq:Err1}
,\ee
where $T^{(\varphi)\,\mu\nu}$ and $T^{(\tx{m})\,\mu\nu}$ are the energy-momentum tensors that originate respectively from the scalar field and matter sector. In the outer region, and as long as one follows the branch $\varphi'=0$, $T^{(\varphi)\,\mu\nu}$ identically vanishes, or at most brings a cosmological constant contribution $-\Mp^2\,\Lambda\,g^{\mu\nu}$\footnote{In general, this contribution corresponds to a constant in the quadratic Horndeski sector: $G_2(X)=-\Mp^2\Lambda$.}. Therefore, noting $P= T^{(\tx{m})\,rr}/f$ and $\beta=G_{4X}(0)$, the function $\mc{J}$ takes the following form:
\be
\mc{J}=f\left(\dfrac{2\beta}{\Mp^2} P-2\beta\Lambda-1\right)
\label{eq:curlyJGR}
.\ee
$P$ should be thought of as the pressure due to matter fields. It can now happen that the pressure reaches the critical value $P_1$, defined as
\be
P_1=\dfrac{\Mp^2}{2 \beta}(1+2\beta\Lambda).
\label{eq:Pc}
\ee
Let us assume this indeed happens at some radius $r=r_1$. Then, nothing a priori forbids the solution to jump from the first branch where $(\varphi'=0,\,\mc{J}\neq0)$ to the second one, with $(\varphi'\neq0,\,\mc{J}=0)$. In this case, the solution would differ from general relativity in the ball $r<r_1$. As we are going to see now, this is forbidden by the regularity of the solution when taking into account the other field equations. In the outer region $r>r_1$, the solution corresponds to general relativity and one can show by using Eq.~(\ref{eq:curlyJGR}) that
\be
\underset{r>r_1}{\underset{r\to r_1}{\tx{lim}}}\mc{J}'=\dfrac{2\beta f_1 P_1'}{\Mp^2},
\label{eq:limitright}
\ee
where $f_1=f(r_1)$ and $P'_1=P'(r_1)$. On the other hand, since $\mc{J}$ would be trivial for $r<r_1$, one has immediately 
\be
\underset{r<r_1}{\underset{r\to r_1}{\tx{lim}}}\mc{J}'=0.
\label{eq:limitleft}
\ee
Such a transition would happen inside the star and not at its surface, since the pressure $P$ would reach the value $P_1$ already. It is therefore natural to ask that $P'_1\neq0$ (there may exists modified gravity models where $P'=0$  at some non-zero radius,  but there is absolutely no reason why this would happen precisely when $P=P_1$). Finally, in the inner region, $\mc{J}$ can be put in the form $\mc{J}(f,h,\varphi',P)$ by use of the $(rr)$ equation (\ref{eq:Err}) in presence of matter, instead of its defining form $\mc{J}(f,h,h',\varphi')$ ---~that is, one can trade $h'$ for $P$. Thus, a priori, $\mc{J}'$ depends on $f$, $f'$, $h$, $h'$, $\varphi'$, $\varphi''$, $P$ and $P'$. All these functions are continuous at $r_1$ by assumption, except maybe $\varphi''$. However, one can also replace this quantity in terms of the other functions as well as the matter energy density $\rho$, by use of the $(tt)$ equation (\ref{eq:Ett}). One ends up with $\mc{J}'(f,h,P,\rho,f',h',\varphi',P')$ which must itself be continous at $r_1$, in contradiction with Eqs.~(\ref{eq:limitright})-(\ref{eq:limitleft}). In summary, even when $G_{4X}(0)\neq0$, no physically sensible solution exists apart from general relativity. The scalar field cannot jump from the trivial branch to another one.

Let us summarize what was proven, and under which conditions. Consider any shift-symmetric Horndeski or beyond theory with a minimally coupled matter sector, as in (\ref{eq:bHSSstar}). Assuming that:
\begin{enumerate}
\item the metric and scalar field are $C^1$ everywhere, spherically symmetric and static, and there exists no metric horizon,
\item spacetime is asymptotically flat with $\varphi'\rightarrow 0$ as $r\rightarrow \infty$,
\item there is a canonical kinetic term $X\subseteq G_2$ in the action and the $G_i$-$F_i$ functions are analytic at the point $X=0$,
\end{enumerate} 
one can conclude that $\varphi$ is constant, and that the only solutions are given by general relativity. In particular, star solutions are identical to their general relativistic counterparts.

As detailed above, it was not necessary to assume that the norm of the current is regular, as opposed to the no-hair theorem for black holes of Sec.~\ref{sec:theorem}. As an alternative proof, one can decide to keep the assumption that $J^2$ is regular everywhere. This is slightly more restrictive, but allows a nice comparison with the black hole case, through Eq.~(\ref{eq:Jsquared}). Indeed, in this equation, there can be two reasons for $J^2$ to blows up. Either $h$ vanishes, and this indicates the presence of a horizon, or $r$ vanishes, that is, one approaches the origin of coordinates. The latter case always happen in a regular geometry as assumed here, because no singularity or horizon shields the origin of coordinates. Hence, from this perspective, it is either the presence or absence of a horizon that forbids the presence of scalar hair.

A number of ways out exist for the black hole no-hair theorem in Horndeski and beyond theories, that we will extensively detail in the forthcoming part, Part \ref{part:2}. Since the theorem proposed for stars relies on very similar assumptions, the same possibilities are offered to escape from it. We explore or indicate theses possibilities in Secs.~\ref{sec:starhairtdep} and \ref{sec:starhairstatic}. 

Interestingly, the no-hair result established here is complementary to what Ref.~\cite{Barausse:2015wia} derived. Let us summarize here the main result of this paper. It investigates the emission of gravitational waves for binary stars in shift-symmetric Horndeski theory. Assuming that the dynamics of the system is perturbative, one can approximate the stars as stationary objects, and perform an expansion around a Minkowski background with $\varphi=0$. This expansion is parametrized by $\epsilon$:
\begin{align}
g_{\mu\nu}&\underset{\epsilon\to0}{=}\eta_{\mu\nu}+\epsilon\, h_{\mu\nu}+\mc{O}(\epsilon^2),
\label{eq:gpertbin}
\\
\varphi&\underset{\epsilon\to0}{=}\epsilon\, \delta\varphi+\mc{O}(\epsilon^2).
\label{eq:phipertbin}
\end{align}
The perturbations then obey the following propagation equations:
\begin{align}
\Box_\eta \hbar_{\mu\nu}&\underset{\epsilon\to0}{=}-\dfrac{2}{\Mp^2}\, T^{(\tx{m})\,\mu\nu}+\mc{O}(\epsilon),
\\
\Box_\eta \delta\varphi&\underset{\epsilon\to0}{=}\mc{S}\left(s_\varphi^{(1)},s_\varphi^{(2)}\right)+\mc{O}(\epsilon),
\end{align}
where $\Box_\eta$ is the d'Alembertian operator in Minkowski spacetime and $\hbar_{\mu\nu}=h_{\mu\nu}-1/2\,h\,\eta_{\mu\nu}$; $\mc{S}$ is a linear function of the so-called sensitivities of the stars 1 and 2, defined as
\be
s_\varphi^{(i)}=\left.\dfrac{1}{m_i}\,\dfrac{\pd m_i}{\pd \varphi}\right|_{N_i,\Sigma_i}
\ee
with $m_i$ the mass of the star $i$, $N_i$ its total baryon number and $\Sigma_i$ its entropy. Reference \cite{Barausse:2015wia} further assumes assumption 3 of the above theorem for stars, and that the scalar field decays as $1/r$ while the metric also deviates from Minkowski spacetime as $1/r$ for large $r$. Under these conditions, sensitivities $s_\varphi^{(i)}$ actually vanish; thus, the expansion in Eqs.~(\ref{eq:gpertbin})-(\ref{eq:phipertbin}) is identical to general relativity at leading order.

This result concerns wave propagation, while the no-hair theorem focuses on the static structure of stars. However, they both combine to point at the absence of differences between general relativity and shift-symmetric (beyond) Horndeski theories. The proof of Ref.~\cite{Barausse:2015wia} comes however with a caveat. The $1/r$ decay of the scalar field and non-homogeneous metric part is valid only outside the Vainshtein radius $r_\tx{V}$ of the theory if it exhibits a Vainshtein screening (see Sec.~\ref{sec:starhairtdep} for more details about this mechanism). Therefore, the above analysis is valid only for wavelenghts $\lambda_\tx{GW}$ greater than the Vainshtein radius. For binary pulsars, $\lambda_\tx{GW}$ is of order $10^9$ km, while for the late inspiral stage of binary neutron stars, it is of order $10^3$ km. In the latter case, it is clear that we cannot be in the regime where the sensitivities vanish. Thus, a more refined analysis may be necessary.

In conclusion, we proved that a shift-symmetric scalar-tensor theory cannot accommodate a non-vanishing Noether current as soon as spacetime is required to be regular, static and spherically symmetric. Using this result, some complementary assumptions allow us to conclude that stars cannot develop scalar hair in Horndeski and beyond theories. Star solutions are therefore identical to general relativity solutions for these theories. This extends what was known previously in the case of black hole geometries. Most of the ways out known for black holes, and presented in Chapters \ref{ch:tdep}-\ref{ch:static}, still allow to circumvent the theorem for stars, as detailed in Secs.~\ref{sec:starhairtdep} and \ref{sec:starhairstatic}. 

\newpage
\thispagestyle{empty}
\cleardoublepage
\part{Compact objects in Horndeski theory and beyond}
\label{part:2}

\newpage
\thispagestyle{empty}
\cleardoublepage
\noindent In this part, we will systematically explore the paths allowing to circumvent the no-hair theorems for black holes (presented in Chapter \ref{ch:nohair}) and for stars (presented in Chapter \ref{ch:nohairstars}). Almost all procedures that allow to construct black holes with scalar hair also provide stars with non-trivial scalar profile, with one exception however. There actually exist many such paths; indeed, one can play with the assumptions of the theorem, and with a huge class of models fitting in the Horndeski and beyond theory. This also means that the no-hair theorems for the shift-symmetric Horndeski theory are relatively weak: it is enough to break only one or two assumptions at a time to obtain solutions that are not physically extravagant. This is summarized in Table \ref{tab:hair}. The second line of the table refers to some solutions that we will describe in the forthcoming part. The table should be read as follows. For instance, the solution proposed by Rinaldi in Ref.~\cite{Rinaldi:2012vy}, fulfills all the assumptions of the theorem, but asymptotic flatness of the metric (it has de Sitter asymptotics). One may refer to Table \ref{tab:hair} if lost in the diversity of solutions we will encounter in this part.

\begin{table}[ht]
 \setlength\extrarowheight{2pt}
 \centering
 \begin{tabular}{|c|m{2.1cm}|m{1.3cm}|m{2.1cm}|m{1.7cm}|m{2.2cm}|m{2.6cm}|}
  \cline{3-7}
\multicolumn{1}{c}{}& &\multicolumn{5}{c|}{Solutions}   \\ 
	\cline{3-7}
 \multicolumn{1}{c}{}& &\centering Rinaldi \cite{Rinaldi:2012vy} &\centering Stealth Schwarz-schild \cite{Babichev:2013cya}  &\centering Sotiriou-Zhou \cite{Sotiriou:2014pfa}  &\centering $\sqrt{-X}\subset G_4$ \cite{Babichev:2017guv} &\centering Purely quartic theory \cite{Babichev:2017guv}
\tabularnewline
	\hline
\multirow{11}{*}{\rotatebox{90}{Assumptions of the theorem}} & \centering Asymptotic flatness & \centering\textcolor{red}{\xmark}  & \centering\textcolor{green}{\cmark}  & \centering\textcolor{green}{\cmark} & \centering\textcolor{green}{\cmark} & \centering\textcolor{green}{\cmark}
	\tabularnewline
	\cline{2-7} 
    & \centering Decay of $\pd_\mu\varphi$ & \centering\textcolor{green}{\cmark} & \centering\textcolor{red}{\xmark} & \centering\textcolor{green}{\cmark} & \centering\textcolor{green}{\cmark} & \centering\textcolor{red}{\xmark}
	\tabularnewline
	\cline{2-7} 
    & \centering Spherical symmetry & \centering\textcolor{green}{\cmark} & \centering\textcolor{green}{\cmark} & \centering\textcolor{green}{\cmark} & \centering\textcolor{green}{\cmark} & \centering\textcolor{green}{\cmark}
	\tabularnewline
	\cline{2-7} 
    & \centering Static $g_{\mu\nu}$ & \centering\textcolor{green}{\cmark} & \centering\textcolor{green}{\cmark} & \centering\textcolor{green}{\cmark} & \centering\textcolor{green}{\cmark} & \centering\textcolor{green}{\cmark}
	\tabularnewline
	\cline{2-7}
		& \centering Static $\varphi$ & \centering\textcolor{green}{\cmark} & \centering\textcolor{red}{\xmark} & \centering\textcolor{green}{\cmark} & \centering\textcolor{green}{\cmark} & \centering\textcolor{green}{\cmark}
	\tabularnewline
	\cline{2-7}
		& \centering Finite $J^2$ & \centering\textcolor{green}{\cmark} & \centering\textcolor{green}{\cmark} & \centering\textcolor{red}{\xmark} & \centering\textcolor{green}{\cmark} & \centering\textcolor{green}{\cmark}
	\tabularnewline
	\cline{2-7}
		& \centering Analytic $G_i-F_i$ & \centering\textcolor{green}{\cmark} & \centering\textcolor{green}{\cmark} & \centering\textcolor{green}{\cmark} & \centering\textcolor{red}{\xmark} & \centering\textcolor{green}{\cmark}
	\tabularnewline
	\cline{2-7}
		& \centering $X \subseteq G_2$ & \centering\textcolor{green}{\cmark} & \centering\textcolor{red}{\xmark} & \centering\textcolor{green}{\cmark} & \centering\textcolor{green}{\cmark} & \centering\textcolor{red}{\xmark}
	\tabularnewline
	\hline
 \end{tabular}
 \caption{Different possibilities to circumvent the no-hair theorem of Chapter \ref{ch:nohair}. This table reads as follows: every column corresponds to a solution. A green checkmark in the associated column means that the solution fulfills the corresponding hypothesis; a red cross means that it does not respect the assumption. The list of solutions presented here is certainly not exhaustive. The purpose is just to show the known minimal violations of the theorem.}
 \label{tab:hair}
\end{table}

Note that there does not exist a physically sensible solution whenever breaking only a single assumption of the theorem. In some cases, one needs to break two assumptions at least to obtain a non-trivial solution that deviates from general relativity. Let us also remark that no solution was built exploiting the lack of spherical symmetry and/or staticity of the metric. Of course, one should not expect that a no-hair theorem exists if the background has no symmetry at all. However, it might be interesting to know if abandoning, say, staticity and spherical symmetry for stationarity and axisymmetry is enough to obtain solutions with scalar hair in Horndeski and beyond theory (as opposed to usual scalar-tensor theories, where the theorems of Chapter \ref{ch:nohairGR} are known to hold). We already discussed this question in the slow rotation limit in Sec.~\ref{sec:slowrot}. Another interesting track would be to abandon the staticity assumption for the metric, while keeping spherical symmetry. In the corresponding case in general relativity, Birkhoff's theorem ensures that no other solution than Schwarzschild's exists in the vacuum; a black hole cannot ``breath'', contracting or inflating while maintaining spherical symmetry. We will discuss in Paragraph \ref{sec:soundness} how this is modified in the context of scalar-tensor theories.

Before going into the detail of these solutions, let us also mention another characteristic of the theorem that can be seen as a weakness. Its formulation is parametrization-dependent. Indeed, one needs to invoke the analyticity of the $G_i$ and $F_i$ functions in order to conclude. As we have seen, there exist several parametrizations of Horndeski and beyond theory, see Appendix \ref{ap:otherparam}. Therefore, it would be much better to be able to formulate the theorem in a way that does not depend on the parametrization. At first, it seems that requiring the analyticity of the $G_i$ and $F_i$ functions is simply equivalent to requiring the analyticity of the Horndeski and beyond action. Strictly speaking, this is true; however, one does not really care about the action itself, but rather about the associated field equations. It can happen that a theory with a non-analytic action is equivalent, up to boundary terms, to an action that is perfectly analytic. One such example is the non-analytic Lagrangian with $G_5\propto \ln|X|$. It is actually equivalent, at the level of field equations, to a linear coupling between the scalar field and the Gauss-Bonnet invariant. Details are given in Sec.~\ref{sec:GB}.

The plan of the part is as follows. In Chapter \ref{ch:tdep}, we introduce a linear time-dependence for the scalar field, while keeping a static metric. We first motivate such a behavior through cosmology, and then investigate in full detail the solutions of two types of Lagrangians: one with reflection symmetry $\varphi\to-\varphi$, and the other without this symmetry. Chapter \ref{ch:static} is devoted to isolated objects. We thus focus on asymptotically flat solutions, and keep a static scalar field. We explore three different ways to circumvent the no-hair theorem. Sections \ref{sec:starhairtdep} and \ref{sec:starhairstatic} also give details on the construction of star solutions with a non-trivial scalar field.
\newpage
\thispagestyle{empty}
\cleardoublepage
\chapter{Black holes and stars with a time-dependent scalar field}\label{ch:tdep}

The no-hair theorem of Sec.~\ref{sec:theorem} relies crucially on the staticity of the metric. Since the field equations bind the scalar to the metric, it is tempting to assume that the scalar field itself is static. However, there is no mathematical obligation to do so. Only the energy-momentum tensor of the scalar field must inherit the symmetries of the geometry, as we mentioned in Chapter \ref{ch:nohairGR}, in the context of general relativity with a complex scalar field \cite{Herdeiro:2014goa}. In fact, if one has in mind the cosmological motivation for modifying gravity, it is natural to allow for a time dependence of the scalar field. Let us see why on the simple example of a de Sitter geometry, which is generically an exact solution of the field equations when no extra matter field is present. Choosing the flat slicing of de Sitter spacetime, the metric takes its well-known form:
\be
\mathrm{d}s^2 = -\, \mathrm{d}\tau^2 + \text{e}^{2 H \tau} (\mathrm{d}\rho^2 + \rho^2 \mathrm{d}\Omega^2)
\label{eq:dSflat}
,\ee
with constant Hubble rate $H$. A homogeneous scalar field $\varphi$ can only depend on $\tau$ in these coordinates. One can map this geometry to the static slicing by the following change of coordinates:
\begin{align}
t&=\tau-\dfrac{1}{2H}\,\tx{ln}\left[1-(H \rho\, \tx{e}^{H\tau})^2\right],
\label{eq:changecoord1}
\\
r&= \rho \,\tx{e}^{H\tau}.
\label{eq:changecoord2}
\end{align}
Then, the metric takes a static form:
\be
\mathrm{d}s^2 = - \,(1-H^2r^2)\,\mathrm{d}t^2 + \dfrac{\mathrm{d}r^2}{1-H^2r^2} + r^2 \mathrm{d}\Omega^2.
\label{eq:dSstatic}
\ee
At the same time, inverting the change of coordinates (\ref{eq:changecoord1})-(\ref{eq:changecoord2}), the scalar field $\varphi(\tau)$ becomes a function of $t$ and $r$, $\varphi(t,r)$. It is therefore legitimate to consider static metrics alongside a time-dependent scalar field. If the cosmological evolution is slow, one may expand the scalar field around a given time $\tau_0$: 
\be
\varphi(\tau) \underset{\tau\to\tau_0}{=} \varphi_0+ \dot{\varphi}_0 (\tau-\tau_0)+ \mc{O}[(\tau-\tau_0)^2]
\label{eq:cosmicexp}
,\ee
where $\varphi_0$ and $\dot{\varphi}_0$ are respectively the value of $\varphi$ and its time derivative at $\tau_0$. Thus, a linear time-dependence for $\varphi$ arises as a natural approximation of cosmological evolution. This will also result in a linear time-dependence in the ``static time'' $t$, as can be seen form Eq.~(\ref{eq:changecoord1}). Additionally, $\varphi = q \tau$ was shown to be an exact cosmological solution for various Horndeski models \cite{Babichev:2012re,Babichev:2016fbg,Appleby:2015ysa,Babichev:2013cya}, as well as a cosmological attractor for models that include cubic and quadratic Horndeski terms \cite{Deffayet:2010qz}. Of course, the presence of some massive body makes things more complicated than a mere coordinate transformation; the scalar field will acquire an additional space dependence which does not result of the coordinate change. Let us parametrize this through the following ansatz: 
\be
\varphi(t,r)= q t + \psi(r),
\label{eq:phitdep}
\ee
where $q$ is a free parameter ---~corresponding to $\dot{\varphi}_0$ in Eq.~(\ref{eq:cosmicexp})~--- and $\psi$ is to be determined from the field equations. As explained, this linear ansatz derives from cosmology, but it also has the following nice property: since the considered models depend only on the derivatives of the scalar field, a linear time dependence of the scalar field does not generate any time dependence in the associated energy-momentum tensor. This does not yet guarantee the existence of solutions, but it is a first consistency check.

In the next section, we will establish the consistency of the ansatz (\ref{eq:phitdep}) in light of the field equations, as well as other possible time dependence. Then, we will examine explicit solutions for the simplest terms that arise in cubic and quartic Horndeski sectors (keeping a simple quadratic sector for consistency). In Sec.~\ref{sec:quartictdep}, we present the black hole solutions that were built using the ansatz (\ref{eq:phitdep}) in the framework of quartic Horndeski models. These have nice integrability properties. We will then explore in Sec.~\ref{sec:cubictdep} cosmological as well as black hole solutions in the simplest cubic model, that can arise from the higher dimensional Dvali-Gabadadze-Porrati model. The equations are more difficult to integrate in this case, and we resorted to numerical integration, based on the results of our work \cite{Babichev:2016fbg}. Finally, Sec.~\ref{sec:conclusiontdep} summarizes the results of this chapter.

\section{Consistency and robustness of the ansatz}

\subsection{Mathematical consistency}
\label{sec:consistency}

We justified that solutions with a static metric and a time-dependent scalar field exist. However, one can wonder whether the ansatz (\ref{eq:phitdep}) for the scalar field, together with a static and spherically symmetric metric, Eq.~(\ref{eq:statmetric}), generically provides solvable field equations. In other words, can we prove a priori that the ansatz (\ref{eq:statmetric})-(\ref{eq:phitdep}) comes with solutions in general? This question was answered positively by the authors of \cite{Babichev:2015rva}. Let us consider a generic scalar-tensor theory with Lagrangian density $L[\varphi,\,g_{\mu\nu}]$. Only diffeomorphism invariance is required. The action $S=\int{\tx{d}^4x L}$ can be used to generate the field equations, that we denote as
\begin{align}
\mc{E}^{(\varphi)}&=\dfrac{1}{\sqrt{-g}}\,\dfrac{\delta L}{\delta \varphi}=0,
\\
\mc{E}^{(g)}_{\mu \nu}&=\dfrac{2}{\sqrt{-g}}\,\dfrac{\delta L}{\delta g^{\mu\nu}} = 0,
\end{align}
for the scalar and metric field equations respectively. The ansatz (\ref{eq:statmetric})-(\ref{eq:phitdep}) only switches on some of the components of $\mc{E}^{(g)}_{\mu \nu}$, namely the $(tt)$, $(rr)$, $(tr)$, $(\theta \theta)$ and $(\phi \phi)$ equations. Spherical symmetry additionally imposes that the $(\theta \theta)$ and $(\phi \phi)$ equations are trivially related through $\mc{E}^{(g)}_{\phi \phi} = \sin^2\theta\, \mc{E}^{(g)}_{\theta \theta}$. Together with the scalar field equation, there are in total five non-trivial equations. On the other hand, only three free functions are at disposal: $h$, $f$ and $\psi$. Therefore, a naive counting seems to indicate that the system is overconstrained. In fact, thanks to diffeomorphism invariance, one can prove that two of the five equations can be deduced from the others. Indeed, let us consider the change of coordinates generated by an infinitesimal vector field $\xi^\mu$. In the new system of coordinates $\hat{x}^\mu=x^\mu+\xi^\mu$, the scalar field and metric read
\begin{align}
\hat{\varphi}&=\varphi- \pd_\mu \varphi\, \xi^\mu,
\\
\hat{g}^{\mu\nu}&=g^{\mu\nu}+ 2\nd^{(\mu} \xi^{\nu)}.
\end{align}
The action is invariant under this coordinate redefinition, which translates mathematically as
\begin{align}
0&=\delta S
\\
 &= \displaystyle\int{\tx{d}^4x\sqrt{-g} \left(\dfrac{1}{\sqrt{-g}}\,\dfrac{\delta L}{\delta \varphi} \delta\varphi+ \dfrac{1}{\sqrt{-g}}\,\dfrac{\delta L}{\delta g^{\mu\nu}} \delta g^{\mu \nu}\right)}
\\
&= \displaystyle\int{\tx{d}^4x\sqrt{-g}\left(-\mc{E}^{(\varphi)} \pd_\mu \varphi\, \xi^\mu+ \mc{E}^{(g)}_{\mu \nu} \nd^{\mu} \xi^\nu\right)},
\label{eq:varaction}
\end{align}
for arbitrary $\xi^\mu$. Thus, integrating by parts the second term in the previous integral,
\be
\mc{E}^{(\varphi)} \pd_\nu \varphi + \nd^\mu \mc{E}^{(g)}_{\mu \nu} = 0.
\label{eq:linkeqs}
\ee
Of the above vector equation, two components are non-trivial. They read
\begin{align}
q \mc{E}^{(\varphi)} + \dfrac{1}{\sqrt{-g}} \pd_r\left(\sqrt{-g}\, g_{rr} \,\mc{E}^{(g)}_{tr}\right)&=0,
\label{eq:linkEphi}
\\
\mc{E}^{(\varphi)} \pd_r \varphi+ \dfrac{1}{\sqrt{-g}} \pd_\mu\left(\sqrt{-g}\, g^{\mu \nu} \mc{E}^{(g)}_{\nu r}\right) -\dfrac12\pd_r g_{\mu \nu}\, \mc{E}^{\mu \nu}_{(g)}&=0.
\end{align}
This explicitly gives, for instance, the scalar field and $(\theta \theta)$ metric equations in terms of the other equations. Thus, only three independent equations remain, with three free functions. The system is not overconstrained, and solutions generically exist.

When the scalar field possesses shift symmetry, one can push further the above calculation. In this case, the scalar field equation $\mc{E}^{(\varphi)}$ takes the simpler form
\be
\mc{E}^{(\varphi)} = \nd_\mu J^\mu = 0
,\ee
where $J^\mu$ is the Noether current associated with shift symmetry, defined in Eq.~(\ref{eq:currentdef}). Integrating by parts the first term in Eq.~(\ref{eq:varaction}) yields
\be
\displaystyle\int{\tx{d}^4x\sqrt{-g} \left[J^\nu \nd_\nu\left(\pd_\mu \varphi \,\xi^\mu\right)+ \mc{E}^{(g)}_{\mu \nu} \nd^{\mu} \xi^{\nu}\right]}=0.
\ee
Considering the particular change of coordinates $\xi^\mu = (\xi(r),0,0,0)$, the same equation reads
\be
\displaystyle\int{\tx{d}^4x\sqrt{-g} \left[J^r \pd_r(q \xi)+ 2\mc{E}^{(g)}_{tr} \nd^{(t} \xi^{r)}\right]}=0.
\label{eq:presquereltrJr}
\ee
Additionally, for the specific vector $\xi^\mu$ considered, 
\be
\nd^{(t} \xi^{r)} = \dfrac12\, g^{rr} \pd_r \xi.
\ee
Inserted into Eq.~(\ref{eq:presquereltrJr}), and taking into account that $\xi(r)$ is arbitrary, this gives an interesting relation:
\be
q J^r = - \mc{E}^{(g)}_{tr} g^{rr}.
\label{eq:linkJrEtr}
\ee
This relation corresponds to the integral of Eq.~(\ref{eq:linkEphi}), with the additional information that the integration constant vanishes. Physically, the $(tr)$ metric equation is related to matter accretion (more details are given in the next paragraph). Therefore, forbidding accretion onto the central object (as a consequence of staticity) also forbids a radial flow of the scalar Noether current, quite consistently. We further discuss in the next paragraph the legitimacy of imposing the staticity of the metric. Before doing so, we use Eq.~(\ref{eq:linkJrEtr}) to prove an extension of the no-hair result of Sec.~\ref{sec:theorem}.

\subsubsection*{Extension of the no-hair theorem}

As a side result, the relation (\ref{eq:linkJrEtr}) allows to extend the no-hair theorem to some linearly time-dependent solutions. Indeed, one of the assumptions of the theorem was that the norm of the current, $J^2$, remains finite everywhere. This allowed us to show as a first step that $J^r$ had to vanish. Whenever $q\neq0$, Eq.~(\ref{eq:linkJrEtr}) already tells us that $J^r$ vanishes. Thus, when $\varphi$ is assumed to depend linearly on time, one can get rid of the assumption that $J^2$ is finite. However, at the same time, the second part of the proof is altered. Indeed, for non-zero $q$, the radial component of the Noether current of shift-symmetric Horndeski theory reads:
\be
\begin{split}
\label{eq:Jrqneq0}
J^r=&-f \varphi' G_{2X} +\dfrac{f}{2 r^2h^2}\left[rfh(r h' + 4 h)\varphi'^2-q^2r^2h'\right]G_{3X}
\\
&+ \dfrac{2 f \varphi'}{r^2h} (f h - h + r f h')G_{4X} -\dfrac{2 f^2 \varphi'}{r^2h^2} \left[(h +rh')fh\varphi'^2-q^2rh'\right]G_{4XX} 
\\
&-\dfrac{f h'}{2r^2h} \left[(3 f-1)hf\varphi'^2+q^2(1-f)\right]G_{5X} 
\\
&+\dfrac{h' f^3\varphi'^2}{2r^2h^2}(f\varphi'^2-q^2) G_{5XX}
,\end{split}
\ee
where a prime stands for a radial derivative ---~to compare with Eq.~(\ref{eq:JrHorni}) for the static case. In general, one cannot factor out a $\varphi'$ term in this equation, because of the cubic and quintic Horndeski terms. However, it is still possible to carry the exact same procedure as in the static case when only reflection-symmetric Horndeski terms are present. Therefore, it is possible to formulate an extension of the no-hair result in the following way; let us keep all the assumptions of the theorem in Sec.~\ref{sec:theorem} except that the scalar field is no longer assumed to be static, nor the current $J^2$ to have a finite norm. Instead, the scalar field is required to depend linearly on time, and the theory to have the reflection symmetry $\varphi\rightarrow-\varphi$. Then, the same conclusion holds, and all black holes have a Schwarzschild geometry.

\subsection{Robustness}
\label{sec:soundness}

We established that it is legitimate to consider a time-dependent scalar field in a static geometry, and that a linear time dependence generically allows for solutions. However, choosing a static metric is physically questionable for at least two reasons. The first objection is related to black hole considerations. In general relativity, Birkhoff's theorem ensures that a spherically symmetric solution of Einstein's equations in the vacuum is also static. However, in Horndeski theory, there is no such theorem, and a spherically symmetric spacetime can perfectly evolve with time. There is therefore no objection against spherically symmetric and non-static metrics in principle. Actually, one can formally view the scalar field as a matter component, and this matter could be accreted onto a black hole. The accretion process does take place in Horndeski theory (for a review on the accretion of dark energy, see \cite{Babichev:2014lda}). It can be treated perturbatively, by neglecting the back-reaction on the metric \cite{Babichev:2010kj}. However, the accreted scalar will eventually change the geometry with time. To be more precise, let us formally split the metric field equations as 
\be
\mc{E}^{(g)\,\mu\nu}= -M_{\tx{Pl}}^2 G^{\mu\nu}+T^{(\varphi)\,\mu\nu}
\label{eq:defTmunu}
,\ee 
where $G^{\mu\nu}$ is Einstein's tensor. Dropping the assumption that the metric is static, the line element can be put in the form
\be
\mathrm{d}s^2 = -\left[1-\dfrac{2m(t,r)}{r}\right]\mathrm{d}t^2 + \dfrac{\mathrm{d}r^2}{f(t,r)} + r^2 \mathrm{d}\Omega^2
.\ee
Accretion onto the central object, say a black hole, is related to the $(tr)$ metric equation in the sense that $\pd_t m$ is generically proportional to $T^{(\varphi)\,r}_{\hphantom{(\varphi)\,r}\,t}$ \cite{Babichev:2014lda}. An increase\footnote{The mass can also decrease with absorption of dark energy, for instance in the case of phantom energy \cite{Caldwell:2003vq}, because the model violates the energy conditions.} in the mass function of the black hole necessarily comes from a non-vanishing $T^{(\varphi)\,r}_{\hphantom{(\varphi)\,r}\,t}$ component. On the contrary, in a static geometry, $G^{\,r}_{\:\:t}$ vanishes and therefore so does $T^{(\varphi)\,r}_{\hphantom{(\varphi)\,r}\,t}$. Imposing a static metric, one forbids by hand the natural accretion; it is certain that one misses other solutions by doing so. Clearly, the solutions with a static metric that we describe in the forthcoming sections do not belong to the same branch as accreting ones. This different branch exists thanks to the complex high-order structure of the considered Horndeski models. Although there is no reason for disregarding non-static metrics, there is a huge technical difficulty in studying (and even finding) such solutions. Allowing the metric to depend on time turns all ordinary differential equations into a set of coupled partial differential equations. Still, time-dependent and spherically symmetric metrics could be the key to unlock many interesting solutions, as well as novel aspects specific to scalar-tensor theories.

Another argument against a static metric comes from cosmology. After all, we do not live in a de Sitter static spacetime (yet). Almost thirty percent of the energy content of the Universe is matter, therefore the spacetime we live in is certainly not static. This objection is maybe less fundamental than the previous one, though more obvious. Indeed, one can still use a de Sitter static metric as an approximation for our current and future spacetime. In the perspective of the large cosmological constant problem, the effect one aims at explaining is a discrepancy of at least fifty five orders of magnitude. Therefore, even an extremely rough approximation, such as saying that we live in a de Sitter spacetime, should not spoil a potential explanation to the large cosmological problem. Even if they are very approximative when it comes to describe the Universe, it is still interesting to obtain exact solutions, and to carry the analysis in explicit backgrounds such as de Sitter or even Minkowski spacetimes. 

With these two objections in mind, we shall pursue with a static geometry. One can still ask whether a linearly time-dependent scalar field is a convenient choice, or the only consistent possibility. There exists an interesting case, where the geometry is Minkowski spacetime and the scalar field depends quadratically on time, the so-called ``Fab Four'' theory \cite{Charmousis:2011ea}. The defining property of this subset of Horndeski theory is that it admits self-tuning solutions. Self-tuning has here a precise meaning: the model must admit flat spacetime as a solution whatever the value of the bare cosmological constant in the action. The scalar field adjusts dynamically through potential phase transitions that modify the value of the bare cosmological constant. Fab Four models are obtained by combination of the four following Lagrangian densities: 
\begin{align}
\label{eq:johnf4}
\mc{L}_{\tx{John}} &= \sqrt{-g} \, V_{\tx{John}}(\varphi) G^{\mu\nu} \pd_\mu\varphi\, \pd_\nu \varphi ,
\\
\label{eq:paul}
\mc{L}_{\tx{Paul}} &=\sqrt{-g}\, V_{\tx{Paul}}(\varphi)   P^{\mu\nu\alpha \beta} \pd_\mu \varphi\, \pd_\alpha \varphi\, \nabla_\nu \pd_\beta \varphi ,
\\
\label{eq:george}
\mc{L}_{\tx{George}} &=\sqrt{-g}\, V_{\tx{George}}(\varphi) R ,
\\
\label{eq:ringo}
\mc{L}_{\tx{Ringo}} &= \sqrt{-g}\, V_{\tx{Ringo}}(\varphi) \hat G
,\end{align}
where the $V_{\tx{Beatle}}$ functions are arbitrary, $\hat G$ is the Gauss-Bonnet density, Eq.~(\ref{eq:GB}), and $P_{\mu\nu\alpha \beta}$ is the double dual Riemann tensor:
\be
P^{\mu\nu}{}_{\rho\sigma} = -\dfrac12 \,\epsilon^{\rho\sigma\lambda\kappa}\,R_{\lambda \kappa}{}^{\xi \tau} \, \dfrac12\, \epsilon_{\xi \tau\mu\nu}
\label{eq:Ptensor}
.\ee
The idea behind such models is of course to get rid of the large cosmological constant. Different subsets are obtained in the same fashion when one desires a de Sitter solution (with an effective cosmological constant independent of the bare one) rather than a Minkowski one \cite{Appleby:2012rx,Martin-Moruno:2015bda}. In \cite{Charmousis:2014mia}, a simple Fab Four model was considered (as well as others), with constant $V_{\tx{John}}(\varphi)=\beta$ and vanishing other potentials. It admits the following Minkowski solution:
\be
\begin{split}
\mathrm{d}s^2 &= -\,\mathrm{d}t^2 +\mathrm{d}r^2 + r^2 \mathrm{d}\Omega^2,
\\
\varphi(t,r)&= \varphi_0 + \varphi_1(t^2-r^2)
\label{eq:Minkowsol}
,\end{split}
\ee
where $\varphi_0$ is free and self-tuning imposes $\beta \varphi_1^2 = \Lambda_{\tx{b}}$ with $\Lambda_\tx{b}$ the bare cosmological constant that appears in the action. Therefore, there also exist solutions with non-linear time dependence (similar solutions were also obtained for the Paul Lagrangian in \cite{Appleby:2015ysa}). The apparent inhomogeneity of the scalar field is very similar to what one obtains for de Sitter spacetime in different slicings, Eqs.~(\ref{eq:dSflat})-(\ref{eq:dSstatic}). The Minkowski solution (\ref{eq:Minkowsol}) can be mapped to a hyperbolic Robertson-Walker spacetime with a homogeneous scalar field, thanks to the following change of coordinates:
\be
\begin{split}
T&=\sqrt{t^2-r^2},
\\
\chi&=\tx{Argth}\left(\dfrac{r}{t}\right)
.\end{split}
\ee
The solution then reads as a specific hyperbolic Robertson-Walker metric, known as Milne spacetime:
\be
\begin{split}
\mathrm{d}s^2 &=-\,\mathrm{d}T^2 +T^2(\mathrm{d}\chi^2 + \tx{sh}^2\chi\, \mathrm{d}\Omega^2),
\\
\varphi(T)&=\varphi_0 + \varphi_1 T^2
.\end{split}
\ee
In passing, one could think about using the Lorentz invariance of Minkowski spacetime to generate infinitely many different scalar profiles, by boosting the solution (\ref{eq:Minkowsol}). However, $\varphi$ is proportional to $t^2-r^2$, which is precisely a Lorentz invariant.

In conclusion, even when assuming a static metric, there can exist different behaviors for the evolution of the scalar field with time, as shown by the explicit example of the Fab Four theory. Not all evolutions are permitted though. For instance, Ref.~\cite{Appleby:2015ysa} proved that Minkowski spacetime is a solution of the simple model $V_{\tx{John}}(\varphi)=\beta$ only when $\varphi$ depends quadratically or linearly on time. As checked explicitly in Paragraph \ref{sec:consistency}, a linearly time-dependent scalar field (\ref{eq:phitdep}) always yields solutions. That is why we will stick to this anstaz in order to examine solutions to specific models in the following sections.

\section{Reflection-symmetric Horndeski models}
\label{sec:quartictdep}

Having justified and discussed the spherically symmetric ansatz (\ref{eq:statmetric})-(\ref{eq:phitdep}), let us now turn to the analysis of specific Horndeski\footnote{The models discussed in this section, discovered and studied before the gravitational wave event GW170817, are now ruled out as dark energy candidates. However, we will see in Chapter \ref{ch:wave} that they are very useful to generate solutions in physically viable models.} models with shift-symmetry. In this section, we will impose an additional symmetry to the theory, namely the $\mathbb{Z}_2$ symmetry $\varphi\rightarrow-\varphi$. We saw at the end of Paragraph \ref{sec:consistency} that, when the theory is reflection-symmetric, the no-hair result can be straightforwardly extended to solutions with a linear time-dependence. Therefore, if any non-trivial solutions are to be found, they will have to break one of the other assumptions of the theorem. Only non-analytic Lagrangian densities, or the ones lacking a standard kinetic term, or else asymptotically curved solutions might exhibit deviations from general relativity. The linearly time-dependent ansatz (\ref{eq:phitdep}) was first proposed in \cite{Babichev:2013cya}, and several solutions were given for the following specific model:
\be
S_{\mathbb{Z}_2}=\displaystyle\int{\tx{d}^4x\sqrt{-g}\left[\zeta (R-2\Lambda_\tx{b})-\eta(\pd\varphi)^2 +\beta G_{\mu\nu}\pd^\mu\varphi\, \pd^\nu\varphi\right]},
\label{eq:John}
\ee
where $\zeta=M_\tx{Pl}^2/2$, $\Lambda_\tx{b}$ is a bare cosmological constant, $\eta$ is a constant without mass dimension and $\beta$ is another constant. In the parametrization of Eq.~(\ref{eq:H}) and after integrations by parts, it corresponds to the following choice:
\be
G_2=2(\eta X - \zeta \Lambda_\tx{b}),~~~~G_4=\zeta+\beta X.
\label{eq:GJohn}
\ee
Although this action is an arbitrary choice among all possible models that possess shift and reflection symmetry, it is the simplest one that captures the essential features of the theory we want to investigate. Analytic results were also obtained for arbitrary $G_2$ and $G_4$ functions \cite{Kobayashi:2014eva}, but, in what follows, we will use the model (\ref{eq:John}) for clarity and concision. One can recognize that the term parametrized by $\beta$ is the John Lagrangian density of the Fab Four theory, Eq.~(\ref{eq:johnf4}), with $V_\tx{John}=\beta$. The metric field equations for the model (\ref{eq:John}) read:
\be
\begin{split}
\label{eq:EOMJohn}
0&=(\zeta +\beta X)G_{\mu\nu} -\eta \left[\partial_\mu\varphi \,\partial_\nu\varphi -\dfrac12\,g_{\mu\nu}(\partial\varphi)^2 \right]  +g_{\mu\nu}\,\zeta\Lambda_\tx{b} +\beta \left\{-\dfrac12R\,\pd_\mu\varphi\, \pd_\nu\varphi\right.
\\
&\quad-\Box\varphi\,\nd_\mu\pd_\nu\varphi+\nd_\lambda\pd_\mu\varphi\,\nd^\lambda\pd_\nu\varphi+\dfrac12\left[(\Box\varphi)^2-\nd_\rho\pd_\sigma\varphi\, \nd^\rho\pd^\sigma\varphi\right]g_{\mu\nu}
\\
&\quad\left.+\,2R_{\lambda(\mu}\pd_{\nu)}\varphi\,\pd^\lambda\varphi-g_{\mu\nu}R^{\rho \sigma}\pd_\rho\varphi\,\pd_\sigma\varphi+R_{\mu\rho\nu\sigma}\,\pd^\rho\varphi\,\pd^\sigma\varphi \vphantom{\dfrac12}\right\}
,\end{split}
\ee
while the scalar field equation is
\be
\nabla_\mu J^\mu =0,~~~~ J^\mu = \left(2\eta g^{\mu\nu} -\beta G^{\mu\nu} \right) \partial_\nu\varphi.
\ee
Using the ansatz (\ref{eq:statmetric})-(\ref{eq:phitdep}), these equations boil down to three ordinary differential equations. The $(tr)$ metric equation, according to Eq.~(\ref{eq:linkJrEtr}), takes a particularly simple form:
\be
\varphi'(2\eta g^{rr}-\beta G^{rr})=0.
\ee
One can clearly see in this equation two allowed branches: the general relativity branch with $\varphi'=0$, or a new branch allowed by the higher order structure of the considered theory, in which one has to impose the geometrical condition $2\eta g^{rr}-\beta G^{rr}=0$. The best way to write the field equations is in terms of the kinetic density $X=-(\pd\varphi)^2/2$, as in \cite{Kobayashi:2014eva}. The metric functions $f$ and $h$ are known in terms of $X$, which has to be found as a root of the following cubic equation:
\be
\beta^2 C_0 = (\zeta-\beta X)^2\left[4\beta^2X-2(\eta\zeta-2\eta\beta X+\beta\zeta\Lambda_\tx{b})r^2\right],
\label{eq:algX}
\ee
with $C_0$ a free integration constant for now. In the literature, exact solutions were given only for specific values of $C_0$ \cite{Babichev:2013cya,Kobayashi:2014eva}. The authors of \cite{Charmousis:2015aya} tried to obtain results for small deviations of the $C_0$ parameter with respect to the previously known solutions. Here, in addition to these results, we will find all regular solutions of the model (\ref{eq:John}) when $\Lambda_\tx{b}=0$, with the ansatz (\ref{eq:statmetric})-(\ref{eq:phitdep}). One can sort the solutions in categories that correspond to the following paragraphs\footnote{Additional solutions like Lifshitz spacetimes also exist \cite{Bravo-Gaete:2013dca}, but we do not discuss them because they are not of direct cosmological interest.}.

\subsection{Stealth Schwarzschild black hole}
\label{sec:stealth}

In the case where both $\eta$ and $\Lambda_\tx{b}$ are taken to be zero in (\ref{eq:John}), that is when focusing on a purely quartic Horndeski theory, one finds that the unique solution of the form (\ref{eq:statmetric})-(\ref{eq:phitdep}) is Schwarzschild geometry. However, the scalar field itself does not vanish whenever $q\neq0$. Explicitly, the solution reads
\be
h = f = 1-\dfrac{2m}{r}, ~~~~\varphi(t,r)=qt \pm q\left[2\sqrt{2rm}+2m\ln\left(\dfrac{\sqrt{r}-\sqrt{2m}}{\sqrt{r}+\sqrt{2m}}\right)\right]
\label{eq:stealthJohn}
,\ee
where $m$ is a free integration constant, which represents the mass of the black hole in Planck units. Here, the parameter $q$ is also free. Additionally, the kinetic density $X$ in this case is
\be
X=\dfrac{q^2}{2}.
\label{eq:Xstealth}
\ee 
This solution is asymptotically flat, but does not contradict the no-hair theorem, because the standard kinetic term is absent ($\eta=0$). It constitutes an example of ``stealth'' configuration. Such solutions are defined as non-trivial configurations of a field with vanishing energy-momentum tensor\footnote{In the case of Horndeski theory, one can define this tensor through Eq.~(\ref{eq:defTmunu}) for simplicity.} \cite{Sokolowski:2003rg,AyonBeato:2005tu}. Here, it is clear that Schwarzschild geometry makes Einstein's tensor vanish, so that the energy-momentum tensor of $\varphi$ vanishes as well. Such a stealth field does not contribute to Einstein's equations, and therefore does not gravitate. One could object that such solutions are of academic interest if they predict no deviation with respect to general relativity. However, the energy-momentum tensor of the scalar field does not always vanish. This is true only on shell, for a specific geometry. In particular, as soon as standard matter is present, the geometry deviates from the general relativity solution. Section \ref{sec:starhairtdep} explores this idea more quantitatively, in the case of stars.

\subsection{De Sitter asymptotics and self-tuning}
\label{sec:deSitterJohn}

The most interesting subclass of the solution has de Sitter asymptotics, and allows for self-tuning, that is an effective cosmological constant which does not depend of the bare one, $\Lambda_\tx{b}$ \cite{Babichev:2013cya}. This class is obtained as soon as the constant $C_0$ of Eq.~(\ref{eq:algX}) is non-zero. In this case, one can express $f$ and $h$ in terms of $X$:
\begin{align}
h &= -\dfrac{2m}{r}+\dfrac{1}{r}\displaystyle\int{\tx{d}r\, \dfrac{2q^2}{C_0}(\zeta-\beta X)^2\left(1+\dfrac{\eta}{\beta}r^2\right)},
\label{eq:hJohnX}
\\
f &= \dfrac{C_0}{2q^2(\zeta-\beta X)^2}h,
\end{align}
where again $m$ is free and represents the mass of the black hole. These equations show that, generically,
\be
f \mathop{\sim}_{r \rightarrow \infty} \dfrac{\eta}{3\beta}r^2.
\ee
In other words, the effective cosmological constant is $\Lambda_\tx{eff}=-\eta/\beta$, totally independent of the bare one $\Lambda_\tx{b}$. If one is interested in de Sitter asymptotics, $\eta$ and $\beta$ must have opposite sign. Sensible anti de Sitter solutions are also obtained for the opposite choice. Incidentally, when $m=0$, there is only a cosmological horizon and $f$ tends towards one close to the origin, thus avoiding a conical singularity. Therefore, there exist solitonic solutions in asymptotically de Sitter spacetimes, i.e., solutions that are regular everywhere with a non-trivial scalar field profile. The field equations are fully integrated in the case 
\be
C_0=\dfrac{\zeta^3}{2\beta\eta^3}(\eta-\beta\Lambda_\tx{b})^2(\eta+\beta\Lambda_\tx{b})
\label{eq:CdSexact}
,\ee
although nothing else but integrability singles out this specific value. The solution is then exactly Schwarzschild-de Sitter geometry, while the kinetic density $X$ is constant:
\begin{align}
h &=f=1-\dfrac{2m}{r}+\dfrac{\eta}{3\beta}r^2,
\label{eq:hJohnSdS}
\\
X&=\dfrac{\zeta}{2\eta\beta}(\eta+\beta\Lambda_\tx{b}).
\label{eq:XJohnSdS}
\end{align}
With respect to Eq.~(\ref{eq:hJohnX}), time has been rescaled to absorb the constant factor between $f$ and $h$. This has the effect to fix the value of the velocity parameter $q$:
\be
q^2=\dfrac{\zeta}{\eta\beta}(\eta+\beta\Lambda_\tx{b}).
\label{eq:qdSJohn}
\ee
The exact expression for the scalar field can be deduced from Eq.~(\ref{eq:XJohnSdS}). For a vanishing mass, this solution becomes a de Sitter spacetime, which was first found in a cosmological framework \cite{Gubitosi:2011sg}. 

Let us make two comments about the self-tuning character of this solution. First, one can be puzzled by the fact that the scalar field is able to ``eat up'' as much vacuum energy as required in order to solve the large cosmological constant problem. It seems that, to compensate the immense amount of positive vacuum energy generated by quantum fields, the scalar sector must be able to store an arbitrarily negative energy. This really sounds like a quantum instability of the model. In fact, we will see in Chapter \ref{ch:pert} that the stability of the solution precisely forbids self-tuning of the cosmological constant to arbitrary values. The allowed range for the effective cosmological constant $\Leff$ is restricted to values close to $\Lb$. Therefore, the specific model (\ref{eq:John}) cannot be used to solve the large cosmological constant problem.

Another point one must be careful with is the renormalization of Newton's constant. What we call Newton's constant is the number obtained as a result of, say, Cavendish experiments. In general relativity, this measured number must be equated to the theoretical quantity $1/(8\pi M_\tx{Pl}^2)$. However, this is not the case in Horndeski theory and its extensions \cite{Babichev:2016kdt}. Performing an expansion in the Newtonian limit, one ends up with a theoretical parameter that must be matched with the experimental value of Newton's constant\footnote{It is shown in \cite{Babichev:2016kdt} that the theoretical parameter in question is body-independent, but depends on the cosmological background (especially the cosmological value of the scalar field).}. It happens that setting the parameter in question to the required value generically prevents the self-tuning of the cosmological constant. If the cosmological constant was large originally (as should be from quantum field theory), it is still large after the self-tuning process. In other words, if one tries to use self-tuning to decrease the value of the cosmological constant, one ends up with a Newton's constant that is way too small with respect to observations. This fundamental difficulty can be overcome only when considering beyond Horndeski models \cite{Babichev:2016kdt}. 

\subsubsection*{Full analysis in the case $\Lambda_\tx{b}=0$}

\begin{figure}[ht]
\centering
\includegraphics[width=\textwidth]{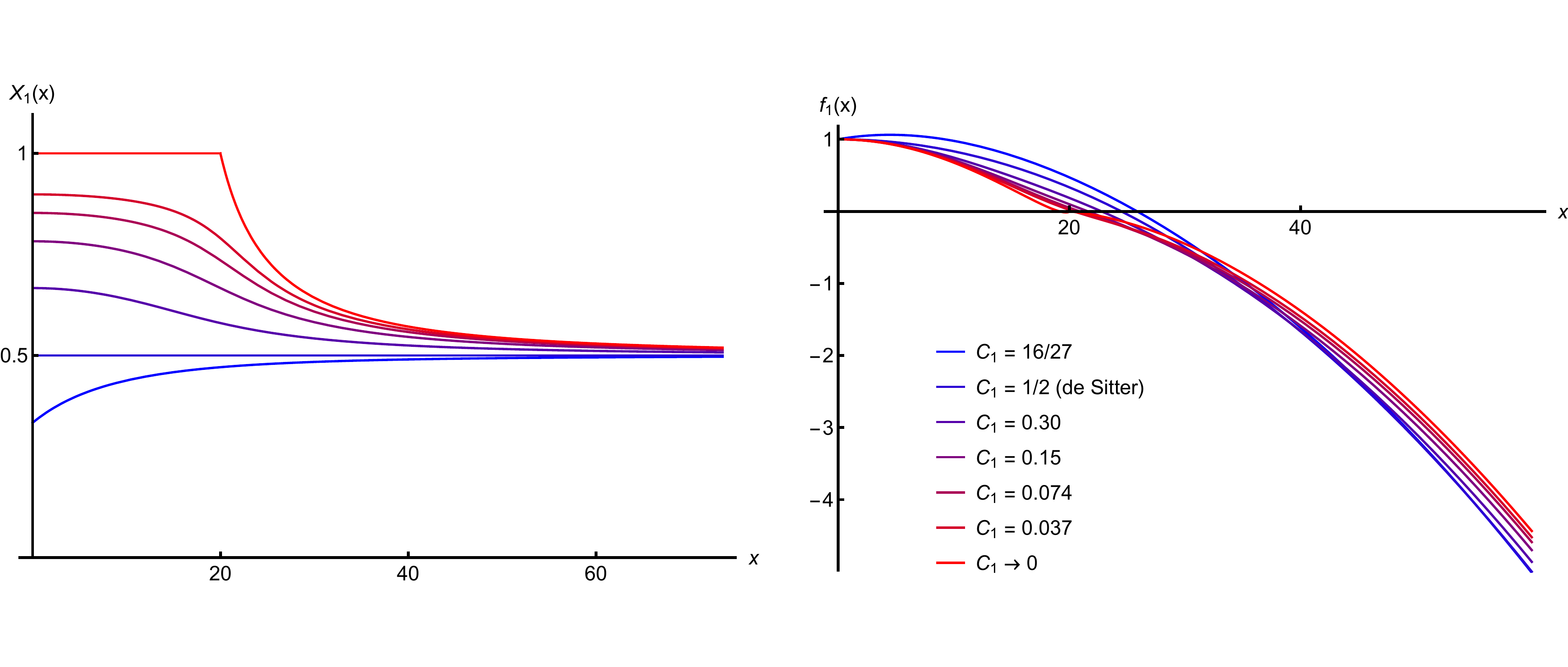}
\caption{Asymptotically de Sitter solution for $\lambda=10^{-2}$. The left panel shows the evolution of the kinetic density with the rescaled radius $x$ for several values of the rescaled integration constant $C_1$. The scalar field is not homogeneous generically, and thus constitutes a soliton. All solutions still have the same asymptotic behavior at spatial infinity. The right panel shows the evolution of the metric function $f$ for the same parameters. The geometry, including the position of the cosmological horizon, is affected by the scalar profile.}
\label{fig:nonconstantX}
\end{figure}

As mentioned above, it is not necessary to fix the $C_0$ parameter to a specific value. In this paragraph, we investigate in full detail the solutions of the cubic equation (\ref{eq:algX}) in the sub-case where $\Lb=0$. We do not derive the full expressions for $h$ and $f$, but they can be obtained straightforwardly by numerical integration of the exact solution for $X$. The solution is most easily written in terms of dimensionless variables. Let us define the following parameters:
\be
C_1=\dfrac{\beta}{\zeta^3}C_0,~~~~Q=\dfrac{\beta}{\zeta}q^2,~~~~\lambda=\dfrac{\eta}{\beta}r_0^2,
\ee
where $r_0$ is some fixed length scale. Note that $\lambda<0$ for de Sitter asymptotics. Let us also redefine the unknown functions according to:
\be
X_1(x)=\dfrac{\beta}{\zeta}X(x\, r_0),~~~~h_1(x)=h(x\,r_0),~~~~f_1(x)=f(x\,r_0).
\label{eq:dlessfunctionsJohn}
\ee
Equation (\ref{eq:algX}) for $X_1$ has three roots, that we will denote as $X_1^{(1)}$, $X_1^{(2)}$ and $X_1^{(3)}$:
\begin{align}
\begin{split}
X_1^{(1)}&=\dfrac{5 \lambda  x^2+4}{6 (\lambda  x^2+1)}+\dfrac{(\lambda x^2+2)^2}{3\cdot2^{2/3} (\lambda  x^2+1)\left[B+\sqrt{B^2-4(\lambda x^2+2)^6}\right]^{1/3}}
\\
&\quad+\dfrac{\left[B+\sqrt{B^2-4(\lambda x^2+2)^6}\right]^{1/3}}{6 \cdot 2^{1/3}(\lambda  x^2+1)},
\end{split}
\\
\begin{split}
X_1^{(2)}&=\dfrac{5 \lambda  x^2+4}{6 (\lambda  x^2+1)}+\dfrac{1+i\sqrt{3}}{2}~\dfrac{(\lambda x^2+2)^2}{3\cdot2^{2/3} (\lambda  x^2+1)\left[B+\sqrt{B^2-4(\lambda x^2+2)^6}\right]^{1/3}}
\\
&\quad-\dfrac{1-i\sqrt{3}}{2}~\dfrac{\left[B+\sqrt{B^2-4(\lambda x^2+2)^6}\right]^{1/3}}{6 \cdot 2^{1/3}(\lambda  x^2+1)},
\end{split}
\\
\begin{split}
X_1^{(3)}&=\dfrac{5 \lambda  x^2+4}{6 (\lambda  x^2+1)}+\dfrac{1-i\sqrt{3}}{2}~\dfrac{(\lambda x^2+2)^2}{3\cdot2^{2/3} (\lambda  x^2+1)\left[B+\sqrt{B^2-4(\lambda x^2+2)^6}\right]^{1/3}}
\\
&\quad-\dfrac{1+i\sqrt{3}}{2}~\dfrac{\left[B+\sqrt{B^2-4(\lambda x^2+2)^6}\right]^{1/3}}{6 \cdot 2^{1/3}(\lambda  x^2+1)},
\end{split}
\end{align}
where
\be
B=-2 \lambda^3 x^6 + 6(9 C_1-2) \lambda^2 x^4+12(9 C_1-2)\lambda  x^2+54 C_1-16.
\ee
An extensive study of the real and imaginary parts of the $X_1^{(i)}$ shows that a real and regular solution for $X_1$ can be found over $0\leq x < \infty$ only when 
\be
0< C_1\leq \dfrac{16}{27}.
\label{eq:C1domain}
\ee
When $C_1$ is in this interval, $X_1=X_1^{(1)}$ for $x<(-1/\lambda)^{1/2}$ and $X_1=X_1^{(3)}$ for $x>(-1/\lambda)^{1/2}$. At spatial infinity, this solution behaves like the exact Schwarzschild-de Sitter solution (\ref{eq:hJohnSdS})-(\ref{eq:XJohnSdS}). Finally, in order for $f$ and $h$ to share the same sign, one must have $Q>0$, that is $\beta>0$. Translated in terms of the original parameters, Eq.~(\ref{eq:C1domain}) reads
\be
0< C_0\leq \dfrac{16\zeta^3}{27\beta}.
\label{eq:C0domain}
\ee
Figure \ref{fig:nonconstantX} shows the various behaviors of $X_1$ in terms of the rescaled radial coordinate $x=r/r_0$. The metric function $f$ is also plotted for a vanishing black hole mass. These solutions can be interpreted as solitons, since the scalar field has a non-homogeneous structure although spacetime is empty and regular.

\subsection{Static universe}

Another class of solutions is obtained when setting $C_0=0$ in Eq.~(\ref{eq:algX}). The solution generically reads:
\begin{align}
X&=\dfrac{\zeta}{\beta},
\label{eq:XEinstein}
\\
h&= -\dfrac{2m}{r}+\dfrac{2\eta}{\eta-2\beta\Lb}-\dfrac{2\sqrt{\beta}(\eta+2\beta\Lb)}{(\eta-2\beta\Lb)^{3/2}\,r}\: \tx{Arctan}\left(\dfrac{r}{2}\sqrt{\dfrac{\eta-2\beta\Lb}{\beta}}\right),
\\
f&= \left(1+\dfrac{\eta-2\beta\Lb}{4\beta}r^2\right)h
,\end{align}
where $m$ is again a mass integration constant, and $q$ was fixed in order to recover $h=f$ at the origin. The solution exists and is regular only provided that $\beta>0$ and $\eta>2\beta\Lb$. When the mass vanishes, the above solution has no horizon and is regular at the origin. At spatial infinity, it corresponds to a static spacetime with negative curvature $K=(2\beta\Lb-\eta)/(4\beta)$, as can be seen from the change of coordinate
\be
\chi=\dfrac{1}{\sqrt{-K}}\,\tx{Argsh}(r\sqrt{-K})
.\ee
Like the asymptotically de Sitter solutions, this solution circumvents the no-hair theorem because it is not asymptotically flat. The particular solution of Ref.~\cite{Babichev:2013cya} was obtained by tuning the parameters of the theory according to $\eta+2\beta\Lb=0$. Except for this specific tuning, the solution again exhibits a solitonic behavior, with a non-homogeneous curvature even when the mass is set to zero.

\subsection{Regularity of the scalar field at the black hole horizon}

Since $\varphi$ itself does not enter the field equations (only $\pd_\mu\varphi$ does), it is not directly measurable. Thus, what really matters is the regularity of the quantities built upon $\pd_\mu\varphi$, such as $X$ or $J^2$. For the solutions described in the above sections, the norm of the current is $J^2= -h (J^t)^2$ since $J^r$ always vanishes. Using again the fact that $J^r=0$, one obtains
\be
J^2=-\,\dfrac{2\beta q^2}{r^3}\left(\dfrac{f}{h}\right)',
\ee
which is generically regular at the horizon. The quantity $X$ is determined through Eq.~(\ref{eq:algX}), where the mass of the black hole $m$ does not appear. There is therefore absolutely no reason for $X$ to be irregular at the black hole horizon. Indeed, for several solutions that we encountered, $X$ is a mere constant ---~see Eqs.~(\ref{eq:XJohnSdS}), (\ref{eq:Xstealth}) or (\ref{eq:XEinstein}). When $X$ depends on the location, we still found that it is regular, as shown in Fig.~\ref{fig:nonconstantX} for instance.

As a by-product of this analysis, the linear time-dependence of $\varphi$ can yield regularity of the scalar field itself close to the black hole horizon \cite{Babichev:2013cya,Kobayashi:2014eva} (a similar result was already noticed earlier in general relativity \cite{Petrich:1988zz} and standard scalar-tensor theories \cite{Jacobson:1999vr}). Let us note $r_\tx{h}$ the position of the horizon where $f$ and $h$ vanish. Then, defining $f'_\tx{h}=f'(r_\tx{h})$ and $h'_\tx{h}=h'(r_\tx{h})$, the regularity of $X$ imposes that
\be
\psi(r) \underset{r\to r_\tx{h}}{\sim} \pm\,\dfrac{q}{\sqrt{h'_\tx{h}f'_\tx{h}}}\,\tx{ln}|r-r_\tx{h}|,
\label{eq:divpsi}
\ee
where $\psi$ is the radial-dependent part of $\varphi$, Eq.~(\ref{eq:phitdep}). This expression clearly diverges when approaching the horizon. However, the logarithmic divergence can precisely be reabsorbed by switching from Schwarzschild to Eddington-Finkelstein coordinates. Let us define the ingoing Eddington-Finkelstein coordinate $v$:
\be
v=t+\displaystyle\int{\dfrac{\tx{d}r}{\sqrt{hf}}}
.\ee
Contrary to $t$, this coordinate does not blow-up when approaching the horizon for an observer who falls into a black hole. In terms of this coordinate, the ansatz (\ref{eq:phitdep}) reads
\be
\varphi(v,r)= qv-q\displaystyle\int{\dfrac{\tx{d}r}{\sqrt{hf}}}+\psi(r).
\ee
The second term in the above equation behaves as 
\be
-q\displaystyle\int{\dfrac{\tx{d}r}{\sqrt{hf}}}\underset{r\to r_\tx{h}}{\sim} -\,\dfrac{q}{\sqrt{h'_\tx{h}f'_\tx{h}}}\,\tx{ln}|r-r_\tx{h}|
\ee
close to the horizon, which exactly cancels the divergence of the plus branch for $\psi$, Eq.~(\ref{eq:divpsi}). Thus, the plus branch for $\varphi$ is regular at the future black hole horizon (while the minus branch is actually regular in the outgoing Eddington-Finkelstein coordinate, i.e., on the past event horizon).

On the other hand, we will see in Chapters \ref{ch:pert} and \ref{ch:wave} that the scalar field associated to the de Sitter solutions of Paragraph \ref{sec:deSitterJohn} becomes homogeneous at spatial infinity only for the minus branch of Eq.~(\ref{eq:divpsi}). Thus, if one selects the cosmologically sensible branch, the scalar field always blows up at the black hole horizon. Again, this is not a real problem as long as the scalar field itself is not observable and all quantities built out of its derivatives remain finite.

\subsection{The case of slow rotation}
\label{sec:slowrotJohn}

As mentioned in Sec.~\ref{sec:slowrot}, the scalar field will not receive any correction from slow rotation at linear order. However, the metric will be modified. This is parametrized by $\omega$, introduced in Eq.~(\ref{eq:HartleThorne}). Reference \cite{Maselli:2015yva} proved that, for the spherically symmetric solutions of the model (\ref{eq:John}) with static metric and linearly time-dependent scalar field, $\omega$ obeys the following equation:
\be
\label{eq:slowrotJohn}
2 (\zeta - \beta X) \left[\omega''+\frac{\omega'}{2} \left(\frac{f'}{f}+\frac{8}{r}-\frac{h'}{h}\right)\right] - 2 \beta X' \omega' =0.
\ee
In \cite{Babichev:2016rlq}, we showed that this equation can be integrated as
\be
\label{eq:slowrotJohn2}
(\zeta- \beta X) \omega'= \frac{C_1}{r^4}\,\dfrac{h}{f},
\ee
with $C_1$ an arbitrary constant. Thus, in general, the equation governing the slow-rotation corrections differs from the general relativistic one. However, many solutions of interest have a constant kinetic density $X$, see for instance Eqs.~(\ref{eq:Xstealth}), (\ref{eq:XJohnSdS}) or (\ref{eq:XEinstein}). The latter solution, Eq.~(\ref{eq:XEinstein}), that describes an Einstein static universe, is even more special, because it has $\zeta= \beta X$; this means that Eq.~(\ref{eq:slowrotJohn}) trivially vanishes. This is a case of strong coupling, where one would have to investigate non-linear terms in order to conclude. In fact, for the Einstein static universe, the lowest order perturbative Lagrangian exactly vanishes, and all perturbations of the solution are strongly coupled, as can be seen from the results of Chapter \ref{ch:pert}. For the stealth Schwarzschild and Schwarzschild-de Sitter solutions though, only $X'$ vanishes. In this case, Eq.~(\ref{eq:slowrotJohn}) boils down to its general relativistic counterpart. When $f=h$, Eq.~(\ref{eq:slowrotJohn2}) shows that $\omega$ exhibits the usual $1/r^3$ decay. One cannot say that this extends the no-hair theorem for shift-symmetric Horndeski theory, because the background solution already differs from general relativity. Thus, when $X$ is constant, one can only say that the way to compute the slow-rotation limit is the same as in general relativity, knowing the non-trivial background solution.

\section{Cubic Lagrangian}
\label{sec:cubictdep} 

In this section, we will investigate the presence of black holes in a model described by the cubic term of Horndeski theory together with a standard quadratic term, based on Ref.~\cite{Babichev:2016fbg}. It is the simplest Horndeski model with higher-order derivatives, and it does not possess reflection symmetry $\varphi\rightarrow-\varphi$. Explicitly, the action we will analyze reads
\be
S_\tx{cubic} = \displaystyle\int{\mathrm{d}^4x \sqrt{-g} \left[\zeta (R -2\Lb)- \eta(\pd \varphi)^2 + \gamma \,\Box \varphi (\pd \varphi)^2  \right]},
\label{eq:DGP}
\ee
where $\zeta$, $\eta$ and $\Lb$ have the same meaning as in Eq.~(\ref{eq:John}), and $\gamma$ is a constant that parametrizes the cubic term. In terms of the Horndeski functions of Eq.~(\ref{eq:H}), this corresponds to
\begin{align}
G_2(X)&=2(\eta X-\zeta\Lb),
\\
G_3(X)&=2\gamma X,
\\
G_4(X)&=\zeta.
\end{align}
The cubic and quadratic term arise in various contexts, e.g., in a certain limit of the Dvali-Gabadadze-Porrati braneworld model \cite{Dvali:2000hr} mentioned in the introduction of this thesis. Explicitly, the action for this model reads:
\be
S_{\rm DGP} = \dfrac{M^3_5}{2} \displaystyle\int_{\cal{M}} \tx{d}^5 x \sqrt{-\Gamma} R_5 
+ \dfrac{M^2_4}{2} \displaystyle\int_{\cal{M}} \tx{d}^4 x \sqrt{-g} R_4 + M^3_5 \displaystyle\int_{\cal{M}} \tx{d}^4 x \sqrt{-g} K,
\label{eq:trueDGP}
\ee
where $\Gamma_{MN}$ is the five-dimensional metric on a manifold $\mc{M}$, $g_{\mu\nu}$ is the induced metric on the boundary $\pd\mc{M}$, $R_5$ and $R_4$ are the corresponding Ricci scalars and $K$ is the Gibbons-Hawking term associated to $g_{\mu\nu}$. $M_5$ and $M_4$ are some defining mass scales, such that $L_\tx{DGP}=M_4^2/M_5^3$ is the scale under which gravity looks four-dimensional. Perturbing $\Gamma_{MN}$ around a Minkowski background, $\Gamma_{MN}=\eta_{MN}+h_{MN}$, and integrating out the fifth dimension, the quadratic terms can be diagonalized into three kinetic terms; one corresponds to the kinetic term of a tensorial degree of freedom, the second one to a vector and the last one to a scalar (all in four dimensions). At this point, however, all degrees of freedom are coupled by cubic and higher order interaction terms. Then, a specific limit, namely
\be
M_4,M_5\rightarrow \infty, ~~~~ \dfrac{M_5^2}{M_4}=\tx{constant},
\ee
allows one to decouple the different degrees of freedom. The scalar sector is then described by a kinetic term and a cubic interaction term \cite{Luty:2003vm,Nicolis:2004qq}, i.e., the resulting action is the scalar part of action (\ref{eq:DGP}), with coefficients $\eta$ and $\gamma$ determined in terms of the mass scales of Eq.~(\ref{eq:trueDGP}). It can be shown that this action allows to account for an accelerated expansion of spacetime \cite{Deffayet:2000uy,Deffayet:2001pu,Nicolis:2004qq} (at least as long as de Sitter metric can be viewed as a small deviation from Minkowski spacetime). On the other hand, the Dvali-Gabadadze-Poratti self-accelerating branch was shown to be plagued by a ghost instability \cite{Koyama:2005tx,Charmousis:2006pn}. However, there is a priori nothing wrong in working with the action (\ref{eq:DGP}) from the start.

When $\Lb=0$, the action (\ref{eq:DGP}) is known as the cubic Galileon, a sub-case of the so called kinetic gravity braiding model. The latter has been studied extensively in the cosmological context \cite{Deffayet:2010qz} as a dark energy candidate. There exists a self-accelerating branch free from ghost instability, with however vanishing sound speed \cite{Deffayet:2010qz,Appleby:2011aa}. It was further shown in the context of local Solar System physics that the presence of a massive body generates stricly positive corrections to the sound speed \cite{Babichev:2012re}, at least inside the Vainshtein radius. Therefore, in this regime, the Cauchy problem is well posed for the cubic Galileon. Additionally, the model (\ref{eq:DGP}) passes the very important test of multi-messenger astronomy. Indeed, arbitrary $G_2$ and $G_3$ Horndeski functions do not affect the propagation of gravitational waves (this is true in an arbitrary background, as shown in \cite{Babichev:2012re,Bettoni:2016mij}). Therefore, the model (\ref{eq:DGP}) predicts an equal speed for electromagnetic and gravitational waves\footnote{The cubic Galileon model is however in clear conflict with integrated Sachs-Wolfe effect data \cite{Renk:2017rzu}. This tension might be alleviated by the presence of $\Lb$ in (\ref{eq:DGP}), or by the consideration of more involved $G_3(X)$ functions.}. The metric field equations read:
\be
\begin{split}
0&=\zeta (G_{\mu \nu} +\Lb \, g_{\mu \nu}) +\eta \left[\dfrac{1}{2}\, g_{\mu \nu} (\partial \varphi)^2 - \partial_\mu \varphi \, \partial_\nu \varphi \right]
\\
 &\quad-\gamma \left\{-\, \Box \varphi \, \partial_\mu \varphi \, \partial_\nu \varphi + \partial_{(\mu} \varphi \, \partial_{\nu)} \left[(\partial \varphi)^2\right] \vphantom{\left[(\partial \varphi)^2\right]} 
-\dfrac{1}{2}\, g_{\mu \nu} \partial^\rho \varphi \, \partial_\rho \left[(\partial \varphi)^2\right] \right\}.
\label{eq:EOM1}
\end{split}
\ee
The Noether current associated with the shift symmetry of the action is
\be
J^\mu =2 \partial_\nu \varphi \left[g^{\mu\nu}(\gamma \,\Box \varphi-\eta) -\gamma \nabla^\mu \pd^\nu \varphi \right],~~~~\nd_\mu J^\mu=0.
\label{eq:currentDGP}
\ee
For numerical integration purposes, we write the scalar field ansatz in a slightly different (but equivalent) way with respect to (\ref{eq:phitdep}), still using the ansatz (\ref{eq:statmetric}) for the metric:
\be
\varphi(t,r) = q t + \displaystyle\int{\mathrm{d}r \, \dfrac{\chi(r)}{h(r)}}.
\label{eq:ansatzDGP}
\ee
With this ansatz, the scalar and metric field equations reduce to a system of three ordinary differential equations (as explained in Paragraph \ref{sec:consistency}). They read:
\begin{align}
\gamma q\, (r^4 h)' \dfrac{f}{h} \chi^2 - \gamma q^3 r^4 h' - 2 \eta q\, r^4 h \chi &= 0,
\label{eq:trDGP}
\\
\eta r^2 \left(\dfrac{f}{h} \chi^2 - q^2 \right) + 2 \zeta r f h' + 2\zeta h (f-1 + \Lb r^2)  &= 0,
\label{eq:mix1DGP}
\\
\left(\dfrac{f}{h} \chi^2 - q^2 \right) \left[ \eta r^2 \sqrt{\dfrac{h}{f}} - \gamma \left(r^2 \sqrt{\dfrac{f}{h}} \chi \right)' \right] &= 2 \zeta r h^2 \left(\sqrt{\dfrac{f}{h}} \right)'.
\label{eq:mix2DGP}
\end{align}
Equation (\ref{eq:trDGP}) is the $(t r)$ metric equation (or, equivalently, $J^r=0$).  Equations (\ref{eq:mix1DGP}) and (\ref{eq:mix2DGP}) are combinations of the $(t r)$, $(t t)$ and $(r r)$ metric equations. 
Again, these field equations have a trivial solution, identical to general relativity with a bare cosmological constant. However, thanks to their higher order structure, they also possess other branches of solution, that we will study in detail in the following paragraphs. Paragraph \ref{sec:cosmoDGP} is devoted to exact cosmological solutions. We will see that several branches exist according to the parameters of the theory, and discuss the various regimes. In Paragraph \ref{sec:analyticDGP}, we explore analytic properties of black hole solutions. Finally, we integrate the field equations numerically in Paragraph \ref{sec:numericsDGP}, and subsequently analyze the solutions. 

\subsection{Cosmological solutions}
\label{sec:cosmoDGP}

\begin{figure}[ht]
\begin{center}
\includegraphics[width=10cm]{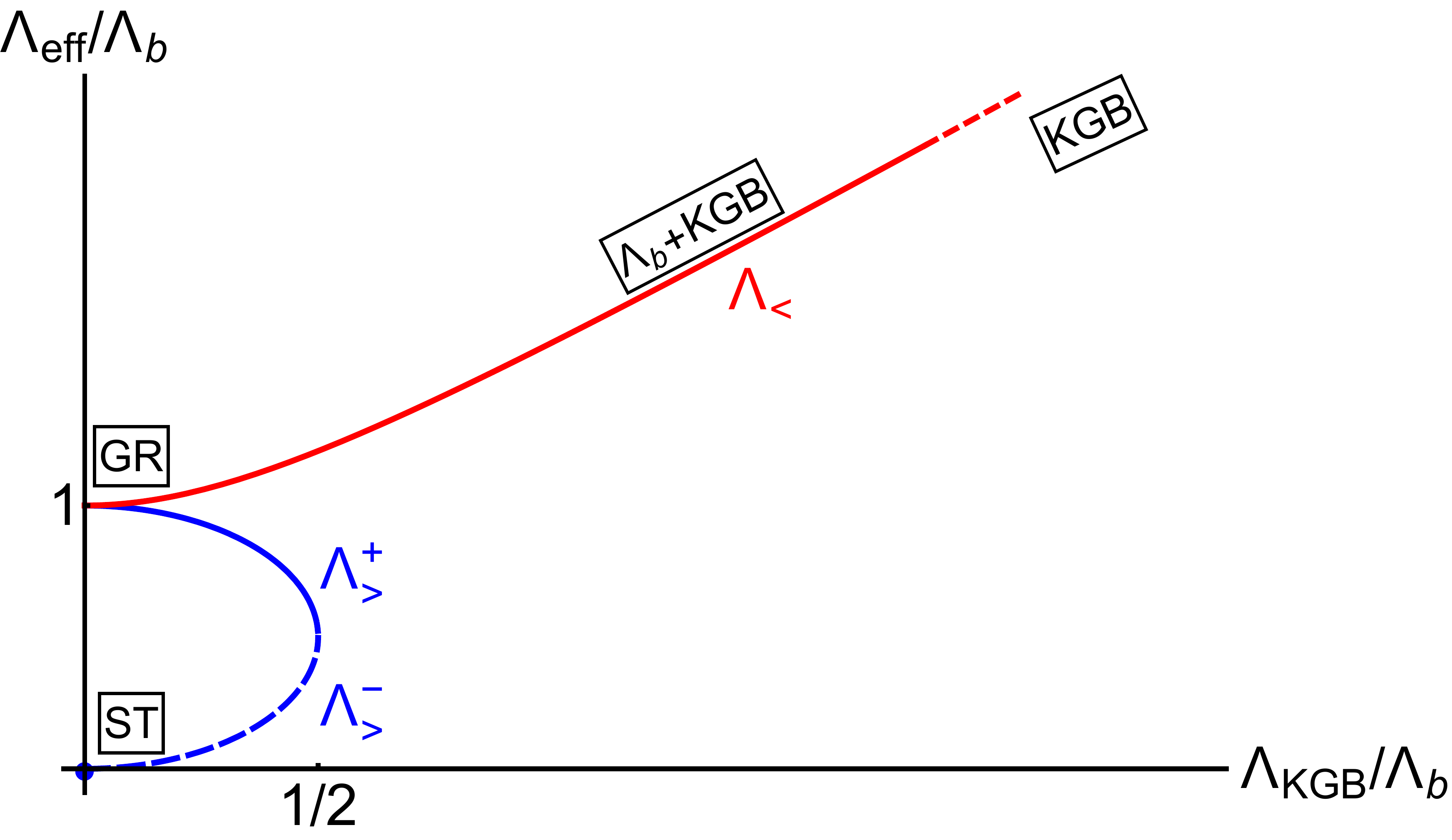}
\caption{Effective cosmological constant for the cubic model (\ref{eq:DGP}). When $\Lambda_\mathrm{eff} \simeq \Lambda_\tx{b}$, the solution becomes identical to general relativity with a bare cosmological constant $\Lb$. The other end of the red branch corresponds to a self-accelerated spacetime, approaching the kinetic gravity braiding solution. The dashed blue branch represents solutions where the bare cosmological constant can be screened, $\Lambda_\mathrm{eff} \ll \Lb$.}
\label{fig:lambdaDGP}
\end{center}
\end{figure}

Let us first consider homogeneous cosmological solutions for the model (\ref{eq:DGP}). These homogeneous solutions describe the far away asymptotics for the black hole solutions we will search for with numerical integration later on. This analysis is complementary to \cite{Babichev:2012re}, which studied the model (\ref{eq:DGP}) with $\Lb=0$ in the cosmological context. The solutions are de Sitter spacetimes:
\begin{align}
f(r) &= h(r)= 1 - \dfrac{\Lambda_\mathrm{eff}}{3} r^2,
\label{eq:cosmoDGP}
\\
\chi(r) &=\dfrac{\eta r}{3 \gamma},
\end{align}
where $\Lambda_\mathrm{eff}$ is to be determined from Eqs.~(\ref{eq:trDGP})--(\ref{eq:mix2DGP}). Note also that the velocity parameter $q$ of Eq.~(\ref{eq:ansatzDGP}) is not free; it is fixed to some particular value $q_0$, given in Eq.~(\ref{eq:qDGP}) below. One can see the homogeneity of the previous solution by using the change of coordinates (\ref{eq:changecoord1})-(\ref{eq:changecoord2}). Then the metric (\ref{eq:cosmoDGP}) reads
\be
\mathrm{d}s^2 = - \,\mathrm{d}\tau^2 + \text{e}^{2 H \tau} (\mathrm{d}\rho^2 + \rho^2 \mathrm{d}\Omega^2),
\ee
with a Hubble rate $H$ such that $H^2=\Leff/3$. It is clear that the scalar field is homogenous when, in the Fridemann-Lemaître-Robertson-Walker coordinates, $\varphi(\tau,\rho) = q_0\tau$. Equations (\ref{eq:trDGP})--(\ref{eq:mix2DGP}) then show that $q_0$ must be a solution of the equation
\be
\dfrac{\eta^2}{3 \gamma^2 q_0^2} = \dfrac{2 \zeta \Lb - \eta q_0^2}{2 \zeta}.
\ee
There are two possible solutions for $q_0$ ($q$ appears only as $q^2$ in the equations of motion, so its overall sign is irrelevant):
\be
q_0^\pm \equiv \left[\dfrac{\zeta \Lb}{\eta} \pm \sqrt{\left(\dfrac{\zeta \Lb}{\eta} \right)^2 - \dfrac{2 \eta \zeta}{3 \gamma^2}}\; \right]^{1/2},
\label{eq:qDGP}
\ee
corresponding to two possible effective cosmological constants:
\be
\Lambda_\mathrm{eff} = \dfrac{\eta^2}{3 \gamma^2 (q_0^\pm)^2} = \dfrac{2 \zeta\Lb - \eta (q_0^\pm)^2}{2 \zeta}
\label{eq:LeffDGP}
.\ee
For $\Lb = 0$, $\eta$ has to be negative and
\begin{equation}
\Lambda_\mathrm{eff} = \left(\dfrac{|\eta|^3}{6 \zeta \gamma^2}\right)^{1/2} \equiv \Lambda_\mathrm{KGB},
\end{equation}
reducing to the kinetic gravity braiding case \cite{Babichev:2012re}, hence the subscript KGB. To summarize, depending on the parameters of the model, the cosmological solutions are given by Eqs.~(\ref{eq:cosmoDGP})--(\ref{eq:qDGP}), with the following values of the effective cosmological constant $\Lambda_\mathrm{eff}$, depending on the parameters of the model: 
\be
\label{eq:casesLeffDGP}
\Lambda_\mathrm{eff}=
\begin{cases}
\Lambda_< = \dfrac{1}{2} \left(\Lb + \sqrt{\Lb^2+4\Lambda_\mathrm{KGB}^2}\right) \ \text{if} \ \eta <0,
\\
\\
\Lambda_>^+ = \dfrac{1}{2} \left(\Lb + \sqrt{\Lb^2-4\Lambda_\mathrm{KGB}^2}\right) \ \text{if} \ \eta >0 \ \text{and} \ \Lb > 2 \Lambda_\mathrm{KGB},
\\
\\
\Lambda_>^- = \dfrac{1}{2} \left(\Lb - \sqrt{\Lb^2-4\Lambda_\mathrm{KGB}^2}\right) \ \text{if} \ \eta >0 \ \text{and} \ \Lb > 2 \Lambda_\mathrm{KGB}.   
\end{cases}
\ee
The different branches are represented in Fig.~\ref{fig:lambdaDGP}. The general relativity solution is restored for $\Lambda_\mathrm{eff}= \Lb$, in which case the presence of the cubic Galileon does not affect the metric. This limit can be attained either via $\Lambda_{>}^+$ or via $\Lambda_{<}$. The latter branch attains, at the other end, the kinetic gravity braiding limit as $\Lb\rightarrow 0$, thus corresponding to self-accelerating solutions \cite{Deffayet:2010qz} with $\Lambda_\mathrm{eff} = \Lambda_\mathrm{KGB}$. Tuning of a large bare cosmological constant to a small effective one is possible in the lower branch $\Lambda_{>}^-$, where $\Lambda_\mathrm{eff} < \Lb$. For this branch the scalar field partially screens the bare value $\Lb$, yielding a cosmological constant of lesser magnitude.

\subsection{Analytic approximations and asymptotics}
\label{sec:analyticDGP}

No exact black hole solution is known for the field equations (\ref{eq:trDGP})--(\ref{eq:mix2DGP}). Therefore, we will mostly resort to numerical integration of these equations. It is possible, however,  to get some analytic insights about the solutions in different limits. 

\subsubsection{Test field limit}
\label{sec:test}

Before solving the full system of equations (\ref{eq:trDGP})--(\ref{eq:mix2DGP}) in the case of black holes, it is instructive to look into a particular limit, where the scalar field does not back-react onto the metric (the ``test field'' approximation). Formally, this approximation can be obtained by setting $\eta = \epsilon \eta_0$ and $\gamma = \epsilon \gamma_0$, and then letting $\epsilon \rightarrow 0$. It is then easy to note that the metric is determined solely by the Einstein-Hilbert part of the action. The Schwarzschild-de Sitter metric with cosmological constant $\Lb$ solves the two equations, Eqs.~(\ref{eq:mix1DGP})-(\ref{eq:mix2DGP}) in this limit. The third equation (\ref{eq:trDGP}), the $(t r)$ metric equation, decouples from the two first ones and gives the scalar field equation on a fixed background metric. Explicitly, one has
\begin{equation}
f(r) = h(r) = 1- \dfrac{2m}{r} - \dfrac{\Lb}{3} r^2,
\end{equation}
with free $m$. Plugging the above expression into the $(t r)$ component of the metric equation yields
\begin{equation}
\chi = \dfrac{\eta}{\gamma}\, \dfrac{r h \pm \sqrt{\Delta}}{(4 h+ r h')},
\label{eq:chitest}
\end{equation}
where 
\be
\begin{split}
\Delta (r)  &=  4m^2+r^4 \left(\frac{4 \gamma ^2 \Lb ^2 q^2}{3 \eta ^2}-\frac{2 \Lb }{3}\right)+r^2 \left(1-\frac{8 \gamma ^2 \Lb  q^2}{3 \eta ^2}\right)
\\
 &\quad - \frac{12 \gamma ^2 m^2 q^2}{\eta ^2 r^2}+\frac{8 \gamma ^2 m q^2}{\eta ^2 r}+ \frac{\Lb ^2 r^6}{9} +\frac{4 \Lb m\, r^3}{3}-4m\,  r .
\end{split}
\ee
Depending on the parameters of the Lagrangian, $\Delta (r)$ may become negative for some range of $r$, rendering the scalar field imaginary. One can check however that, in the case of physical interest, $1/\sqrt{\Lb} \gg m$, $\Delta$ can be positive everywhere outside the horizon by requiring that
\begin{equation}
\left(\dfrac{\gamma q}{\eta} \right)^2 < \dfrac{1}{3 \Lb}
.\end{equation}
The scalar field becomes imaginary for $r \lesssim 3m/2$, i.e., in the interior region of the horizon, so that the exterior solution is well-behaved everywhere in this regime. At this point, we would like to draw a parallel with the results of \cite{Babichev:2010kj}. In this reference, a process of scalar field accretion onto a static spherically symmetric black hole was studied, where the back-reaction of the scalar field on the black hole was neglected ---~exactly the situation considered in this paragraph, the test approximation. The key difference of the accreting solution in \cite{Babichev:2010kj} is an integration constant, which vanishes for the solution presented here. As discussed above, the solution in the test field approximation follows from the $(tr)$ Einstein equation (\ref{eq:trDGP}), which is equivalent to the equation $J^r=0$. This last equation can be obtained from the scalar field equation (\ref{eq:currentDGP}) by integrating along the radial coordinate and setting to zero the integration constant. In the case of accretion, this integration constant (the primary hair of the black hole) is not set to zero; instead, it is chosen in such a way that the solution for the scalar field describes a so called transonic flow, so that it is smooth and free of singularities (at least for radii larger than the radius of the scalar sound horizon). The above considerations illustrate the fact that, when back-reaction is taken into account, other solutions ---~with non-zero flux~--- may exist, as already mentioned in Paragraph \ref{sec:soundness}. Again, the ansatz (\ref{eq:phitdep}) is not unique, but rather corresponds to a special case of zero scalar flow. 

\subsubsection{Asymptotic behavior at small and large \texorpdfstring{$r$}{}}

Solving the system of equations (\ref{eq:trDGP})--(\ref{eq:mix2DGP}) near the origin, $r\to 0$, one finds the following asymptotic behavior:
\begin{align}
\label{as1}
h(r) &\underset{r \rightarrow 0}{=}- b \,r^{-4} + c\, r^{-8/3}+ o(r^{-8/3}),
\\
f(r) &\underset{r \rightarrow 0}{=} - \dfrac{1}{3} + a\, r^{4/3} + o(r^{4/3}),
\\
\chi(r) &\underset{r \rightarrow 0}{=} d \, r^{-13/3}+ o(r^{-13/3}),
\label{as3}
\end{align}
where $a, b, c$ and $d$ depend on the parameters of the theory and are fixed by the field equations (their exact expressions are not interesting for us here). Note that unlike general relativity black holes, the $f(r)$ component of the metric is finite at the origin. Furthermore, one can actually show analytically that the behavior of the solutions near the black hole singularity depends only on the radial part of the scalar field and not on the time dependent part. Indeed, imposing a static $(q=0)$ scalar field, and further setting $\eta=0$, one can find an exact solution for all $r$ which has the same behavior as (\ref{as1})--(\ref{as3}) in the $r\rightarrow 0$ region. Therefore, we can also conclude that, for $r\to 0$, the leading order behavior of the solution is determined by the higher-order Galileon term  $\Box \varphi\, (\partial \varphi)^2$, rather than by $(\partial \varphi)^2$ or the $\Lb$ term. This is expected as, close to the singularity, the higher order Dvali-Gabadadze-Porrati term contains in total more derivatives than the $\eta$ and $\Lb$ terms. The numerical integration presented below confirms the behavior (\ref{as1})--(\ref{as3}), see in particular  Figs.~\ref{fig:pureDGP} and \ref{fig:bhDGP}.

Let us now look for the large $r$ asymptotic behavior of the solution to Eqs.~(\ref{eq:trDGP})--(\ref{eq:mix2DGP}). We assume that, at spatial infinity, the solution has the following power expansion in $1/r$:
\begin{equation}
h(r) = \sum\limits_{n=-2}^\infty \dfrac{c^{(n)}_h}{r^n}, ~~~~
f(r) = \sum\limits_{n=-2}^\infty \dfrac{c^{(n)}_f}{r^n}, ~~~~
\chi(r) = \sum\limits_{n=-1}^\infty \dfrac{c^{(n)}_\chi}{r^n}.
\end{equation}
One can always rescale time so that $c^{(-2)}_h = c^{(-2)}_f$, in order for the speed of light to be asymptotically equal to unity. Then, the asymptotic expansion reads
\begin{align}
\label{hfcfar1}
h(r) &\underset{r \rightarrow \infty}{=} - \dfrac{\Lambda_\mathrm{eff}}{3} r^2 + 1 + \mathcal{O} \left(\dfrac{1}{r}\right),
\\
f(r) &\underset{r \rightarrow \infty}{=} - \dfrac{\Lambda_\mathrm{eff}}{3} r^2 + c^{(0)}_f + \mathcal{O} \left(\dfrac{1}{r}\right),
\\
\chi(r) &\underset{r \rightarrow \infty}{=} \dfrac{\eta r}{3 \gamma} + \dfrac{c^{(-1)}_\chi}{r} + \mathcal{O} \left(\dfrac{1}{r^2}\right),
\label{hfcfar3}
\end{align}
where $c^{(0)}_f$ and $c^{(-1)}_\chi$ are particular functions of the Lagrangian parameters (we do not give their exact expression here, since they are cumbersome). It is important to stress that the metric in the expansion (\ref{hfcfar1})--(\ref{hfcfar3}) asymptotically approaches the metric of the homogeneous cosmological solution, since $\Lambda_\mathrm{eff}$ in (\ref{hfcfar1})--(\ref{hfcfar3}) is given by (\ref{eq:LeffDGP}).   
Note that in this expansion, the velocity parameter $q$ remains arbitrary; it may not coincide with $q_0$, which is fixed by the cosmological solution. The question then arises whether the asymptotic solution (\ref{hfcfar1})--(\ref{hfcfar3}) is homogeneous, since in the time-dependent part of the scalar field enters an arbitrary velocity $q$, which does not necessarily match the cosmological solution. To check the homogeneity of the scalar field, one can find explicitly the solution for $\varphi$ by integration of (\ref{hfcfar1})--(\ref{hfcfar3}):
\begin{equation}
\varphi(t,r) \underset{r \rightarrow \infty}{=} q t - \dfrac{\eta}{\Lambda_\mathrm{eff} \gamma} \: \text{ln} \left(\sqrt{\dfrac{\Lambda_\mathrm{eff}}{3}} r \right) 
+ \mathcal{O}\left(\dfrac{1}{r} \right),
\end{equation}
and then by the change of coordinates (\ref{eq:changecoord1})-(\ref{eq:changecoord2}), one has:
\begin{equation}
\varphi(\tau , \rho) \underset{\rho \rightarrow \infty}{=} q_0\: \tau + (q_0-q) \sqrt{\dfrac{3}{\Lambda_\mathrm{eff}}} \: \text{ln} \left(\sqrt{\dfrac{\Lambda_\mathrm{eff}}{3}} \rho \right) + \mathcal{O}\left(\dfrac{1}{\rho} \right)
\label{eq:corrqq0}
.\end{equation}
Although $\varphi(\tau, \rho)$ appears to be inhomogeneous when $\rho\to\infty$, one should keep in mind that the value of $\varphi$ itself is not a physical observable, because of the shift symmetry of the problem. Only derivatives of $\varphi$ enter equations of motion. One can easily conclude from (\ref{eq:corrqq0}) that $\pd_\rho\varphi\sim \rho^{-1}$ when $\rho\to\infty$, which becomes negligible with respect to $\pd_\tau\varphi\sim q_0$. $\pd_\mu\varphi$ behaves like the associated cosmological solution from Paragraph \ref{sec:cosmoDGP}, i.e., the one with $q = q_0$. 
In the case $q=q_0$, the previous expansion gets simplified as follows:
\begin{align}
h(r) &\underset{r \rightarrow \infty}{=} 1 - \dfrac{2m}{r} - \dfrac{\Lambda_\mathrm{eff}}{3}r^2   + \mathcal{O} \left(\dfrac{1}{r^6}\right),
\\
f(r) &\underset{r \rightarrow \infty}{=} 1- \dfrac{2m}{r}  - \dfrac{\Lambda_\mathrm{eff}}{3} r^2  + \mathcal{O} \left(\dfrac{1}{r^4}\right),
\\
\chi(r) &\underset{r \rightarrow \infty}{=} \dfrac{\eta r}{3 \gamma} + \dfrac{\gamma q_0^2 m}{\eta r^2} + \mathcal{O} \left(\dfrac{1}{r^5}\right),
\end{align}
where $q_0$ and $\Lambda_\mathrm{eff}$ are given correspondingly in (\ref{eq:qDGP}) and (\ref{eq:LeffDGP}), and 
$m$ is a free constant. Here, we see effectively the important role played by the time dependent part of the scalar field which determines the asymptotic behavior of the black hole solution as well as the modified value of the effective cosmological constant. 
The asymptotic solution for $\varphi$ in Friedmann coordinates reads, in this case,
\begin{equation}
\varphi(\tau , \rho) \underset{\rho \rightarrow \infty}{=} q_0 \tau + \mathcal{O} \left( \dfrac{1}{\rho^3} \right).
\label{eq:corrbh}
\end{equation}
Note the much faster decay of the inhomogeneous part, $\rho^{-3}$, in the case $q=q_0$ with respect to the case $q\neq q_0$, Eq.~(\ref{eq:corrqq0}).

\subsection{Numerical integration}
\label{sec:numericsDGP}

In this section, we will perform the numerical integration of the system of ordinary differential equations (\ref{eq:trDGP})--(\ref{eq:mix2DGP}), which is a consequence of the equations of motion of the model (\ref{eq:DGP}) with the ansatz (\ref{eq:ansatzDGP}). It is convenient to introduce dimensionless quantities in order to integrate numerically this system of equations. The theory (\ref{eq:DGP}) contains four dimensionful parameters $\zeta, \eta, \gamma, \Lb$. Besides, the ansatz for the scalar field introduces an extra dimensionful quantity $q$. Thus, in total, there are five dimensionful parameters, which have to be combined in a number of dimensionless quantities. Let us define first the dimensionless radius $x = r/r_0$ with $r_0$ some fixed length scale (which we later choose to be the horizon radius). Then, let us define three dimensionless constants as combinations of the parameters of the Lagrangian, the velocity $q$, and the length scale $r_0$ as follows:
\begin{equation}\label{eq:dlessconst}
\alpha_1 = - \dfrac{\gamma q}{r_0 \eta},~~~~ \alpha_2 = - \dfrac{\eta q^2 r_0^2}{\zeta},~~~~ \alpha_3 = \Lb r_0^2.
\end{equation}
Let us also redefine the functions as in Eq.~(\ref{eq:dlessfunctionsJohn}):
\begin{equation}\label{eq:dlessfunctionsDGP}
h_\tx{n} (x)= h(x\,r_0),~~~~ f_\tx{n} (x)= f(x\,r_0),~~~~ \chi_\tx{n}(x) = \dfrac{\chi(x\,r_0)}{q}.
\end{equation}
The above redefinition implies that the scalar field is measured in units of $q$. The equations of motion (\ref{eq:trDGP})--(\ref{eq:mix2DGP}) can then be rewritten in terms of the dimensionless quantities:
\begin{align}
\label{eq:tralpha}
\alpha_1 (x^4 h_\tx{n})' \dfrac{f_\tx{n}}{h_\tx{n}}\chi_\tx{n}^2 + 2x^4 h_\tx{n} \chi_\tx{n} - \alpha_1x^4 h_\tx{n}' &= 0,
\\
\label{eq:mix1alpha}
\alpha_2 x^2 \left(1 - \dfrac{f_\tx{n}}{h_\tx{n}} \chi_\tx{n}^2 \right) + 2 x f_\tx{n} h_\tx{n}' + 2h_\tx{n} (-1+f_\tx{n}+ \alpha_3 x^2) &=0,
\\
\left(1 - \dfrac{f_\tx{n}}{h_\tx{n}} \chi_\tx{n}^2 \right) \left[\alpha_2 x^2 \sqrt{\dfrac{h_\tx{n}}{f_\tx{n}}} + \alpha_1 \alpha_2 \left(x^2 \sqrt{\dfrac{f_\tx{n}}{h_\tx{n}}} \chi_\tx{n}\right)' \right] &= 2 x h^2 \left(\sqrt{\dfrac{f_\tx{n}}{h_\tx{n}}} \right)',
\label{eq:mix2alpha}
\end{align}
where a prime throughout this section denotes a derivative with respect to the dimensionless radius $x$. Henceforth, we choose the length scale $r_0$ to be the radius of horizon, i.e., in terms of $x$ the black hole horizon is at $x=1$. Note that the first two equations (\ref{eq:tralpha}) and (\ref{eq:mix1alpha}) are algebraic equations on $f_\tx{n}$ and $\chi_\tx{n}$. Thus, one can resolve Eqs.~(\ref{eq:tralpha}) and (\ref{eq:mix1alpha}) to find $f_\tx{n}$ and $\chi_\tx{n}$ in terms of $h_\tx{n}$ and $h'_\tx{n}$. By substituting the obtained expressions in the third equation of the system, Eq.~(\ref{eq:mix2alpha}), one arrives at a second order ordinary differential equation on $h_\tx{n}$. To find the unique solution, two boundary conditions should be supplemented. Let us impose one boundary condition at the black hole horizon: we will require that the radial function $h_\tx{n}$ vanishes at $x = 1$ (which can be simply thought as a definition of the black hole horizon). As a second boundary condition, we specify (arbitrarily) the derivative of $h_\tx{n}$ at the point $x =1$, $h'_\tx{n}|_1$. By integration from $x=1$, we then select the value of $h'_\tx{n}|_1$ such that the solution has the desired cosmological behavior at large $x$. In other words, we use the numerical shooting method. 

\subsubsection{The case \texorpdfstring{$\eta=0$, $\Lb=0$}{}}
\label{sec:eta0Lambda0}

\begin{figure}[ht]
\begin{center}
\includegraphics[width=10cm]{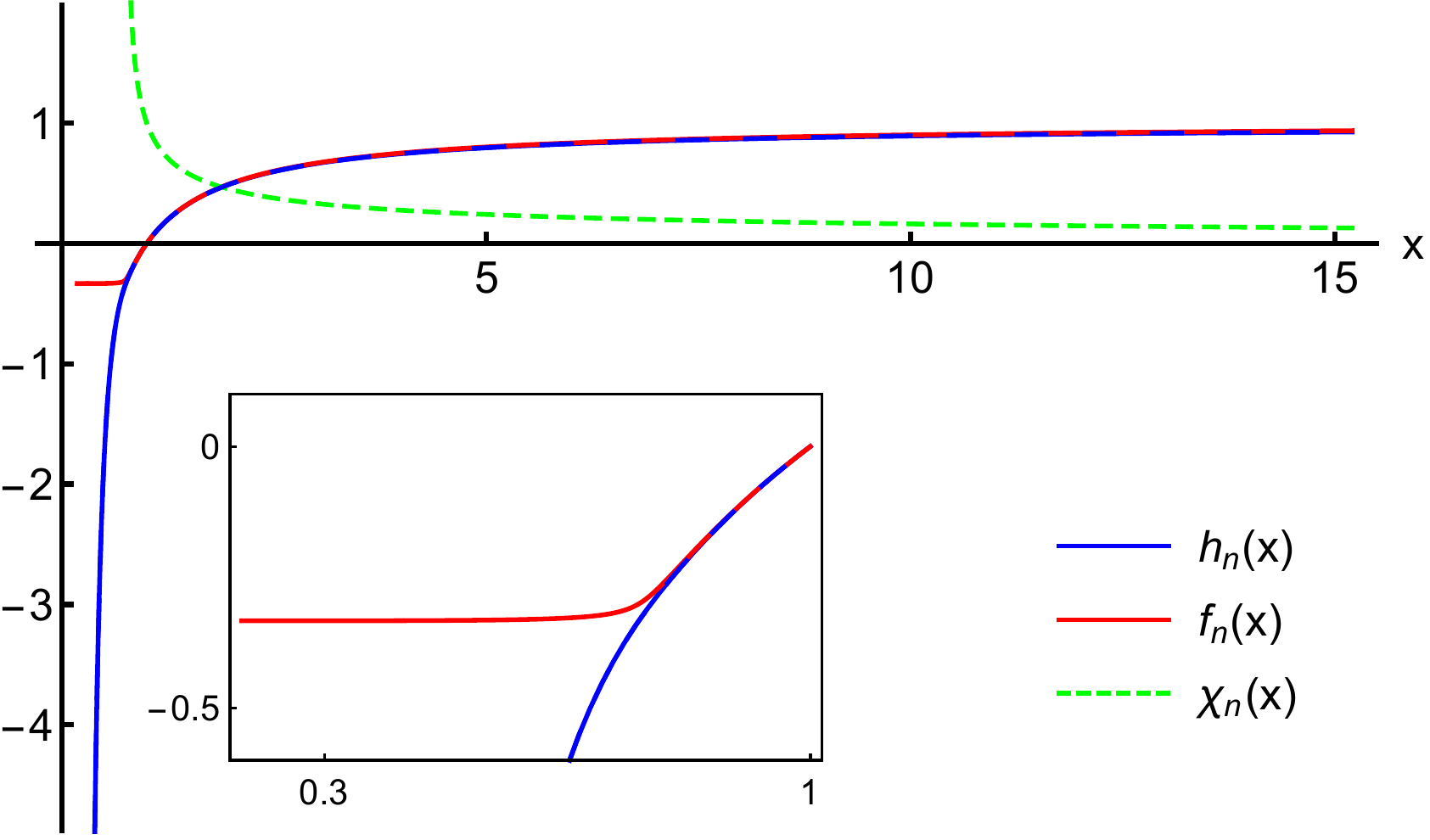}
\caption{Asymptotically flat black hole in the $\eta=0$, $\Lb=0$ case. For this solution,  $\alpha_1 \cdot \alpha_2 = 10^{-3}$, and the solution stops at $x=0.18$ for a numerical precision of 14 digits. The zoomed plot shows the black hole region with more details; in particular, there is no cusp in $f_\tx{n}$.}
\label{fig:pureDGP}
\end{center}
\end{figure}

First, let us consider the case of vanishing $\eta$ and $\Lb$. The action (\ref{eq:DGP}) contains only the Einstein-Hilbert and the cubic terms in this case. The only relevant dimensionless parameter is $\alpha_1 \cdot \alpha_2$. In the absence of a black hole, the corresponding cosmological solution is Minkowski spacetime, represented by the blue dot at the origin in Fig.~\ref{fig:lambdaDGP}. Solving numerically the system of equations (\ref{eq:tralpha})--(\ref{eq:mix2alpha}), one gets asymptotically flat black holes, as shown in Fig.~\ref{fig:pureDGP}.
For general boundary condition of the equations, $f_\tx{n}$ and $h_\tx{n}$ approach different constants at infinity. However, they can be matched by adjusting the numerical value of the derivative of $h_\tx{n}$ at the location of the event horizon, $h_\tx{n}'|_1$. The numerical solutions are always well-behaved in the direction of increasing $x$ (note that the numerical integration is performed from the event horizon). However, when the numerical precision is increased, the numerical integration cannot be continued below some radius inside the horizon, because the numerical code breaks down there. It should be stressed that this is a generic feature of all the simulations we carried out (i.e., including the ones described in the paragraph below), and not specific to the $\eta=0$, $\Lb=0$ case. We could not conclude on the origin of this numerical singularity: it can be either a numerical artifact or a physical pathology at that point. However, the presence of a physical singularity inside the horizon in the test field limit, as shown in Paragraph \ref{sec:test}, suggests that the breakdown of the numerical code may indeed indicate a singular behavior of the solutions, rather than a numerical glitch.

\subsubsection{Generic case}

\begin{figure}[!ht]
\centering
   \begin{subfigure}[b]{0.75\textwidth}
   \includegraphics[width=1\linewidth]{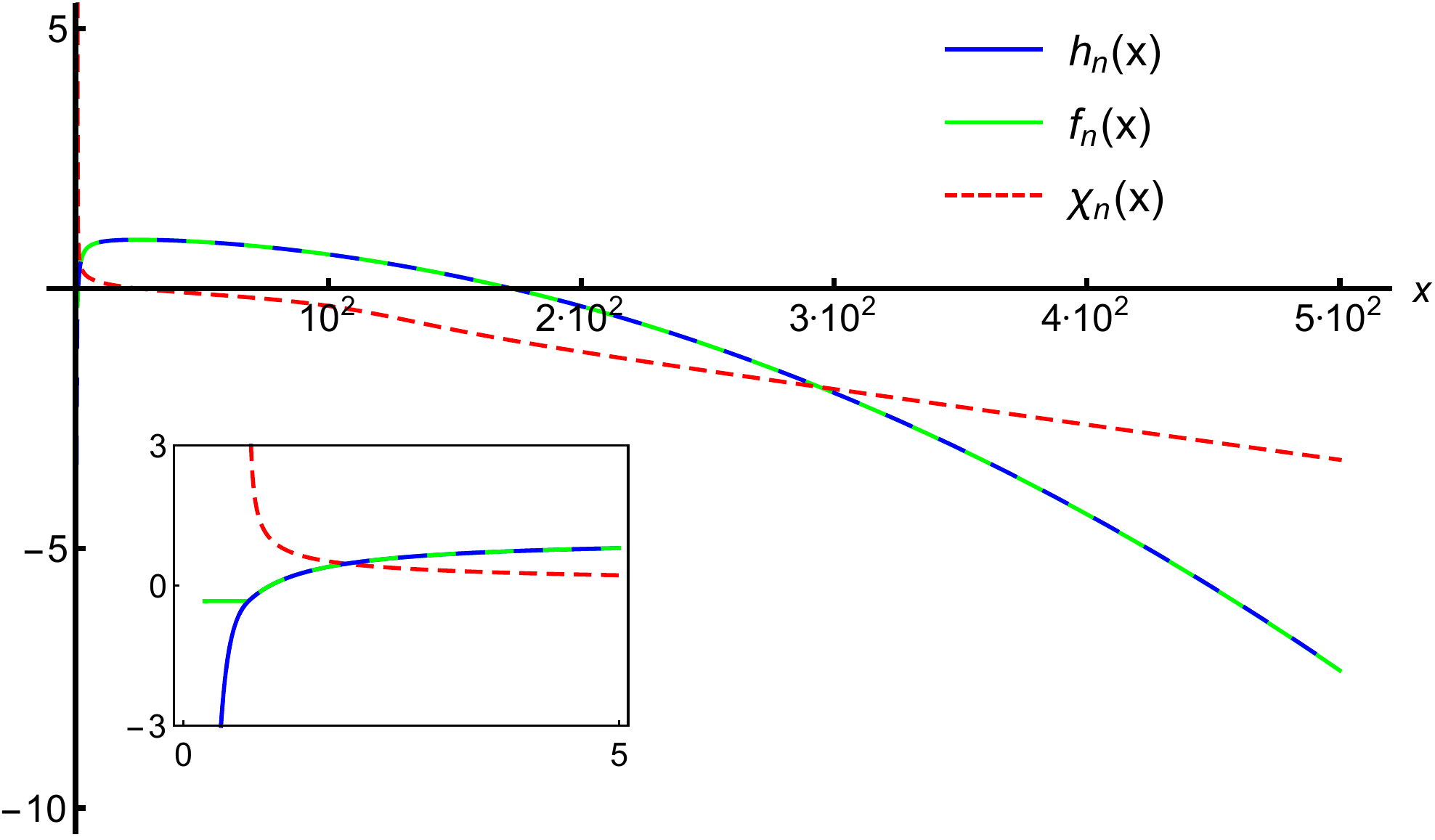}
   \caption{$\eta<0$}
   \label{fig:bhDGPetaneg} 
\end{subfigure}
\begin{subfigure}[b]{0.75\textwidth}
   \includegraphics[width=1\linewidth]{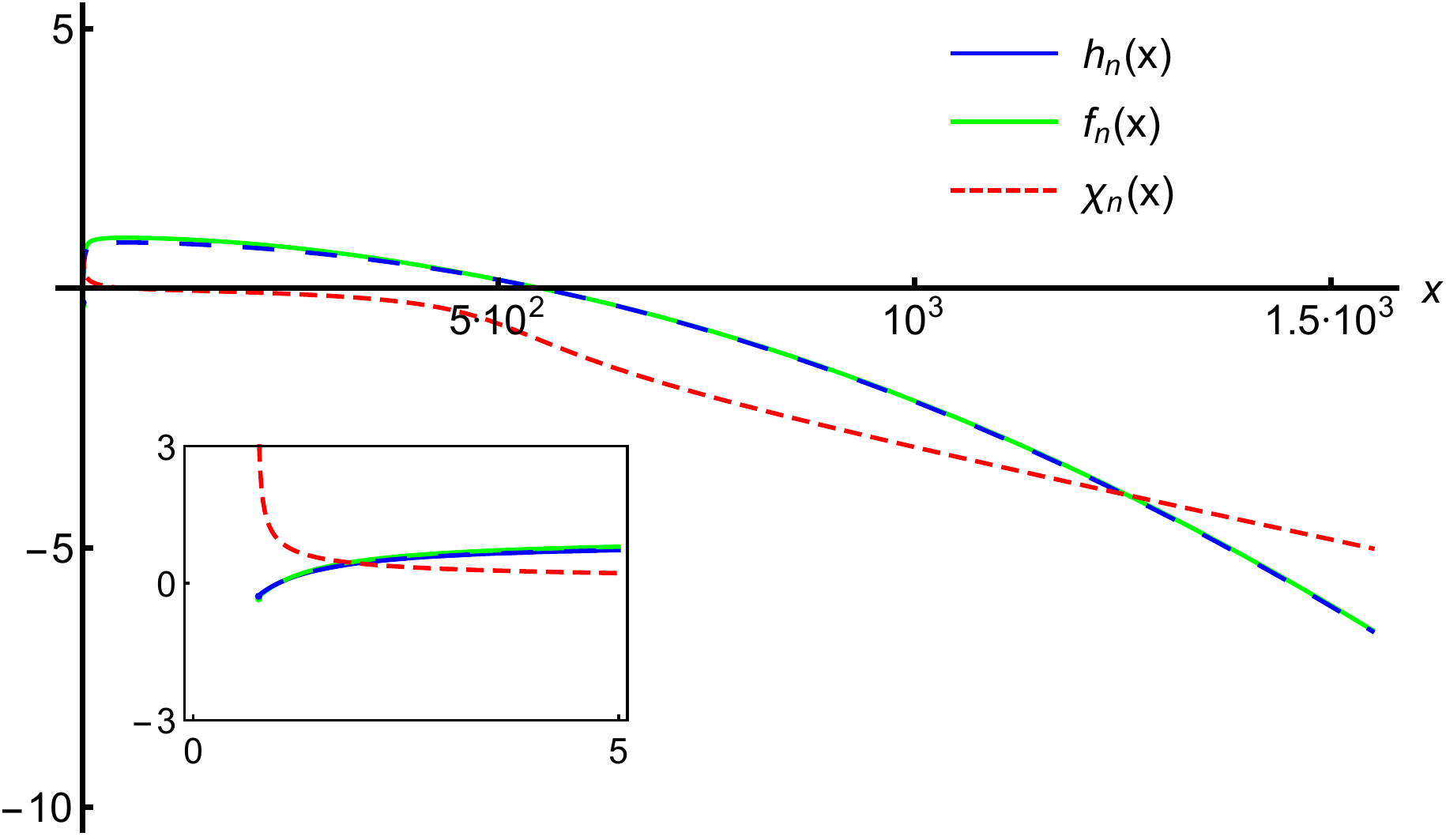}
   \caption{$\eta>0$}
   \label{fig:bhDGPetapos}
\end{subfigure}
\caption{(a) Typical black hole in a de Sitter universe for the action (\ref{eq:DGP}). The parameters of this solution are $\alpha_1 = 50$, $\alpha_2 = 2.5 \cdot 10^{-7}$ and $\alpha_3 = 10^{-4}$. For this choice of parameters, $\eta<0$, the velocity is $q \simeq 0.87\, q_0$ and the bare cosmological constant $\Lb$ is about 25 times greater than the kinetic gravity braiding one, $\Lambda_\mathrm{KGB}$. This solution is therefore in the $\Lambda_<$ branch, close to the general relativistic regime (see Fig.~\ref{fig:lambdanum}). The framed plot shows a zoom on the black hole region. (b) Another solution sitting in the $\Lambda_>^+$ branch, with $\alpha_1 = 10^2$, $\alpha_2 = -3 \cdot 10^{-7}$ and $\alpha_3 = 10^{-5}$; $q \simeq 0.53 \,q_0$ and $\Lb \simeq 5 \,\Lambda_\mathrm{KGB}$.}
\label{fig:bhDGP}
\end{figure}

In this subsection, we will consider general non-zero values of $\eta$ and $\Lb$. The typical behavior of such solutions is presented in Fig.~\ref{fig:bhDGP}. In contrast to the case $\eta=0$, $\Lb=0$, the asymptotic solutions are no longer flat. One expects de Sitter asymptotics, according to the study of Paragraph \ref{sec:cosmoDGP}.

Let us comment at this point on some details of the numerical solutions presented here. We use the shooting method, starting from the location of the black hole horizon, i.e., $h_\tx{n}|_1=0$ in the rescaled quantities. The value of the derivative $h'_\tx{n}|_1$ is not, however, fixed by the condition at the event horizon. Whenever we find some numerical solution, $f_\tx{n}$ and $h_\tx{n}$ always behave like $x^2$ at large $x$, but with $f_\tx{n}/h_\tx{n} \neq 1$ asymptotically, in general. We therefore use the freedom of choosing  $h'_\tx{n}|_1$ so that $f_\tx{n}\sim h_\tx{n}\sim -C_1x^2$ as $x\to \infty$, where $C_1$ is some constant related to  the effective cosmological constant. It is, however, only possible to do for some range of $q$ (assuming the parameters of the Lagrangian are fixed) such that $q$ does not deviate too much from the value $q_0$. In this case, there is a unique choice of $h'_\tx{n}|_1$ so that $f_\tx{n}$ and $h_\tx{n}$ coincide at large $x$. On the contrary, for the values of $q$ that are far from $q_0$ it is impossible to do so, whichever boundary conditions are chosen. In what follows, we will focus on the solutions for which $f_\tx{n}/h_\tx{n}=1$ asymptotically at large $x$, and we will discard other solutions. These numerical solutions have a de Sitter-like asymptotic behavior,
\begin{align}
\label{numass1}
h_\tx{n}(x) &\underset{x\rightarrow \infty}{\sim} f_\tx{n}(x) \underset{x \rightarrow \infty}{\sim} - C_1x^2,
\\
\chi_\tx{n}(x) &\underset{x \rightarrow \infty}{\sim} - C_2 x,
\label{numass2}
\end{align}
with some positive constants $C_1$, $C_2$, c.f. Eq.~(\ref{hfcfar1})--(\ref{hfcfar3}). In addition, we checked that the norm of the derivative of the scalar, $(\partial \varphi)^2$, approaches a constant value at infinity, a further consistency check with the analytic cosmological solution (\ref{eq:cosmoDGP}). 

Depending on the choice of parameters and therefore on the particular branch, one may expect that all the black hole solutions fall in 
one of the three families (\ref{eq:casesLeffDGP}), with a corresponding asymptotic value of $\Lambda_\mathrm{eff}$. It was indeed possible to find black hole solutions for both positive and negative $\eta$. 
For positive $\eta$, however, which gives two cosmological branches, $\Lambda_>^\pm$, we only found numerical solutions which approach one of the branches, the $\Lambda_>^+$ one.

\begin{figure}[ht]
\begin{center}
\includegraphics[width=12cm]{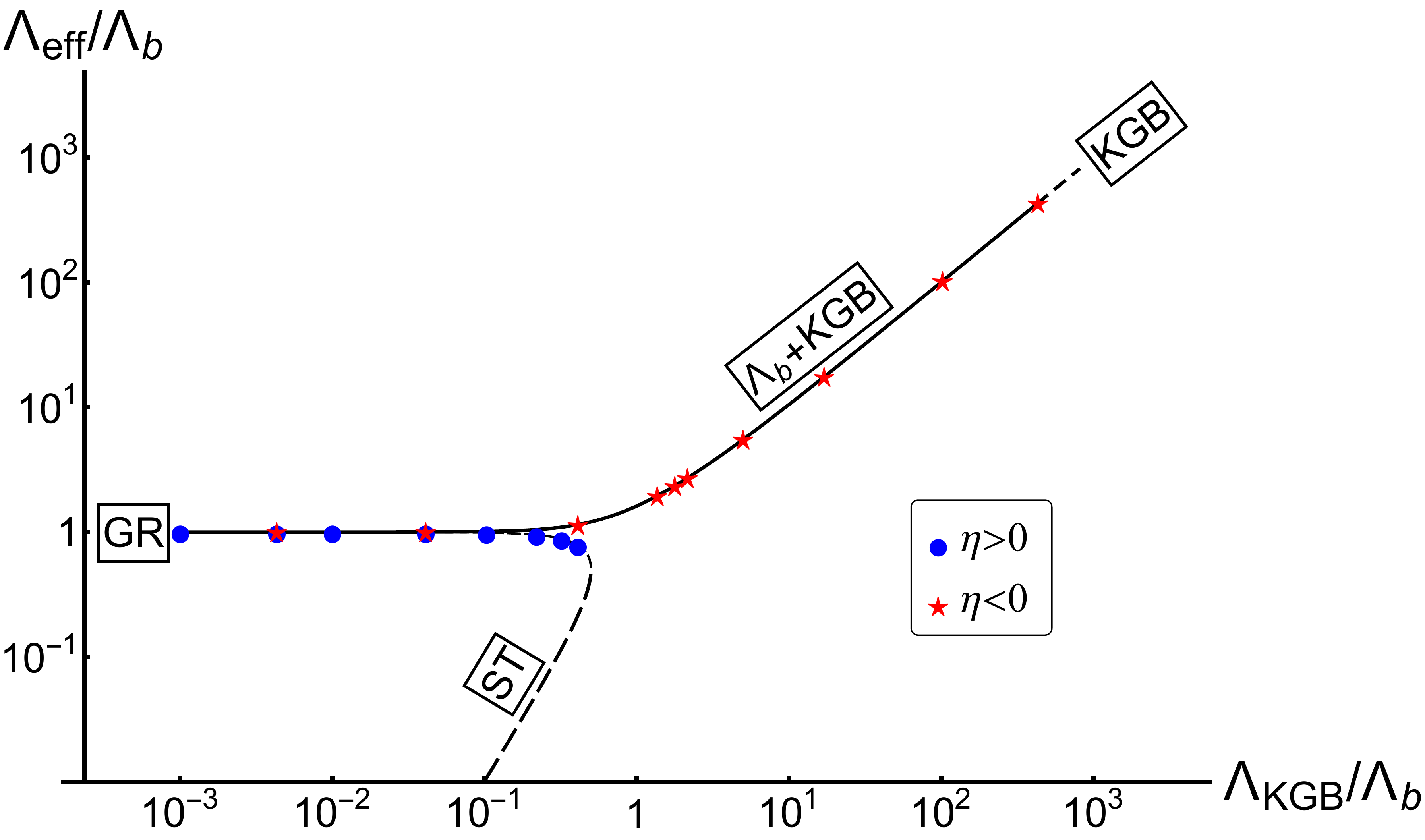}
\caption{Comparison between the far away metric of black hole solutions and their associated cosmological solutions. The black solid line is the value of $\Lambda_\mathrm{eff}/\Lb$ expected from cosmology as a function of $\Lambda_\mathrm{KGB}$. The blue dots (red stars) represent the values of the same quantity obtained from numerical simulations with $\eta>0$ ($\eta<0$). There is a perfect agreement between the numerical results and the theoretical prediction.}
\label{fig:lambdanum}
\end{center}
\end{figure}

For a set of parameters $\zeta$, $\eta$, $\Lb$, $\gamma$, the cosmological solution is given by (\ref{eq:cosmoDGP}) with $\Lambda_\mathrm{eff}$ along one of the branches of Eq.~(\ref{eq:casesLeffDGP}) and $q_0$ given by (\ref{eq:qDGP}). Remarkably, all numerical solutions for a fixed set $\zeta$, $\eta$, $\Lb$, $\gamma$ asymptotically approach the $\Lambda_<$ cosmological solution for $\eta<0$ and the $\Lambda_>^+$ cosmological solution for $\eta>0$. This means that the constants $C_1$, $C_2$ in Eqs.~(\ref{numass1})-(\ref{numass2}) are respectively $\Lambda_\mathrm{eff}/3$ and $\eta/(3 \gamma)$. In Fig.~\ref{fig:lambdanum}, we show the (normalized) cosmological constant which is read off the numerical solutions, versus the analytical results for the homogeneous cosmological solutions.

\begin{figure}[!ht]
\begin{center}
\includegraphics[width=12cm]{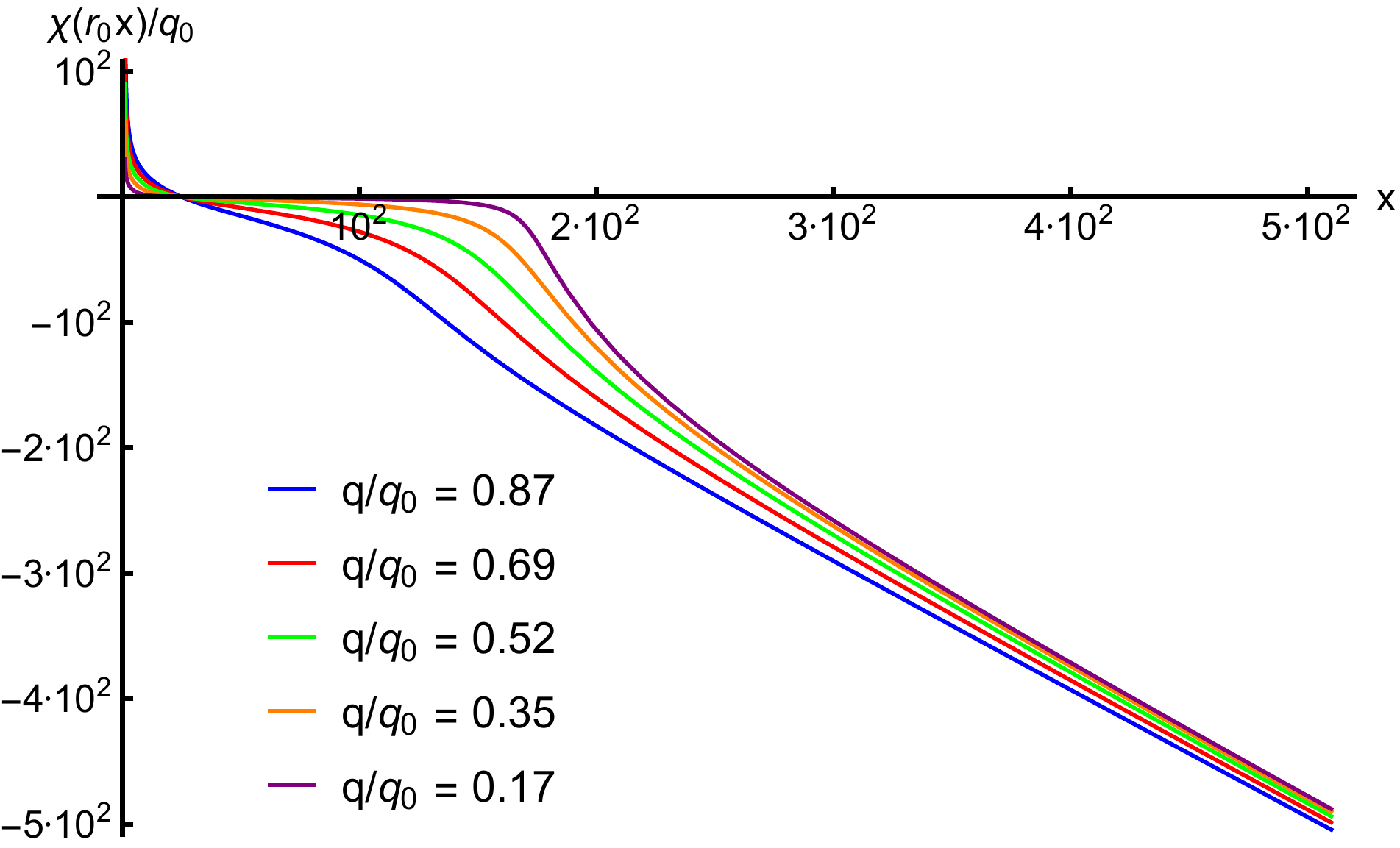}
\caption{The scalar field function $\chi$ for different values of the velocity $q$. The parameters of the Lagrangian are kept constant (they are the same as in Fig.~\ref{fig:bhDGPetaneg}). Here, we plotted $\chi/q_0=\chi_\tx{n}\,q/q_0$ rather than $\chi_\tx{n}$, so that for all solutions, the scalar field is measured in units of $q_0$. The solutions all behave identically far away from the black hole.}
\label{fig:primhair}
\end{center}
\end{figure}

It is important to stress here that the numerical values $C_1$, $C_2$ do not depend on a particular value of $q$, which is a free parameter entering the scalar field ansatz and eventually the definition of the dimensionless parameters $\alpha_i$ via Eq.~(\ref{eq:dlessconst}). The value of $q$ determines the details of the black hole solutions, but not the faraway behavior, as expected from 
the discussion in Sec.~\ref{sec:analyticDGP}. Therefore, as shown in Fig.~\ref{fig:primhair}, there exists a whole family of solutions parametrized by $q$ for a given set of parameters in the Lagrangian and, more importantly, for a given black hole mass. Indeed, in Fig.~\ref{fig:primhair} for instance, the (black hole) horizon location is kept fixed. The velocity parameter $q$ thus has the characteristics of primary hair.

It is worth mentioning that from the numerical solutions described above, with $f\sim h$ at large $r$, 
one can construct physically equivalent solutions with $f \neq h$ at infinity.
Indeed, changing the time parametrization according to $t'= t/\sqrt{C}$ yields:
\begin{align}
\varphi(t',r) &= q \sqrt{C} t' + \displaystyle\int{\mathrm{d}r \, \dfrac{\chi(r)}{h(r)}},
\\
\mathrm{d}s^2 &= - C h(r) \,\mathrm{d}t'^2 + \dfrac{\mathrm{d}r^2}{f(r)} + r^2 \mathrm{d}\Omega^2.
\end{align}
instead of (\ref{eq:ansatzDGP}). Defining $\tilde{h} = C h$ and $\tilde{\chi} = C \chi$, one gets back the old ansatz (\ref{eq:ansatzDGP}) with $q \to q \sqrt{C}$. Note that if the solution in the coordinates $(t,r)$is such that $f\sim h$ at large $r$, the same solution in the coordinates $(t',r)$ has the asymptotic behavior 
$\tilde{h} \sim C f$.
In terms of dimensionless parameters, this corresponds to replacing $(\alpha_1, \: \alpha_2)$ by $(\sqrt{C} \alpha_1, \: C \alpha_2)$. 
It is clear, however, that all these solutions with arbitrary $C$ are physically equivalent.

\section{Star solutions}
\label{sec:starhairtdep}

So far, we have been focusing on black hole solutions. It is natural to ask what happens if a smooth matter source is placed
instead of a black hole, so that the solution now describes a star configuration. Most of the ways to circumvent the black hole no-hair theorem, listed in Table \ref{tab:hair}, may be extended to the case of star solutions. 

On the other hand, it is known that screening mechanisms generically operate in Horndeski theory (in particular Vainshtein's mechanism \cite{Vainshtein:1972sx,Babichev:2013usa}). When such a mechanism operates, deviations with respect to general relativity exist, but they remain very small in dense environments. In the case of Vainshtein's mechanism, considering a massive and spherical body, there exists a region where the scalar field is heavily suppressed and the general relativistic solution is restored. This region is delimited by the Vainshtein radius $r_\tx{V}$, function of the mass of the object and the coupling constants parametrizing the theory. The suppression of the scalar background is achieved either through non-linear self-interactions of the scalar degree of freedom, or interactions between the spin-2 and scalar degrees of freedom. This screening mechanism is believed to operate generically in Horndeski theory \cite{Narikawa:2013pjr,Koyama:2013paa,Kase:2013uja}. 
Therefore, one could legitimately expect that star solutions are very close to general relativity, even if the scalar field is not exactly trivial. 

\begin{figure}[ht]
\begin{center}
\includegraphics[width=\textwidth]{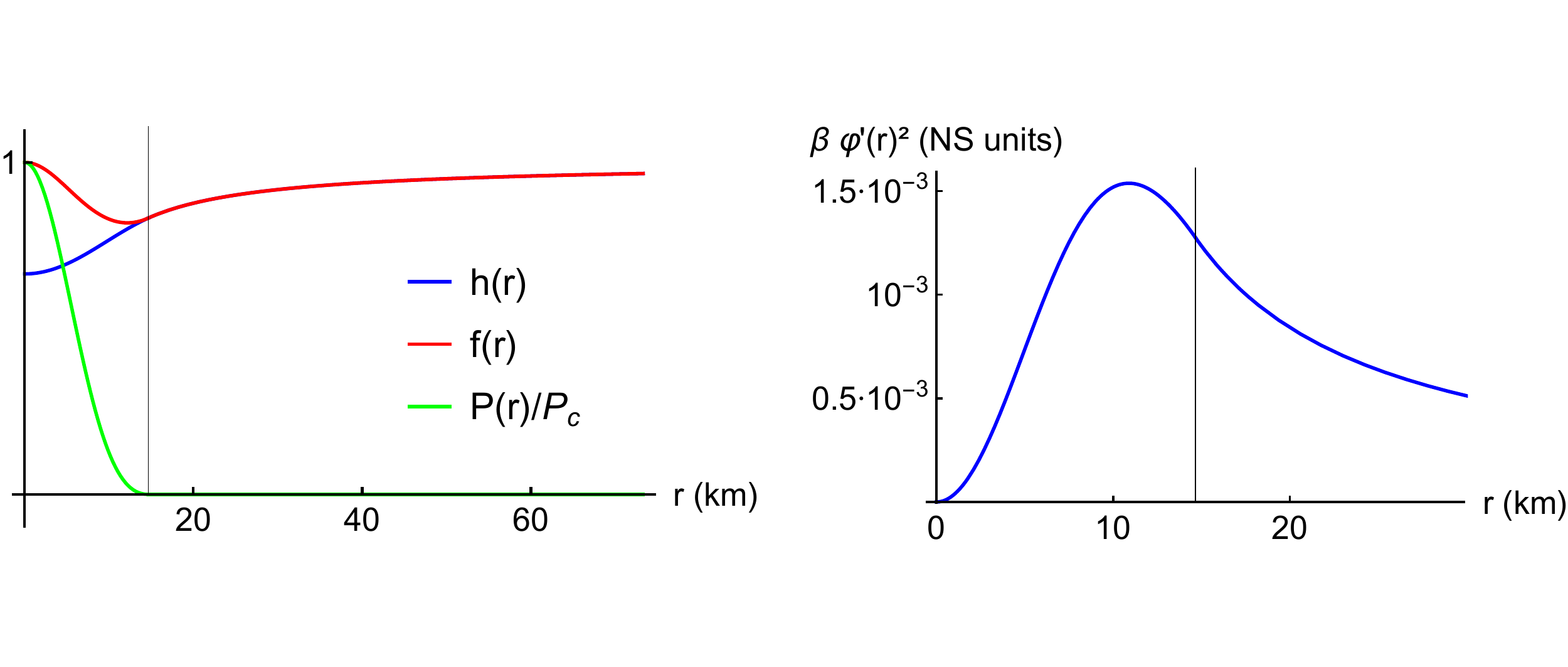}
\caption{Star solution for the model (\ref{eq:Johnstar}). The left panel shows the metric functions, and the pressure normalized to its central value $P_\tx{c}$. The right panel shows the non-trivial behavior of the scalar field through the quantity $\beta\varphi'^{\,2}$, in the $c=1$, $G=1$, $M_\odot=1$ units. In both panels, the vertical black line corresponds to the surface of the star. The central density is $\rho_\tx{B;c}=3\cdot10^{17}$~kg/m$^3$, and the parameter $\beta q^2$ is chosen to be $5.3\cdot10^{-3}$ in the units used here (one can actually determine exact bounds on this quantity). This star has gravitational mass $m=0.83\; M_\odot$ and radius $R_\ast=15$~km. With the same equation of state and central density, one gets $m=1.1~M_\odot$ and $R_\ast=18$~km in general relativity.}
\label{fig:profileJohnq}
\end{center}
\end{figure}

However, Vainshtein's mechanism relies on some assumptions. In particular, if some terms in the action (\ref{eq:H}) are absent, the argument may break down. It is notably the case if the linear coupling of the scalar to Einstein's tensor dominates \cite{Kase:2013uja}. The simplest example that exhibits this behavior is the following action:
\be
S_\tx{J}=\displaystyle\int{\tx{d}^4x \sqrt{-g}(\zeta R+\beta G^{\mu\nu}\pd_\mu\varphi\,\pd_\nu\varphi)}+S_\tx{m},
\label{eq:Johnstar}
\ee
where $S_\tx{m}$ is the action that describes ordinary matter. This corresponds to the action $S_{\mathbb{Z}_2}$, Eq.~(\ref{eq:John}), with $\eta=0$ and $\Lb=0$. Star solutions for this model were first studied in \cite{Cisterna:2015yla}. We saw in Paragraph \ref{sec:stealth} that the model (\ref{eq:Johnstar}) possesses a stealth Schwarzschild solution, Eq.~(\ref{eq:stealthJohn}). The idea is to match this exact exterior solution with a regular star solution. To find the solution inside the star, one must first specify which type of matter it is made of. The simplest model is a perfect fluid, with the following energy-momentum tensor\footnote{A non-minimal coupling of matter to the metric might actually prove interesting in light of the gravitational wave event GW170817, see Part \ref{part:3} for more details.}:
\be
T^{(\tx{m})}_{\mu\nu}=(\rho+P)u_\mu u_\nu+P g_{\mu\nu},
\ee
where $\rho$ and $P$ are respectively the energy density and the pressure of the perfect fluid, and $u^\mu$ is its normalized four velocity. We will use a static and spherically symmetric metric, Eq.~(\ref{eq:statmetric}). The pressure and energy density are also assumed to depend on $r$. At hydrostatic equilibrium, only the time component of $u^\mu$ is non-vanishing. On the other hand, we will keep the linear time dependence of the scalar field, Eq.~(\ref{eq:phitdep}). There are thus five unknown functions: $f$, $h$, $\varphi$, $P$ and $\rho$. As in the black hole case, the first three equations are the $(tt)$ and $(rr)$ metric equations, and the condition $J^r=0$. Another equation is obtained by requiring that the energy-momentum tensor of matter is conserved:
\be
\nd_\mu T^{(\tx{m})\,\mu\nu}=0.
\ee
Finally, the fifth equation required is the equation of state that links $P$ to $\rho$. One can work either with simplified equations of state or tabulated ones (see e.g., \cite{Akmal:1998cf,Haensel:2004nu}). Here, we will use a polytropic equation of state as a naive model. The pressure is then given by
\be
P=K \rho_\tx{B}^{1+1/n},
\ee
where $\rho_\tx{B}$ is the baryonic mass density, $n$ the polytropic index and $K$ is some constant. The first law of thermodynamics allows us to relate $\rho$ and $\rho_\tx{B}$, and we arrive at the following equation of state:
\be
\rho=nP+\left(\dfrac{P}{K}\right)^{n/(n+1)}.
\ee
We now have a closed system of ordinary differential equations. For practical purposes, units of length, time and mass are chosen so that $c=1$, $G=1$ and $M_\odot=1$; the values $n=2$ and $K=123$ (in the chosen units) give sensible results for the mass and radius of neutron stars in general relativity. As a boundary condition, one can provide the baryonic mass density at the center of the star $\rho_\tx{B;c}$; it will be chosen of order $5\cdot10^{17}$~kg/m$^3$ (i.e., $4.9\cdot10^{-4}$ in the adapted units) to fit with realistic equations of state. The equations are then integrated numerically from the center of the star to the point where the pressure practically vanishes. This point corresponds to the surface of the star. The solution is matched with the exterior Schwarzschild metric in a $C^1$ way. From this matching, one can extract the gravitational mass of the star. A typical result is shown in Fig.~\ref{fig:profileJohnq}.

\begin{figure}[ht]
\begin{center}
\includegraphics[width=10cm]{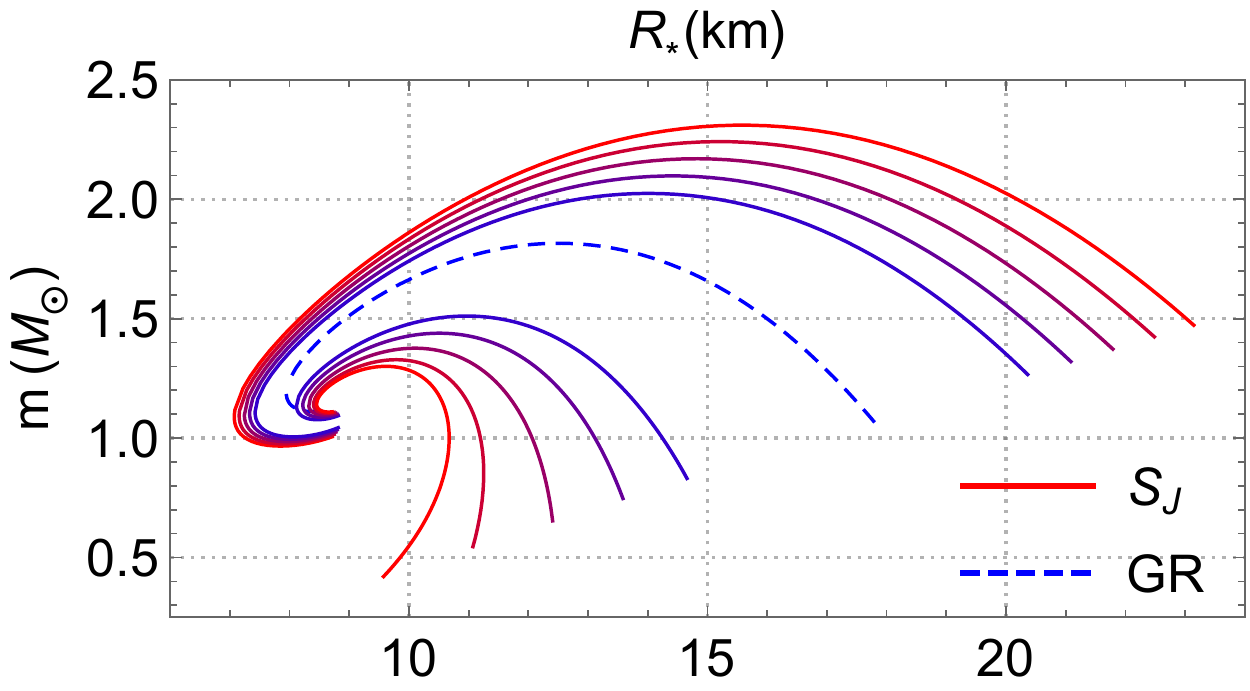}
\caption{Mass-radius relation for the polytropic equation of state. The central density varies from $3\cdot10^{17}$ to $3\cdot10^{21}$~kg/m$^3$. The dashed blue curve is obtained in general relativity. The other reddish curves are all obtained for the model (\ref{eq:Johnstar}), for various values of the combination $\beta q^2$, ranging from $5.3\cdot10^{-3}$ to $1.2\cdot10^{-2}$ in absolute value (and in the system of units used here). When $\beta>0$ ($\beta<0$ respectively), stars tend to have smaller (larger) mass with respect to general relativity. At least in general relativity, only the portion of curve where $\tx{d}m/\tx{d}R_\ast<0$ represents stable solutions. Deviations with respect to general relativity in principle allow one to put some bounds on the solution. For instance, $\beta>0$ seems disfavored because the model could then not accommodate stars of mass $2~M_\odot$, which are known to exist \cite{Antoniadis:2013pzd}. However, one should take into account the poor knowledge we have on the internal structure of neutron stars, and thus on the correct equation of state.}
\label{fig:MRq}
\end{center}
\end{figure}

As an experimental observable, one can plot the mass-radius relation for various values of the central density. For the polytropic equation of state described above, the mass-radius relation has the shape presented in Fig.~\ref{fig:MRq}. The curve is also plotted for general relativity with the same equation of state, in order to allow comparison. For a given star radius, $\beta<0$ implies heavier neutron stars, while $\beta>0$ implies lighter stars compared to general relativity. The observation of stars with mass higher than, say, 1.5~$M_\odot$ can thus rule out the solutions with positive $\beta$. One should keep in mind that the equation of state that we used is very idealized. Thus, the obtained mass-radius relations are very approximate.

Let us stress again that, although Schwarzschild solution is recovered outside the star, the situation is different from usual screening mechanisms. First, there is a huge deviation from general relativity inside the star. Second, the scalar field is not screened up to some radius $r_\tx{V}$ as in the case of Vainshtein's mechanism. It does not back-react \textit{at all} on the metric, arbitrarily far from the source.

Star solutions of the model (\ref{eq:Johnstar}) were further studied in \cite{Silva:2016smx,Cisterna:2016vdx,Maselli:2016gxk}. These references took into account more realistic equations of state, and examined the relation between mass and moment of inertia for neutron stars. These solutions should account for most neutron stars, except millisecond pulsars \cite{Cisterna:2016vdx}. Requiring the theory to reproduce the heaviest known pulsar, PSR J0348+0432, which has mass $2.01 \pm 0.04 \; M_\odot$ \cite{Antoniadis:2013pzd}, puts a bound on $\beta$ when it is positive \cite{Cisterna:2016vdx}. Gamma ray bursts from the pulsar SGR0526-066 provide an additional test for redshift \cite{Higdon:1990dv}. The authors of \cite{Cisterna:2016vdx} checked that their solutions are consistent with the allowed redshifts $z = 0.23 \pm 0.07$ for masses between 1 and 1.5~$M_\odot$. The redshift analysis does not provide sharper constraints than the maximal mass test, though. 

Alternatively, one can find significant deviations with respect to general relativity in beyond Horndeski theory. Indeed, Ref.~\cite{Kobayashi:2014ida} proved that Vainshtein's mechanism generically breaks down inside matter sources for these models (as opposed to Horndeski theory). This result was checked by a fully non-linear and numerical calculation in \cite{Babichev:2016jom}. In this case, the scalar field is assumed to play a role at cosmological scales (as dark energy). Matching between the local and the cosmological solution crucially generates the breaking of Vainshtein's mechanism inside the star. Again, this work was extended to slow rotation and realistic equations of state, giving some bounds on the parameters of the considered model \cite{Sakstein:2016oel}.

It remains though to elucidate which configurations would be energetically favored, between general relativity and Horndeski and beyond models. In standard scalar-tensor theory (\ref{eq:STaction}), it is well known that solutions with a non-trivial scalar profile can be energetically favored with respect to general relativity, a process named ``scalarization''. Damour and Esposito-Far\`ese exhibited some non-minimal couplings between matter and the metric for which it is indeed the case \cite{Damour:1993hw}. Such a study in Horndeski and beyond theory would require a better understanding of the concept of energy.

\section{Conclusions}
\label{sec:conclusiontdep}

In this chapter, we introduced solutions where the scalar field does not only depend on space, but also on time. This is actually imposed by the boundary conditions if the scalar field is assumed to play a significant role as dark energy. We saw in detail how this can be a consistent procedure when the scalar field depends linearly on the cosmological time. 
We also saw that, in order to be fully general, one should allow the metric to depend on time as well. Still, it is possible to build solutions with scalar hair when keeping a static metric.

We studied solutions in two significant subclasses of Horndeski theory. The first subclass possesses reflection symmetry, $\varphi\to-\varphi$. We focused on the specific action (\ref{eq:John}) because it is the simplest model that is representative of both the quartic and quadratic Horndeski sectors. The solutions presented for this model include some exact Schwarzschild and Schwarzschild-de Sitter solutions, as well as an Einstein static universe and solitons. The stealth Schwarzschild and Schwarzschild-de Sitter solutions are particularly interesting for self-tuning purposes. We will further analyze them from the point of view of stability in Chapter \ref{ch:pert}, and use them as well in Chapter \ref{ch:wave} to generate solutions for other interesting models. We also discussed the regularity of the scalar field and invariants built upon it close to the horizon, and examined the slowly rotating limit. We showed that the equation which governs slow-rotation corrections is often similar to its general relativistic counterpart.

The second subclass of Horndeski theory that we studied does not possess reflection symmetry. It is the simplest model with higher order derivatives one can consider: an Einstein-Hilbert term, a cosmological constant, together with the simplest quadratic and cubic Horndeski terms (\ref{eq:DGP}). Although it was not possible to integrate the full system of equations exactly, we could still study analytically various asymptotic regimes and some specific limits, obtaining insight about the full solutions. Additionally, we performed numerical integration of the full system of equations for different parameter ranges. It is important to stress that solutions exist for a range of the scalar field velocity $q$ entering the scalar field ansatz (\ref{eq:ansatzDGP}) (for fixed parameters of the theory $\zeta$, $\eta$, $\Lb$ and $\gamma$). $q$ is a free parameter, independent on the mass of the black hole. Thus, $q$ can be treated as a parameter corresponding to primary hair. At the same time, the asymptotic behavior at large distances is controlled by the fixed ---~in terms of the Lagrangian parameters~--- value $q_0$, Eq.~(\ref{eq:qDGP}). $q$ determines the behavior of a solution at intermediate distances, while the cosmological homogeneous configuration is restored at large radii, independently on $q$. This interpretation of $q$ should be taken with the following reservation: the solution for $\varphi$ in fact depends on $q$ even asymptotically, Eq.~(\ref{eq:corrqq0}); however, the value of $\partial_\mu\varphi$ approaches the cosmological homogeneous solution.

As an outlook, it would be interesting to incorporate other $G_2$ or $G_3$ functions in the study. A straightforward extension of our work is to examine the stability of the above black hole solutions; as already mentioned, this was discussed in some perturbative regime in \cite{Babichev:2012re}. All it takes to get the full result is to combine the upcoming stability analysis, Chapter \ref{ch:pert}, with the numerical solutions presented in Sec.~\ref{sec:cubictdep}. 

Finally, we discussed star solutions, also in the case where the scalar field depends on time. For some models, stars significantly differ from those of general relativity. Neutron star data allow to put some bounds on the parameters of these models. These bounds will be improved with a better knowledge of the equation of state, which should follow from gravitational wave observations \cite{Takami:2014zpa}. We also saw that the interplay between the cosmological time dependence and local physics is crucial, notably in the case of beyond Horndeski theory.
\newpage
\thispagestyle{empty}
\cleardoublepage
\chapter{Black holes and stars with a static scalar field}\label{ch:static}

This thesis is devoted to the study of compact astrophysical objects. To describe them, we would like to define the notion of an isolated system. Of course, realistic objects cannot literally be isolated from the rest of the Universe. Distant stars surround us, and at cosmological scale, the presence of various energy components makes the Universe curved. In spite of these objections, it seems reasonable that the local properties of, say, a star, should be well approximated by assuming that spacetime behaves likes Minkowski spacetime far away from the source of the gravitational field. Additionally, if a scalar field is invoked to account for dark energy, one could expect that its cosmological evolution is slow with respect to black hole time scales. Thus, we will consider in this section asymptotically flat spacetimes, as defined in Chapter \ref{ch:nohairGR}, with a static scalar field.

On the side of the scalar field, it might appear in contradiction with the arguments given in the previous chapter to consider a static scalar. Indeed, for the solutions presented above, the radial profile of the scalar field is directly proportional to its time derivative $q$. Therefore, it does not make sense to approximate these particular solutions as static, even when the scalar field evolution at the cosmological level is slow. However, the scalar field does not necessarily roll with time in general; if a potential term is present, it can stabilize the field to the minimum value of the potential ---~breaking shift-symmetry at the same time, though. What is sure is that other branches of solutions exist, and that the scalar field does not always need to depend on time in static coordinates, even when used for cosmological purposes (see the solutions presented in \cite{Rinaldi:2012vy,Anabalon:2013oea,Minamitsuji:2013ura} for instance). Furthermore, one can look at scalar-tensor models from a different perspective, simply assuming that a scalar field is present in the fundamental theory of gravity. Even if this field plays no role at cosmological level, the resulting theory is in the class (\ref{eq:H})-(\ref{eq:bH4})-(\ref{eq:bH5}) (or at least is part of the degenerate higher-order scalar-tensor theories). It differs from general relativity, and can be tested against observations in strong curvature regimes. In passing, from this perspective, the scalar field vanishes over cosmological scales; thus, the simultaneous observation of gravitational and electromagnetic waves does not constrain these models any more (or at least very weakly).
With this in mind, we are now going to see how to implement solutions that deviate from general relativity, based on our work \cite{Babichev:2017guv}. We will respect the assumptions of the theorem concerning asymptotic flatness and staticity, especially for the scalar field. That is, we will consider a scalar field that depends on the radial coordinate $r$ only, and such that
\be
\varphi(r) \underset{r \rightarrow \infty}{=} \varphi_\infty+\mc{O}\left(\dfrac1r\right),~~~~\varphi'(r) \underset{r \rightarrow \infty}{=} \mc{O}\left(\dfrac{1}{r^2}\right),
\ee
where $\varphi_\infty$ is an arbitrary constant, that one can decide to set to zero because of shift symmetry. Since we will respect the first assumptions of the theorem, it is necessary to break at least one of the remaining hypotheses in order to find non-trivial solutions. One can either use Horndeski densities that are not analytic around Minkowski vacuum, allow the norm of the current to diverge, or remove the standard kinetic term from the action. Each of the three following sections is accordingly devoted to explicit examples illustrating these three possible ways to circumvent the theorem.

\section{Non-analytic Horndeski functions}

The first possibility to get round the theorem is to examine Lagrangian densities that are not analytic when the spacetime becomes flat and the scalar field approaches a constant. In general, such models will not possess physical solutions; however, there exist exceptions where the spacetime can be asymptotically flat, with an asymptotically trivial scalar field. The loophole in this case is in the second step of the no-hair theorem proof. For certain (non-analytic) models, a vanishing radial component for the current does not lead to the trivial solution $\varphi'=0$. In order to find these models, the idea is to select $G_i$ Horndeski functions (or beyond Horndeski $F_i$) that yield a $\varphi'$-independent piece in $J^r$, Eq.~(\ref{eq:staticJrbH})\footnote{It could also occur that $J^r$ contains negative powers of $\varphi'$. However, such solutions would acquire an infinite current when approaching Minkowski vacuum.}. At the same time, we will keep the standard kinetic term $X\subseteq G_2$, so we do not break this assumption of the theorem. This way, $\varphi'$ (appearing in the standard kinetic term) will be forced to a non-trivial value from the condition $J^r=0$. A careful examination of Eq.~(\ref{eq:staticJrbH}) reveals that for each function $G_i$ and $F_i$, there exists an appropriate choice:
\be
\begin{split}
G_2 &\supseteq \sqrt{-X},
\\
G_3 &\supseteq \tx{ln}|X|,
\\
G_4 &\supseteq \sqrt{-X},
\\
G_5 &\supseteq \tx{ln}|X|,
\\
F_4 &\supseteq (-X)^{-3/2},
\\
F_5 &\supseteq X^{-2}.
\label{eq:nanalytic}
\end{split}
\ee
An action involving the standard kinetic term, $X\subseteq G_2$,  additionally to one of the above Lagrangians, has the potential to possess a non-trivial, static and asymptotically flat black hole solution, with also regular behavior for the current. It suffices a priori for one of the $G_i$ or $F_i$ functions to have such a form. However, although this imposes that $\varphi'$ is not trivial, it does not guarantee the existence of a black hole solution. It is only a necessary condition. Indeed, we will see in Paragraph \ref{sec:regularGB} that the model $G_5 \propto \tx{ln}|X|$, together with a standard kinetic term and a Ricci scalar, does not possess any black hole solution with regular $J^2$. In this section, we present exact black hole solutions in Horndeski and beyond Horndeski theories, for models with $G_4 \supset \sqrt{-X}$ and $F_4 \propto (-X)^{-3/2}$ respectively. Both solutions admit secondary hair and are asymptotically flat while the scalar field asymptotically decays. On the contrary, we show that no black hole solutions with a regular norm of the current $J^2$ can be found for the model with $G_5 \propto \tx{ln}|X|$.

\subsection{Quartic Horndeski Lagrangian}
\label{sec:sqrtX}

Following the method stemming from (\ref{eq:nanalytic}), let us first consider the following action:
\be
\begin{split}
S_\tx{n.a.} &=\displaystyle\int{\mathrm{d}^4x \sqrt{-g} \left\{ \left[\zeta + \beta \sqrt{(\pd \varphi)^2/2}\right] R- \eta (\pd \varphi)^2 \vphantom{\dfrac{\beta}{\sqrt{2(\pd \varphi)^2}}}\right.}
\\
&\quad\left.- \dfrac{\beta}{\sqrt{2(\pd \varphi)^2}} \left[(\Box \varphi)^2-\nd_\mu \pd_\nu \varphi\nd^\mu \pd^\nu \varphi\right] \right\}.
\label{eq:sqrtX}
\end{split}
\ee
where ``n.a.'' stands for non-analytic. Equivalently, one can set
\be
G_2 =2 \eta X,~~~~G_4 = \zeta + \beta \sqrt{-X},~~~~G_3 =0,~~~~G_5=0,
\ee
with the usual meaning for $\eta$ and $\zeta$. Here, $\beta$ is a new coupling constant that is dimensionless as $\eta$. Note that $\eta$ or $\beta$ could be absorbed in a redefinition of the scalar field. We will not do so, in order to keep track of the origin of the various terms. The $G_2$ term is simply a canonical kinetic term, and the coefficient $\zeta$ in $G_4$ yields an Einstein-Hilbert piece in the action. The $\beta \sqrt{-X}$ term is in the class defined by (\ref{eq:nanalytic}) and gives a $\varphi$-independent contribution to the current. It is interesting to note in passing that the above action for $\zeta=0$ admits global scale invariance, as was shown in \cite{Padilla:2013jza}. Using the ansatz (\ref{eq:statmetric})-(\ref{eq:statphi}) for the metric and scalar field, one obtains for the radial component of the current, Eq.~(\ref{eq:staticJrbH}):
\be
J^r= \dfrac{\beta  \sqrt{2 f}}{r^2} \text{sgn}(\varphi')-2\eta  \varphi' f.
\label{eq:currentsqrtjohn}
\ee
The first term does not depend on $\varphi'$; it depends on its sign, but as we will see below, all solutions keep a fixed sign in the static region of the black hole. Solving Eq.~(\ref{eq:currentsqrtjohn}), one gets
\be
\varphi'=\pm \dfrac{\beta}{\sqrt{2} \eta  r^2 \sqrt{f}}.
\label{eq:phiprime}
\ee
This expression for $\varphi'$ is real for $f>0$, i.e., outside of the black hole horizon. Applying the sgn function to the $J^r=0$ equation, one finds that $\beta$ and $\eta$ necessarily share the same sign. Two other equations remain to be solved, namely the $(tt)$ and $(rr)$ components of Einstein equations. They can be found by specializing the equations of Appendix \ref{ap:fieldeqs} to the specific model (\ref{eq:sqrtX}). The $(tt)$ equation is particularly simple to solve once Eq.~(\ref{eq:phiprime}) has been used. It is actually a first order differential equation on $f$. The $(rr)$ equation then imposes that $h$ is equal to $f$ (up to an overall constant that simply amounts to a redefinition of time). The solution takes the following form:
\be
f(r) = h(r) = 1 -\dfrac{2m}{r}-\dfrac{\beta ^2}{4 \zeta  \eta  r^2},
\label{eq:fsol1}
\ee
where $m$ is a free integration consant. Additionally, the kinetic density $X$ reads:
\be
X(r) = -\dfrac{\beta^2}{4\eta^2 r^4}
\label{eq:Xstat}
,\ee
from which one can compute the scalar field. Because of shift symmetry, the scalar field is determined up to some constant. This freedom can be used to impose that $\varphi$ vanishes at spatial infinity. Then, the solution depends on the sign of the parameters $\eta$ and $\beta$:
\be
\begin{split}
\varphi(r) &= \pm \sqrt{\dfrac{2\zeta}{\eta}} \left\{\tx{Arctan} \left[\dfrac{\beta ^2+4\zeta  \eta\, m\,  r}{\beta  \sqrt{4 \zeta  \eta \, r (r-2m)-\beta ^2}}\right]- \tx{Arctan}\left(\dfrac{2m}{\beta}  \sqrt{\zeta \eta}\right) \right\}
\\
&\quad\tx{if}~\beta>0~\tx{and}~\eta>0,
\\
\varphi(r) &= \pm \sqrt{\dfrac{2\zeta}{-\eta}} \left\{\tx{Argth}\left[\dfrac{\beta ^2+4\zeta  \eta\, m\,r}{\beta  \sqrt{\beta ^2-4 \zeta  \eta \, r (r-2m)}}\right]+\tx{Argth}\left(\dfrac{2m}{\beta}  \sqrt{-\zeta  \eta}\right)\right\}
\\
&\quad\tx{if}~\beta<0~\tx{and}~\eta<0.
\label{eq:phisol}
\end{split}
\ee
The above solution describes a black hole with mass $m$. Note that the non-trivial scalar field back-reacts on the metric in an interesting way: the spacetime solution is of the Reissner-Nordstr\"om form. This is possibly related to the remnant of global conformal invariance shared by the action (\ref{eq:sqrtX}), as the spacetime metric solution has zero Ricci scalar curvature (as does Reissner-Nordstr\"om spacetime). Positive $\eta$ formally corresponds to an imaginary charge of the Reissner-Nordstr\"om metric. In this case, there exists an event horizon for any value of $m$ including that of zero (unlike Reissner-Nordstr\"om spacetime). On the other hand, when $\eta$ is negative, the scalar field manifests itself in an electric-like contribution where $\sqrt{-\beta^2/(4 \zeta \eta)}$ plays a role similar to that of electric charge for spacetime. This ``electric charge'' is however not an integration constant; it depends entirely on the parameters of the theory, which are fixed. Any such black hole experiences the exact same correction to the Schwarzschild metric. Choosing a negative $\eta$ significantly affects the inner structure of the black hole, but the solution is not to be trusted beyond the event horizon, as can be seen from the fact that $\varphi'$ becomes imaginary there. For negative $\eta$, there exists a lower bound on $m$:
\be
m_\tx{min} = \dfrac12\,\sqrt{\dfrac{\beta^2}{-\zeta \eta}},
\ee
which, when saturated gives an extremal black hole. Whenever $m<m_\tx{min}$, the solution does not describe a black hole any more, but rather a naked singularity. In terms of stability, positive $\eta$ corresponds to the ``correct'' sign in the standard kinetic term. 
The stability, however, also depends on the quartic Horndeski term. Therefore, one cannot conclude on the stability of the solutions only by the sign of $\eta$, see e.g., \cite{Deffayet:2010qz,Babichev:2012re}. The solution presented above is asymptotically flat. It then fulfills all assumptions of the no-hair theorem but the analyticity of the Horndeski densities $G_i$, due to the presence of $\sqrt{-X}$ in $G_4$. The metric features a Newtonian fall-off at spatial infinity. At $r\to \infty$, the scalar field decays as:
\be
\varphi(r) \mathop{=}_{r \rightarrow \infty} \pm \dfrac{\beta}{\sqrt{2}\eta r} + \mathcal{O}(r^{-2}).
\label{eq:farphi}
\ee
This solution does not have primary hair, as no integration constant other than $m$ appears in (\ref{eq:farphi}). The black hole manifestly has secondary hair due to the non-trivial scalar-tensor mixing. To conclude about faraway asymptotics, let us remark that a cosmological constant can be added to the initial action. The solution is modified in the same way as it is in general relativity, and acquires anti-de Sitter or de Sitter asymptotics. Explicitly, setting $G_2 = 2(\eta X-\zeta\Lb)$ and $G_4 =\zeta + \beta \sqrt{-X}$, one gets 
\be
f(r) =h(r)= 1-\dfrac{2m}{r} -\dfrac{\beta ^2}{4 \zeta  \eta  r^2} -\dfrac{\Lb}{3} r^2,
\ee
and the scalar field can still be computed from Eq.~(\ref{eq:phiprime}).

Let us now examine the near-horizon asymptotics. As a direct consequence of Eq.~(\ref{eq:phiprime}), the derivative of the scalar field diverges at the horizon. This is however a coordinate-dependent statement, which ceases to be true using the tortoise coordinate, for instance; the divergence is absorbed in the coordinate transformation. On the other hand, it is easy to check from Eq.~(\ref{eq:phisol}) that the scalar itself is finite at the horizon. Crucially, $X$ does not diverge either close to the horizon, Eq.~(\ref{eq:Xstat}). Also, since the metric is identical to the Reissner-Nordstr\"om solution, it is clearly regular. Therefore, all physically meaningful quantities are well behaved when approaching the horizon.

Finally, in the interior of the black hole, $f<0$ and Eq.~(\ref{eq:phiprime}) would imply that $\varphi'$ becomes imaginary. This feature is not specific to the solution presented here: all known static solutions possess it \cite{Rinaldi:2012vy, Minamitsuji:2013ura, Anabalon:2013oea}. The solution can therefore not be trusted beyond the event horizon.

\subsection{Quartic beyond Horndeski Lagrangian}
\label{sec:quarticbH}

A very similar analysis can be carried out for the beyond Horndeski quartic function $F_4$. Following Eq.~(\ref{eq:nanalytic}), one may consider the Lagrangian defined by:
\be
G_2 = 2\eta X,~~~~G_4= \zeta,~~~~F_4 = \gamma(-X)^{-3/2},~~~~G_3 =0,~~~~G_5=0, ~~~~F_5=0.
\ee
The new constant $\gamma$ parametrizes the beyond Horndeski term. One can follow the same steps as in the previous paragraph. The $J^r=0$ equation provides an expression for the kinetic density $X$:
\be
X=-\left[\dfrac{2 \gamma}{\eta} \; \dfrac{f(rh)'}{r^2 h}\right]^2
\label{eq:XbH}
.\ee
Then, using the $(rr)$ equation (\ref{eq:Err}), it is possible to determine a particular combination of $f$ and $h$:
\be
\dfrac{f(rh)'}{r^2 h}=\dfrac{\eta}{24\gamma^2} \left[-\zeta+\sqrt{\zeta^2+\dfrac{48 \zeta \gamma^2}{\eta r^2}}\right]
\label{eq:combinationbH}
.\ee
Substituting this into the $(tt)$ equation, one ends up with a first order differential equation on $f$, the solution of which is
\be
f=\dfrac{1}{r\left(4\gamma\sqrt{-X}-\zeta\right)^2}\left[C-\displaystyle\int{\tx{d}r(\zeta+\eta r^2 X)\left(4\gamma \sqrt{-X}-\zeta\right)}\right]
,\ee
with $C$ a free integration constant. $X$ is known in terms of $r$, by combining Eqs.~(\ref{eq:XbH}) and (\ref{eq:combinationbH}). One can also compute $h$ from Eq. (\ref{eq:combinationbH}). The explicit expression for $f$ reads
\be
\begin{split}
 f(r)&= \vphantom{\left[\dfrac{r}{\gamma \zeta  \left(24 \gamma+\sqrt{6} \sqrt{96 \gamma^2+\zeta  \eta  r^2}\right)}\right]} \dfrac{1}{144 \gamma ^2 \eta  r \left\{2 \zeta +\sqrt{\zeta [\zeta + 48 \gamma ^2/(\eta r^2)]}\right\}^2 \left(48\gamma ^2+\zeta  \eta  r^2\right)}
\\
&\quad \times\left\{\vphantom{\left[ \dfrac{r}{\gamma \zeta  \left(24 \gamma+\sqrt{6} \sqrt{96 \gamma^2+\zeta  \eta  r^2}\right)}\right]}24 \gamma^2 r^2 \zeta \eta \left[10\zeta \sqrt{\zeta \eta (48 \gamma^2+\zeta  \eta  r^2)}+\eta (27 C+16 \zeta ^2 r)\right]\right.
\\
&\quad\vphantom{\left[ \dfrac{r}{\gamma \zeta  \left(24 \gamma+\sqrt{6} \sqrt{96 \gamma^2+\zeta  \eta  r^2}\right)}\right]}+r^4 \zeta^3 \eta^2\left[\sqrt{\zeta \eta (48 \gamma^2+\zeta  \eta  r^2)}-\zeta \eta r\right]
\\
&\quad\vphantom{\left[ \dfrac{r}{\gamma \zeta  \left(24 \gamma+\sqrt{6} \sqrt{96 \gamma^2+\zeta  \eta  r^2}\right)}\right]} +1152 \gamma^4 \left[8 \zeta \sqrt{\zeta \eta (48\gamma^2+\zeta  \eta  r^2)}+27 C \eta +18 \zeta ^2 \eta  r\right]
\\
&\quad\vphantom{\left[ \dfrac{r}{\gamma \zeta  \left(24 \gamma+\sqrt{6} \sqrt{96 \gamma^2+\zeta  \eta  r^2}\right)}\right]} +288 \sqrt{6} \gamma^3 \zeta^{3/2} \sqrt{\eta } (48 \gamma^2+\zeta  \eta  r^2) 
\\
&\quad\left.\times\tx{ln}\, \left[ \dfrac{r}{2\gamma \zeta  \left(12 \gamma+\sqrt{3} \sqrt{48 \gamma^2+\zeta  \eta  r^2}\right)}\right]\right\}
,\end{split}
\ee
where we assumed that $\eta$ is positive ($\eta$ and $\gamma$ must have opposite sign). Again, this solution is asymptotically flat, with a Newtonian fall-off. Taking for instance a positive $\eta$, and defining the quantity
\be
m= \dfrac{-9C\sqrt{\eta}+4\sqrt{3} \gamma \zeta^{3/2} \,\tx{ln}\, (12\gamma^2 \zeta^3 \eta)}{18\zeta^2 \sqrt{\eta}}
,\ee
one can expand $f$ at spatial infinity and get
\be
f(r) \mathop{=}_{r \rightarrow \infty} 1-\dfrac{2m}{r}+\dfrac{20 \gamma^2}{\zeta  \eta  r^2} + \mathcal{O}(r^{-3})
.\ee
Therefore, $m$ should be interpreted as the gravitational mass of the black hole. The solution is very similar to the one obtained when considering $G_4\propto\sqrt{-X}$.

\subsection{Quintic Horndeski Lagrangian}
\label{sec:regularGB}

\begin{figure}[ht]
\begin{center}
\includegraphics[width=10cm]{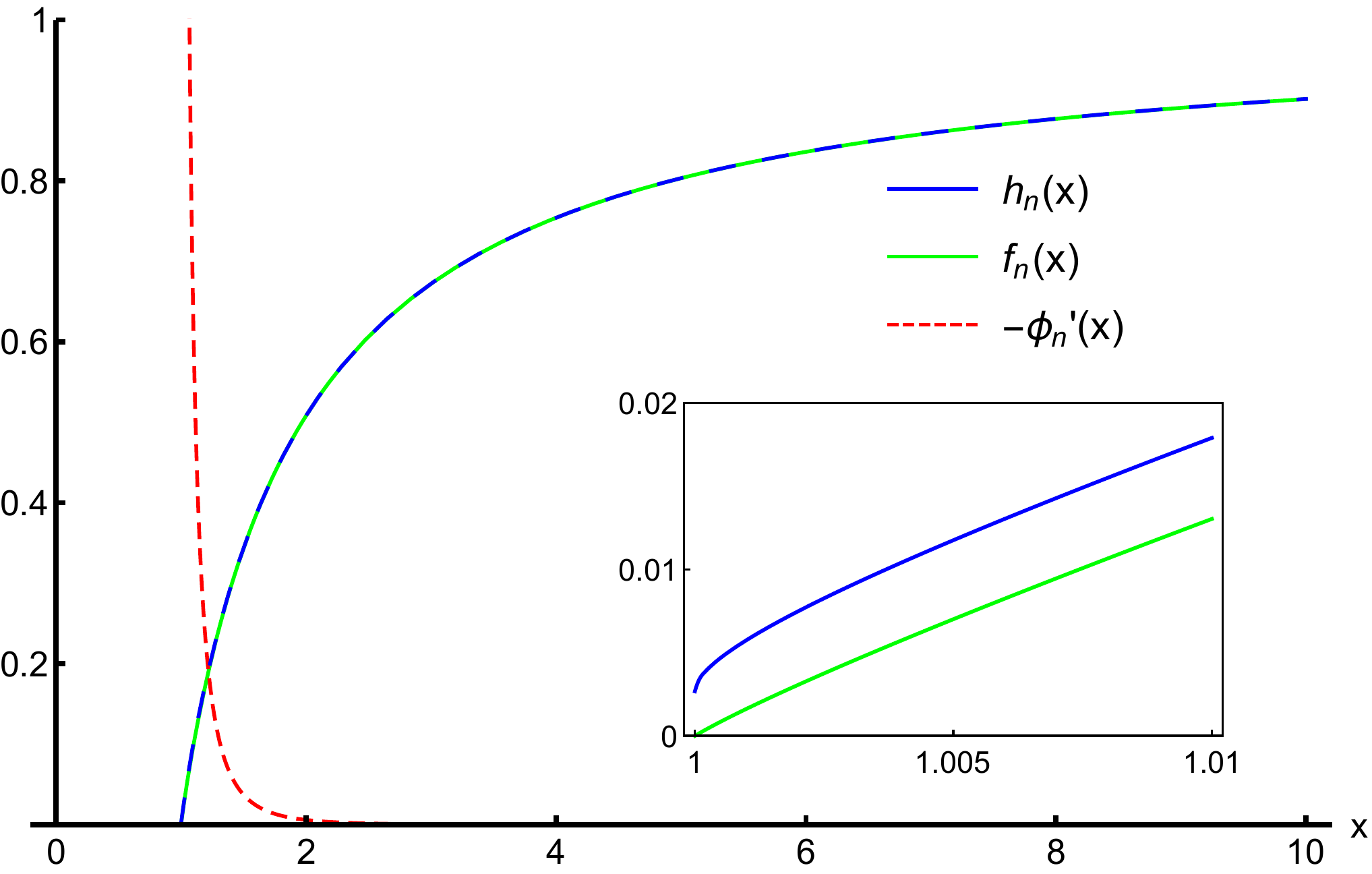}
\caption{Typical numerical solution, obtained for $\alpha_\tx{n}=1/10$. $f_\tx{n}$ and $h_\tx{n}$ are in very good agreement with the spatial infinity expansion (\ref{eq:farGBh})--(\ref{eq:farGBphi}) at large values of $x$. However, the zoomed plot reveals a pathological behavior close to the point where $f_\tx{n}$ vanishes. It is clear that $h_\tx{n}$ does not vanish at the same time.}
\label{fig:coordsing}
\end{center}
\end{figure}

\begin{figure}[ht]
\begin{center}
\includegraphics[width=10cm]{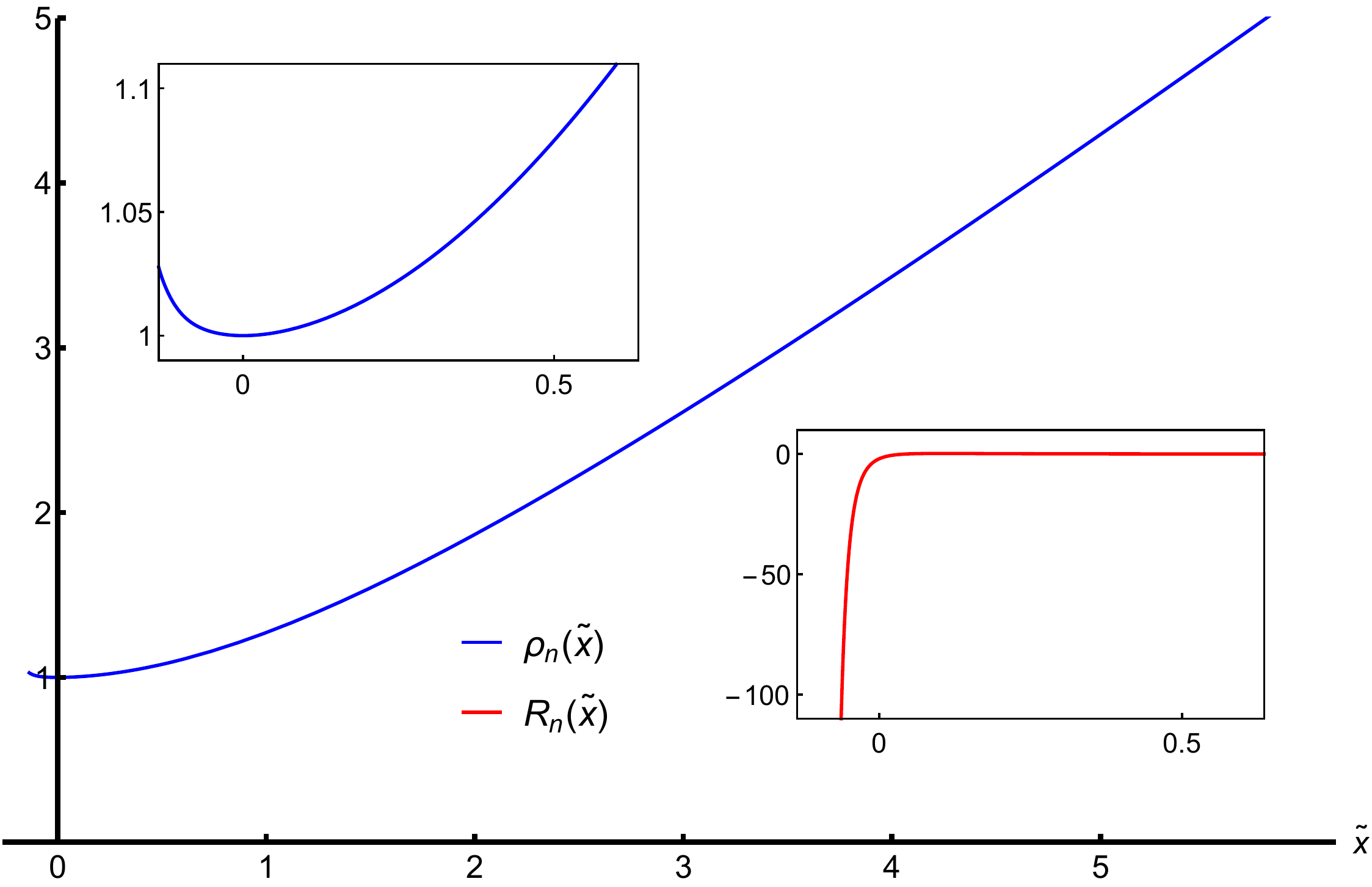}
\caption{Numerical solution obtained for $\alpha_\tx{n}=1/10$. At large $\tilde{x}$, $\rho_\tx{n}(\tilde{x}) \sim \tilde{x}$. At $\tilde{x}=0$, the zoomed plot shows that $\rho_\tx{n}$ starts increasing again. However, the simulation breaks down at $\tilde{x} \simeq -0.13$. The second framed plot shows that the Ricci scalar $R_\tx{n}$, in red, diverges at this point, indicating a curvature singularity.}
\label{fig:nakedsing}
\end{center}
\end{figure}

In this section, we will examine the quintic Horndeski Lagrangian built of the following elements:
\be
G_2 =2\eta X,~~~~G_5 = \alpha\, \tx{ln}|X|,~~~~G_4=\zeta,~~~~G_3 =0,
\label{eq:GBlag}
\ee
while the beyond Horndeski sector is assumed to vanish, and where $\alpha$ is a constant. We discuss in more detail the origin and the properties of this model in Sec.~(\ref{sec:GB}). For now, it is sufficient to see that the function $G_5$ is not analytic at the point $X=0$. We shall impose that the radial component of the current vanishes: $J^r=0$. First, let us take a look at the spatial infinity expansion of the solution, assuming that it can be expanded in a $1/r$ series:
\begin{align}
h &\mathop{=}_{r\rightarrow\infty} 1 -\dfrac{2m}{r} - \dfrac{8\alpha ^2 m^3}{7\zeta  \eta  r^7} + \mathcal{O}(r^{-8}),
\label{eq:farGBh}
\\
f &\mathop{=}_{r\rightarrow\infty} 1 -\dfrac{2m}{r} - \dfrac{4\alpha ^2 m^3}{2\zeta  \eta  r^7} + \mathcal{O}(r^{-8}),
\label{eq:farGBf}
\\
\varphi' &\mathop{=}_{r\rightarrow\infty} -\dfrac{2\alpha m}{2\eta r^5} + \mathcal{O}(r^{-6}),
\label{eq:farGBphi}
\end{align}
with $m$ a free integration constant. The corrections with respect to general relativity are therefore very mild far away from the source. These corrections are in agreement with the post-Newtonian corrections for a distributional source found in \cite{Amendola:2007ni}. The scalar field $\varphi$ decays as $1/r^4$ and the only free parameter is the mass of the central object, $m$. There is no tunable scalar charge as expected, since $J^r=0$ is already an integral of the scalar equation of motion. The above expansion cannot be trusted whenever the $\alpha^2$ corrections become of the same order as the mass term, i.e., when
\be
r \lesssim \left(\dfrac{\alpha^2 m^2}{\zeta \eta}\right)^{1/6}
.\ee
To go further, we resort to numerical integration, because we could not integrate analytically the field equations. The radial component of the current reads
\be
J^r= f \left[\dfrac{\alpha  (f-1) h'}{r^2 h}-2\eta \varphi' \right],
\label{eq:GBcurrent}
\ee
and one can still use the $(tt)$ and $(rr)$ equations of Appendix \ref{ap:fieldeqs}. First, imposing $J^r=0$, it is possible to extract $h'/h$ as a function of $\varphi'$ and $f$:
\be
\dfrac{h'}{h}=\dfrac{2\eta r^2 \varphi'}{\alpha(f-1)}
\label{eq:hGB}
.\ee
Using Eq.~(\ref{eq:hGB}), the $(rr)$ equation becomes a second-order algebraic equation on $\varphi'$; the solution for $\varphi'$ in terms of $f$ is:
\be
\varphi'=\dfrac{-2 \zeta  \eta  r^3 f \pm\sqrt{4 \zeta ^2 \eta ^2 r^6 f^2- 2 \alpha^2 \eta \zeta r^2 f (1 -f)^2(5 f-  1)}}{\alpha \eta r^2 f (5 f-1)}
\label{eq:phiGB}
.\ee
Two branches exist for $\varphi'$. Relying on the numerical analysis, we select the plus branch of the above two; the minus branch gives pathological solutions that extend only to a finite radius. Equation (\ref{eq:phiGB}) also fixes the sign of $\eta$. Indeed, taking the limit $f \rightarrow 0$, as expected for a black hole, one can check that the sign of the term under the square root is determined by the sign of $\eta$ at leading order ($\zeta$ is positive by convention). If $\eta$ was negative, the scalar field would become imaginary before reaching the assumed horizon. Therefore, we will restrict the analysis to positive $\eta$. One is then left with a single master equation on $f$, which turns out to be a first-order ordinary differential equation. To write it in a form adapted to numerical resolution, let us introduce a length scale $r_0$, and consider functions of $x = r/r_0$, rather than $r$. Then, the master equation depends merely on one dimensionless parameter, that we will call $\alpha_\tx{n}$:
\be
\alpha_\tx{n} = \dfrac{\alpha}{\sqrt{2\eta \zeta} r_0^2}
.\ee
We use the following dictionary between dimensionless and dimensionful quantities:
\be
f_\tx{n} (x) = f(x \,r_0),~~~~ h_\tx{n} (x) = h(x \,r_0),~~~~\varphi_\tx{n} (x) = \sqrt{\dfrac{2\eta}{\zeta}} \varphi(x \,r_0).
\ee
The master equation in terms of $f_\tx{n}$ reads:
\be
\begin{split}
\label{eq:masterf}
\vphantom{\dfrac12}&4 f_\tx{n}^3 \left(40 x \alpha_\tx{n} ^4 f_\tx{n}'+5 x^4 \alpha_\tx{n} ^2+53 \alpha_\tx{n} ^4\right)+x \alpha_\tx{n} ^2 \left(-12 x \Sigma + x^4+\alpha_\tx{n} ^2\right) f_\tx{n}'
\\
\vphantom{\dfrac12}&-2 f_\tx{n} \left[-x \alpha_\tx{n} ^2 \left(10 x \Sigma+x^4+4 \alpha_\tx{n} ^2\right) f_\tx{n}'+6 x \alpha_\tx{n} ^2 \Sigma+x^8-8 x^4 \alpha_\tx{n} ^2-13 \alpha_\tx{n} ^4\right]
\\
\vphantom{\dfrac12}&+\alpha_\tx{n} ^2 f_\tx{n}^2 \left[\left(5 x^5-94 x \alpha_\tx{n} ^2\right) f_\tx{n}'+10 x \Sigma-34 x^4-116 \alpha_\tx{n} ^2\right]
\\
\vphantom{\dfrac12}&-5 \alpha_\tx{n} ^4 f_\tx{n}^4 \left(15 x f_\tx{n}' +34\right)+2 \left(x^4+\alpha_\tx{n} ^2\right) \left(x \Sigma-\alpha_\tx{n} ^2\right)+50 \alpha_\tx{n} ^4 f_\tx{n}^5 = 0,
\end{split}
\ee
where
\be
\Sigma=\sqrt{x^2 f_\tx{n} \left\{f_\tx{n} \left[\alpha_\tx{n} ^2 f_\tx{n} (11-5 f_\tx{n})+x^4-7 \alpha_\tx{n} ^2\right]+\alpha_\tx{n} ^2\right\}}.
\ee
Since this is a first-order differential equation, one needs to specify a single initial condition. Because we are a priori looking for a black hole, we will impose that $f_\tx{n}$ vanishes at $x=1$. Then one can proceed to numerical integration. A typical result of the numerical integration is displayed in Fig.~\ref{fig:coordsing}. Far away, the metric and scalar field fit very well the expansion given in Eqs.~(\ref{eq:farGBh})--(\ref{eq:farGBphi}). However, taking a closer look at the black hole region itself, one remarks that $h_\tx{n}$ does not vanish when $f_\tx{n}$ does, as should be the case for a black hole. This behavior is shown in Fig.~\ref{fig:coordsing}, and we also confirmed this by an analytical expansion close to the point where $f$ vanishes. All curvature invariants being finite there, this suggests the presence of a coordinate singularity.
To go further, one must remark that there is no way to extend the solution in the region $x<1$ because $\varphi_\tx{n}'$ becomes imaginary there. Therefore, one has to change coordinates. As the coordinate singularity arises from the $g_{rr}$ part of the metric, let us define a new radial coordinate $\tr$ as
\be
\tx{d}\tr = \dfrac{\tx{d}r}{\sqrt{f(r)}}
.\ee
The ansatz for the metric now takes the following form:
\be
\tx{d}s^2 = -h(\tr)\, \tx{d}t^2 + \tx{d}\tr^2+\rho(\tr)^2\tx{d}\Omega^2
\label{eq:ansatz2}
,\ee
where $\rho$ is a new unknown function, interpreted as the areal radius, i.e., the radius that measures the area of constant $\tr$ 2-spheres. Repeating the same procedure as above, one can eliminate $h$ and $\varphi'$ in the equations and obtain a master equation on $\rho$ only, which is second order\footnote{The higher order of the equation translates the fact that the new ansatz (\ref{eq:ansatz2}) involves an additional reparametrization freedom $\tr \rightarrow \tr +$constant.}. With the same convention as before for the units and the index n, let us define $\tilde{x}=\tilde{r}/r_0$. The master equation then reads:
\be
\begin{split}
&\vphantom{\dfrac12}50 \alpha_\tx{n} ^4 \rho_\tx{n}'^{\:11}-170 \alpha_\tx{n} ^4 \rho_\tx{n}'^{\:9}+\rho_\tx{n}^2 \tilde{\Sigma}^3-2 \alpha_\tx{n} ^2 \left(\rho_\tx{n}^4+\alpha_\tx{n} ^2\right) \rho_\tx{n}'+4 \left(5 \alpha_\tx{n} ^2 \rho_\tx{n}^4+53 \alpha_\tx{n} ^4\right) \rho_\tx{n}'^{\:7}
\\
&\vphantom{\dfrac12}-2 \left(17 \alpha_\tx{n} ^2 \rho_\tx{n}^4+58 \alpha_\tx{n} ^4\right) \rho_\tx{n}'^{\:5}+2 \left(8 \alpha_\tx{n} ^2 \rho_\tx{n}^4-\rho_\tx{n}^8+13 \alpha_\tx{n} ^4\right) \rho_\tx{n}'^{\:3}
\\
&\vphantom{\dfrac12}+2 \alpha_\tx{n} ^2 \rho_\tx{n} \rho_\tx{n}' \rho_\tx{n}'' \left[\rho_\tx{n}^4 \left(5 \rho_\tx{n}'^{\:4}+2 \rho_\tx{n}'^{\:2}+1\right)\vphantom{\left(\rho_\tx{n}'^2-1\right)^2}-\alpha_\tx{n} ^2 \left(\rho_\tx{n}'^{\:2}-1\right)^2 \left(75 \rho_\tx{n}'^{\:4}-10 \rho_\tx{n}'^{\:2}-1\right)\right]
\\
&\vphantom{\dfrac12}+\rho_\tx{n}^2 \tilde{\Sigma} \left\{\rho_\tx{n}'^{\:2} \left[\alpha_\tx{n} ^2 \left(15 \rho_\tx{n}'^{\:4}-23 \rho_\tx{n}'^{\:2}+9\right)+8 \alpha_\tx{n} ^2 \rho_\tx{n} \left(5 \rho_\tx{n}'^{\:2}-3\right) \rho_\tx{n}''+\rho_\tx{n}^4\right]-\alpha_\tx{n} ^2\right\}=0
,\end{split}
\ee
where
\be
\tilde{\Sigma}=\sqrt{\rho_\tx{n}'^{\:2} \left(\alpha_\tx{n} ^2 \left(-5 \rho_\tx{n}'^{\:4}+11 \rho_\tx{n}'^{\:2}-7\right)+\rho_\tx{n}^4\right)+\alpha_\tx{n} ^2}
.\ee
To proceed with numerical integration, one needs to specify two initial conditions. The configuration we will impose is equivalent to the one of the previous analysis: we will require that $\rho_\tx{n}(0)=1$ and $\rho_\tx{n}'(0)=0$. In terms of the old ansatz, this would translate as $f_\tx{n}(1)=0$, and $x=1$ is mapped to $\tilde{x}=0$. The result is shown in Fig.~\ref{fig:nakedsing}.
The solution can be continued in the $\tilde{x}<0$ range, i.e., beyond the point where the old coordinate system becomes singular. In the new coordinates, the areal radius of 2-spheres $\rho$ decreases with $\tr$ up to $\tr=0$, and then starts increasing again when $\tr$ goes to negative values. However, the solution can be continued only in a very short range of negative $\tr$. When the integration fails, the Ricci scalar explodes; thus, this solution describes a curvature singularity which is not shielded by any horizon.

A second possibility though, is to interpret this type of solution as a wormhole. For a similar solution, although for the different theory (\ref{eq:string}) presented in the next section, this interpretation has been suggested \cite{Kanti:2011jz}. There are two branches of the solution for $\varphi'$, and the idea of \cite{Kanti:2011jz} amounts to gluing these two branches at $\tr=0$, so that the solution on the left is symmetrical to the solution on the right: $\varphi(-\tr)=\varphi(\tr)$, $\varphi'(-\tr)=-\varphi'(\tr)$, etc. One obtains two copies of the same asymptotically flat universe, glued together at $\tr=0$, where the areal radius $\rho$ is minimal. Doing this, one creates a throat that relates two universes, i.e., a wormhole. The price to pay for this is that certain quantities, namely $\varphi'$ and $h'$, become discontinuous at $\tr=0$. This can be fixed, as proposed in \cite{Kanti:2011jz}, by adding some matter located on the throat. If one tunes this matter to the right density and pressure, one can account for the discontinuities at the throat.

Thus, there are no black hole solutions for the Lagrangian (\ref{eq:GBlag}), if one requires that the norm of the current is finite. Far away from the curvature singularity though, the metric can describe the exterior of a star. A similar behavior was found in \cite{Charmousis:2015txa} in a different but close set-up.

\section{Infinite norm of the current}
\label{sec:GB}

In this section, we wish to present briefly another specific solution that circumvents the no-hair theorem, discussed in \cite{Sotiriou:2013qea,Sotiriou:2014pfa}. It is based on an action that includes a standard kinetic term and a linear coupling between the scalar field and the Gauss-Bonnet density $\hat{G}$, defined in Eq.~(\ref{eq:GB}). Explicitly, the action is
\be
S_\tx{GB}=\displaystyle\int{\tx{d}^4x \sqrt{-g}\left[\zeta R - \eta (\pd\varphi)^2-\dfrac{\alpha}{4}\varphi\,\hat{G}\right]},
\label{eq:actionGB}
\ee
with usual $\zeta$ and $\eta$, and coupling constant $\alpha$. The black hole solutions of a very similar action were studied previously in \cite{Campbell:1990fu} and \cite{Kanti:1995vq}. In these references, the coupling of the scalar was inspired by the dilaton of string theory, under the form 
\be
\tx{e}^{\alpha\varphi/4} \hat{G}.
\label{eq:string}
\ee
The characteristic features of the solutions in this theory are similar to those of the model (\ref{eq:actionGB}), which actually results from (\ref{eq:string}) in the limit $\alpha\, \varphi \ll1$. At first sight, one could object that the action (\ref{eq:actionGB}) does not fit in the framework of shift-symmetric Horndeski theory. First, the presence of $\varphi$ itself appears in contradiction with the shift symmetry. However, the field equations derived from (\ref{eq:actionGB}) are shift-symmetric. Indeed, under the transformation $\varphi\to\varphi+C$ with consant $C$, one ends up with an additional term $\alpha C \hat{G}$. Knowing that the Gauss-Bonnet density is a total derivative, this additional term is merely a boundary term. Thus, one can omit it and recover the original action: the action is indeed shift-symmetric. Another objection is that such a term does not seem to be present in the generic formulation of Horndeski theory (\ref{eq:H}). However, the action (\ref{eq:actionGB}) certainly generates second-order field equations, and Horndeski theory is precisely the most generic scalar-tensor theory with second-order field equations. Therefore, the model (\ref{eq:actionGB}) must be describable in terms of Horndeski theory. Reference \cite{Kobayashi:2011nu} actually proved that it is equivalent to the model (\ref{eq:GBlag}) with $G_5=\alpha \ln |X|$, by comparing the field equations of both theories. We already studied this model in the previous section, because of its non-analytic character that a priori allows for non-trivial solutions. We showed there that no black hole solutions exist. However, we assumed a finite norm for the Noether current $J^\mu$. Reference \cite{Sotiriou:2014pfa} found solutions in this framework by allowing the scalar quantity $J^2$ to diverge near the horizon of the black hole. Therefore, the solutions described in this section will break two of the assumptions of the theorem: the $G_5$ function is not analytic at $X=0$, and the norm of the current $J^2$ is not finite everywhere. It should be underlined that the action (\ref{eq:actionGB}) has an interesting distinctive property with respect to other non-analytic models (\ref{eq:nanalytic}). It is the only one that has analytic field equations around $\varphi=0$ at the covariant level \cite{Sotiriou:2014pfa} (for the other models, it is only true in the case of spherical symmetry). 

The model (\ref{eq:actionGB}) has the following scalar field equation:
\be
\Box\varphi=-\dfrac{\alpha}{8\eta}\hat G
.\ee
Since $\hat G$ only vanishes in flat spacetime, this equation actually says that $\varphi$ cannot be trivial in a curved background. The no-hair theorem therefore does not apply, and Schwarzschild metric is not even present among the solutions to the field equations. One can get a perturbative or numerical solution to these field equations \cite{Kanti:2011jz,Campbell:1990fu,Sotiriou:2014pfa}. The scalar field can be made regular at horizon. To do so, one must tune the integration constant $Q$ of the scalar field equation $\nd_\mu J^\mu=0$ accordingly with the mass of the black hole. $Q$ is defined in Eq.~(\ref{eq:scalarcharge}). Note that for solutions with regular $J^2$, $J^r$ vanishes and $Q=0$. $Q$ is called a ``scalar charge'', because when it does not vanish,
\be
\varphi \underset{r \rightarrow \infty}{\sim} \dfrac{Q}{2\eta r}
\label{eq:phifarSZ}
.\ee
The solution with regular $\varphi$ has secondary hair, because $Q$ depends on the mass of the black hole. The (numerical) solution appears to have a singularity at non-zero radius, which can be shielded by a horizon for a large enough mass of the black hole. The norm of the current diverges as
\be
J^2\underset{r \rightarrow r_\tx{h}}{\sim} \dfrac{\alpha f'_\tx{h}}{r_\tx{h}^2(r-r_\tx{h})},
\ee
where $r_\tx{h}$ is the radius of the horizon and $f'_\tx{h}$ the derivative of $f$ at this point. The fact that $J^2$ is not bounded is unusual and puzzling. First, $Q\neq0$ implies that there is a continuous flux of scalar current towards the black hole, while the metric is assumed to remain static. In the case of a star, this is impossible because of the regularity of the solution at the origin. Here, the problem is hidden behind the black hole horizon. However, there is no obvious contradiction in starting from a static star with $Q=0$, and collapsing it down into a black hole with $Q\neq0$; indeed, there is in between a dynamical phase where $Q$ can evolve. Some work has been initiated for studying collapses in the model (\ref{eq:actionGB}), in \cite{Benkel:2016kcq}. However, this work only studied the decoupling limit in which the scalar field does not back-react on the metric. In this study, the scalar indeed has the behavior (\ref{eq:phifarSZ}), but this is automatic as soon as $Q\neq0$. To obtain significant results, one must study the problem in full generality, in particular at the horizon of the black hole, where regularity must be ensured.  


Concerning perturbations, we checked that the solution presented in \cite{Sotiriou:2014pfa} is linearly stable, as presented in Chapter \ref{ch:pert}. Finally, the slow rotation limit of the solution presented in this section also differs from general relativity. Indeed, the equation that governs the evolution of the rotation speed $\omega$, Eq.~(\ref{eq:HartleThorne}), reads \cite{Maselli:2015yva}:
\be
\begin{split}
&(2\alpha f \varphi'+r)\omega''+\left[3\alpha\varphi'f'+2\alpha f\varphi''+\dfrac{6\alpha}{r}f\varphi'-\alpha f\varphi'\dfrac{h'}{h}\right.
\\
&\left.+\dfrac{r}{2}\left(\dfrac{f'}{f}-\dfrac{h'}{h}\right)+4\right]\omega'=0,
\end{split}
\ee
where the ansatz (\ref{eq:statmetric})-(\ref{eq:statphi}) was used. As a consequence, frame-dragging experiments can a priori help distinguishing this theory from general relativity.

\section{No standard kinetic term}
\label{sec:quartic}

The third possibility to look for non-trivial black hole solutions is to give up the presence of a standard kinetic term, and to consider a model which only involves the other Horndeski and beyond terms. In this section, we exhibit a family of Lagrangian densities with a general relativistic metric, but a non-trivial scalar field. These black hole solutions are therefore stealth, and as such similar to the Schwarzschild solution discussed in Sec.~\ref{sec:stealth} in the context of the model (\ref{eq:John}). In contrast to this solution, however, the solutions presented here have time-independent scalar field configurations. For concreteness and simplicity, let us set 
\be
G_2=0,~~~~G_3=0,~~~~G_5=0,~~~~F_5=0,
\ee
with arbitrary (regular) $G_4$ and $F_4$. Doing so, one automatically gets rid of the canonical kinetic term, so that the no-hair theorem does not apply any more. The scalar field and the metric are still assumed to be static, Eqs.~(\ref{eq:statmetric})-(\ref{eq:statphi}). The equations of motion actually involve the density $X$ only, as one can see from the equations of Appendix \ref{ap:fieldeqs}. Extracting the combination $(rh)'f/h$ from both the $(rr)$ equation and the $J^r=0$ equation and equating the two expressions, 
one is left with
\be
\dfrac{G_{4X}}{G_{4X}+2XG_{4XX}+4X^2F_{4X}+8XF_4} =\dfrac{G_4}{G_4-2XG_{4X}-4X^2F_4} 
\label{eq:Xconst}
.\ee
It is remarkable that the above equation does not involve the radial parameter $r$. Equation (\ref{eq:Xconst}) should be understood as an equation on $X$ for a fixed choice of  $G_4$ and $F_4$. Let us assume for now that Eq.~(\ref{eq:Xconst}) has a solution, $X=X_0$. 
The fact that Eq.~(\ref{eq:Xconst}) does not involve $r$ means that $X$ is constant everywhere. This greatly simplifies the $(tt)$ equation, which can be immediately integrated. The solution reads
\begin{align}
h(r)&= 1-\dfrac{2m}{r},
\nn
f(r) &=\left(1-\dfrac{2m}{r}\right) \dfrac{G_4(X_0)}{G_4(X_0)-2 X_0 G_{4X}(X_0)-4X_0^2F_4(X_0)},
\end{align}
where $m$ is an integration constant and keeping in mind that $X_0$ must be a solution of Eq.~(\ref{eq:Xconst}). Therefore, the static and spherically symmetric solutions of a fully general quartic Horndeski theory boil down to a simple Schwarzschild metric, up to a solid angle deficit (corresponding to the constant in front of $f$). One can avoid a solid angle deficit (which would lead to a curvature singularity even for $m=0$) by requiring an extra condition on the functions $G_4$ and $F_4$, such that the factor in front of $f$ is 1. The combination of this condition with Eq.~(\ref{eq:Xconst}) gives
\begin{align}
0&=G_{4X}(X_0)+2X_0F_4(X_0),
\label{eq:constrquartic1}
\\
0&=G_{4XX}(X_0)+4F_4(X_0)+2X_0F_{4X}(X_0),
\label{eq:constrquartic2}
\end{align}
for some value $X=X_0$. Thus, infinitely many theories possess a stealth Schwarzschild black hole solution. Namely, all those which fulfill the constraints given in Eqs.~(\ref{eq:constrquartic1})-(\ref{eq:constrquartic2}) at some point $X_0$. It is very interesting to notice that the condition (\ref{eq:constrquartic1}) is exactly the one that allows for the propagation of gravitational waves at the speed of light, see Part \ref{part:3}. The models that pass this gravitational wave test are therefore also free from conical singularity in spherical symmetry.

There is a subclass among these models that has interesting properties in order to find exact stealth solutions; it is the subspace of $\{G_4,\:F_4\}$ theories where $F_4=0$. In this subclass, the models that possess such a stealth black hole are the theories with $G_{4X}(X_0)=0$ and $G_{4XX}(X_0)=0$. Any theory of the type
\be
G_4(X)=\zeta+\sum_{n>2} \beta_n (X-X_0)^n
\label{eq:examplestealth}
\ee
will allow for a Schwarzschild metric with a non-trivial scalar field. A more general examination of theories having $X=X_0$ with $G_{4X}(X_0)=0$ and $G_{4XX}(X_0)=0$ shows that any such theory allows for all Ricci-flat solutions, with a non-vanishing hidden scalar field. For instance, these theories admit as a solution the Kerr metric (here in Boyer-Lindquist coordinates):
\be
\begin{split}
\tx{d} s^2 = &- \left(1-\dfrac{2 m r}{r^2+a^2 \cos^2 \theta}\right)\tx{d} t^2 -\dfrac{4 m r a \sin^2 \theta}{r^2+a^2 \cos^2 \theta} \tx{d}t \,\tx{d}\phi +\dfrac{r^2+a^2 \cos^2\theta}{r^2-2 m r +a^2}\tx{d} r^2
\\
&+ \tx(r^2+a^2 \cos^2\theta)\,\tx{d}\theta^2 + \left(r^2+a^2+\dfrac{2 m r a^2 \sin^2\theta}{r^2+a^2 \cos^2\theta}\right)\sin^2 \theta \,\tx{d}\phi^2
,\end{split}
\label{eq:Kerr}
\ee
with a scalar field given by
\be
\begin{split}
\varphi (r,\theta)= &\sqrt{-2 X_0} \left[ a\sin \theta- \sqrt{a^2-2 mr+r^2}\right.
\\
&\left.-m~\tx{ln} \left(\sqrt{a^2-2 m r+r^2}-m+r\right)\right],
\end{split}
\ee
$a$ being the rotation parameter and $m$ the mass of the black hole. This scalar field is regular everywhere outside of the event horizon of the Kerr black hole. A remarkable characteristic of this class of solutions is that, even though the geometry is asymptotically flat, the scalar does not vanish at spatial infinity: its derivative $\varphi'$ tends towards a finite constant. This violates another assumption of the no-hair theorem; it is required that $\varphi'\rightarrow 0$ at spatial infinity. Therefore, the class of solutions discussed in this paragraph breaks two hypotheses. 

The black hole solutions found in this section are reminiscent of the properties of the ghost condensate in the field of a black hole \cite{Mukohyama:2005rw}. Indeed, for this theory, which contains only a non-trivial function $G_2(X)$ (while other functions are zero) with a minimum at some $X=X_0$, the situation is very similar. At the point $X=X_0$, the energy-momentum tensor for this theory becomes equivalent to that of the cosmological term. Adjusting $G_2(X)$ in such a way that the cosmological term is zero, one gets a stealth black hole solution, similar to the solutions presented in this paragraph. In the case of the $G_2(X)$ theory, there is a pathology though ---~the theory becomes non-dynamical at the point $X=X_0$. A way to overcome this pathology is to introduce higher-order terms. Therefore it is still to be understood whether a theory that satisfies (\ref{eq:constrquartic1})-(\ref{eq:constrquartic2}) is healthy at the point $X=X_0$.

\section{Camouflaged stars}
\label{sec:starhairstatic}

In this paragraph, we would like to exhibit a class of solutions that we call ``camouflaged'' stars, because the solution outside the star is the Schwarzschild metric with a vanishing scalar field, while the interior solution deviates from general relativity. We will still consider a static ansatz for both the metric and scalar field, as in the rest of this chapter. Such solutions circumvent the Cauchy-Kowalewski theorem because of the higher order branch structure of Eq.~(\ref{eq:branches}). Let us recall that it was proven in Sec.~\ref{sec:Jr} that the radial component of the scalar current must vanish for stars, and this can be achieved either through setting $\varphi'=0$ or $\mc{J}=0$, where $\mc{J}$ is defined in Eq.~(\ref{eq:curlyJ}). Here, we are interested in the non-trivial branch $\mc{J}=0$. We thus want to circumvent the argument of Sec.~\ref{sec:nohairstar}. Rewriting Eq.~(\ref{eq:curlyJGR}) without normalizing the kinetic term, i.e., considering a term $\eta X\subseteq G_2$ instead of $X\subseteq G_2$, one has:
\be
\mc{J}=f\left(\dfrac{2\beta}{\Mp^2} P-2\beta\Lb-\eta\right)
\label{eq:curlyJGR2}
.\ee
If the transition from the outer branch $\varphi'=0$ to the inner branch $\mc{J}=0$ happens precisely at the surface of the star, it is natural to expect that the derivative of the pressure vanishes at the same point. In this case, the argument about the continuity of $\mc{J}'$ given in Sec.~\ref{sec:nohairstar} does not work any more, and one can a priori find solutions that jump from one branch to the other. The transition happens at the surface of the star if the critical pressure $P_1$, Eq.~(\ref{eq:Pc}), vanishes:
\be
P_1=\dfrac{\Mp^2}{2\beta}(\eta+2\beta\Lb)=0.
\ee
Apart from a fine tuning of the parameters, this is possible only if $\eta=0$ and $\Lb=0$. A vanishing bare cosmological constant is not problematic if one is examining asymptotically flat solutions. Setting $\eta=0$ removes the standard kinetic term and breaks an assumption of the no-hair theorem, which is exactly what we want. Therefore, in the rest of this paragraph, we will set $\eta=0$ and $\Lb=0$. Then, outside of the star, since $P=0$, Eq.~(\ref{eq:curlyJGR2}) shows that $\mc{J}$ automatically vanishes. This means that one does not even have to change of branch at the surface of the star: one just follows the branch $\mc{J}=0$ everywhere, which happens to coincide with the trivial branch $\varphi'=0$ in the outer region. Therefore, camouflaged star solutions should exist for any Horndeski or beyond theory without standard kinetic term and bare cosmological constant.

\begin{figure}[ht]
\begin{center}
\includegraphics[width=\textwidth]{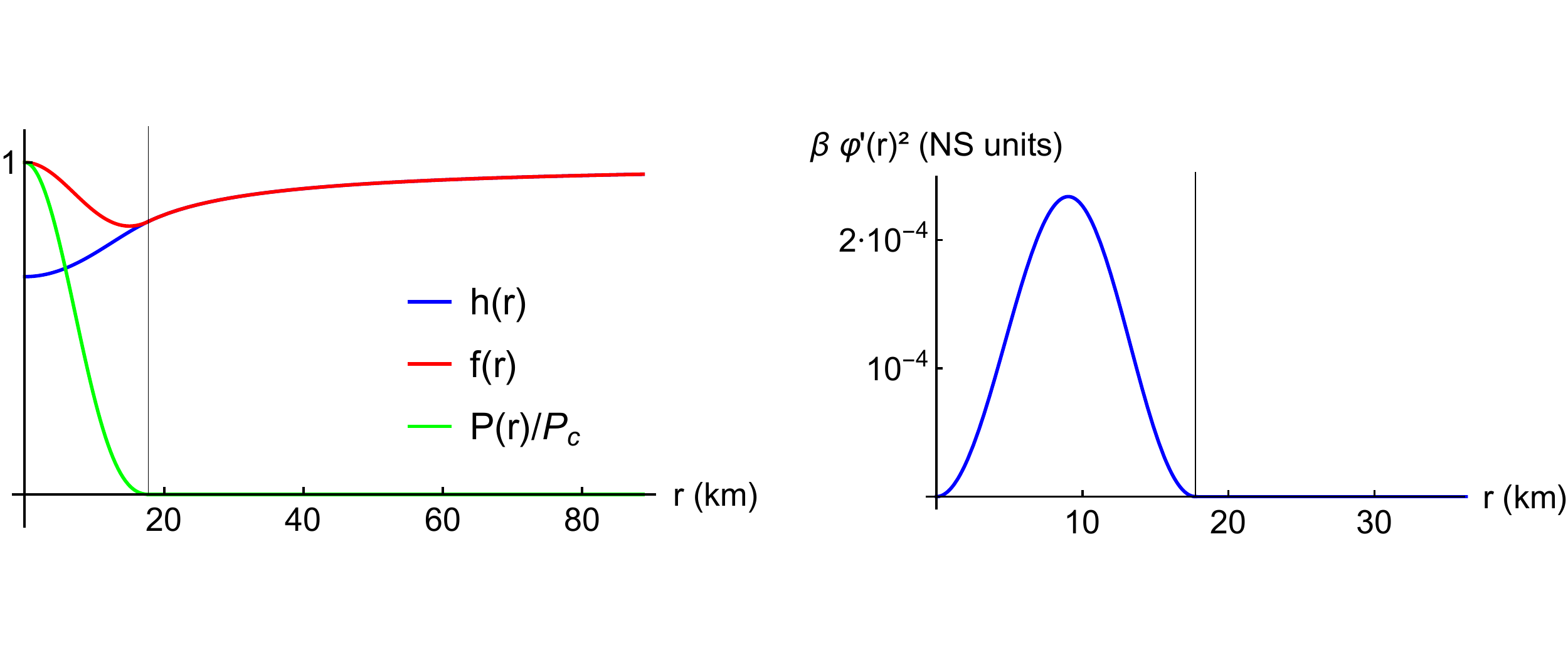}
\caption{Camouflaged star for the model (\ref{eq:Johnstar}). As in Fig.~\ref{fig:profileJohnq}, the left panel shows the metric and normalized pressure. To the right is plotted the quantity $\beta\varphi'^{\,2}$. Again, the vertical black line corresponds to the surface of the star. The central density is still $\rho_\tx{B;c}=5\cdot10^{17}$ kg/m$^3$. This star has gravitational mass $m=1.1\; M_\odot$ and radius $R_\ast=18$ km.}
\label{fig:profileJohnqzero}
\end{center}
\end{figure}

\begin{figure}[!ht]
\begin{center}
\includegraphics[width=10cm]{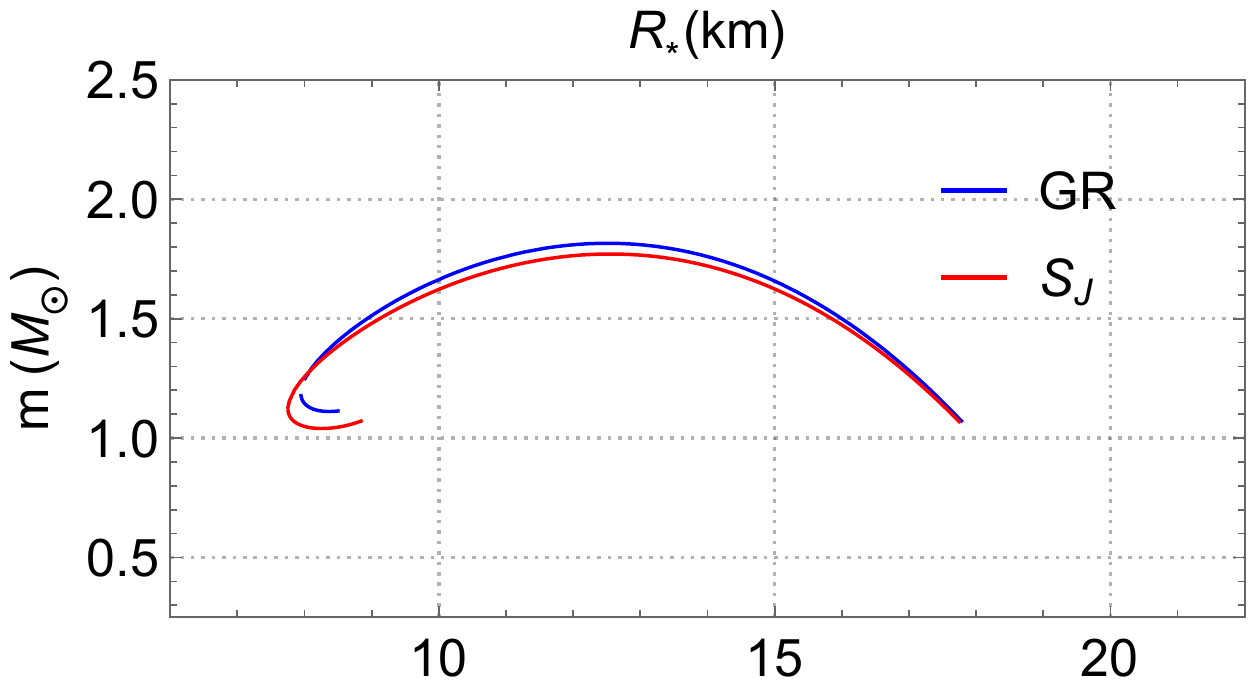}
\caption{Mass-radius relation for the polytropic equation of state. The central density varies again from $3\cdot10^{17}$ to $3\cdot10^{21}$ kg/m$^3$. The blue curve corresponds to general relativity, while the red one is derived from the action $S_\tx{J}$. The difference between the two theories is very mild. Given the uncertainties on the internal structure of neutron stars, the two solutions are virtually indistinguishable.}
\label{fig:MRqzero}
\end{center}
\end{figure}

Such a solution was exhibited by Cisterna et al. in \cite{Cisterna:2015yla}, again for the action (\ref{eq:Johnstar}), that we studied in detail in Sec.~\ref{sec:starhairtdep}. The notable difference between Sec.~\ref{sec:starhairtdep} and the solution presented here is that the scalar field does not depend on time any more ($q=0$). The characteristics of this solution are detailed in Figs.~\ref{fig:profileJohnqzero} and \ref{fig:MRqzero}.

Let us remark that one could straightforwardly construct similar solutions in the quartic Horndeski and beyond sector, for the theories that fulfill the constraints (\ref{eq:constrquartic1})-(\ref{eq:constrquartic2}). It suffices to solve numerically the field equations inside the star. This would be a way to check explicitly that theories with $G_{4X}(X_0)=0$, $G_{4XX}(X_0)=0$ ---~in the notations of Eqs.~(\ref{eq:constrquartic1})-(\ref{eq:constrquartic2})~--- are not equivalent to general relativity.

In passing, nothing opposes to the existence of solutions with non-analytic $G_i$--$F_i$ functions, as was done for black holes in \cite{Babichev:2017guv}. On the other hand, the black hole solution presented in \cite{Sotiriou:2014pfa} for the linear coupling between $\varphi$ and the Gauss-Bonnet density cannot describe the exterior of a star. Indeed, it possesses a non-vanishing current, and we showed this is impossible in Sec.~\ref{sec:Jr}. Rather, for this theory, the faraway behavior of the star solutions is the one derived in \cite{Amendola:2007ni,Babichev:2017guv}, with a scalar decaying as $\varphi\underset{r\to\infty}{\sim}\varphi_\infty/r^4$, and $1/r^7$ corrections to the Schwarzschild metric.

\section{Conclusions}

This chapter was devoted to finding static, spherically symmetric and asymptotically flat solutions with a static scalar field. Such configurations were the main target of the no-hair theorem given in Chapter \ref{ch:nohair}. As a result, it is difficult to find non-trivial solutions, and only specific models can accommodate such solutions ---~as opposed to the solutions of Chapter \ref{ch:tdep}, where hair was generically non-trivial, and we only examined the simplest models. 

A first way to generate scalar hair is to work with models that are built using specific functions $G_i$ and $F_i$ as in Sec.~(\ref{eq:nanalytic}). In fact, for this class of solutions, $\varphi$ has to be non-trivial in order to achieve a finite norm of the current. We found six different models in the Horndeski and beyond theory (four if restricted to the Horndeski theory) which possibly admit hairy black holes, see Eq.~(\ref{eq:nanalytic}). As two illustrative examples, we found black hole solutions with secondary hair in a subclass of quartic Horndeski and beyond Horndeski theory. All observable quantities made of the metric and the scalar field are well-behaved, both at spatial infinity and at the horizon. The scalar field decays like $1/r$, and back-reacts on the metric through a rapidly damped contribution in $1/r^2$. In the same spirit, we also investigated in detail the quintic theory which is in the family (\ref{eq:nanalytic}) under study. Interestingly, it is equivalent to a linear coupling between the scalar field and the Gauss-Bonnet density, whereupon the scalar field is sourced by the Gauss-Bonnet curvature scalar away from the trivial configuration. Within this subclass, no black hole solutions  were found with a regular norm of the current $J^2$, although the solutions asymptotically agree with Dirac sourced star solutions found previously in this theory \cite{Amendola:2007ni}. This might mean that the finiteness of the norm is not relevant for black hole solutions. Indeed, for the same theory, solutions with a diverging norm of the current have been obtained in \cite{Sotiriou:2014pfa}, thus violating another assumption of the no-hair theorem. They exhibit secondary scalar hair, where the scalar charge is fixed according to the mass in order for the scalar field not to diverge at horizon. This solution, although puzzling, exhibits no pathology. A third way to build black holes with hair, presented in Sec.~\ref{sec:quartic}, is to remove the canonical kinetic term from the action. Although this is not very natural from an effective field theory point of view, this corresponds to breaking another assumption of the theorem. We allowed arbitrary $G_4$ and $F_4$ quartic Horndeski and beyond functions. We obtained the generic solution, which is described by a Schwarzschild metric. In the case of $G_4$ alone, inspection of the field equations shows that regularity conditions actually allow for any Ricci-flat solution with a constant density $X$. The stationary Kerr metric is indeed solution to the field equations, with a non-trivial scalar field profile. Finally, we briefly examined star solutions in models which exactly recover general relativity as a vacuum solution, but allow for two branches when matter is present. Again, it would be interesting to study the energetic properties of such objects.

The study of the perturbations of such solutions, however, could reveal hidden pathologies of the theories or/and solutions. For instance, an unpleasant feature of the family of Lagrangians (\ref{eq:nanalytic}) is that they are non-analytic around Minkowski vacuum. The solutions of Sec.~\ref{sec:quartic} with $F_4=0$, on the other hand, are strongly coupled around a Ricci-flat background. These peculiarities, unseen at the level of solutions, would probably play a role at the level of perturbations.

\newpage
\thispagestyle{empty}
\cleardoublepage
\part{Linear perturbations}
\label{part:3}

\newpage
\thispagestyle{empty}
\cleardoublepage
\noindent As mentioned in the introduction, the existence of some solution with scalar hair does not make it a physically relevant black hole candidate. One must in principle check that it can result from gravitational collapse, and first that it is stable. A fully non-linear treatment of stability is a mathematically monumental task. So far, in general relativity, only Minkowski spacetime was proven to be fully stable \cite{Christodoulou:1993uv}, which goes to show the difficulty of the problem. A slightly easier task is to establish linear stability, i.e., to show that initial data close enough to some given solution are not evolved too far by the linearized field equations, and eventually asymptote the solution in question. This remains very difficult, and, in general relativity, it was only established that Schwarzschild solution is linearly stable in this sense \cite{Dafermos:2016uzj}.

We will restrict our analysis to a simplified version of linear stability, based on the quadratic expansion of the Horndeski and beyond action. When expanded at second order, this action contains a massless spin-2 degree of freedom, with two polarizations, and a scalar degree of freedom. Each of them obeys a wave equation, and thus propagates in its own effective metric. There are three pathologies one wants to avoid:
\begin{itemize}
\item \textit{gradient instabilities}. This happens when the signature of (one of) the effective metric(s) is not Lorentzian, and thus some degree of freedom has exponentially growing modes;
\item \textit{tachyonic instabilities}. Again, some degree of freedom has exponentially growing modes, but this time because the associated mass term in the quadratic action has a wrong sign;
\item \textit{ghosts}. A ghost is a degree of freedom which can acquire an arbitrarily negative energy; as soon as it interacts with a healthy degree of freedom, an instability appears. The energy of the ghost decreases while the energy of the healthy degree of freedom increases of the same amount. Then, the modes of these two fields are filled by an infinity of particles, which immediately destabilizes the solution.
\end{itemize}
In this thesis, we focus only on gradient instabilities and ghosts, because they are more dangerous than tachyonic instabilities. To understand why, let us consider some field $\varphi$ obeying a Klein-Gordon equation (in two-dimensional flat spacetime for simplicity). Let us decompose it in modes of given momentum $k$, $\varphi_k\propto \tx{e}^{i(\omega_kt-kx)}$. If the sign in front of $\varphi''$ is the wrong one in the field equation (which corresponds to a gradient instability):
\begin{align*}
\ddot{\varphi}+c^2\varphi''+M^2\varphi=0
,\end{align*}
then the modes with $k>M/c$ will grow exponentially on a time-scale $\tau=(k^2/c^2-M^2)^{-1/2}\to0$ when $k\to\infty$. There is no lower bound on this time scale, i.e., the solution is immediately destabilized. On the other hand, if the sign in front of $\varphi''$ is correct while the sign of the mass term is wrong (tachyonic instability):
\begin{align*}
\ddot{\varphi}-c^2\varphi''-M^2\varphi=0
,\end{align*}
modes with $k<M/c$ will grow exponentially on a time-scale of $\tau=(M^2-k^2/c^2)^{-1/2}<1/M$. Thus, the instability might be kept under control if $M$ is small enough. This is why the study of the kinetic operator must come before the study of the mass terms, although the latter is important also. Note that a ghost degree of freedom also leads to an infinitely rapid instability, at the quantum level. For instance, if the ghost field is coupled to some positive energy degree of freedom through gravity, a virtual graviton may decay into ghost particles and healthy ones. The production rate is infinite if no cutoff is imposed in the ultraviolet sector of the theory \cite{Cline:2003gs}.

In common lore, the presence of a ghost is often deduced from the absence of lower bound on the Hamiltonian density. We discuss this criterion in Chapter \ref{ch:pert}, and show that we must abandon it for a more generic one, as we argued in \cite{Babichev:2018uiw}. In the same chapter, we compute the effective metric for gravitational waves and scalar perturbations of some solutions presented in Chapter \ref{ch:tdep}. We apply the new criterion to establish the stability domain of these solutions. In parallel, the linear perturbations tell us at which speed gravitational waves propagate; therefore, thanks to the tools developed in Chapter \ref{ch:pert}, it is possible to examine the speed of gravitational waves in light of the recent binary neutron star merger, GW170817. This is the aim of Chapter \ref{ch:wave}, based on our work \cite{Babichev:2017lmw}.
\newpage
\thispagestyle{empty}
\cleardoublepage
\chapter{Black hole stability}\label{ch:pert}

The notion of causal cone and effective metric will be essential to the stability arguments, so let us dwell on this point before entering details in the body of the chapter. Since no mass term is associated with the rank-2 tensor mediating gravity, general relativity is dynamically described by a massless spin-2 degree of freedom. The effective action describing this degree of freedom can be found by expanding the action (\ref{eq:GRaction}) up to the second order, around some given background solution. According to this effective action, the spin-2 degree of freedom obeys a second-order differential equation (see Appendix \ref{ap:diagonal} for more details). This kinetic operator might be encoded in an effective metric, which defines then a causal cone of propagation. In the case of general relativity, this effective metric is the metric $g_{\mu\nu}$ itself. A second and independent assumption of general relativity is that all matter fields universally couple to this metric, in order to satisfy the weak equivalence principle. These postulates imply that electromagnetic and gravitational waves propagate on the same causal cones, i.e., with the same speed. 

However, modified gravity degrees of freedom ---~including spin-2 ones~--- propagate in an effective metric which can be different from that of general relativity. Each degree of freedom now a priori comes with its own kinetic operator, or equivalently its own effective metric\footnote{We do not consider here Lorentz-breaking theories \cite{Horava:2009uw,Jacobson:2008aj}, where equations of motion can be of higher order. Also, in beyond Horndeski theory and degenerate higher-order scalar-tensor theories, one may get Euler-Lagrange equations of third order; at least for beyond Horndeski models that can be disformally related to Horndeksi theory, the equations can be cast in a form that involves second-order time derivatives only \cite{Gleyzes:2014qga}.}. Provided that the equation of motion is hyperbolic, each effective metric defines a causal cone of propagation.
These causal cones are inherently different for different spins ---~scalar, vector, or tensor~--- and their structure determine whether the degrees of freedom are healthy or not. Note that for Horndeski and beyond theory, the scalar and tensor degrees of freedom mix together, and it is in general only for the most symmetric backgrounds that one manages to demix them. Furthermore, in modified gravity theories, matter is still assumed to couple universally to a single metric in order to pass stringent fifth-force experiments. This introduces the matter causal cone, in addition to gravity cones, and the physical metric, associated to geodesic free-fall, which matter couples to. 

Brans-Dicke theory, Eq.~(\ref{eq:BDaction}), or more generally scalar-tensor theories, Eq.~(\ref{eq:STaction}), constitute simple examples of theories with different metrics for matter and spin-2 perturbations. When one works with the variable $\tilde g_{\mu\nu}$ of Eq.~(\ref{eq:STaction2}), one is said to work in the Jordan (or physical) frame; $\tilde g_{\mu\nu}$ is the metric which defines the causal cone of matter. On the other hand, the scalar-tensor action may be rewritten in terms of $g_{\mu\nu}$ as defined in Eq.~(\ref{eq:gEinstein}). If one uses $g_{\mu\nu}$ as a variable, one works in the so-called Einstein frame, because the dynamical part of the action becomes a mere Ricci scalar for the spin-2 degree of freedom. As a consequence, the causal cone of this degree of freedom is defined by $g_{\mu\nu}\neq \tilde g_{\mu\nu}$. However, in this case, the two cones still coincide because $\tilde g_{\mu\nu}$ and $g_{\mu\nu}$ are conformally related. As we will see, this is no longer the case for Horndeski and beyond theory. 

The effect of multiple causal cones and mixing is that different species can now have subluminal or superluminal propagation\footnote{These multiple possibilities have actually been recently constrained by the gravitational wave event GW170817, but we will not say more for the moment (this constraint is treated in Chapter \ref{ch:wave}).} (luminality being defined with respect to the matter causal cone, the one felt by electromagnetic waves).  In this chapter, we will see how, starting from the causal cone structure of propagating degrees of freedom, one can infer if the perturbations in question are healthy ---~in other words that they do not generate ghost or gradient instabilities. In particular, the sign of the determinant of the effective metric defines the hyperbolicity condition, which if satisfied, means that a particular solution is safe from imaginary speeds of propagation and therefore gradient instabilities. On the other hand, the local orientation of the cone tells us about the absence/presence of a ghost degree of freedom.

A complementary way to find the good or sick nature of propagating degrees of freedom is often described via the Hamiltonian density of the degrees of freedom in question. Once the effective action of some degree of freedom is known, one defines the conjugate momentum and writes down the Hamiltonian density of the associated field. It is known that if the Hamiltonian density is bounded from below, then there exists a stable ground state. The contrary is often assumed to be true: if a Hamiltonian density is unbounded from below, then the system is unstable and admits a ghost instability. One of the main aims of this chapter is to explicitly show that this inverse statement is not always true. In other words, if a Hamiltonian density is unbounded from below, this does not necessarily signify that it generates a ghost instability. 

Section \ref{Sec2} explains in detail why the reciprocal Hamiltonian criterion may fail. A sounder criterion, based on causal cones, is proposed and used to re-derive the stability conditions of k-essence, as an application. We will then move on, in Sec.~\ref{sec:Johnstab}, to apply the causal cone criterion to the stability of a Horndeski model, namely the action (\ref{eq:John}), and the associated Schwarzschild-de Sitter solution, Eqs.~(\ref{eq:CdSexact})--(\ref{eq:qdSJohn}). The mixed combination of space and time dependence for the scalar, as well as the higher order nature of the theory, leads to causal scalar and tensor cones which are quite complex. This is the reason why the Hamiltonian-based analysis proposed in \cite{Ogawa:2015pea} (as well as Refs.~\cite{Takahashi:2015pad,Takahashi:2016dnv,Kase:2018voo} that use the same arguments\footnote{Reference \cite{Maselli:2016gxk} also uses these arguments, but it proves the stability of the odd-parity modes outside neutron stars, and this is correct.} as in \cite{Ogawa:2015pea}) gave the wrong conclusion, stating the instability of such black holes. Although the Hamiltonian density associated with the spin-2 degree of freedom is unbounded from below in Schwarzschild coordinates, we will see that it is bounded from below in an appropriate coordinate system. The spin-2 and matter causal cones indeed keep compatible orientations. We will complete this analysis by deriving the scalar causal cone, and by showing that it also has a compatible orientation with the two previous cones for a certain range of parameters of the model. We present conclusions in Sec.~\ref{sec:concpert}.

\section{Hamiltonian vs stability}
\label{Sec2}

The reason why the Hamiltonian criterion for instability may fail is simple, although it goes against standard lore originating from particle physics or highly symmetric backgrounds associated to Friedmann-Lema\^{\i}tre-Robertson-Walker cosmology. The Hamiltonian density is not a scalar quantity, and therefore depends on the coordinate system it is associated with. As such, we will explicitly see that Hamiltonian densities can be unbounded from below, but under a coordinate transformation can be transformed to a bounded density. The key point will be the coordinate system on which the Hamiltonian is to be defined in relation to the effective causal cones. The coordinate system will have to be of a certain ``good'' type in order for the Hamiltonian density to be conclusive. 

For our purposes, we will restrict ourselves to configurations where essentially the problem is mathematically 2-dimensional. This includes the case for planar, cylindrical or spherical symmetry, for example. A ``good'' coordinate system will first involve the existence of a common timelike direction for all causal cones. Secondly, it will involve the existence of a common spacelike direction exterior to all causal cones\footnote{A causal cone represents an open set whose interior is bounded by the characteristics of the cone. The complementary of this set with boundary is an open set which is the exterior of the cone.}. If such a coordinate system exists, then we will show that the total Hamiltonian density is bounded from below in this coordinate system, and the solution is stable. If such a coordinate system does not exist, on the contrary, then the Hamiltonian density is always unbounded from below. The relevant criteria emerging from the causal cones will inevitably lead to the knowledge of ghost or gradient instabilities present in the system.

\subsection{Causal cones and Hamiltonian in a general coordinate system}

Let us consider any possible effective metric $\mathcal{S}_{\mu\nu}$\footnote{We use the notation $\mathcal{S}_{\mu\nu}$ for the effective metric, because we will first apply this discussion to k-essence, where the degree of freedom is a $\mc{S}$calar. $\mathcal{G}_{\mu\nu}$ will be used in Sec.~\ref{sec:effmetrics} to denote the effective metric in which the spin-2 degree of freedom ($\mc{G}$ravitational waves) propagates.} in which some field $\chi$ propagates. $\chi$ is to be viewed as the perturbation of some background field, and $\mc{S}_{\mu\nu}$ is derived from the second-order (in terms of $\chi$) action of the theory under consideration. Locally, $\mc{S}_{\mu\nu}$ defines a causal cone through $\mc{S}_{\mu\nu} \tx{d}x^\mu \tx{d}x^\nu = 0$, or equivalently through $\mc{S}^{\mu\nu} k_\mu k_\nu = 0$ for a wave vector $k_\mu$ ($\mc{S}^{\mu\nu}$ denotes the inverse of $\mc{S}_{\mu\nu}$). Additionally, let us consider the physical metric $g_{\mu\nu}$, to which matter fields are universally coupled. To simplify,
we shall assume that this standard metric $g_{\mu\nu}$ is flat. This is not restrictive, since one can always chose a local system of coordinates where this is the case. If the Lagrangian defining the dynamics of $\chi$ reads
\be
\mathcal{L}^{(2)} =-\dfrac{1}{2}\,\mathcal{S}^{\mu\nu} \partial_\mu\chi\partial_\nu\chi,
\label{eq:chiLag}
\ee
then the \textit{conjugate momentum} of $\chi$ is defined as
\begin{equation}
p = \frac{\partial \mathcal{L}^{(2)}}{\partial \dot\chi}
=-\mathcal{S}^{00}\dot\chi - \mathcal{S}^{0i}\partial_i\chi,
\label{Eq:p}
\end{equation}
and the \textit{Hamiltonian density} associated to $\chi$ is
\begin{equation}
\mathcal{H}^{(2)} = p\dot\chi - \mathcal{L}^{(2)} =
-\,\frac{1}{2\mathcal{S}^{00}}
\left(p+\mathcal{S}^{0i}\partial_i\chi\right)^2
+\frac{1}{2}\,\mathcal{S}^{ij} \partial_i\chi \partial_j\chi.
\label{Eq:H2}
\end{equation}
Note that its positiveness depends only on $\mathcal{S}^{00}$ and
$\mathcal{S}^{ij}$, but not on the mixed components
$\mathcal{S}^{0i}$, although we shall see that they are actually
crucial for the stability analysis. Stability is indeed a
physical (observable) statement, which should be coordinate
independent, whereas the Hamiltonian density is not a
scalar and depends thus on the coordinate system.

\begin{figure}
\centering
\includegraphics[width=\textwidth
]{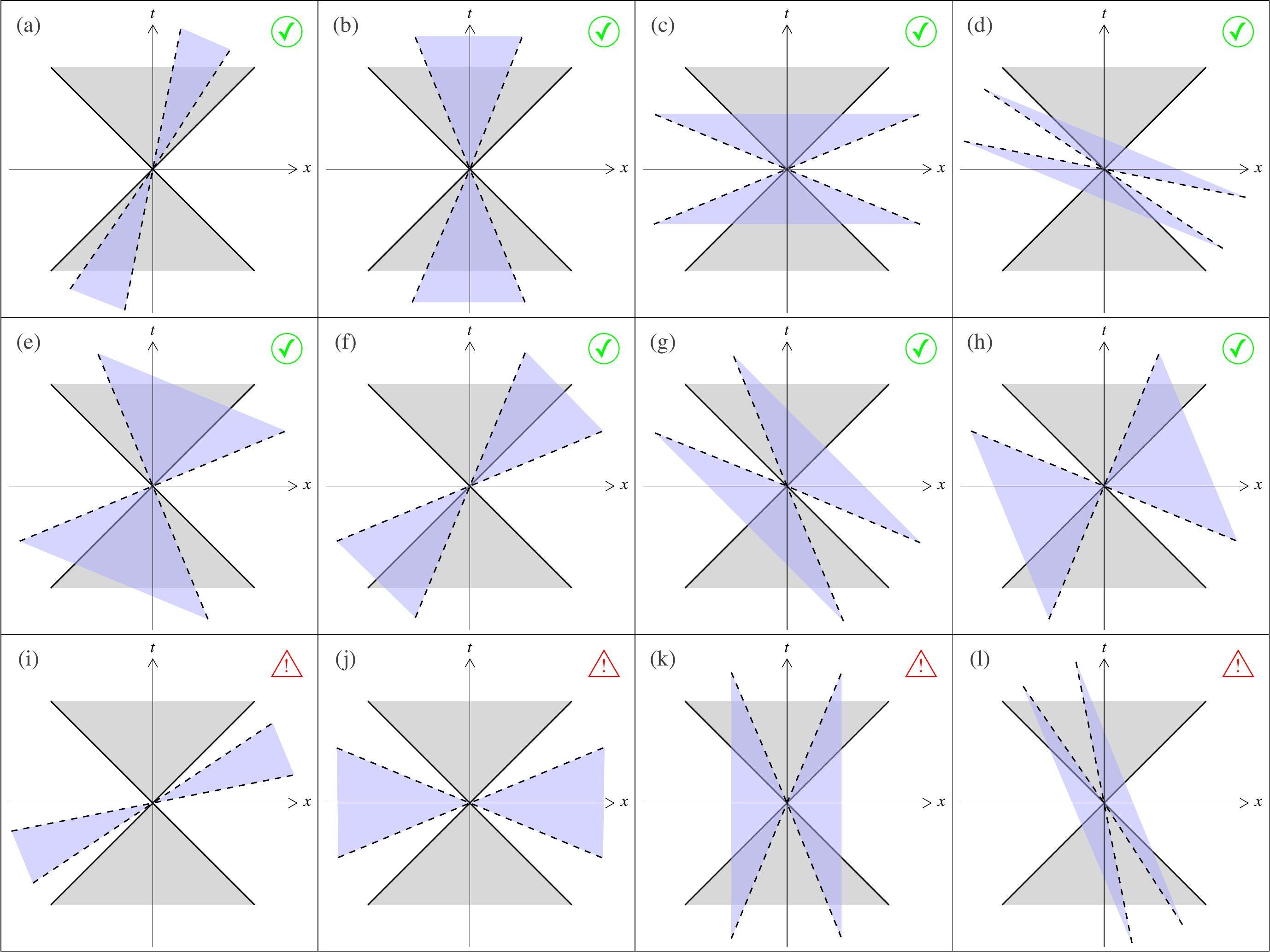}
\caption{All possible relative orientations of two causal cones, defined by $\mc{S}_{\mu\nu}$ (dashed blue) and $g_{\mu\nu}$ (solid gray). The coordinate system is chosen so that $g_{\mu\nu}$ is flat, and thus its characteristics appear at $\pm45^\circ$. We do not plot the equivalent configurations exchanging left and right, and do not consider the
limiting cases where some characteristics coincide. The first row
(a)--(d) are safe cases in which the two metrics can be
diagonalized simultaneously by an appropriate choice of
coordinates ---~corresponding then to panels (b) or (c). Although
the kinetic contribution to their Hamiltonian density is
unbounded from below in cases (a) and (d), it is positive in (b)
and (c). The second row (e)--(h) are again safe cases, for which
the kinetic contribution to the Hamiltonian density can be proven
to be positive in an appropriate coordinate system, actually
corresponding to case (e), but the two metrics cannot be
simultaneously diagonalized ---~two quadratic forms can always be simultaneously diagonalized if at least one of them is positive (or negative) definite; here both metrics have hyperbolic signature, and this is the reason why the non-simultaneously diagonalizable cases (e)--(h) are possible. The third row (i)--(l) are unstable cases, for which the two metrics can be simultaneously diagonalized as in (j) and (k), but they have then opposite signatures in this $(t,x)$ subspace. Their total Hamiltonian density remains unbounded from below in all coordinate systems.
}
\label{fig:CausalCones}
\end{figure}

To simplify even further the discussion, let us assume that
$\mathcal{S}^{\mu\nu}$ is of the form
\begin{equation}
\begin{bmatrix}
\mathcal{S}^{00} & \mathcal{S}^{01} & 0 & 0 \\
\mathcal{S}^{01} & \mathcal{S}^{11} & 0 & 0 \\
0 & 0 & \mathcal{S}^{22} & 0 \\
0 & 0 & 0 & \mathcal{S}^{33}
\label{Eq:matrixG}
\end{bmatrix},
\end{equation}
with $\mathcal{S}^{22}\geq 0$ and $\mathcal{S}^{33}\geq 0$, and
let us focus on the $(t,x)$ subspace as shown in
Fig.~\ref{fig:CausalCones}.
In the neighborhood of a spherical
body, for instance, it is natural to choose spherical coordinates
where $\mathcal{S}^{\theta\theta} = 1/r^2$ and
$\mathcal{S}^{\phi\phi} = 1/(r^2\sin^2\theta)$, the
difficulties being restricted to the $(t,r)$ subspace. 
In order
for this metric to define a cone, with non-empty interior and
exterior, it is necessary that its determinant be negative:
\begin{equation}
\mc{D}= \mathcal{S}^{00}\mathcal{S}^{11}
- (\mathcal{S}^{01})^2 < 0.
\label{Eq:Det}
\end{equation}
Note that this hyperbolicity condition depends on the
off-diagonal component $\mathcal{S}^{01}$, contrary to the sign
of Hamiltonian density (\ref{Eq:H2}) above. The inverse
$\mathcal{S}_{\mu\nu}$ of matrix (\ref{Eq:matrixG})
(this is \textit{not} the tensor $\mathcal{S}^{\mu\nu}$ with indices lowered by $g_{\mu\nu}$, i.e., $\mathcal{S}_{\mu\nu}\neq g_{\mu\lambda}g_{\nu\rho}\mathcal{S}^{\lambda\rho}$)
reads in the $(t,x)$ subspace
\begin{equation}
\dfrac{1}{\mc{D}}\begin{bmatrix}
\mathcal{S}^{11} & -\mathcal{S}^{01} \\
-\mathcal{S}^{01} & \mathcal{S}^{00}
\end{bmatrix}.
\label{Eq:Ginverse}
\end{equation}
One can thus conclude that when $\mathcal{S}^{\mu\nu}$ indeed
defines a cone, then $\mathcal{S}_{00}$ has the opposite sign of
$\mathcal{S}^{11}$, and $\mathcal{S}_{11}$ the opposite sign of
$\mathcal{S}^{00}$.

Let us now consider the exhaustive list of possible cone orientations of
Fig.~\ref{fig:CausalCones}. In the situation of panel (a), the
time axis is outside the dashed blue cone defined by
$\mathcal{S}^{\mu\nu}$. This means that $\mathcal{S}_{00}\, \tx{d}t\,
\tx{d}t > 0$, and therefore $\mathcal{S}^{11} < 0$. This implies that
the Hamiltonian density (\ref{Eq:H2}) is unbounded from below
because of the contribution of $\mathcal{S}^{ij}
\partial_i\chi \partial_j\chi/2$, when $\partial_1\chi$ is large
enough (and $p$ is chosen to compensate
$\mathcal{S}^{0i}\partial_i\chi$). This conclusion remains the
same for all panels of this figure in which the time axis is
outside the dashed cone, namely (f), (h), (i), (j), and (k). On
the contrary, when the time axis is within the dashed cone (in
all other panels of Fig.~\ref{fig:CausalCones}), this corresponds
to $\mathcal{S}^{11} > 0$, and the second term of the Hamiltonian density
(\ref{Eq:H2}) is thus positive.

Similarly, in the situation of panel (d), the $x$ axis is within
the dashed cone, therefore $\mathcal{S}_{11}\, \tx{d}x\, \tx{d}x < 0$,
which implies $\mathcal{S}^{00} > 0$. In this case, the
Hamiltonian density (\ref{Eq:H2}) is unbounded from below because
of the contribution of its first term
$-\left(p+\mathcal{S}^{0i}\partial_i\chi\right)^2/
\left(2\mathcal{S}^{00}\right)$. This conclusion remains the same
for all panels in which the $x$ axis is inside the dashed cone,
namely (g), (h), (j), (k), and (l). In all other panels, the $x$
axis is outside the dashed cone, therefore $\mathcal{S}^{00} < 0$
and the first term of the Hamiltonian density (\ref{Eq:H2}) is thus positive.

Note that panels (h), (j) and (k) have both their time axis
outside the dashed cone and their $x$ axis within it. This means
that the Hamiltonian density (\ref{Eq:H2}) is always negative,
while that corresponding to matter (coupled to $g_{\mu\nu}$ and
propagating thereby in the solid grey cone) is always positive.
It thus naively seems that any coupling between matter and $\chi$,
or any indirect coupling via another field (for instance
gravity), will lead to deadly instabilities. This is indeed the
case for panels (j) and (k), but not for panel (h). Indeed, if
one chooses another coordinate system such that the new time $t'$
lies within the intersection of both cones (superposition of the
grey and blue regions), and the new spatial direction $x'$ is
outside both cones (white region), then one gets simultaneously
the four conditions $g'^{00} < 0$, $g'^{11} > 0$,
$\mathcal{S}'^{00} < 0$, and $\mathcal{S}'^{11} > 0$. Therefore,
both the Hamiltonian density (\ref{Eq:H2}) for the degree
of freedom $\chi$ and its analogue for matter are positive in
this coordinate system. This suffices to prove that no
instability can be caused by the kinetic terms in the situation
of panel (h).

Let us recall that when a total Hamiltonian density (including
all interacting fields) is bounded from below, then the
lowest-energy state is necessarily stable. It is indeed
impossible to reach a higher energy state (for any field) without violating
energy conservation. But note that the converse theorem does not
exist, as underlined by the reasoning above: a Hamiltonian
density which is unbounded from below does not always imply an
instability. 

Although the Hamiltonian density is not bounded by
below in the situations corresponding to panels (a), (d), (f),
(g) and (h) of Fig.~\ref{fig:CausalCones}, there exists a choice
of coordinates mapping them to panels (b), (c) or (e), where the
new total Hamiltonian density is bounded from below. This suffices
to guarantee the stability of the lowest-energy state, as
computed in this new coordinate system. The only generically
unstable cases correspond to the third row of
Fig.~\ref{fig:CausalCones}, panels (i) to (l), because their
total Hamiltonian density is never bounded from below in any
coordinate system. They are such that the matrix
$\mathcal{S}^{\mu\lambda} g_{\lambda\nu}$ is diagonalizable and
possesses two negative eigenvalues. Conversely, it is
easy to write the inequalities needed on the components of the
effective metric $\mathcal{S}^{\mu\nu}$ to be in the eight safe
cases corresponding to the first two rows, panels (a) to (h): in
addition to the hyperbolicity condition (\ref{Eq:Det}), one just
needs
\begin{equation}
\mathcal{S}^{00} < \mathcal{S}^{11}\quad\text{and/or}\quad
|\mathcal{S}^{00}+\mathcal{S}^{11}| < 2 |\mathcal{S}^{01}|,
\label{eq:condS}
\end{equation}
when focusing on the $(t,x)$ subspace in a coordinate system such
that $g_{\mu\nu} = \text{diag}(-1,1)$.
But these inequalities are less enlightening than
Fig.~\ref{fig:CausalCones} itself, in which it is immediate to
see whether the two causal cones have both a common exterior
(when one should specify initial data) and a common interior.
When one chooses new coordinates such that time lies within the
cone intersection, and space is outside both cones, then the
total Hamiltonian density caused by kinetic terms becomes
positive.

\subsection{An illustration: the effect of boosts}

To understand this better, let us just consider the boosts of
special relativity in flat spacetime, instead of the arbitrary
coordinate transformations allowed in general relativity. Under these boosts, 
the metric $g^{\mu\nu}$ in which matter propagates is unchanged and still reads
$\text{diag}(-1,1,1,1)$. In the simple cases of panels (b) and
(c) of Fig.~\ref{fig:CausalCones}, the components of $\mathcal{S}^{\mu\nu}$ in the $(t,x)$
subspace read $\text{diag}(-1/c_\tx{s}^2,1)$ (up to an overall positive constant), where $c_\tx{s}$ is the speed corresponding to
the characteristics of the dashed blue cone. The wave equation for the field
$\chi$ reads $\mathcal{S}^{\mu\nu}
\partial_\mu\partial_\nu\chi = 0$, and it admits as solutions
arbitrary functions of $(x\pm c_\tx{s} t)$. Panel (b) corresponds to
$c_\tx{s}^2 < 1$ while panel (c) corresponds to $c_\tx{s}^2 > 1$. If one now performs a boost
of speed $-v$, one finds that the components of
$\mathcal{S}'^{\mu\nu}$ in the new coordinate system read
\begin{equation}
\frac{1}{c_\tx{s}^2(1-v^2)}
\begin{bmatrix}
-1 + v^2 c_\tx{s}^2\quad & v (1-c_\tx{s}^2) \\
v (1-c_\tx{s}^2)\quad & c_\tx{s}^2-v^2
\end{bmatrix}.
\label{Eq:BoostedG}
\end{equation}
One thus immediately sees that $\mathcal{S}'^{11} < 0$ (with
$\mathcal{S}'^{00}$ still negative) when one chooses $|c_\tx{s}|<|v|<1$, i.e., that panel (b) is mapped to panel (a). Although
we started from the stable situation of panel (b), in which the
total Hamiltonian density is positive, the
contribution of the field $\chi$ is no longer bounded
from below in this boosted frame corresponding to panel (a). This
is an illustration of what we stated in the previous paragraph. The fact that the Hamiltonian density is unbounded from below is a mere coordinate effect in the present situation, and
it has no physical meaning. The model is stable, but one is not
computing the ``right'' quantity in the boosted frame of panel
(a) (we shall come back to this ``right'' quantity below).

Note that a negative value of $\mathcal{S}'^{11}$ in the boosted
frame of panel (a) always comes together with a significant
non-zero value of $|\mathcal{S}'^{01}| = |\mathcal{S}'^{10}| >
\sqrt{-\mc{D}}$, where $\mc{D}$ is the determinant (\ref{Eq:Det}). The
reason is that this determinant must remain negative in all
coordinate systems ---~and actually remains strictly equal to $\mc{D}$
when one considers only special-relativistic boosts as here.
These non-zero off-diagonal components of $\mathcal{S}'^{\mu\nu}$
are crucial for the existence of an inverse boost taking us back
to the situation of panel (b), where the total Hamiltonian
density is positive. If they were absent, then the metric
$\text{diag}(\mathcal{S}'^{00}, \mathcal{S}'^{11})$ would be
negative definite, it would not define any causal cone, and the
Cauchy problem would be ill-posed. Note also that the magnitude
of these off-diagonal components of $\mathcal{S}'^{\mu\nu}$ is crucial. For instance, panel (i) of
Fig.~\ref{fig:CausalCones} corresponds to $\mathcal{S}'^{00} < 0$
and $\mathcal{S}'^{11} < 0$ like panel (a), and it does satisfy
$|\mathcal{S}'^{01}| > \sqrt{-\mc{D}}$, but also the inequality
$|\mathcal{S}'^{01}| <
\left|\mathcal{S}'^{00}+\mathcal{S}'^{11}\right|/2$, contrary to Eq.~(\ref{eq:condS}). This leads
to the situation of panel (j) when diagonalizing
$\mathcal{S}^{\mu\nu}$ by an appropriate boost. In this case (j),
the two metrics $g^{\mu\nu}$ and $\mathcal{S}^{\mu\nu}$ have
opposite signatures in the $(t,x)$ subspace, so that the field $\chi$ behaves as a ghost in this subspace, and
the model is unstable as soon as $\chi$ is somehow coupled to
matter (including indirectly, e.g., via gravity).

Let us now apply a boost to the case of panel (c) of
Fig.~\ref{fig:CausalCones}. If one chooses $|c_\tx{s}|^{-1} < |v| < 1$,
then one finds from Eq.~(\ref{Eq:BoostedG}) that $\mathcal{S}'^{00}
> 0$ (with $\mathcal{S}'^{11}$ still positive), i.e., one obtains
the situation of panel (d). Here again, as described above, one finds that the contribution of the field $\chi$
to the Hamiltonian density is no longer bounded from below in this
boosted frame, whereas it was positive in the initial frame
corresponding to panel (c). The fact that the first term of
(\ref{Eq:H2}), proportional to $\dot\chi^2$, becomes negative is
related to the fact that the null direction $N^\mu$ (with respect to $\mathcal{S}'_{\mu\nu}$) can have a ``wrong'' time-orientation in the boosted frame. Indeed, from panel (d), when this null vector $N^\mu$ points towards positive values of $x'$, it seems to go
backwards with respect to time $t'$\footnote{On the other hand,
the possible negative value of Hamiltonian (\ref{Eq:H2}) in the
previous case of panel (a) is less obvious, since the null
vectors $N^\mu$ (with respect to $\mathcal{S}'_{\mu\nu}$)
always remain future-oriented. In that case,
negative values are caused by the second term of (\ref{Eq:H2})
involving the spatial derivative $\partial_1\chi$, i.e., by a specific spatial dependence of the initial data.}. As underlined above, the
hypersurface $t'=0$ cannot be consistently used to specify
initial data in this case, since it is not spacelike with respect
to $\mathcal{S}'_{\mu\nu}$. Therefore, the sign of the Hamiltonian density
(\ref{Eq:H2}) at $t'=0$ does not have much meaning anyway. The
conclusion is the same as before: the fact that the Hamiltonian density is unbounded by
below in the boosted frame of panel (d) is a mere coordinate
effect, without any physical meaning, and the model is actually
stable, as proven by the positive total Hamiltonian density in
the frame of panel (c).

\subsubsection*{Mixing of energy and momentum}

It is also instructive to compute the energy of a system in a
boosted frame (still in flat spacetime, to simplify the
discussion). Although it differs from $g^{\mu\nu}$, the effective
metric $\mathcal{S}^{\mu\nu}$ is a tensor. Therefore, the Lagrangian $\mathcal{L}^{(2)}$, Eq.~(\ref{eq:chiLag}), is diffeomorphism invariant, and this implies
that four Noether currents are conserved. They read \cite{Wald:1984rg}:
\begin{equation}
T_\mu^\nu =
\frac{\delta \mathcal{L}^{(2)}}{\delta (\partial_\nu \chi)}\,
\partial_\mu \chi - \delta_\mu^\nu\, \mathcal{L}^{(2)},
\end{equation}
where the index $\mu$ specifies which of the four currents is considered, $\nu$ denotes its components, and $\delta_\mu^\nu$ is the Kronecker symbol. The current conservation reads as usual
\be
\partial_\nu T^\nu_\mu = 0 \Leftrightarrow \partial_0
T^0_\mu + \partial_i T^i_\mu = 0.
\ee
When integrating this identity
over a large spatial volume $V$ containing the whole physical
system under consideration, the spatial derivatives become
vanishing boundary terms, and one gets the standard conservation
laws for total energy and momentum, $\partial_t P_\mu = 0$, with
$P_\mu = \int\!\!\!\int\!\!\!\int_V T_\mu^0\, \tx{d}^3 x$. For
$\mu = 0$, the energy density $T_0^0$ coincides with the on-shell
value of the Hamiltonian density (\ref{Eq:H2}). As recalled
above, if it is bounded from below, then the lowest-energy state
must be stable. But it should be underlined that the three
components of the total momentum $P_i$ are also conserved, and
that the components $T_i^0 = p\,\partial_i\chi$ ---~with $p$ still
given by Eq.~(\ref{Eq:p})~--- have no preferred sign, since there
is no privileged spatial direction. When changing coordinates,
the total 4-momentum of the system becomes $P'_\lambda =
(\partial x^\mu/\partial x'^\lambda)P_\mu$, and in particular,
the energy gets mixed with the initial 3-momentum, $P'_0 =
(\partial x^\mu/\partial x'^0)P_\mu$, or simply $P'_0 = (P_0+ v
P_1)/\sqrt{1-v^2}$ for a mere boost of velocity $v$ in the $x$
direction. Of course, $g^{\mu\nu}P_\mu P_\nu$ (as well as
$\mathcal{S}^{\mu\nu}P_\mu P_\nu$) is a scalar quantity, and it
remains thus invariant under coordinate transformations. However,
it is not always negative, contrary to the standard ``minus rest
mass squared'' in special relativity, therefore the magnitude of
the spatial components $P_i$ is not always bounded by $P_0$. For
instance, in panels (c) or (d) of Fig.~\ref{fig:CausalCones}, when $\chi$ propagates outside the solid grey
cone, it obviously corresponds to a positive $g^{\mu\nu}P_\mu P_\nu$,
i.e., a spacelike $P_\mu$ with respect to $g^{\mu\nu}$. It is
thus clear that a negative value of $P'_0 = (P_0+ v
P_1)/\sqrt{1-v^2}$ is reachable for a large enough boost velocity
$|v| < 1$. The fact that $P'_0$ can also become negative in the
case of panel (a) is much less obvious, but it can be checked
that it coincides with the spatial integral of the on-shell expression
of the Hamiltonian density (\ref{Eq:H2}) with the boosted effective metric
(\ref{Eq:BoostedG}). In such a case, a large enough boost
velocity $|c_\tx{s}|<|v|<1$ generates a negative $\mathcal{S}'^{11}$,
and thereby a possibly negative Hamiltonian density (\ref{Eq:H2}), when
initial data on the $t'=0$ hypersurface are chosen with a large
spatial gradient $\partial_{x'}\chi$ (but a small
$\partial_{t'}\chi$). 

With this slightly different viewpoint of momentum and energy conservation, one can also understand why situations like
panels (a) or (d) of Fig.~\ref{fig:CausalCones} are stable in
spite of their Hamiltonian density (\ref{Eq:H2}) which is
unbounded from below. Indeed, not only their
total energy $P'_0$ is conserved, but also their 3-momentum
$P'_i$. Thus, for any boost speed $v$, $(P'_0- v
P'_1)/\sqrt{1-v^2}$ is also conserved. In the case of panel (a), a boost of speed $v$ such that $|c_\tx{s}|<v<1$ brings us to panel (b), where the energy $P_0$ is obviously bounded from below. Similarly, a boost of speed $v$ such that $|c_\tx{s}|^{-1}<v<1$ maps panel (d) to panel (c), making the energy positive in the new frame for the same reason. In
other words, stability is not ensured by the fact that the Hamiltonian density is bounded from below in the case of panels (a) and (d), but by the fact that the linear combination $T'^0_0 - v T'^0_1$ is positive. 

In more general
situations involving arbitrary coordinate transformations, the
initial energy $P_0$ which is bounded from below is
again a linear combination of conserved quantities in the
new frame, $P_0 = (\partial x'^\mu/\partial x^0)P'_\mu$.

\subsection{Application to k-essence}
\label{sec:kessence}

The above discussion is very generic. The simplest example of a theory with different causal cones for a spin-0 degree of freedom and matter is k-essence \cite{Bekenstein:1984tv,ArmendarizPicon:1999rj,%
Chiba:1999ka,ArmendarizPicon:2000ah}. Again, to simplify, we will consider a flat spacetime, i.e., without
any metric perturbation. In terms of the parametrization of Horndeski theory that we used, k-essence corresponds to a non-linear function $G_2$ of the standard kinetic term:
$\mathcal{L}_2 = G_2(X)$, where as usual $X$ stands for $-g^{\mu\nu}\partial_\mu\varphi\,\partial_\nu\varphi/2$. If
one writes the scalar field as $\varphi = \bar\varphi+\chi$,
where $\bar\varphi$ denotes the background solution and $\chi$ a
small perturbation, one finds that the second-order expansion of
this Lagrangian reads \cite{ArmendarizPicon:1999rj,Babichev:2006vx,Bruneton:2006gf,%
Bruneton:2007si,Babichev:2007dw}
\be
\mathcal{L}^{(2)}_2 =-\dfrac{1}{2} \,\mathcal{S}^{\mu\nu} \partial_\mu\chi \partial_\nu\chi,
\label{eq:kessenceLag}
\ee
where $\mathcal{S}^{\mu\nu}$ is given by:
\be
\mathcal{S}^{\mu\nu} = G_{2X}(\bar X) g^{\mu\nu} - G_{2XX}(\bar X)\pd^\mu\bar\varphi \,\pd^\nu\bar\varphi.
\label{Eq:GmunuKessence}
\ee
In the above equation, it should be noted that the differential operator $\pd^\mu$ stands for $g^{\mu\nu}\pd_\nu$. $\mathcal{S}^{\mu\nu}$ is not proportional to $g^{\mu\nu}$
as soon as $G_{2XX}(\bar X) \neq 0$ and the background solution has a
non-vanishing gradient $\partial_\mu\bar\varphi$. Thus, $\mathcal{S}^{\mu\nu}$ and $g^{\mu\nu}$ define
different causal cones. 

As an application of the discussion in the previous paragraphs, let us re-derive the
stability conditions of k-essence. These conditions have been written several times in
the literature \cite{Aharonov:1969vu,ArmendarizPicon:1999rj,%
Babichev:2006vx,Bruneton:2006gf,Bruneton:2007si,Babichev:2007dw}. They read $G_{2X}(\bar X) > 0$ and
$2 \bar X G_{2XX}(\bar X) + G_{2X}(\bar X) > 0$. Note that no
condition is imposed on $G_{2XX}(\bar X)$ alone.

When the background scalar gradient $\partial_\mu\bar\varphi$ is timelike with respect to $g^{\mu\nu}$,
the causal cones can be represented as panels (a), (b), (c) or (d)
of Fig.~\ref{fig:CausalCones}, where the gray cone (with solid
lines) is defined by $g^{\mu\nu}$ and the blue one (with dashed
lines) by $\mathcal{S}^{\mu\nu}$. Panel (a) is actually
transformed into (b), and (d) into (c), if one chooses a
coordinate system such that the spatial gradients
$\partial_i\bar\varphi$ vanish, the vector
$\partial_\mu\bar\varphi$ pointing then exactly in the time
direction. Panels (a) and (b) correspond to $G_{2XX}(\bar X) > 0$, and mean that
the spin-0 degree of freedom $\chi$ propagates slower than light. Panels (c) and (d)
correspond to $G_{2XX}(\bar X)< 0$, and describe a superluminal
scalar field, but this does not lead to any causality problem as
soon as this dashed cone remains always a cone, with a non-empty
exterior where one may define Cauchy surfaces to specify initial
data. This has already been discussed in detail 
\cite{Aharonov:1969vu,Babichev:2006vx,%
Bruneton:2006gf,Bruneton:2007si,Babichev:2007dw}.
Paradoxes only occur
when one wants to specify initial data on the $t=0$ hypersurface
in the situation of panel (d): this is forbidden because this
hypersurface is not spacelike with respect to the dashed cone.

Let us choose some coordinates such that $g_{\mu\nu} =
\text{diag}(-1,1,1,1)$. Then, if $\partial_\mu \bar\varphi$ is
timelike with respect to $g_{\mu\nu}$, it is always possible to
boost this coordinate system such that $\partial_i\bar\varphi =
0$. One thus gets $\mathcal{S}^{\mu\nu} =
\text{diag}(-G_{2X} - \dot{\bar\varphi}^2 G_{2XX},
G_{2X},G_{2X},G_{2X})$. To be in the situation of panels (b) or (c) of
Fig.~\ref{fig:CausalCones}, it is necessary to have
$\mathcal{S}^{00}<0$ and $\mathcal{S}^{xx}>0$, therefore one needs
$-G_{2X} - \dot{\bar\varphi}^2 G_{2XX} < 0$ and $G_{2X} > 0$. Since $\bar
X =-\frac12 g^{\mu\nu}\partial_\mu\bar\varphi \partial_\nu\bar\varphi =
\dot{\bar\varphi}^2/2$ in this specific coordinate system, the
covariant expressions of these conditions are necessarily
$G_{2X}(\bar X)> 0$ and $G_{2X}(\bar X) + 2\bar X G_{2XX}(\bar X) > 0$, and we recover the correct result. 

The result remains
the same when the background scalar gradient $\partial_\mu
\bar\varphi$ is spacelike (still with respect to $g_{\mu\nu}$).
Then one may choose the $x$ coordinate in its direction, so that
its only non-vanishing component is $\bar\varphi' =
\partial_1\bar\varphi$. In this coordinate system, the components
of the effective metric read $\mathcal{S}^{\mu\nu} =
\text{diag}(-G_{2X}, G_{2X} -\bar\varphi'^{\,2}G_{2XX},G_{2X},G_{2X})$, while $\bar X = -\bar\varphi'^{\,2}/2$,
therefore one recovers strictly the same covariant inequalities.

Finally, when $\partial_\mu \bar\varphi$ is a null vector (again
with respect to $g_{\mu\nu}$, i.e., $\bar X = 0$), it is possible
to choose a coordinate system in which $\partial_\mu \bar\varphi
= \left( \dot{\bar\varphi}, \dot{\bar\varphi}, 0,0\right)$, and
the non-vanishing components of the effective metric read
$\mathcal{S}^{00} = -G_{2X}-\dot{\bar\varphi}^2 G_{2XX}$,
$\mathcal{S}^{11} = G_{2X}-\dot{\bar\varphi}^2 G_{2XX}$,
$\mathcal{S}^{01} = \mathcal{S}^{10} = \dot{\bar\varphi}^2 G_{2XX}$, and $\mathcal{S}^{22} = \mathcal{S}^{33} = G_{2X}$. Then, one of the characteristics defined by
$\mathcal{S}^{\mu\nu}$ coincides with one of those defined by
$g^{\mu\nu}$, corresponding to a velocity $-1$ for spin-0
perturbations. This is thus a limiting case of those plotted in
Fig.~\ref{fig:CausalCones}. But when $G_{2X}(\bar X) > 0$,
consistently with the same covariant inequalities as above, one
finds that the causal cones defined by $g^{\mu\nu}$ and
$\mathcal{S}^{\mu\nu}$ have both a common interior and a common
exterior, and the background solution is thus stable.

\section{Stable black hole solutions in Horndeski theory}
\label{sec:Johnstab}

Let us now illustrate these findings with a specific example. We will discuss some solutions of the shift and reflection symmetric action (\ref{eq:John}), that we recall here for convenience:
\be
S_{\mathbb{Z}_2}=\displaystyle\int{\tx{d}^4x\sqrt{-g}\left[\zeta (R-2\Lambda_\tx{b})-\eta(\pd\varphi)^2 +\beta G_{\mu\nu}\pd^\mu\varphi\, \pd^\nu\varphi\right]}.
\label{eq:John2}
\ee
Regarding stability, Appleby and Linder examined the action (\ref{eq:John2}) with vanishing $\Lambda_\text{b}$ in a cosmological framework \cite{Appleby:2011aa}. From the study of scalar perturbations, they found that there always exists either a gradient instability or a ghost. This pathology can however be cured by the introduction of a bare cosmological constant $\Lambda_\text{b}$, as we will see below. The stability of the black hole solutions with a static scalar field was discussed in \cite{Kobayashi:2012kh,Kobayashi:2014wsa,Ganguly:2017ort}. Stable parameter regions were exhibited. Then, Ogawa et al. tackled the case where the scalar field acquires time-dependence \cite{Ogawa:2015pea}. They claimed that the solutions were always unstable, whatever the coupling parameters of the theory. However, their argument made use of the fact that the Hamiltonian is unbounded from below; as argued in Sec.~\ref{Sec2}, this criterion can lead to erroneous conclusions. We show in Paragraph \ref{sec:bhstab} that there indeed exist stable black hole solutions for given parameters. These are stable against simultaneous scalar, gravitational and matter perturbations. We will first derive the effective metrics in which spin-2 and scalar perturbations respectively propagate.

\subsection{The effective metrics for spin-2 and scalar perturbations}
\label{sec:effmetrics}


The analysis will be mostly focused on the spherically symmetric Schwarzschild-de Sitter solutions presented in Paragraph \ref{sec:deSitterJohn}, Eqs.~(\ref{eq:hJohnSdS})--(\ref{eq:XJohnSdS}). Let us recall this solution in a compact form:
\begin{align}
\label{eq:hJohnSdS2}
\tx{d}s^2 &= -h(r)\, \tx{d}t^2+\dfrac{\tx{d}r^2}{f(r)}+ r^2\left(\tx{d}\theta^2 +\sin^2\theta\, \tx{d}\phi^2\right),
\\
h(r) &= f(r)= 1- \frac{2m}{r} - \frac{\Lambda_\text{eff}}{3}\, r^2,
\\
\Lambda_\text{eff} &= -\frac{\eta}{\beta},
\label{EqLambdaEff}
\\
\varphi(t,r) &= q\left[ t - \int\frac{\sqrt{1-h(r)}}{h(r)} \tx{d}r\right],
\label{Eqphi}
\\
q^2 &= \frac{\eta+ \beta\Lambda_\text{b}}{\eta\beta}\,\zeta,
\label{Eqq2}
\end{align}
where $q$ parametrizes the linear time-dependence of the scalar field\footnote{There actually exist two branches for the scalar field, corresponding to a plus or minus sign in front of the $r$ integral. We keep only the minus branch, so that this solution is mapped to a homogeneous and expanding one in Friedmann-Lemaître-Robertson-Walker coordinates, see \cite{Babichev:2016kdt}.}. Again, the constant $\Lambda_\tx{eff}$ plays the role of an effective cosmological constant, and is a priori independent of the bare one $\Lambda_\text{b}$, with the velocity $q$ playing the role of a tuning integration constant to $\Lambda_\text{b}$ via Eq.~(\ref{Eqq2}). For consistency, the right-hand side of Eq.~(\ref{Eqq2}) should be positive; since $\zeta$ is always positive, one must therefore have
\be
(\eta+ \beta\Lambda_\text{b}) \eta \beta>0,
\label{eq:exist1}
\ee
for this solution. Since this solution is meant to describe the present acceleration of the expansion of the Universe, $\Lambda_\tx{eff}$ should be positive, which translates as
\be
\eta \beta <0.
\label{eq:Lambdapos}
\ee
Let us now proceed with the perturbative analysis. We will actually perform a mode analysis\footnote{A covariant analysis to find the effective metrics would be more powerful, but it is a hard task to demix spin-0 and spin-2 perturbations without assuming any symmetry on the background. This result has been obtained in the case of quadratic and cubic Horndeski theory, but not for the quartic sector. More details are given in Appendix \ref{ap:diagonal}.}. Indeed, one can exploit the spherical symmetry of the background to decompose perturbations on the basis of spherical harmonics $Y_{\ell m}(\theta,\phi)$. Since modes of given orbital numbers $\ell$ and $m$ do not interact at linear level, one can study stability mode by mode. This mode formalism was first developed by Regge and Wheeler \cite{Regge:1957td} in the framework of general relativity. Their work was completed by Vishveshwara \cite{Vishveshwara:1970cc} and Zerilli \cite{Zerilli:1970se}. The metric and scalar field are perturbed according to:
\begin{align}
g_{\mu \nu}&=\bar{g}_{\mu \nu} +h_{\mu \nu},
\\
\varphi &= \bar{\varphi}+\chi,
\end{align}
where a bar denotes the background solution (\ref{eq:hJohnSdS2})--(\ref{Eqq2}). A priori, $h_{\mu \nu}$ and $\chi$ are arbitrary (small) functions of $t$, $r$, $\theta$ and $\phi$. As shown in \cite{Regge:1957td}, one can then decompose these perturbations on the base of spherical harmonics:
\begin{align}
h_{tt}&=h(r)\sum_{\ell, m}H_{0,\ell m}(t,r)Y_{\ell m}(\theta,\phi), 
\label{htt}
\\
h_{tr}&=\sum_{\ell, m}H_{1,\ell m}(t,r)Y_{\ell m}(\theta,\phi),
\\
h_{rr}&=\frac{1}{f(r)}\sum_{\ell, m}H_{2,\ell m}(t,r)Y_{\ell m}(\theta,\phi),
\\
h_{ta}&=\sum_{\ell, m}\left[\beta_{\ell m}(t,r)\partial_{a}Y_{\ell m}(\theta,\phi)+\mathfrak{h}_{0,\ell m}(t,r)E_{ab}\,\partial^{b}Y_{\ell m}(\theta,\phi)\right],
\label{hta} 
\\
h_{ra}&=\sum_{\ell, m}\left[\alpha_{\ell m}(t,r)\partial_{a}Y_{\ell m}(\theta,\phi)+\mathfrak{h}_{1,\ell m}(t,r)E_{ab}\,\partial^{b}Y_{\ell m}(\theta,\phi)\right],
\label{hra} 
\\ 
\begin{split}
h_{ab}&=\sum_{\ell, m} \left\{K_{\ell m}(t,r) g_{ab} Y_{\ell m}(\theta,\phi)+\sum_{\ell, m} G_{\ell m}(t,r) \nabla_a \pd_b Y_{\ell m}(\theta,\phi)\right.
\\
&\quad\left.+\mathfrak{h}_{2,\ell m}(t,r)\left[E_{a}^{~c}\nabla_{c}\pd_{b}Y_{\ell m}(\theta,\phi)+E_{b}^{~c}\nabla_{c}\pd_{a}Y_{\ell m}(\theta,\phi)\right]\vphantom{\sum_{\ell, m}}\right\},
\end{split}
\label{eq:hab}
\\
\chi&=\sum_{\ell, m}\chi_{\ell m}(t,r)Y_{\ell m}(\theta,\phi),
\label{eq:chimodes}
\end{align}
where indices $a$ and $b$ run over angular coordinates ($\theta$ and $\phi$), and $E_{ab}=\sqrt{\det\gamma}~\epsilon_{ab}$ with $\gamma_{ab}$ the two-dimensional metric on the sphere and $\epsilon_{ab}$ the totally antisymmetric symbol ($\epsilon_{\theta \varphi}=1$). In the above sums, $\ell$ runs from 0 to infinity, and $m$ from $-\ell$ to $\ell$. There are then eleven free functions parametrizing a given mode: $H_0$, $H_1$, $H_2$, $\alpha$, $\beta$, $K$, $G$, $\mathfrak{h}_0$, $\mathfrak{h}_1$, $\mathfrak{h}_2$ and $\chi$ (corresponding to the ten components of a two-by-two symmetric tensor, and to the scalar; of course, because of gauge invariance, the number of free functions may be reduced). Each of the terms in Eqs.~(\ref{htt})--(\ref{eq:chimodes}) behaves in a specific way under the parity transformation $(\theta,\phi)\to (-\theta,-\phi)$; either it picks a $(-1)^\ell$ or a $(-1)^{\ell+1}$ coefficient. In the former case, the perturbation is said to have even parity and in the latter, odd parity. The terms associated with Gothic letters $\mathfrak h$ correspond to the odd parity perturbations, while all others are even parity perturbations. 

At the same time, in scalar-tensor theories, a mode of given $(\ell,m)$ should describe the two polarizations of the spin-2 degree of freedom and the scalar degree of freedom. Generically, the odd part of some mode corresponds to one polarization of the spin-2 degree of freedom, while the even part corresponds to the other polarization and the scalar degree of freedom (see \cite{Kobayashi:2012kh,Kobayashi:2014wsa} for instance). The cases $\ell=0$ and $\ell=1$ constitute exceptions however. For instance, in the case of the monopole mode $\ell=0$, the odd terms in Eqs.~(\ref{hta})--(\ref{eq:hab}) are simply absent. Indeed, it is a spherically symmetric mode, and it is expected that the spin-2 degree of freedom does not propagate with this symmetry. Thus, if the effective action which describes this mode is non-trivial, it necessarily corresponds to the spin-0 field. We will now use this fact in order to determine the effective metric for the scalar degree of freedom, in the background given by Eqs.~(\ref{eq:hJohnSdS2})--(\ref{Eqq2}). Let us thus focus on a spherically symmetric perturbation. In this case, Eqs.~(\ref{htt})--(\ref{eq:hab}) boil down to:
\be
h_{\mu \nu}=\begin{bmatrix}
h(r) H_0(t,r) & H_1(t,r) & 0 & 0 \\
H_1(t,r) & H_2(t,r)/f(r) & 0 & 0 \\
0 & 0 & K(t,r) r^2 & 0\\
0 & 0 & 0 & K(t,r) r^2 \sin^2 \theta
\end{bmatrix}
,\ee
with free functions $H_i$ and $K$ (which correspond to $H_{i,00}$ and $K_{00}$ in the previous notations). Inserting these perturbations into the action, one can isolate the terms which are quadratic in $h_{\mu \nu}$ and $\chi$. This gives the second order perturbed action:
\be
\delta^{(2)}_\tx{s}S_{\mathbb{Z}_2}= \displaystyle\int{\tx{d}t\, \tx{d}r\, 4\pi r^2\mc{L}^{(2)}_\tx{s}},
\ee
where the factor $4\pi r^2$ corresponds to the trivial angular integration, and $\mc{L}^{(2)}_\tx{s}$ is the Lagrangian density from which one can extract the causal structure of the perturbations. The subscript ``s'' stands for scalar, since only a spherically symmetric mode is excited. The calculations can be simplified by using the diffeomorphism invariance generated by an infinitesimal vector $\xi^\mu$. In some new system of coordinates $\hat{x}^\mu=x^\mu+\xi^\mu$, the metric and scalar transform according to
\begin{align}
\hat{g}_{\mu \nu} &= g_{\mu \nu} -2\nd_{(\mu}\xi_{\nu)},
\\
\hat{\varphi}&=\varphi-\pd_\mu\varphi\, \xi^\mu.
\end{align}
With a well-chosen $\xi^\mu$, one can in fact set $K$ and $\chi$ to zero. This completely fixes the gauge. Explicitly,
\be
\xi^\mu=\left(\dfrac1q\left(\chi+\varphi'\dfrac{Kr}{2}\right),-\dfrac{Kr}{2},0,0\right).
\ee
Note that this gauge fixing is possible only when $q\neq0$, i.e., the scalar field is not static. In this gauge, $\mc{L}^{(2)}_\tx{s}$ reads, after numerous integrations by parts and using the background field equations,
\be
\begin{split}
\label{eq:H012Lag}
\mc{L}^{(2)}_\tx{s} &= c_1 H_0 \dot{H}_2+c_2H_0'H_1+c_3H_0'H_2+c_4H_1\dot{H}_2+c_5H_0^2+c_6H_2^2+c_7H_0H_2
\\
&\quad+c_8H_1H_2.
\end{split}
\ee
Here a dot represents a time derivative, and all $c_i$ are background coefficients with radial (but no time) dependence, the detailed expression of which can be found in Appendix \ref{sec:appmonop}. This three-field Lagrangian should boil down to a Lagrangian depending on a single dynamical variable. As a first step in this direction, it is easy to eliminate $H_2$ since the associated field equation is algebraic in $H_2$:
\be
H_2 = -\dfrac{1}{2c_6}(-c_1\dot{H}_0-c_4 \dot{H}_1+c_3 H_0'+c_7 H_0+c_8H_1).
\ee
Inserting back this expression in $\mc{L}^{(2)}_\tx{s}$, one obtains
\be
\begin{split}
\mc{L}^{(2)}_\tx{s} &= \tilde{c}_1 \dot{H}_0^2+\tilde{c}_2H_0'^2+\tilde{c}_3H_0'\dot{H}_0+\tilde{c}_4\dot{H}_1^2+\tilde{c}_5\dot{H}_0 H_1'+\tilde{c}_6 \dot{H}_0 \dot{H}_1+\tilde{c}_7H_0'H_1
\\
&\quad+\tilde{c}_8\dot{H}_0H_1+\tilde{c}_9H_0^2+\tilde{c}_{10}H_0H_1+\tilde{c}_{11}H_1^2,
\label{eq:H0H1Lag}
\end{split}
\ee
where the $\tilde{c}_i$ coefficients are again given in Appendix \ref{sec:appmonop} in terms of the $c_i$. A trickier step is to trade $H_1$ and $H_0$ for a single variable, since the associated field equations are differential equations, not algebraic ones. To this end, let us introduce an auxiliary field $\pi_\tx{s}$ as a linear combination of $H_0$, $H_1$ and their first derivatives:
\be
\pi_\tx{s}=\dot{H}_0 + a_2 H_0' + a_3 \dot{H}_1 + a_4 H_1' + a_5 H_0 + a_6 H_1,
\label{eq:pilinear}
\ee
with some $a_i$ coefficients to be determined soon. The idea is to introduce $\pi_\tx{s}$ at the level of the action, group all the derivatives inside $\pi_\tx{s}$, and then to solve for the algebraic equations giving $H_0$ and $H_1$ in terms of $\pi_\tx{s}$. Therefore, let us rewrite the Lagrangian as
\be
\begin{split}
\mc{L}^{(2)}_\tx{s} &= a_1[-\pi_\tx{s}^2+2\pi_\tx{s} (\dot{H}_0 + a_2 H_0' + a_3 \dot{H}_1 + a_4 H_1' + a_5 H_0 + a_6 H_1)]
\\
&\quad+ a_7 H_0^2+a_8 H_1^2+a_9 H_0 H_1.
\label{eq:piH0H1Lag}
\end{split}
\ee
Variation of (\ref{eq:piH0H1Lag}) with respect to $\pi_\tx{s}$ ensures Eq.~(\ref{eq:pilinear}). Now, a simple identification with Lagrangian (\ref{eq:H0H1Lag}) allows one to determine the $a_i$ in terms of the $\tilde{c}_i$. Again, these coefficients are given in Appendix \ref{sec:appmonop}. Variation of (\ref{eq:piH0H1Lag}) with respect to $H_0$ and $H_1$ gives a system of two linear equations, which can be easily solved to write these two fields in terms of $\pi_\tx{s}$ and its derivatives. We do not write down their expression here because of their consequent length, but the procedure is straightforward\footnote{The same procedure for the stealth Schwarzschild black hole (\ref{eq:stealthJohn})-(\ref{eq:Xstealth}) breaks down at this point. Section \ref{sec:stealthstab} of the corresponding appendix is devoted to this particular case.}. At this point, the Lagrangian density depends on a single variable $\pi_\tx{s}$. We will examine its kinetic part only, neglecting the potential associated to this degree of freedom and thereby focusing on the causal structure. This kinetic part reads
\be
\mc{L}_\tx{s;\:Kin}^{(2)}= -\dfrac12(\mc{S}^{tt}\dot{\pi_\tx{s}}^2 + 2\mc{S}^{tr} \dot{\pi_\tx{s}} \pi_\tx{s}' + \mc{S}^{rr} \pi_\tx{s}'^2),
\label{eq:piLag}
\ee
with
\begin{align}
\mc{S}^{tt} &= \dfrac{c_1^2 c_3^2 c_4^2}{4c_2 \Delta}(-2c_4^2c_5+c_2c_1c_4'-c_2c_4c_1'-c_1c_4c_2'),
\\
\mc{S}^{rr} &= -\dfrac{c_1^2 c_3^2 c_4^2}{2 \Delta}(-c_3c_8+c_2c_6),
\\
\mc{S}^{tr} &= -\dfrac{c_1^2 c_3^2 c_4^2}{4 \Delta}(-c_4c_7+c_1c_8),
\\
\begin{split}
\label{Ddef}
\Delta &= c_6^2 \Bigl\{2(-c_3 c_8+c_2c_6)(- c_2c_4c_1'+c_1c_2c_4'-c_1c_4c_2')
\\
&\quad+\bigl[4c_3c_8c_5+c_2(c_7^2-4c_6c_5)\bigr]c_4^2-2c_2c_4c_7c_1c_8+c_2c_1^2c_8^2\Bigr\}.
\end{split}
\end{align}
Alternatively, one can remark that the scalar mode propagates to linear order in the given black hole background (\ref{eq:hJohnSdS2}) with an effective two-dimensional metric $\mc{S}_{\mu \nu}$:
\be
\mc{L}_\tx{s;\:Kin}^{(2)}= - \dfrac12\, \mc{S}^{\mu \nu} \pd_\mu \pi_\tx{s} \pd_\nu \pi_\tx{s}.
\ee
One can read from Eq.~(\ref{eq:piLag}) the inverse metric:
\be
\mc{S}^{\mu \nu}=\begin{bmatrix}
\mc{S}^{tt} & \mc{S}^{tr} \\
\mc{S}^{tr} & \mc{S}^{rr}
\end{bmatrix}
,\ee
and the metric itself:
\be
\mc{S}_{\mu \nu}=\dfrac{1}{\mc{S}^{tt}\mc{S}^{rr}-(\mc{S}^{tr})^2}\begin{bmatrix}
\mc{S}^{rr} & -\mc{S}^{tr} \\
-\mc{S}^{tr} & \mc{S}^{tt}
\end{bmatrix}.
\ee
From this last object, one can determine the hyperbolicity condition, the propagation speeds, and all the information needed for the causal structure of the scalar mode. The hyperbolicity condition for instance reads
\be
(\mc{S}^{tr})^2-\mc{S}^{tt}\mc{S}^{rr}>0.
\label{eq:hypscal}
\ee
The speed of a wave moving towards or away from the origin is then given by
\be
c_\tx{s}^\pm=\dfrac{\mc{S}^{tr} \pm \sqrt{(\mc{S}^{tr})^2-\mc{S}^{tt}\mc{S}^{rr}}}{\mc{S}^{tt}}.
\ee
The hyperbolicity condition ensures that these propagation speeds are well defined. At any given point, $c_\tx{s}^+$ and $c_\tx{s}^-$ generate the scalar causal cone. Finally, one needs to know where the interior of the cone is located. This can be easily determined by checking whether a given direction (for instance the one generated by the vector $\pd_t$) is time or space-like with respect to the metric $\mc{S}_{\mu \nu}$.

A similar analysis must be carried out for the spin-2 mode. It was actually already realized by Ogawa et al. in \cite{Ogawa:2015pea}. They studied odd-parity perturbations which, as we mentioned, generically correspond to one of the spin-2 polarizations. These perturbations are obtained by keeping only $\mathfrak{h}_0$, $\mathfrak{h}_1$and $\mathfrak{h}_2$ in Eqs.~(\ref{htt})--(\ref{eq:hab}). For the solution (\ref{eq:hJohnSdS2})--(\ref{Eqq2}), the gravity perturbations propagate in a two-dimensional effective metric $\mc{G}_{\mu \nu}$ such that
\begin{align}
\mc{G}^{tt}&= \dfrac{-2(2-\beta q^2/\zeta)}{2\zeta+\beta q^2} \dfrac{1}{h}\left(\zeta+\beta q^2 \dfrac{h-2}{2h}\right) ,
\label{eq:Gtt}
\\
\mc{G}^{rr}&= \dfrac{2(2-\beta q^2 /\zeta)}{(2\zeta+\beta q^2)} h\left(\zeta-\beta q^2 \dfrac{h+2}{2h}\right) ,
\label{eq:Grr}
\\
\mc{G}^{tr}&= \dfrac{-2(2-\beta q^2/\zeta)}{(2\zeta+\beta q^2)} \beta q \varphi'.
\label{eq:Gtr}
\end{align}
The hyperbolicity condition coming from $\mc{G}_{\mu\nu}$ reads:
\be
(\mc{G}^{tr})^2-\mc{G}^{tt}\mc{G}^{rr}>0,
\label{eq:hypgrav}
\ee
and the speeds of inwards/outwards moving gravitational waves are given by
\be
c_\tx{g}^\pm=\dfrac{\mc{G}^{tr} \pm \sqrt{(\mc{G}^{tr})^2-\mc{G}^{tt}\mc{G}^{rr}}}{\mc{G}^{tt}}.
\ee
In a nutshell, we have found the effective metrics in which the spin-2 and scalar degrees of freedom propagate, and they are given by $\mc{G}^{\mu \nu}$ and $\mc{S}^{\mu \nu}$ respectively.

\subsection{Homogeneous solutions: a stability window}

We will first apply the above analysis to de Sitter solutions, that is solution (\ref{eq:hJohnSdS2})--(\ref{Eqq2}) with $m=0$. Of course, in this case, the analysis presented above is not strictly necessary, but it allows us to cross check our results with cosmological perturbation theory. In particular, we arrive at the same conclusion as \cite{Appleby:2011aa} for the model (\ref{eq:John2}) with vanishing $\Lambda_\text{b}$: there is no stable homogeneous configuration. However, switching on a non-trivial $\Lambda_\text{b}$, the hyperbolicity conditions (\ref{eq:hypscal}) and (\ref{eq:hypgrav}) read respectively:
\begin{align}
(3 \beta\Lambda_\text{b}+\eta)(\eta-\beta\Lambda_\text{b})&<0,
\label{eq:hypscal2}
\\
(3\eta+\beta\Lambda_\text{b})(\eta-\beta\Lambda_\text{b})&>0.
\label{eq:hypgrav2}
\end{align}
These two conditions must be supplemented with the fact that the spin-2, matter and scalar cones have a non-empty intersection and a common exterior. It is enough to check the orientation of the cones at $r=0$, since the solution under analysis is homogeneous. If the cones have compatible orientations at $r=0$, this will remain true everywhere else. The calculation is then particularly simple, since $\mc{S}^{tr}$ and $\mc{G}^{tr}$ vanish at $r=0$, meaning that the cones are either aligned (and symmetric around the $t$ axis) or inclined at ninety degrees. The $t$ axis is of course always in the interior of the matter causal cone. Therefore, the spin-2 and scalar causal cones are compatible if $\mc{S}_{tt}(r=0)<0$ and $\mc{G}_{tt}(r=0)<0$. In terms of the parameters of the theory, this translates as
\begin{align}
\eta (\eta+3\beta \Lambda_\text{b})<0,
\label{eq:goodorientscal}
\\
\eta(3 \eta+\beta \Lambda_\text{b})>0,
\label{eq:goodorientgrav}
\end{align}
respectively. Thus, there are in total six conditions to fulfill for stability and existence of the solution: Eqs.~(\ref{eq:exist1}), (\ref{eq:Lambdapos}), (\ref{eq:hypscal2}), (\ref{eq:hypgrav2}), (\ref{eq:goodorientscal}) and (\ref{eq:goodorientgrav}). They actually define an non-empty subspace of the parameter space. The cosmological solution is stable if and only if
\begin{align}
\tx{either}~\eta&>0,~\beta<0~\tx{and}~\dfrac{\Lambda_\text{b}}{3}<-\dfrac{\eta}{\beta}<\Lambda_\text{b},
\label{eq:range1}
\\
\tx{or}~\eta&<0,~\beta>0~\tx{and}~\Lambda_\text{b}<-\dfrac{\eta}{\beta}<3\Lambda_\text{b}.
\label{eq:range2}
\end{align}
In the following section, we give an example of parameters that fulfill this criterion. Let us stress that the above restrictions prevent one from using the theory (\ref{eq:John2}) as a self-tuning model. Indeed, the above equations tell us that the effective cosmological constant has to be of same magnitude as the bare one. Thus, in the case of the model (\ref{eq:John2}), stability prevents the scalar field from absorbing a huge amount of vacuum energy (present through a large $\Lb$ here). In Chapter \ref{ch:wave}, we will see some models that allow for an actual self-tuning.

\subsection{Black holes in de Sitter: example of a stable configuration}
\label{sec:bhstab}

The causal cone analysis is fully relevant when the solution no longer describes a homogeneous cosmology, but rather a black hole embedded in such a cosmology. The two conditions for the background solution to exist and have de Sitter asymptotics, Eqs.~(\ref{eq:exist1})-(\ref{eq:Lambdapos}), do not depend on the presence of a mass $m\neq0$. Therefore, they remain identical when a black hole is present.

\begin{figure}[!ht]
\centering
\includegraphics[width=\textwidth]{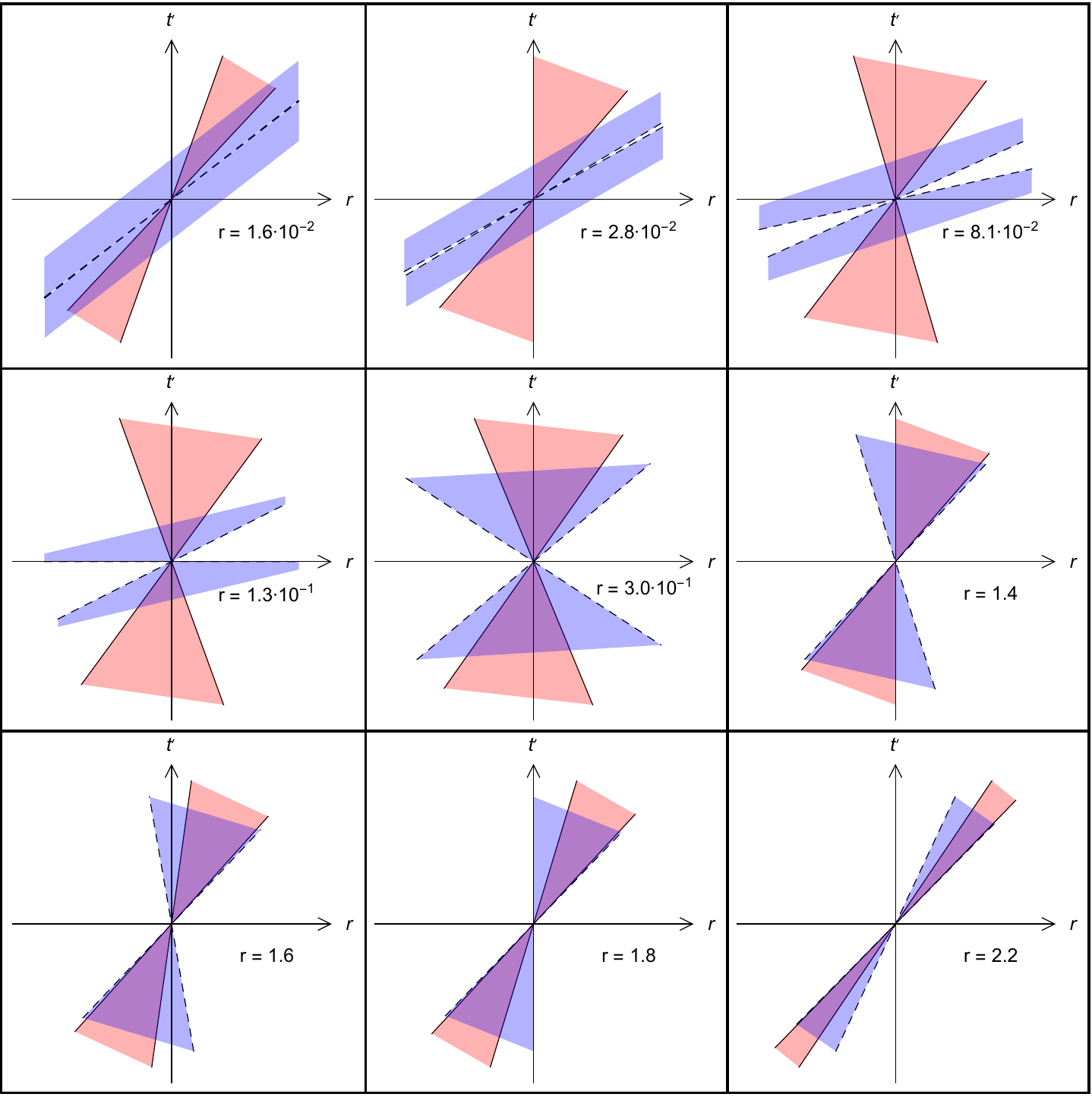}
\caption{The scalar and spin-2 causal cones in Schwarzschild-de Sitter geometry, respectively in dashed blue and plain red. In this plot, the time coordinate $t'$ has been rescaled with respect to the original one, so that the matter causal cone, associated to $g_{\mu\nu}$, corresponds to lines at $\pm45^\circ$. The parameters of the Lagrangian are chosen so that the associated cosmological solution is stable: $\eta=1/2$, $\beta=-1$, $\zeta=1$, $\Lambda_\text{b}=1$ in Planck units. The radius $r$ varies between the black hole horizon located at $r\simeq 9.4\cdot 10^{-3}$ and the cosmological horizon at $r\simeq2.4$. For this set of parameters, there always exist a hypersurface which is spacelike with respect to the three metrics $g_{\mu\nu}$, $\mc{G}_{\mu\nu}$ and $\mc{S}_{\mu\nu}$, and a time direction which is timelike for these three metrics.}
\label{fig:bhsoundcone}
\end{figure}

On the other hand, the expressions of $\mc{G}_{\mu \nu}$ and $\mc{S}_{\mu \nu}$ become very complicated with a non-vanishing black hole mass. It is still possible to prove that the hyperbolicity conditions for both $\mc{S}_{\mu \nu}$ and $\mc{G}_{\mu \nu}$ are not modified with respect to the de Sitter case. They are again given by Eqs.~(\ref{eq:hypscal2})-(\ref{eq:hypgrav2}). To ensure the compatibility of orientation between the scalar, matter and spin-2 cones is however more tricky. We checked numerically that the conditions (\ref{eq:goodorientscal}) and (\ref{eq:goodorientgrav}) for these three cones to be compatible in the de Sitter case lead to compatible cones also when the mass parameter $m$ is switched on. That is, for parameters in the range (\ref{eq:range1})-(\ref{eq:range2}), the three cones seem to have a compatible orientation even close to the black hole horizon. This remains true for arbitrary mass of the black hole (as long as the black hole horizon remains smaller than the cosmological horizon). Figure \ref{fig:bhsoundcone} provides an illustrative example of this numerical check, for a given set of parameters that falls in the range (\ref{eq:range1}).
In this case, the cones have compatible orientations everywhere, meaning that the background solution is stable against gravitational, scalar and matter perturbations. Remarkably, the scalar causal cone entirely opens up when approaching the black hole horizon, without becoming pathological.

Let us stress here why Ref.~\cite{Ogawa:2015pea} would have claimed that the situation exhibited in Fig.~\ref{fig:bhsoundcone} is unstable, in the light of the discussion of Sec.~\ref{Sec2}. In this paper, the spin-2 effective metric components $\mc{G}^{tt}$ and $\mc{G}^{rr}$ were required to be negative and positive respectively. It was proven, however, that the product  $\mc{G}^{tt}\mc{G}^{rr}$ is always positive in the vicinity of a horizon. Figure \ref{fig:bhsoundcone} shows that, indeed, the $t$ axis ``leaves'' the causal cone of the spin-2 degree of freedom (red cone), close to the event and cosmological horizon, while the $r$ axis remains in the exterior of the cone. This makes the quantity $\mc{G}^{tt}\mc{G}^{rr}$ positive close to horizons, and the associated Hamiltonian density unbounded from below. However, our analysis so far clearly shows that it does not signal an instability in any way.

\section{Conclusions}
\label{sec:concpert}

In this chapter, we have studied stability criteria for solutions in modified gravity theories, without particular restriction to scalar-tensor theories. We then applied these criteria to establish the stability of certain black holes with non-trivial scalar hair, that we encountered in Chapter \ref{ch:tdep}.

The tools developed, as well as the stability criteria concerning Hamiltonian densities, are generically applicable in modified gravity theories. The starting ingredients for the applicability of these tools are multiple gravitational degrees of freedom, a clear characteristic of theories going beyond general relativity. In order to treat the problem consistently, we used the notion of causal cone, each associated to a healthy propagating degree of freedom. Indeed, the local existence of well-defined causal cones allows to determine the healthy propagation of modes about an effective background solution. We saw that, unlike standard lore, the Hamiltonian densities associated to each of the modes do not suffice to exhibit an instability. The failure of the Hamiltonian criterion, in more complex background metrics, is due to the fact that it is not a scalar quantity. Each Hamiltonian density, associated to a propagating mode, depends on the particular coordinate system one is using. So although a Hamiltonian density which is bounded from below signals that the mode is stable, the converse is not true. Namely, a Hamiltonian density found to be unbounded from below in some coordinate system is inconclusive on instability. One may find a coordinate transformation rendering the Hamiltonian bounded from below as we saw explicitly in Sec.~\ref{Sec2}.

Standard lore is recovered only for a class of ``good'' coordinate systems that are defined with respect to all the causal cones present in the system: scalar, spin-2, matter, etc. Namely, these ``good'' coordinate systems exhibit a timelike coordinate common to all causal cones and spacelike coordinates for all causal cone exteriors. If such a coordinate system exists, then the Hamiltonian is indeed bounded from below and the degrees of freedom are well behaved, propagating in a timelike direction with a hyperbolic operator. If not, then indeed the Hamiltonian for at least one of the degrees of freedom is always unbounded from below, and it presents a ghost instability.

The subtlety arises due to the complexity of the background solution. Indeed, the key point for the examples here is that the background scalar is space and time dependent. Then the causal cones can tilt and open up when approaching the horizon (event or cosmological). As a result, the original time (space) coordinate of the background metric may ``leave'' the interior (exterior) of a causal cone associated to some degree of freedom. This can lead to a misinterpretation of the Hamiltonian density associated to the initial coordinates, which ``leave'' the causal cone of the degree of freedom in question.
It should be emphasized that the failing Hamiltonian stability criterion is not due to the mixing of degrees of freedom, as illustrated by the simple k-essence example of Sec.~\ref{Sec2}. One may also consider the case of a theory including a $G_3$ Horndeski term, as in Sec.~\ref{sec:cubictdep}, where the de-mixing of modes has been achieved \cite{Babichev:2012re} for an arbitrary background\footnote{This result is not known for $G_4$ theories, as discussed in Appendix \ref{ap:diagonal}}. We saw in Sec.~\ref{sec:cubictdep} that in such a model were found self-accelerating vacua.
In a standard Friedmann-Lemaître-Robertson-Walker coordinate system, where the dark energy scalar depends purely on cosmological time, such self-accelerating solutions generically give (depending on the coupling constants of the theory) a stable vacuum \cite{Babichev:2012re}, with an associated Hamiltonian which is bounded from below. When one considers the precise same stable vacua in a spherical coordinate system, where the metric is static but the scalar field now depends both on space and time, the same Hamiltonian density can be found to be unbounded from below.
This, as we emphasized, is an artifact of a bad use of a coordinate system (here static) whereas the Friedmann-Lemaître-Robertson-Walker coordinates are indeed ``good'' (satisfying the causal cone criteria, its cosmological time remaining notably within the causal cones). This example demonstrates that misinterpretations related to Hamiltonian densities are not due to mixing of degrees of freedom but, crucially, to the background depending (or not) on multiple coordinates. It is for this reason that one does not encounter problems with the Hamiltonian in Friedmann-Lemaître-Robertson-Walker coordinates for example or when the scalar field is static, as in Chapter \ref{ch:static}. The analysis presented in this chapter is therefore relevant for backgrounds with lesser symmetry, for example stationary backgrounds involving rotating black holes. For stationary backgrounds, the $\theta$ and $r$ dependent effective metrics may again tilt and open up as one approaches a horizon. Clearly, the analysis can also be used in vector-tensor theories, where black holes similar to the one studied here have been found \cite{Chagoya:2016aar,Minamitsuji:2016ydr,Babichev:2017rti}.
In any case, let us emphasize that there always exist ``bad'' coordinates in which the Hamiltonian density of a stable solution appears unbounded from below. It suffices that its time direction is outside at least one of the causal cones, or that some spatial direction is inside one of them. For instance, a mere exchange of $t$ and $x$ creates such a spurious pathology, whereas the physics is obviously unchanged. As usual in general relativity, one should never trust coordinate-dependent quantities.

We also underlined that, when there exists a ``good'' coordinate system in which the total Hamiltonian density is bounded from below, then it may also be computed in other coordinate systems, but it no longer corresponds to the mere energy. It becomes a linear combination of the energy and momentum densities, whose spatial integrals over the whole system are all conserved. In other words, the stability of the solution is still guaranteed by the bounded from below character of a conserved quantity, but it is no longer the Hamiltonian which plays this role.

Using the above tools, we have corrected a misinterpretation \cite{Ogawa:2015pea,Takahashi:2015pad,Takahashi:2016dnv,Kase:2018voo} in the literature about the said instability of a class of hairy black holes. It is true, as stated in \cite{Ogawa:2015pea}, that the Hamiltonian density for the spin-2 degree of freedom is always unbounded from below in Schwarzschild coordinates when approaching a horizon. However, at the same time, the spin-2 causal cone remains compatible with the matter causal cone under some conditions on the parameters defining the model. In other words, there exist coordinate systems where the Hamiltonian density for the spin-2 degree of freedom is bounded from below. We completed this stability analysis by computing the scalar causal cone, thanks to the study of $\ell=0$ perturbations. Again, there exists a domain of parameters where the three causal cones share a common time and a common spacelike hypersurface. Hence, the class of hairy black holes studied here is free of ghost and gradient instabilities for a given range of parameters of the model. This is an important result considering the rarity of stable hairy black holes in gravitational physics (see for example \cite{Bronnikov:1978mx, Harper:2003wt} and \cite{Volkov:1998cc} for the instability of two celebrated cases). 

However, the reader might be aware that the model (\ref{eq:John2}) (as well as most of the quartic and quintic Horndeski sector) were recently ruled out as dark energy candidates, by a single gravitational event, GW170817. We are going to see however, that the work presented in this chapter is not vain, and that both stability and consistency with GW170817 can be fulfilled easily.
\newpage
\thispagestyle{empty}
\cleardoublepage
\chapter{The speed of gravitational waves}\label{ch:wave}

On August 17, 2017 were detected almost simultaneously gravitational waves and a gamma-ray burst, as well as an electromagnetic signal in more or less all possible observation channels \cite{TheLIGOScientific:2017qsa,Monitor:2017mdv,GBM:2017lvd}. This event, baptized with the beautiful name of GW170817, was interpreted as the merger of two neutron stars. The gravitational wave signal before the merger is clean and corresponds to general relativistic expectations for the final inspiral of two neutron stars. It also allows to estimate roughly the masses of the two components, which were between 0.86 and 2.26 $M_\odot$. Although it was possible to detect the signal approximately 100~s before the merger, the post-merger phase did not provide much information, and the properties of the remnant object remain elusive. This experiment provides priceless information on the strong field regime of gravity. At the same time, it also constitutes an entirely new cosmological test. First, it provides an independent measurement of the current Hubble rate $H$ (it is found to be $70^{+12}_{-8}$~km.s$^{-1}$.Mpc$^{-1}$). This might help to reconcile the values obtained from supernovae data and cosmic microwave background, for which a discrepancy has been observed.

Second, and this is what is the most constraining for generalized scalar-tensor theories, this event proves that light and gravitational waves propagate at the same speed in the intergalactic medium (we will note their respective speeds as $c_\text{light}$ and $c_\text{grav}$ in the rest of this chapter). Indeed, these waves traveled over a distance of around 40~Mpc, and arrived with a delay of 1.7~s. This constrains the
speeds of light and gravity to differ by no more than a few parts
in $10^{15}$. The constraint is probably even stronger, since the delay might be due to the time that is necessary for a jet to form during the merger and emit the electromagnetic radiation\footnote{Note that the absence of gravitational Cherenkov radiation for high-energy astroparticles already imposed an indirect lower bound on the speed of gravitational waves, $1-c_\tx{grav}/c_\tx{light}<2\cdot10^{-15}$ \cite{Moore:2001bv}.}. On the other hand, most Horndeski and beyond theories predict that, if the scalar field plays a significant role in cosmology, especially in the late Universe, its presence should affect the speed of propagation of gravitational waves (pretty much as the presence of matter slows down the propagation of light in material media).

References \cite{Creminelli:2017sry,Ezquiaga:2017ekz}
(see also \cite{Lombriser:2015sxa,Lombriser:2016yzn,Bartolo:2017ibw,Baker:2017hug,Sakstein:2017xjx}) have characterized which
Horndeski models and their generalizations satisfy exactly $c_\text{grav} = c_\text{light}$ in a homogeneous
Universe. This chapter will focus on this particular subclass of the Horndeski and beyond theory. In order to ensure
$c_\text{grav} = c_\text{light}$, no quintic Horndeski or beyond
term is allowed. For example, interactions involving the Gauss-Bonnet term, like the model (\ref{eq:actionGB}), are excluded\footnote{By excluded, throughout this chapter, we refer to the cases where the extra mode is a dark energy field, giving an effective acceleration to the Universe at late times. If the extra mode, say a scalar, is not varying at cosmological scales, but only locally, it may not influence gravitational waves in their 40~Mpc journey to Earth detectors.} \cite{Ezquiaga:2017ekz, Creminelli:2017sry}. Furthermore, the
function $F_4$ must be related to $G_4$ by
\cite{Creminelli:2017sry,Ezquiaga:2017ekz}\footnote{Note
that there is a sign mistake in Eq.~(11) of
Ref.~\cite{Creminelli:2017sry}, and that
Ref.~\cite{Ezquiaga:2017ekz} defines $F_4$
with a sign oppposite to Eq.~(\ref{eq:bH4}).}
\begin{equation}
F_4(\varphi,X) = -\dfrac{G_{4X}}{2X}.
\label{EqBigF4}
\end{equation}
It is however easy to map some quartic Horndeski model into a Horndeski and beyond one that fulfills the condition (\ref{EqBigF4}), through a disformal transformation that depends only on $X$, i.e., by writing the initial Horndeski action in terms of
\be
\tilde g_{\mu\nu}= g_{\mu\nu} + D(X)\pd_\mu\varphi\,\pd_\nu\varphi,
\label{eq:disformal2}
\ee 
with a suitably chosen $D$. This is a subcase of the generic disformal transformation, Eq.~(\ref{eq:disformal}). Then, at least in homogeneous backgrounds, the new theory fulfills $c_\text{grav}=c_\text{light}$. Note that such a transformation cannot be operated on quintic Horndeski theories. For completeness, let us give the most general degenerate higher-order scalar-tensor theory 
that passes the gravitational wave test (at least at the cosmological level). Its Lagrangian reads \cite{Creminelli:2017sry}:
\be
\begin{split}
\mc{L}_\tx{DHOST}&=\mc{L}_2+\mc{L}_3+C(\varphi,X)B_4(\varphi,X)R-\dfrac{4CB_{4X}}{X}\pd^\mu\varphi\,\pd^\nu\varphi\nd_\mu\pd_\nu\varphi\,\Box\varphi
\\
&\quad+ \left(\frac{4C B_{4X} }{X} +\frac{6 B_4 C_X^2}{C}+  8C_X B_{4X} \right) \pd^\mu\varphi \nd_\mu\pd_\nu\varphi\, \pd_\lambda\varphi \nd^\lambda\pd^\nu\varphi
\\
&\quad-\frac{8 C_X B_{4X}}{X} \left(\pd_\mu\varphi \nd^\mu\pd^\nu\varphi\, \pd_\nu\varphi\right)^2,
\label{eq:DHOSTcT1}
\end{split}
\ee
where $\mc{L}_2$ and $\mc{L}_3$ are unchanged with respect to Eqs.~(\ref{eq:L2})-(\ref{eq:L3}), and $C$ and $B_4$ are new free functions of $\varphi$ and $X$. 

The condition (\ref{EqBigF4}) ensures that the speeds of
gravitational and electromagnetic waves coincide at least in a
homogeneous cosmological background. However, the waves of the
GW170817 event did pass nearby massive bodies during their 40~Mpc
journey, and if their speeds slightly differed in such
inhomogeneous situations, this would a priori suffice
to increase the delay between their detections. It is thus
important to check that these speeds remain equal even in
inhomogeneous backgrounds. Actually,
Ref.~\cite{Ezquiaga:2017ekz} claims that the condition
(\ref{EqBigF4}) also suffices around arbitrary backgrounds, and we will confirm so below
for a specific exact solution. But this reference uses the results of
\cite{Bettoni:2016mij}, which needed to neglect scalar-spin-2
mixing terms in order to extract the spin-2 propagation speed.
Generically, the separation of the spin-2 and spin-0 degrees of
freedom is background dependent and highly non-trivial, as shown in Appendix \ref{ap:diagonal}. In the same spirit, Ref.~\cite{Creminelli:2017sry} computed the Arnowitt-Deser-Misner decomposition of the Lagrangian (\ref{eq:DHOSTcT1}). In the unitary gauge, it is very similar to general relativity (at least when $C=1$): 
\be
\mc{L}_\tx{DHOST} = \mc{L}_2+\mc{L}_3+B_4 \left(^{(3)}R +  K_{ij} K^{ij} - K^2  \right),
\ee
where $^{(3)}R$ is the three dimensional Ricci scalar on constant $\varphi$ hypersurfaces, and $K_{ij}$ the associated extrinsic curvature. However, it is not completely obvious from this expression that spin-2 perturbations propagate at $c_\text{grav}=c_\text{light}$, because the Lagrangian contains for instance mixing terms proportional to $\dot{h}_{ij}D_iN_j$, in usual Arnowitt-Deser-Misner notation; the shift $N_i$ cannot be eliminated as in general relativity, because the gauge was already fixed in order to absorb the scalar field (see also a related discussion in \cite{Langlois:2017dyl}). This decomposition might allow one to conclude generically about the speed of gravitational waves, but a careful justification is still needed. 

The present chapter is based on our work \cite{Babichev:2017lmw}, and its aim is twofold. First, in Sec.~\ref{sec:GWspeed}, we check that the speed equality remains satisfied even in a very inhomogeneous situation, namely in the vicinity of a black hole, where gradients are large and where the separation of spin-2 and spin-0 degrees of freedom is difficult. This will be done for an exact Schwarzschild-de Sitter solution of the specific model (\ref{eq:John2}) ---~more precisely for its disformally transformed version, as in Eq.~(\ref{eq:disformal2}). We also report that this solution is ghost-free and has no gradient instability for some ranges of the parameters defining the theory. Then, in Sec.~(\ref{sec:selft}), we show that an infinity of self-tuning cosmological models still exist while taking into account the $c_\text{grav} = c_\text{light}$ constraint. In such models, the energy-momentum tensor of the scalar field almost perfectly counterbalances the very large bare cosmological constant assumed to be present in the Lagrangian, so that the observable accelerated expansion of the Universe is consistent with a tiny effective cosmological constant.

\section{The speed of gravitational waves in strongly curved backgrounds}
\label{sec:GWspeed}

\subsection{From the Horndeski to the physical frame}

The model (\ref{eq:John2}) is a non-trivial quartic Horndeski model. Thus, as such, it has $c_\tx{grav}\neq c_\tx{light}$ in cosmological backgrounds. This is however true only as long as matter is assumed to be minimally coupled to $g_{\mu\nu}$. As mentioned in the introduction of this chapter, it suffices to couple matter to a different metric $\tilde g_{\mu\nu}$, disformally related to
$g_{\mu\nu}$, to change the matter
causal cone so that $c_\text{grav} = c_\text{light}$ is ensured,
at least in a homogeneous Universe. In the present model (\ref{eq:John2}), the
disformal transformations given in
\cite{
Bettoni:2013diz,Zumalacarregui:2013pma,%
Crisostomi:2016czh,Achour:2016rkg} or the gravity speed derived
in \cite{Bettoni:2016mij,Ezquiaga:2017ekz} might be used to prove that
this physical metric must read:
\begin{equation}
\tilde g_{\mu\nu} = g_{\mu\nu} - \frac{\beta}{\zeta-\beta X}\, \pd_\mu\varphi\, \pd_\nu\varphi.
\label{Eqgtilde}
\end{equation}
Of course, any metric proportional to this $\tilde{g}_{\mu \nu}$
would also be allowed, since it would not change the causal cone,
even if the conformal factor depends on $X$. In standard nomenclature for Brans-Dicke gravity, the non-physical $g_{\mu\nu}$ would be called the ``Einstein frame'' metric. However, its perturbations do not describe a pure spin-2 degree of freedom in the present case, because of the kinetic mixing introduced by the $G^{\mu \nu} \pd_\mu\varphi\,\pd_\nu\varphi$ term of action (\ref{eq:John2}). Therefore, $g_{\mu\nu}$ will be called the ``Horndeski frame'' metric rather than the ``Einstein frame'' one. On the other hand, $\tilde{g}_{\mu \nu}$ is still called the ``Jordan frame'' metric, or physical metric. As in standard Brans-Dicke, it is easier to work in the non-physical frame because the metric sector is simpler there. One should keep in mind that the analogy is to be taken with caution, because the frames of the higher order theories are related disformally (\ref{eq:disformal2}), and not conformally as in Brans-Dicke. One may also rewrite the action (\ref{eq:John2}) in terms of this
$\tilde g_{\mu\nu}$, and one finds that it becomes of the Horndeski and beyond form, Eqs.~(\ref{eq:L2})--(\ref{eq:L5}) and (\ref{eq:bH4})--(\ref{eq:bH5}), with rather complicated functions $\tilde
G_4(\tilde X)$ and $\tilde F_4(\tilde
X)$ (involving nested square roots), which now do
satisfy the constraint (\ref{EqBigF4}) in terms of the variable
$\tilde X = -\tilde g^{\mu\nu}\pd_\mu\varphi\,\pd_\nu\varphi/2$ ($\tilde g^{\mu\nu}$ denoting of course the inverse of $\tilde g_{\mu\nu}$). This guarantees that the speeds of light and
gravity coincide at least in the asymptotic homogeneous Universe,
far away from any local massive body.

Before proceeding to the causal cone analysis (which will inform us about both the speed of propagation of spin-2 waves and stability), let us find the form of the background solution (\ref{eq:hJohnSdS2})--(\ref{Eqq2}) in the physical frame. For this family of solutions, the quantity $X$ is constant, which simplifies the disformal transformation (\ref{Eqgtilde}). To simplify notation, let us set
\be
D=-\dfrac{\beta q^2}{\zeta-\beta q^2/2},
\ee
with the notations of Eqs.~(\ref{eq:John2}) and (\ref{eq:hJohnSdS2})--(\ref{Eqq2}). The disformed metric $\tilde g_{\mu\nu}$ acquires off-diagonal terms in the original $(t,r)$ coordinates due to the $t$ and $r$ scalar field dependence. We then diagonalize the physical metric using the following coordinate redefinition\footnote{The sign in front of the $r$ integral can actually be either a plus or minus, corresponding to the fact that two branches exist for the scalar solution, Eq.~(\ref{Eqphi}). However, as in the previous chapter, only the minus branch for the scalar field is mapped to a homogeneous and expanding one in the Friedmann-Lemaître-Robertson-Walker coordinates of the physical frame.}:
\be
\tilde t=\sqrt{1-D}\left\{t+ \int\!\!\tx{d}r \,\frac{D\sqrt{1-h(r)}}{h(r)[h(r)-D]} \right\},
\label{Eq:ttilde}
\ee
Note that, for this coordinate transformation to be well defined, one needs $D<1$. For the solution (\ref{eq:hJohnSdS2})--(\ref{Eqq2}), this bound reads
\be
(3\eta+\beta \Lambda_\tx{b})(\eta-\beta \Lambda_\tx{b})>0.
\label{eq:Bconstraint2}
\ee
This bound must be kept in mind for the upcoming stability analysis, that will constrain the range of the parameters. The background solution in the physical frame then recovers the same form as the original background, namely:
\begin{align}
\label{eq:ansatzphys}
\tx{d}\tilde s^2 &= -\tilde h(r)\, \tx{d}\tilde t^2 +\frac{\tx{d}r^2}{\tilde h(r)} + r^2\left(\tx{d}\theta^2 +\sin^2\theta\, \tx{d}\phi^2\right),
\\
\tilde h(r) &= 1- \frac{2\tilde m}{r} - \frac{ \tilde \Lambda_\text{eff}}{3}\, r^2,
\\
\varphi(\tilde t, r) &= \tilde q\left[\tilde t - \int\!\frac{\sqrt{1-\tilde h(r)}}{ \tilde h(r)}\, \tx{d}r\right],
\label{Eqphiphys}
\end{align}
where the rescaled parameters of the solution are defined as
\begin{align}
\tilde q &=\frac{q}{\sqrt{1-D}},
\\
\tilde m &= \frac{m}{1-D},
\\
\tilde \Lambda_\text{eff} &= \frac{\Lambda_\text{eff}}{1-D}=\left(\dfrac{\Lambda_\text{eff} + \Lambda_\text{b}}{3\Lambda_\text{eff}-\Lambda_\text{b}}\right) \Lambda_\text{eff},
\label{EqLambdaEffphys}
\end{align}
still with $\Leff=-\eta/\beta$ and $q$ fulfilling Eq.~(\ref{Eqq2}). The solution in the physical frame is asymptotically de Sitter only for positive $\tilde \Lambda_\tx{eff}$. In terms of the Lagrangian parameters, this translates as:
\be
\eta \beta(\eta-\beta\Lambda_\tx{b})(3\eta+\beta\Lambda_\tx{b})<0,
\label{eq:exist2}
\ee
to be combined with constraints (\ref{eq:exist1}) and (\ref{eq:Bconstraint2}). It is easy to check that the three conditions together imply that the solution was also asymptotically de Sitter in the original Horndeski frame, i.e., that $\Lambda_\tx{eff}>0$. Note that the above transformation can be trivially extended to the stealth solution (\ref{eq:stealthJohn})-(\ref{eq:Xstealth}). Hence, the physical metrics are again black hole solutions. This is not a trivial result, as a disformal transformation may change the nature of solutions, rendering them even singular upon going from one frame to the other.

At this stage, it seems from Eq.~(\ref{EqLambdaEffphys}) that a very small $\tilde
\Lambda_\text{eff}$ remains possible, for instance if
$\Lambda_\text{eff} = -\eta/\beta$ is chosen to almost compensate
$\Lambda_\text{b}$. However, the field equations written in
the physical frame $\tilde g_{\mu\nu}$ actually always imply
$\tilde \Lambda_\text{eff} \simeq \Lambda_\text{b}$
\cite{Babichev:2016kdt}. Moreover, we will see below that the
stability of the solution forces the observable $\tilde \Lambda_\text{eff}$ to be of the same
order of magnitude as $\Lambda_\text{b}$ (or even larger).
Therefore, in the simple model (\ref{eq:John2}), the small
observed cosmological constant cannot be explained by the
self-tuning mechanism, and some other reason must be invoked,
like in standard general relativity. It remains that this model
is observationally consistent if the constant
$\Lambda_\text{b}$ entering (\ref{eq:John2}) is small enough.

As mentionned in Chapter \ref{ch:pert}, the odd-parity perturbations of solution
(\ref{eq:ansatzphys})--(\ref{EqLambdaEffphys}) have been analyzed in
\cite{Ogawa:2015pea}. The causal structure carries through upon disformal field redefinitions (\ref{eq:disformal2}), as long as these are not singular. Thus, in the physical frame, spin-2 perturbations propagate in the effective metric $\mathcal{G}_{\mu\nu}$, as given in Eqs.~(\ref{eq:Gtt})--(\ref{eq:Gtr}). One can compare it with the metric $\tilde g_{\mu\nu}$,
Eq.~(\ref{Eqgtilde}), to which matter (including photons) is
assumed to be coupled. The result is remarkably simple:
\be
\tilde{g}_{\mu\nu}= \left(1+
\frac{\Lambda_\text{b}}{\Lambda_\text{eff}}\right)
\mathcal{G}_{\mu\nu}.
\label{EqGmunueff}
\ee
The two metrics are related by a constant conformal factor. Therefore, they have identical causal structure at any point of spacetime, provided that the conformal factor is positive. When this is the case, matter and spin-2 perturbations propagate exactly the same way (on the contrary, if the above conformal factor is negative, the cone of the spin-2 degree of freedom is exactly complementary to the matter one and they have no overlap nor common exterior). One is therefore led to impose that
\be
\Lambda_\tx{eff}(\Lambda_\tx{b}+\Lambda_\tx{eff})>0,
\label{eq:samemetric1}
\ee
i.e., in terms of the Lagrangian parameters:
\be
\eta(\eta-\beta\Lambda_\tx{b})>0.
\label{eq:samemetric2}
\ee
Under this condition, even close to the black hole, the causal cones of spin-2 and matter perturbations
exactly coincide. In other words, the universal coupling of
matter to the disformal metric (\ref{Eqgtilde}) suffices to
ensure $c_\text{grav} = c_\text{light}$ even in a very
inhomogeneous configuration. 

\subsection{Stability of the solution}

We are now going to study the stability of the above solution, based on the results of Chapter \ref{ch:pert}. We do so in the Horndeski frame because we already know the causal structure of the spin-2 and scalar degrees of freedom from Chapter \ref{ch:pert}. In this chapter, three causal cones had to be considered: the matter causal cone and the cones associated to scalar and gravitational perturbations (with their associated effective metrics). Now, if matter couples to the physical metric $\tilde g_{\mu\nu}$, the effective metric for the spin-2 degree of freedom is identical to the effective metric for light, provided Eq.~(\ref{eq:samemetric2}) is fulfilled. This effectively reduces the number of causal cones under scrutiny from three to two. Stability can then be ensured only if the
scalar causal cone shares a common interior direction and a
common exterior hypersurface with that of $\tilde g_{\mu\nu}$ (or equivalently $\mc{G}_{\mu\nu}$).

Again, an exact calculation is possible in the case of a homogeneous spacetime, i.e., when $\tilde m=0$. In comparison to the case of Chapter \ref{ch:pert}, where matter was minimally coupled to $g_{\mu\nu}$, there is one more condition to fulfill (corresponding to the fact that the change of frame must be well-defined). Thus, there are in total seven conditions. These include Eqs.~(\ref{eq:exist1}), (\ref{eq:Bconstraint2}), (\ref{eq:exist2}) and (\ref{eq:samemetric2}), as well as two hyperbolicity conditions and one orientation compatibility condition. The two hyperbolicity conditions are actually unchanged, and are given by Eqs.~(\ref{eq:hypscal2})-(\ref{eq:hypgrav2}). Finally, the scalar and matter/spin-2 causal cones have compatible orientations provided that 
\be
\eta(3\eta+\beta\Lb)>0.
\ee
It is straightforward to check that this seven inequalities again define the range of Eqs.~(\ref{eq:range1})-(\ref{eq:range2}). Rewriting these conditions in terms of the observed $\tilde\Lambda_\text{eff}$, one obtains
\begin{align}
\tx{either}~\eta&>0,~\beta<0~\tx{and}~\Lambda_\text{b} < \tilde\Lambda_\text{eff},
\label{eq:rangeLphys1}
\\
\tx{or}~\eta&<0,~\beta>0~\tx{and}~\Lambda_\text{b} < \tilde\Lambda_\text{eff} <\frac{3}{2}\,\Lambda_\text{b}.
\label{eq:rangeLphys2}
\end{align}
Again, when setting $\Lambda_\text{b}=0$, the interval of
stability disappears. This is more clearly seen using the ranges in the form (\ref{eq:range1})-(\ref{eq:range2}). In
other words, it is the presence of vacuum energy which allows for
a window of stability for the black hole solution. As stressed below Eq.~(\ref{EqLambdaEffphys}) though, Eqs.~(\ref{eq:rangeLphys1})-(\ref{eq:rangeLphys2}) mean that
self-tuning is impossible in this specific model, since the
observed cosmological constant must always be larger than the
bare one.

Taking $\tilde m \neq0$ again makes the stability analysis very difficult technically, but the results are essentially unchanged with respect to Paragraph \ref{sec:bhstab}. Indeed, the effect of coupling matter to $\tilde g_{\mu\nu}$ rather than $g_{\mu\nu}$ is merely to transform the matter causal cone at $\pm 45^\circ$ into the red causal cone of Fig.~\ref{fig:bhsoundcone}. Thus, it even simplifies the analysis. Again, for parameters that yield a stable cosmological configuration, the black hole solutions seem to be always stable as well.

\section{Self-tuning models with unit speed for gravitational waves}
\label{sec:selft}

Although the model (\ref{eq:John2}) (considered as the Horndeski-frame theory) is not able to self-tune a bare cosmological constant, there still exists an infinite class of other Horndeski and beyond models which provide self-tuning, as shown in Ref.~\cite{Babichev:2016kdt}. In this section, we prove that a subclass of them also satisfies the $c_\text{grav} = c_\text{light}$
constraint. From now on, we will consider the physical frame only. Thus, let us change notations slightly to avoid putting tildes everywhere. Matter will be assumed to couple minimally to $g_{\mu\nu}$, and we no
longer consider any disformal transformation such as
(\ref{Eqgtilde}). Additionally, in order to avoid hiding several different scales in the functions of
$X$, it is convenient to work with the
dimensionless quantity
\begin{equation}
\mc{X}= \dfrac{-\pd_\mu\varphi\,\pd^\mu\varphi}{M^2},
\label{EqX}
\end{equation}
$M$ being the only mass scale entering the Lagrangian of the
scalar field $\varphi$, itself chosen dimensionless (beware not
to confuse $M$ with the Planck mass $M_\text{Pl}$). All the
coefficients entering dimensionless functions of $\mc{X}$ will also
be assumed to be of order 1. Up to a total
derivative, the shift-symmetric Horndeski and beyond Lagrangian\footnote{The quintic sector is excluded here, because it cannot play a cosmological role. The cubic sector is omitted, because it must anyway be passive for the self-tuning solutions derived in 
\cite{Babichev:2016kdt}.} may then be rewritten
as
\be
\begin{split}
\mathcal{L}&=\dfrac{M_\text{Pl}^2}{2} \left(R-2\Lambda_\text{b}\right) -M^4 \mc{X} f_2(\mc{X}) - 4 s_4(\mc{X}) G^{\mu\nu} \pd_\mu \varphi\,\pd_\nu\varphi
\\
&\quad-\,\frac{f_4(\mc{X})}{M^2}\, \varepsilon^{\mu\nu\rho\sigma} \varepsilon^{\alpha\beta\gamma}_{\hphantom{\alpha\beta\gamma}\sigma}\,
\pd_\mu\varphi\, \pd_\alpha\varphi\,\nd_\nu\pd_\beta\varphi\, \nd_\rho\pd_\gamma\varphi.
\label{EqLf}
\end{split}
\ee
The full translation of the $s_i$--$f_i$ functions in terms of the $G_i$--$F_i$ that we used so far is provided in Appendix \ref{ap:otherparam}. The action (\ref{EqLf}) is supplemented by a matter Lagrangian, minimally coupled to the variable $g_{\mu\nu}$ in terms of which Eq.~(\ref{EqLf}) is written. For instance, the model (\ref{eq:John2}) corresponds to constant values $\zeta = M_\text{Pl}^2/2$,
$f_2 = -\eta/M^2$, $s_4 = -\beta/4$, and $f_4 = 0$. In terms of these notations, the $c_\text{grav} = c_\text{light}$ constraint (\ref{EqBigF4})
becomes then
\begin{equation}
f_4(\mc{X}) = -\frac{4s_4(\mc{X})}{\mc{X}}.
\label{EqSmallf4}
\end{equation}
This fixes $f_4(\mc{X})$, while $f_2(\mc{X})$ and $s_4(\mc{X})$ are arbitrary. For monomials,
this means that one needs
\begin{equation}
f_2 = k_2 \mc{X}^\alpha,\quad s_4 = \kappa_4 \mc{X}^\gamma,\quad
f_4 = -4\kappa_4 \mc{X}^{\gamma-1}.
\label{Eqf4}
\end{equation}
where $k_2$, $\kappa_4$, $\alpha$ and $\gamma$ are dimensionless
constants of order 1.
Note that negative exponents $\alpha$ and $\gamma$ are perfectly
allowed and consistent in this cosmological context, where the
background solution corresponds to a strictly positive value of
$\mc{X}$. Small perturbations are thus well-defined around such a background, because the density $\mc{X}$ will never cross 0 and thus will not run into the pole of one the functions of the Lagrangian.

Particular self-tuning models respecting $c_\text{grav} =
c_\text{light}$ are thus easily obtained from (\ref{Eqf4}).
However, we already mentioned in Paragraph \ref{sec:deSitterJohn}
that one cannot impose $\Lambda_\text{eff} \ll \Lambda_\text{b}$ and be satisfied with this result. This is because making $\Lambda_\text{eff}$ very small in general has a side effect: it renormalizes Newton's constant to an accordingly small value, way to small to fit the observed value \cite{Babichev:2016kdt}. Luckily however,
a subclass among the models (\ref{Eqf4}) is such that $M_\text{Pl}$ (or equivalently Newton's constant)
remains unrenormalized, and it is thus possible to get
$\Lambda_\text{eff} \ll \Lambda_\text{b}$ in a consistent way. The self-tuning $\Lambda_\text{eff} \ll \Lambda_\text{b}$ is achieved by choosing an appropriate value of $M$.
This subclass corresponds to the exponent $\gamma =
-3/2$, i.e., $s_4 = \kappa_4 \mc{X}^{-3/2}$ and $f_4 =
-8\kappa_4 \mc{X}^{-5/2}$. In terms of the $G_i$ notation
of Eq.~(\ref{eq:H}), this reads
\begin{eqnarray}
G_2(X) &=& - M_\text{Pl}^2 \Lambda_\text{b}
- k_2 M^4 \left(\frac{2X}{M^2}\right)^{\alpha+1},
\label{EqG2}\\
G_4(X) &=& \frac{1}{2} M_\text{Pl}^2
- 2 \kappa_4 M^3 \left(2X\right)^{-1/2},
\label{EqG4b}
\end{eqnarray}
while $F_4$ is given by Eq.~(\ref{EqBigF4}).
One then finds that the Schwarzschild-de Sitter equations of
Ref.~\cite{Babichev:2016kdt} can be solved provided $\alpha \neq
-1$ and $\alpha\neq -1/2$, and they imply
\begin{equation}
(H^2)^{\alpha+1}(M^2)^{\alpha+2} \propto
(M_\text{Pl}^2 \Lambda_\text{b})^{\alpha+3/2},
\label{EqMagnitude}
\end{equation}
where $H$ is the Hubble rate, defined as $H^2=\Leff/3$.
The proportionality factor depends on the dimensionless constants $k_2$, $\kappa_4$ and
$\alpha$, and is thus itself of order 1.
Therefore, if $\alpha\neq -2$, it suffices to choose $M$
appropriately to get $H$ equal to the observed value, whatever
the large $\Lambda_\text{b}$ entering the action. Note that
all these models (with $\alpha \not\in\{-2, -1,
-1/2\}$ and $\gamma = -3/2$) admit exact
Schwarzschild-de Sitter solutions such that $\Lambda_\text{eff}$ is consistent with its small observed value,
and they also satisfy $c_\text{grav} = c_\text{light}$, at least
in the asymptotic homogeneous Universe.

As underlined at the very end of \cite{Babichev:2016kdt}, if
the bare cosmological constant happens to take the huge value
$\Lambda_\text{b} \simeq M_\text{Pl}^2$, then the particular
case $\alpha=-5/4$ needs a rather natural value of the
scale $M \simeq 100~\text{MeV}$, similar to usual elementary
particle masses.

Another interesting particular case is $\alpha=-3/2$, for
instance $f_2 = s_4 = -\mc{X}^{-3/2}$ and $f_4 = 4 \mc{X}^{-5/2}$ (choosing
here $k_2 = \kappa_4 = -1$ to simplify, the signs being imposed
by the field equations). This corresponds to
\begin{eqnarray}
G_2(X) &=& - M_\text{Pl}^2 \Lambda_\text{b}
+ M^5 \left(2X\right)^{-1/2},
\label{EqG2c}\\
G_4(X) &=& \frac{1}{2} M_\text{Pl}^2
+ 2 M^3 \left(2X\right)^{-1/2},
\label{EqG4c}\\
F_4(X) &=&
2M^3 \left(2X\right)^{-5/2}.
\label{EqF4c}
\end{eqnarray}
Then, the exact version of Eq.~(\ref{EqMagnitude}) implies that
one must choose $M = 2\sqrt{3}\, H$, i.e., the very small
observed Hubble expansion rate $H$ must actually be put by hand
in the action via the scale $M$. But this drawback comes
with the great bonus that this observed $H$ now depends only on
$M$, and no longer on the bare vacuum energy density
$M_\text{Pl}^2 \Lambda_\text{b}$. Therefore, even if
$\Lambda_\text{b}$ happens to change because of a phase
transition during the cosmological evolution of the Universe, the
effective $\Lambda_\text{eff} = (M/2)^2 = 3 H^2$ remains constant
and small.

The important conclusion is that elegant self-tuning cosmological
models are still allowed, even when taking into account the
experimental constraint $c_\text{grav} = c_\text{light}$. Note
that for these models, we did not prove that the speed equality
remains valid in the vicinity of massive bodies. However, the
result of Sec.~\ref{sec:GWspeed} for the simple model (\ref{eq:John2}) and the argument
of Refs.~\cite{Ezquiaga:2017ekz,Langlois:2017dyl} show that it
may remain true, at least for Schwarzschild-de Sitter black-hole
solutions. The stability of these self-tuning models should also
be analyzed, as we did above for the model (\ref{eq:John2}). Aside of this, it would be of great interest to study a
more realistic cosmological evolution for these self-tuning
models, as in \cite{Martin-Moruno:2015lha,Starobinsky:2016kua}, where certain branches of solutions
were shown to screen matter as well as the cosmological
constant.

\newpage
\thispagestyle{empty}
\cleardoublepage
\chapter*{Summary and outlook}
\addcontentsline{toc}{chapter}{Summary and outlook}
\markboth{Summary and outlook}{Summary and outlook}

\section*{Summary}
\addcontentsline{toc}{section}{Summary}

Scalar-tensor theories (in a broad sense) have interesting cosmological properties. Aside from this, it is important to know whether they can accommodate compact objects, and if so, to what extent these objects are similar to those encountered in general relativity. My thesis elaborated on these questions. The first essential tools for this analysis are no-hair theorems. We discussed a previously established no-hair theorem in Horndeski theory, aimed at static, spherically symmetric and asymptotically flat black holes. We extended it to star configurations under very similar assumptions. 

At the same time, Horndeski and beyond theory is complex and many assumptions are required to prove the no-hair theorems. Accordingly, there also exist many ways to arrive at solutions which do possess hair. We explored these paths in the second part of the thesis. One of the essential results of this part is that, when the scalar field plays the role of dark energy, the black hole (or star) solutions generically possess hair. We saw so by analyzing the effect of the simplest cubic and quartic models in Horndeski theory. The quintic sector has no interest in this context, since it is ruled out by gravitational wave experiments. When a $\mathbb{Z}_2$ symmetry is imposed on the scalar field (i.e., in the quadratic plus quartic case), exact solutions are easy to find. Some reproduce exactly general relativistic solutions, including a Schwarzschild-de Sitter spacetime that has simple self-tuning properties. In the cubic case, one has to use numerical integration to find black hole solutions, but there exist some with similar self-tuning properties. 

This part was also the occasion to study which models possess asymptotically flat solutions with a non-trivial static scalar field (as opposed to the case where its time evolution is imposed by cosmology). It is more difficult to find solutions in this case, and actually only few models circumvent the no-hair theorem. It is the presence (or absence) of specific terms in the Lagrangian that allows for solutions that deviate from general relativity. We examined in detail these terms in the case of the quartic and quintic sector of Horndeski and beyond theory. It is still legitimate to consider quintic models in this framework, where the scalar field does not have to play a role at cosmological scale. In the shift-symmetric case, the quintic Lagrangian that forces a non-trivial scalar field amounts to a linear coupling between the scalar field and the Gauss-Bonnet invariant. It does not admit regular black hole solutions, unless one also allows the norm of the current to diverge at the horizon.

Once background solutions with hair are known, a further step is to study their stability (and eventually their formation through collapse). It was the goal of the third part to study the stability of some solutions with hair presented in the rest of the thesis. We focused on the solutions where the scalar is a dark energy field. In this case, the intricate time and space dependence makes it impossible to use some standard stability criteria. In particular, we proved that when the Hamiltonian is unbounded from below (which would usually correspond to a ghost) in some coordinate system, the solution can still be stable. Instead, we established the right stability criterion: the causal cones associated to each degree of freedom (scalar, spin-2 and matter) must share a common time and a common spacelike Cauchy hypersurface.
This result is not limited to a specific scalar-tensor theory, and might prove interesting in various modified gravity models. As an example, we applied this criterion to the self-tuned Schwarzschild-de Sitter solution presented earlier in the thesis. There exists a window of stability for the parameters of the theory (which does not seem to depend on the presence of a black hole). The stability conditions actually prevent one  from downgrading a large bare cosmological constant to a small effective value compatible with observation.

As a by-product of the linear stability analysis, we were able to compute the speed of gravitational waves in a strongly curved spherically symmetric background (as opposed to weakly curved cosmological backgrounds, where the result was already known). When the speed of light and of gravitational waves is the same over large cosmological scales, this remains true in the strongly curved environment of a black hole. We finally gave a class of Horndeski and beyond models that pass gravitational wave tests as well as local tests, and provide an actual self-tuning of the cosmological constant.

\section*{Outlook}
\addcontentsline{toc}{section}{Outlook}

Further developments in the field of scalar-tensor theories can arise either from observation, or from yet unsolved theoretical questions. As to observations, gravitational wave interferometers will bring one the most important contributions in the forthcoming years\footnote{Note that the large-scale structure at low redshifts will also soon be scrutinized by various experiments \cite{Bose:2018orj}, while other projects are proposed to increase the precision of the local tests in the Solar System \cite{Sakstein:2017pqi}.}. Concerning black holes, the accumulation of events will statistically allow us to put very narrow bounds on the post-Newtonian parameters and to check general relativistic predictions (as illustrated by Figs.~7 and 8 of Ref.~\cite{TheLIGOScientific:2016pea}). On the other hand, we will learn a lot about the equation of state and core structure of neutron stars, for instance through their tidal deformability (see Fig.~5 of Ref.~\cite{TheLIGOScientific:2017qsa} for instance). It would be extremely useful to have at hand viable modified gravity models, as benchmarks in the alternative theory space. Of course, modeling a binary merger in modified gravity is an extraordinarily difficult task. An interesting and fairly unexplored starting point would be the study of gravitational collapse in Horndeski and beyond theory, in a spherically symmetric situation. It would be worth extending the numerical code of Ref.~\cite{OConnor:2009iuz} to investigate such situations.

On the cosmological side, another crucial issue is the effect of self-tuning on matter or radiation. Indeed, the aim of self-tuning mechanisms is to tune away a huge bare cosmological constant, but matter and radiation should not be equally suppressed by this mechanism (for instance, the existence of a radiation dominated era is necessary for Big Bang nucleosynthesis). At least for some specific self-tuning models, it is impossible to have both a coherent cosmological history and stability at all epochs \cite{Linder:2013zoa}. References \cite{Martin-Moruno:2015lha} and \cite{Starobinsky:2016kua} went further on this topic, but it is not clear yet whether some models can have a sensible cosmological history, together with self-tuning and stability. It would be particularly interesting to study the self-tuning Horndeski and beyond models that pass the gravitational wave test from this perspective. As we saw in this thesis, stability seems to compete with self-tuning. If this is also true in the case of the models presented in Sec.~\ref{sec:selft}, the ability of scalar-tensor theories to provide a realistic self-tuning would be seriously jeopardized.

Finally, a theoretical issue that was raised recently is the well-posed character of the field equations. References \cite{Papallo:2017qvl,Papallo:2017ddx} proved that quartic and quintic Horndeski theory are not ``strongly hyperbolic'' ---~a stronger requirement than the ``weak'' hyperbolicity we used in this thesis~--- around a flat background in any harmonic gauge. This can be interpreted as the impossibility to determine an estimate of the energy at a given time after the moment when initial conditions were specified. Deciding whether this constitutes an actual flaw of the theory requires further study.



\appendix

\newpage
\thispagestyle{empty}
\cleardoublepage
\chapter{Alternative parametrizations of Horndeski and beyond theory}
\label{ap:otherparam}

In this appendix, we provide and discuss other parametrizations of Horndeski theory. In the literature, $\varphi_\mu$ often stands for $\pd_\mu\varphi$, as well as $\varphi_{\mu\nu}$ for $\nd_\mu\pd_\nu\varphi$. Using these notations, let us define the following building blocks:
\begin{align}
L_{(2,0)} &= -\frac{1}{3!}\, \varepsilon^{\mu\nu\rho\sigma}\varepsilon^\alpha_{\hphantom{\alpha}\nu\rho\sigma}\,\varphi_\mu\varphi_\alpha
= \varphi_\mu\varphi^\mu,
\label{eqL2Gal}
\\
L_{(3,0)} &= -\frac{1}{2!}\, \varepsilon^{\mu\nu\rho\sigma}\varepsilon^{\alpha\beta}_{\hphantom{\alpha\beta}\rho\sigma}\,\varphi_\mu \varphi_\alpha\varphi_{\nu\beta}
= \varphi_\mu\varphi^\mu \,\Box\varphi - \varphi^\mu\varphi_{\mu\nu}\varphi^\nu,
\label{eqL3Gal}
\\
L_{(4,0)} &=-\,\varepsilon^{\mu\nu\rho\sigma}\varepsilon^{\alpha\beta\gamma}_{\hphantom{\alpha\beta\gamma}\sigma}\,\varphi_\mu \varphi_\alpha\varphi_{\nu \beta}\varphi_{\rho\gamma}
\label{eqL4Gal}
\\
&=\varphi_\mu\varphi^\mu\left(\Box \varphi\right)^2-2 \,\varphi^{\mu} \varphi_{\mu\nu}\varphi^{\nu} \,\Box \varphi-\varphi_\mu\varphi^\mu \varphi_{\nu\rho}\varphi^{\nu\rho}+2 \,\varphi^{\mu}\varphi_{\mu\nu}\varphi^{\nu\rho}\varphi_{\rho},
\\
L_{(5,0)} &=-\,\varepsilon^{\mu\nu\rho\sigma}\varepsilon^{\alpha\beta\gamma\delta}\varphi_\mu \varphi_\alpha\varphi_{\nu\beta} \varphi_{\rho\gamma}\varphi_{\sigma\delta}
\label{eqL5Gal}
\\
\begin{split}
&=\varphi_\mu\varphi^\mu\left(\Box \varphi\right)^3-3\, \varphi^{\mu} \varphi_{\mu\nu}\varphi^{\nu}\left(\Box \varphi\right)^2-3 \,\varphi_\mu\varphi^\mu\varphi_{\nu\rho}\varphi^{\nu\rho}\,\Box \varphi
\\
&\quad +6 \,\varphi^{\mu}\varphi_{\mu\nu}\varphi^{\nu\rho}\varphi_{\rho}\, \Box \varphi+2 \,\varphi_\mu\varphi^\mu\varphi_{\nu}^{\hphantom{\nu}\rho}\varphi_{\rho}^{\hphantom{\rho}\sigma}\varphi_{\sigma}^{\hphantom{\sigma}\nu}
\\
&\quad +3 \,\varphi_{\mu\nu}\varphi^{\mu\nu}\varphi^{\rho}\varphi_{\rho\sigma}\varphi^{\sigma}-6\, \varphi^{\mu}\varphi_{\mu\nu}\varphi^{\nu\rho}\varphi_{\rho\sigma}\varphi^{\sigma},
\end{split}
\\
L_{(4,1)} &= -\,\varepsilon^{\mu\nu\rho\sigma}\varepsilon^{\alpha\beta\gamma}_{\hphantom{\alpha\beta\gamma}\sigma}\,\varphi_\mu \varphi_\alpha R_{\nu\rho\beta\gamma}
= - 4\, G^{\mu\nu} \varphi_\mu \varphi_\nu,
\label{eqL41Gal}
\\
L_{(5,1)} &= -\,\varepsilon^{\mu\nu\rho\sigma}\varepsilon^{\alpha\beta\gamma\delta} \varphi_\mu\varphi_\alpha\varphi_{\nu\beta}R_{\rho\sigma\gamma\delta}
\label{eqL51Gal}
\\
\begin{split}
&=2\, \varphi_\mu \varphi^\mu R\,\Box \varphi-2\,\varphi^{\mu} \varphi_{\mu\nu}\varphi^{\nu}R-4\, \varphi^{\mu} R_{\mu\nu}\varphi^\nu\, \Box \varphi
\\
&\quad-4\, \varphi_\mu \varphi^\mu \varphi^{\nu\rho} R_{\nu\rho}+8\, \varphi^{\mu}\varphi_{\mu\nu}R^{\nu\rho}\varphi_{\rho}+4\,\varphi^{\mu} \varphi^{\nu} \varphi^{\rho\sigma}R_{\mu\rho\nu\sigma}.
\end{split}
\end{align}
These definitions are those of Ref.~\cite{Babichev:2016kdt}. Horndeski and beyond theories are obtained by multiplying the above building blocks by arbitrary functions of $\varphi$ and $X$. However, these functions differ from the $G_i$ and $F_i$ we used in the body of the thesis, although they are of course related. Before introducing these functions, let us stress that the kinetic density of the scalar field is designated by many different notations, each author having its own preference in the literature. In the body of this thesis, we stuck to the notation $X=-\pd_\mu\varphi \,\pd^\mu\varphi/2$. We also used the notation $\mc{X}$ in the last chapter to designate the same quantity with a different normalization, in terms of a mass scale $M$, $\mc{X}=-\pd_\mu\varphi\,\pd^\mu\varphi/M^2$, see Eq.~(\ref{EqX}). This quantity was also designated by the letter $X$ in \cite{Babichev:2016kdt,Babichev:2017lmw}. At the same time, the authors of \cite{Gleyzes:2014dya,Gleyzes:2014qga,Langlois:2015cwa,Langlois:2015skt,Achour:2016rkg,BenAchour:2016fzp,Crisostomi:2016tcp,Crisostomi:2016czh} use the letter $X$ to designate $\pd_\mu\varphi\, \pd^\mu\varphi$.

The Horndeski and beyond functions that are used in Chapter \ref{ch:wave} are more easily defined in terms of $\mc{X}$. This is the quantity we will use to define alternatively the Horndeski and beyond action:
\begin{equation}
S_\tx{bH} = \dfrac{\Mp^2}{2}\int \tx{d}^4x \sqrt{-g}\,(R-2\Lb)+ \sum_{(n,p)} \int \tx{d}^4x \sqrt{-g}\, \mathcal{L}_{(n,p)} 
\label{eqAction}
\end{equation}
where the Lagrangians $\mathcal{L}_{(n,p)}$ are related to the $L_{(n,p)}$ ones by
\begin{align}
\mathcal{L}_{(2,0)} &= M^2 f_2(\varphi,\mc{X}) L_{(2,0)} = -M^4 \mc{X} f_2(\varphi,\mc{X}),
\label{eqL2}\\
\mathcal{L}_{(3,0)} &= f_3(\varphi,\mc{X}) L_{(3,0)},
\label{eqL3}\\
\mathcal{L}_{(4,0)} &=\frac{1}{M^2} f_4(\varphi,\mc{X}) L_{(4,0)},
\label{eqL4}\\
\mathcal{L}_{(5,0)} &=\frac{1}{M^4} f_5(\varphi,\mc{X}) L_{(5,0)},
\label{eqL5}\\
\mathcal{L}_{(4,1)} &= s_4(\varphi,\mc{X}) L_{(4,1)},
\label{eqL41}\\
\mathcal{L}_{(5,1)} &=\frac{1}{M^2} s_5(\varphi,\mc{X}) L_{(5,1)}
\label{eqL51}.
\end{align}
One can then relate the two formulations of Horndeski and beyond theory, and for instance translate the $G_i$ and $F_i$ functions in term of the $s_i$ and $f_i$. Packing $\mathcal{L}_{(4,0)}$ and $\mathcal{L}_{(4,1)}$ together, as well as $\mathcal{L}_{(5,0)}$ and $\mathcal{L}_{(5,1)}$, one can show that 
\begin{align}
\mathcal{L}_{(2,0)} &= G_2(\varphi,X),
\label{eqGeneralizedHorndeskiOld2}
\\
\mathcal{L}_{(3,0)} &= -G_3(\varphi,X)\Box\varphi +\text{total derivative},
\label{eqGeneralizedHorndeskiOld3}
\\
\begin{split}
\mathcal{L}_{(4,0)} + \mathcal{L}_{(4,1)} &=G_4(\varphi,X) R + G_{4X} \left[\left(\Box\varphi\right)^2
- \varphi_{\mu\nu}\varphi^{\mu\nu}\right]
\\
&\quad+F_4(\varphi,X) \varepsilon^{\mu\nu\rho\sigma}
\varepsilon^{\alpha\beta\gamma}_{\hphantom{\alpha\beta\gamma}\sigma}\,
\varphi_\mu
\varphi_\alpha \varphi_{\nu\beta} \varphi_{\rho\gamma}
+\text{total derivative},
\label{eqGeneralizedHorndeskiOld4}
\end{split}
\\
\begin{split}
\mathcal{L}_{(5,0)} + \mathcal{L}_{(5,1)} &=
G_5(\varphi,X) G^{\mu\nu} \varphi_{\mu\nu}
-\frac{1}{6} G_{5X}
\left[\left(\Box \varphi\right)^3
- 3\, \Box\varphi\, \varphi_{\mu\nu}\varphi^{\mu\nu}\right.
\\
&\quad\left.+ 2\, \varphi_{\mu\nu}\varphi^{\nu\rho}
\varphi_\rho^{\hphantom{\rho}\mu}\right]+F_5(\varphi,X) \varepsilon^{\mu\nu\rho\sigma}
\varepsilon^{\alpha\beta\gamma\delta}\, \varphi_\mu \varphi_\alpha
\varphi_{\nu\beta} \varphi_{\rho\gamma}\varphi_{\sigma\delta}
\\
&\quad+\text{total derivative},
\label{eqGeneralizedHorndeskiOld5}
\end{split}
\end{align}
where the $G_i$ and $F_i$ are related to the new functions by
\begin{align}
G_2(\varphi,X) &= -\Mp^2\Lb-M^4 \mc{X} f_2(\varphi,\mc{X}),\\
G_3(\varphi,X) &= M^2\left[\mc{X} f_3(\varphi,\mc{X})
+ \frac{1}{2}\int f_3(\varphi,\mc{X}) \tx{d}\mc{X}\right],\\
G_4(\varphi,X) &= \dfrac{\Mp^2}{2}-2 M^2 \mc{X} s_4(\varphi,\mc{X}),
\label{eqG4}\\
F_4(\varphi,X) &= \left[-f_4(\varphi,\mc{X})+4 s_{4\mc{X}}(\varphi,\mc{X})\right]/M^2,
\label{eqF4}\\
G_5(\varphi,X) &= 4\mc{X} s_5(\varphi,\mc{X}) + 2\int s_5(\varphi,\mc{X}) \tx{d}\mc{X},\\
F_5(\varphi,X) &= \left[-f_5(\varphi,\mc{X}) + \frac{4}{3} s_{5\mc{X}}(\varphi,\mc{X})\right]/M^4.
\end{align}
References \cite{Gleyzes:2014dya,Gleyzes:2014qga,Lin:2014jga,Crisostomi:2016tcp} used yet another notation, with functions $A_i$ and $B_i$; they relate to the functions $f_i$ and $s_i$ as follows:
\begin{align}
A_2(\varphi,2X) &= -M^4 \mc{X} f_2(\varphi,\mc{X}),\\
A_3(\varphi,2X) &= M^3 \mc{X}^{3/2} f_3(\varphi,\mc{X}),\\
A_4(\varphi,2X) &= -M^2 \mc{X} \left[\mc{X} f_4(\varphi,\mc{X}) + 2 s_4(\varphi,\mc{X})\right],\\
B_4(\varphi,2X) &= - 2M^2 \mc{X} s_4(\varphi,\mc{X}),\\
A_5(\varphi,2X) &= M \mc{X}^{3/2}\left[\mc{X} f_5(\varphi,\mc{X}) +2 s_5(\varphi,\mc{X})\right],\\
B_5(\varphi,2X) &= -4 M \mc{X}^{3/2} s_5(\varphi,\mc{X}),
\end{align}
assuming that the gradient of $\varphi$ is timelike. This should be true on cosmological backgrounds, where the spatial variations of $\varphi$ are neglected with respect to cosmic time evolution.

\chapter{Horndeski and beyond field equations in spherical symmetry}
\label{ap:fieldeqs}

In this appendix, we provide the useful field equations to treat Horndeski and beyond theory in spherical symmetry. The equations given hereafter assume the parametrization of Eqs.~(\ref{eq:H}) and (\ref{eq:bH4})-(\ref{eq:bH5}), together with shift symmetry, Eq.~(\ref{eq:SS}). The ansatz for the metric is given by Eq.~(\ref{eq:statmetric}), and the scalar field is assumed to be static too\footnote{The same equations including the time dependence of the scalar field, i.e., $q\neq0$, can be found in \cite{Babichev:2016kdt} in the parametrization of Eqs.~(\ref{eqL2})--(\ref{eqL51}).}, Eq.~(\ref{eq:statphi}). In general, it is sufficient to make use of two of the mertic field equations of motion, the $(tt)$ and $(rr)$ ones, together with the fact that the radial component of the Noether current $J^\mu$ vanishes. First, the $(tt)$ equation reads: 
\be
\begin{split}
\label{eq:Ett}
&G_2 -\dfrac12f\varphi'X'G_{3X}+ \dfrac{2}{r} \left(\dfrac{1 - f}{r} - f'\right) G_4 + \dfrac{4}{r}f \left(\dfrac{1}{r} + \dfrac{X'}{X} + \dfrac{f'}{f}\right) X G_{4X} 
\\
& + \dfrac{8}{r} fX X' G_{4XX}+ \dfrac{1}{r^2}f\varphi'\left[(1-3f)\dfrac{X'}{X}-2f'\right]XG_{5X}-\dfrac{2}{r^2}f^2\varphi'XX'G_{5XX}
\\
&+\dfrac{16}{r}fX^2X'F_{4X}+\dfrac{8}{r}f\left(\dfrac{4X'}{X}+\dfrac{f'}{f}+\dfrac{1}{r}\right)X^2F_4
\\
&-\dfrac{12}{r^2}f^2h\varphi'X^2\left(\dfrac{2f'}{f}-\dfrac{5X'}{X}\right)F_5-\dfrac{24}{r^2}f^2h\varphi'X^2X'F_{5X}=0.
\end{split}
\ee
Let us recall that, when $\varphi$ depends on $r$ only, $X=-f \varphi'^2/2$. Then, the radial component of the current $J^r$ is given by:
\be
\begin{split}
\label{eq:staticJrbH}
J^r&=-f \varphi' G_{2X} - f\dfrac{r h' + 4 h}{r h} X G_{3X} + 2 f \varphi' \dfrac{f h - h + r f h'}{r^2 h}G_{4X} 
\\
&\quad+ 4 f^2 \varphi' \dfrac{h +rh' }{r^2h} X G_{4XX} -f h' \dfrac{1 - 3 f}{r^2h} X G_{5X} +2 \dfrac{h' f^2}{r^2h} X^2 G_{5XX}
\\
&\quad+ 8 f^2 \varphi' \dfrac{h +rh' }{r^2h} X (2 F_4+X F_{4X})- 12 \dfrac{f^2 h'}{rh} X^2 (5 F_5+2 X F_{5X})
.\end{split}
\ee
This expression should be equated to zero whenever assuming that $J^2$ does not diverge. Finally, to be precise, rather  than the $(rr)$ metric equation itself, we use a linear combination of it with the $J^r=0$ equation, in the fashion described in \cite{Babichev:2016kdt}. Namely, if $\mc{E}^{rr}$ stands for the $(rr)$ metric field equation, we write $\mc{E}^{rr}-J^r\pd^r\varphi$. The result gives:
\be
\begin{split}
\label{eq:Err}
&G_2 - \dfrac{2}{r^2h}(frh'+fh-h) G_4 + \dfrac{4f}{r^2h} (rh'+h) X G_{4X} - \dfrac{2}{r^2h}f^2h'\varphi'XG_{5X}
\\
&+\dfrac{8f}{r^2h}(rh'+h)X^2F_4-\dfrac{24}{r^2h}f^2h'\varphi'X^2F_5 = 0.
\end{split}
\ee
Interestingly, the cubic Horndeski term $G_3$ disappears from this combination.




\chapter{De-mixing spin-2 and spin-0 degrees of freedom}
\label{ap:diagonal}

In this appendix, we summarize the progresses that have been made in the (covariant) decomposition of the spin-2 and scalar degrees of freedom. 

\section{The case of general relativity}

Let us start by describing what happens in usual general relativity (with or without a cosmological constant $\Lambda$, because we will only focus on the kinetic part of the effective action). Of course, in this case, there is no spin-0 degree of freedom, and thus no de-mixing to achieve; but this will illustrate the generic procedure from which one can extract the effective metric for the spin-2 degree of freedom, $\mc{G}_{\mu\nu}$. The metric is assumed to be perturbed according to:
\be
g_{\mu\nu}=\bar g_{\mu\nu}+h_{\mu\nu}.
\label{eq:pertmetric}
\ee
Then, one can expand the Einstein-Hilbert action $S_\tx{EH}=\int \tx{d}^4x \sqrt{-g} \,\zeta R$ at second order (in terms of $h_{\mu\nu}$). One gets
\be
S_\tx{EH}[g_{\mu\nu}]\underset{h\to0}{=}S_\tx{EH}[\bar g_{\mu\nu}]+\delta^{(1)}S_\tx{EH}+\delta^{(2)}S_\tx{EH}+\mc{O}(h^3).
\ee
The quantity $\delta^{(1)}S_\tx{EH}$ provides the background field equations, while $\delta^{(2)}S_\tx{EH}$ can be used to determine the propagation of perturbations over a given background. One obtains 
\be
\delta^{(1)}S_\tx{EH}=-\zeta \displaystyle\int{ \tx{d}^4x \sqrt{-g} \, G^{\mu\nu}h_{\mu\nu}},
\ee
meaning that the background field equations for the metric are Einstein's equations, $G_{\mu\nu}=0$. The kinetic part (i.e., highest derivatives terms only) of the second-order perturbed Einstein-Hilbert action reads:
\be
\begin{split}
\delta^{(2)}S_\tx{EH;\,Kin}&=\zeta \displaystyle\int \tx{d}^4x \sqrt{-g} \,\left(-\dfrac14 \nd_\rho h_{\mu\nu}\nd^\rho h^{\mu\nu}+\dfrac12 \nd_\rho h_{\mu\nu}\nd^\nu h^{\rho\mu}-\dfrac12 \pd_\mu h \nd_\nu h^{\mu\nu}\right.
\\
&\quad+\left.\dfrac14\pd_\mu h\, \pd^\mu h \right).
\end{split}
\ee
Unless specified otherwise, the geometric quantities such as Einstein's tensor or the covariant derivative are the ones associated with the background metric $\bar g_{\mu\nu}$. The variation of $\delta^{(2)}S_\tx{EH;\,Kin}$ with respect to $h_{\mu\nu}$ yields a linear second-order differential equation on $h_{\mu\nu}$:
\be
\Box h_{\mu\nu}-2\nd_\alpha \pd_{(\mu}h_{\nu)}^{\;\alpha}+\nd_\mu\pd_\nu h -g_{\mu\nu}\left(\Box h-\nd_\alpha\pd_\beta h^{\alpha\beta}\right)=0
\label{eq:prophGR}
\ee
Again, this equation is valid only when neglecting lower derivative terms, but it suffices for instance in order to determine the causal structure and characteristics on some given background. Defining $\hbar_{\mu\nu}=h_{\mu\nu}-h g_{\mu\nu}/2$, and imposing the transverse gauge $\nd_\mu\hbar^{\mu\nu}$, Eq.~(\ref{eq:prophGR}) reads $\Box \hbar_{\mu\nu}=0$, showing that spin-2 excitations propagate in the effective metric $g_{\mu\nu}$ in general relativity.

\section{Quadratic and cubic Horndeski sectors}

The case of quadratic Horndeski theory (supplemented with the usual Ricci scalar) is equally simple to that of general relativity. Namely, considering the action 
\be
S_2=\displaystyle\int{\tx{d}^4x\sqrt{-g}\left[\zeta R+G_2(X)\right]},
\ee
if one perturbs the metric according to Eq.~(\ref{eq:pertmetric}) and the scalar according to 
\be
\varphi=\bar\varphi+\chi,
\ee
one finds that spin-2 and scalar perturbations do not mix (at least at kinetic level). Thus, the spin-2 excitations are identical to those of general relativity, and propagate in $\mc{G}_{\mu\nu}=g_{\mu\nu}$. The scalar perturbations can be shown to propagate in the effective metric described in Paragraph \ref{sec:kessence}:
\be
\mathcal{S}_2^{\mu\nu} = G_{2X} g^{\mu\nu} - G_{2XX}\pd^\mu\varphi \,\pd^\nu\varphi.
\ee
Things get more complicated already for the cubic Horndeski action. Reference \cite{Babichev:2012re} managed to diagonalize spin-2 and spin-0 perturbations for the simple following cubic term which corresponds to $G_3(X)=2\gamma X$:
\be
S_3=\displaystyle\int{\tx{d}^4x\sqrt{-g}\left(\zeta R-2\gamma X \Box\varphi\right)},
\ee
with $\gamma$ a constant\footnote{Let us note that Ref.~\cite{Babichev:2012re} had also included a standard kinetic term $G_2 \propto X$ in its analysis. The full analysis for arbitrary functions $G_2$ and $G_3$ should not be too difficult to obtain using the same procedure.}. One may guess that the spin-2 sector should be unaffected by the presence of the cubic term, but this is not obvious at first. With the same notations as above for the perturbations, one finds the following second-order Lagrangian (keeping again only highest-order derivatives):
\be
\begin{split}
\delta^{(2)}S_\tx{3;\,Kin}&=\displaystyle\int \tx{d}^4x \sqrt{-g} \left\{\zeta\left(-\dfrac14 \nd_\rho h_{\mu\nu}\nd^\rho h^{\mu\nu}+\dfrac12 \nd_\rho h_{\mu\nu}\nd^\nu h^{\rho\mu}-\dfrac12 \pd_\mu h \nd_\nu h^{\mu\nu}\right.\right.
\\
&\quad+\left.\dfrac14\pd_\mu h\, \pd^\mu h\right) +\gamma\left[
2\Box\varphi \left(\partial_\mu\chi\right)^2
- 2 \nabla_\mu\partial_\nu\varphi\,
\partial^\mu\chi \,\partial^\nu\chi\vphantom{\dfrac12}\right.
\\
&\quad+\left.\left. \partial_\mu\varphi\,\partial_\nu\varphi\,
\partial_\lambda\chi\nabla^\lambda h^{\mu\nu}
- 2\partial^\mu\varphi\, \partial^\nu\varphi\,
\partial_\mu\chi\, \left(\nd_\lambda h^\lambda_{\;\nu}
-\frac{1}{2}\pd_\nu h\right)
\right] \right\}.
\label{eq:actionpertDGP}
\end{split}
\ee
Although the first terms are identical to general relativity, one immediately sees that the second part of the action (proportional to $\gamma$) mixes the kinetic terms of $h_{\mu\nu}$ and $\chi$. Thus, one cannot call $h_{\mu\nu}$ the spin-2 and $\chi$ the spin-0 perturbations respectively. One first has to diagonalize the above kinetic structure. This is achieved \cite{Babichev:2012re} by the following change of variables:
\begin{equation}
\hat h_{\mu\nu} = h_{\mu\nu} -\frac{2\gamma}{\zeta}\left[\partial_\mu\varphi\,\partial_\nu\varphi
-\frac{1}{2}g_{\mu\nu}
\left(\partial_\lambda\varphi\right)^2\right] \chi.
\label{eq:changemetric}
\end{equation}
Then, the kinetic part of $\delta^{(2)}S_3$ reads, in terms of $\hat h_{\mu\nu}$ and $\chi$: 
\be
\delta^{(2)}S_\tx{3;\,Kin}[\hat h_{\mu\nu},\chi]=\delta^{(2)}S_\tx{EH;\,Kin}[\hat h_{\mu\nu}]+\displaystyle\int \tx{d}^4x \sqrt{-g} \left(-\dfrac12 \mc{S}_3^{\mu\nu}\pd_\mu\chi\,\pd_\nu\chi\right),
\label{eq:cubicsep}
\ee
with
\be
\mc{S}_3^{\mu\nu}=-\gamma g^{\mu\nu}\left[4\Box\varphi
+\dfrac{\gamma}{\zeta}\left(\partial_\lambda\varphi\right)^4 \right]
+4\gamma\nabla^\mu\partial^\nu\varphi
+4 \frac{\gamma^2}{\zeta}\left(\partial_\lambda\varphi\right)^2
\partial^\mu\varphi\,\partial^\nu\varphi.
\ee
Equation (\ref{eq:cubicsep}) yields the desired result: $\hat h_{\mu\nu}$ and $\chi$ do not talk to each other, and propagate in their own effective metric (respectively $g^{\mu\nu}$ and $\mc{S}_3^{\mu\nu}$).

\section{Quartic Horndeski sector}

For the goals of Chapter \ref{ch:pert}, we tried to carry out a similar analysis and diagonalize the scalar and spin-2 perturbations for the quartic Horndeski action. However, it was not possible to arrive at a conclusion this way. We had to resort to the mode analysis presented in Sec.~\ref{sec:Johnstab}, with the drawback that it is only valid over a spherically symmetric background. In passing, Ref.~\cite{Bettoni:2016mij} tried to do the same calculation with in mind the speed of gravitational waves, but could conclude only by neglecting the mixing terms between spin-0 and spin-2 degrees of freedom. We summarize here the point at which we arrived in this calculation. We analyzed the following action (which corresponds to $G_4=\zeta+\beta X$):
\be
S_4=\displaystyle\int{\tx{d}^4x\sqrt{-g}\left[\zeta R+\beta G^{\mu\nu}\pd_\mu\varphi\,\pd_\nu\varphi\right]}.
\ee
Using again the same notations for perturbations, one obtains
\be
\begin{split}
\delta^{(2)}S_\tx{4;\,Kin}&=\displaystyle\int \tx{d}^4x \sqrt{-g} \left\{\beta G^{\mu\nu}\pd_\mu\chi\,\pd_\nu\chi+(\zeta+\beta X)\left(-\dfrac14 \nd_\rho h_{\mu\nu}\nd^\rho h^{\mu\nu}\right.\right.
\\
&\quad\left.+\dfrac12 \nd_\rho h_{\mu\nu}\nd^\nu h^{\rho\mu}-\dfrac12 \pd_\mu h \nd_\nu h^{\mu\nu}+\dfrac14\pd_\mu h\, \pd^\mu h\right) 
\\
&\quad-\beta\left[\dfrac14\pd^\rho\varphi\,\pd^\sigma\varphi\nd_\rho h_{\mu\nu}\nd_\sigma h^{\mu\nu}+\dfrac12\pd^\nu\varphi\,\pd^\sigma\varphi\nd_\rho h_{\mu\nu}\nd^\rho h^\mu_{\;\sigma}\right.
\\
&\quad-\pd^\nu\varphi\,\pd^\sigma\varphi\nd_\rho h_{\mu\nu}\nd_\sigma h^{\mu\rho}-\dfrac12\pd^\mu\varphi\,\pd^\sigma\varphi\nd_\rho h_{\mu\nu}\nd^\nu h_\sigma^{\;\rho}
\\
&\quad+\dfrac12 \pd^\rho\varphi\,\pd^\sigma\varphi\nd_\mu h_{\rho\sigma}\nd_\nu h^{\mu\nu}+\dfrac12\pd^\rho\varphi\,\pd^\mu\varphi\,\pd_\mu h\nd_\nu h_\rho^{\;\nu}
\\
&\quad+\dfrac12 \pd^\sigma\varphi\,\pd^\nu\varphi\,\pd_\mu h\nd_\nu h^\mu_{\;\sigma}-\dfrac14 \pd^\mu\varphi\,\pd^\nu\varphi\,\pd_\mu h\,\pd_\nu h
\\
&\quad-\dfrac12 \pd^\rho\varphi\,\pd^\sigma\varphi\nd_\mu h_{\rho\sigma}\pd^\mu h +\nd^\mu\pd^\nu\varphi\,\pd^\rho\chi\nd_\rho h_{\mu\nu}
\\
&\quad-2\nd_\mu\pd_\rho\varphi\,\pd^\rho\chi\nd_\nu h^{\mu\nu}+\nd^\mu h_{\mu\nu} \Box\varphi\,\pd^\nu\chi-\pd_\mu h\,\Box\varphi\, \pd^\mu \chi
\\
&\quad\left.\left.+\nd_\mu\pd_\nu\varphi\,\pd^\nu\chi\,\pd^\mu h\vphantom{\dfrac12}\right]\right\}.
\end{split}
\ee
The part which is quadratic in $h_{\mu\nu}$ is much more involved than in the previous case. Indeed, because of the presence of Einstein's tensor in the action, one does not expect to recover that spin-2 perturbations simply propagate in the effective metric $g_{\mu\nu}$. The last five terms in the above expression are kinetic mixing terms between $h_{\mu\nu}$ and $\chi$. They are the terms one would like to get rid of, in order to diagonalize the perturbations. However, we found indications that no linear change of variables ---~in the fashion of Eq.~(\ref{eq:changemetric})~--- allows to eliminate these terms. 

The second-order kinetic action is much simplified in the transverse traceless gauge, $h=0$ and $\nd_\mu h^{\mu\nu}=0$. In this gauge, it boils down to:
\be
\begin{split}
\delta^{(2)}S_\tx{4;\,Kin}&=\displaystyle\int \tx{d}^4x \sqrt{-g} \left\{\beta G^{\mu\nu}\pd_\mu\chi\,\pd_\nu\chi-\dfrac14(\zeta+\beta X)\nd_\rho h_{\mu\nu}\nd^\rho h^{\mu\nu}\right.
\\
&\quad-\beta\left[\dfrac14\pd^\rho\varphi\,\pd^\sigma\varphi\nd_\rho h_{\mu\nu}\nd_\sigma h^{\mu\nu}+\dfrac12\pd^\nu\varphi\,\pd^\sigma\varphi\nd_\rho h_{\mu\nu}\nd^\rho h^\mu_{\;\sigma}\right.
\\
&\quad\left.\left.+\nd^\mu\pd^\nu\varphi\,\pd^\rho\chi\nd_\rho h_{\mu\nu}\vphantom{\dfrac12}\right]\right\},
\end{split}
\ee
in agreement with Ref.~\cite{Bettoni:2016mij}. Even in this gauge, however, it is not possible to diagonalize the perturbations by a mere linear transformation of the variables.

\chapter{Monopole perturbation of some Horndeski black holes}
\label{sec:appmonop}

\section{Coefficients for the Schwarzschild-de Sitter solution}
\label{sec:coeffmonop}

The aim of this appendix is to display the explicit expressions of the
various coefficients used in the analysis of Sec.~\ref{sec:effmetrics}.
Those entering Eq.~(\ref{eq:H012Lag}) and subsequent read
\begin{align}
c_1&=-\dfrac{\beta q}{r} \varphi' \sqrt{\dfrac{f}{h}},
\\
c_2&=2fc_1,
\\
c_3&=- \dfrac{1}{2r}\sqrt{\dfrac{f}{h}} (-2\zeta h + \beta q^2-3 \beta hf \varphi'^2),
\\
c_4 &= \dfrac{2}{h} c_3,
\\
c_5 &= \dfrac{q^2}{4r^2}\dfrac{1}{\sqrt{hf}} [2\beta(1-f-rf')+\eta r^2],
\\
\begin{split}
c_6 &= -\dfrac{1}{4hr^2} \sqrt{\dfrac{f}{h}} \left\{\dfrac12 h \varphi'^2[(2\beta-12 \beta f+\eta r^2)h-12 \beta f r h']+rh'(\beta q^2-2\zeta h)\right.
\\
&\quad-\left.\vphantom{\dfrac12}2\zeta h^2-\beta q^2 h\right\},
\end{split}
\\
c_7 &= \dfrac{q^2}{4hr^2}\dfrac{1}{\sqrt{hf}} [2 \beta f rh'+h(2\beta-2\beta f+\eta r^2)],
\\
c_8 &= \dfrac{q}{2hr^2} \varphi' \sqrt{\dfrac{f}{h}} [-6 \beta f rh'+h(2\beta-6\beta f+\eta r^2)].
\end{align}
The coefficients entering Eq.~(\ref{eq:H0H1Lag}) and subsequent read
\begin{align}
\tilde{c}_1&=-\dfrac{c_1^2}{4c_6},
\\
\tilde{c}_2&=-\dfrac{c_3^2}{4c_6},
\\
\tilde{c}_3&=\dfrac{c_1 c_3}{2c_6},
\\
\tilde{c}_4&=-\dfrac{c_4^2}{4c_6},
\\
\tilde{c}_5&=\dfrac{c_4 c_3}{2c_6},
\\
\tilde{c}_6&=-\dfrac{c_1 c_4}{2c_6},
\\
\tilde{c}_7&=\dfrac{2c_2c_6-c_8c_3}{2c_6},
\\
\tilde{c}_8&=\dfrac{c_8c_1-c_4c_7}{2c_6},
\\
\tilde{c}_9&=-\dfrac{c_6(c_7c_3'+c_3c_7'-c_7^2+4c_5c_6)-c_7c_3c_6'}{4c_6^2},
\\
\tilde{c}_{10}&=-\dfrac{c_7 c_8}{2c_6},
\\
\tilde{c}_{11}&=-\dfrac{c_8^2}{4c_6}.
\end{align}
Finally, the coefficients entering Eq.~(\ref{eq:H01Lag}) read
\begin{align}
a_1 &= \tilde{c}_1,
\\
a_2 &=\dfrac{\tilde{c}_3}{2\tilde{c}_1},
\\
a_3 &=\dfrac{\tilde{c}_6}{2\tilde{c}_1},
\\
a_4 &= 0,
\\
a_5 &= \dfrac{2\tilde{c}_1\tilde{c}_7-\tilde{c}_8 \tilde{c}_3}{\tilde{c}_6\tilde{c}_3},
\\
a_6 &= \dfrac{\tilde{c}_7}{\tilde{c}_3},
\\
a_7 &= \tilde{c}_9-a_1 a_5^2 + (a_1 a_2 a_5)',
\\
a_8 &= \tilde{c}_{11}-a_1 a_6^2,
\\
a_9 &= \tilde{c}_{10}-2a_1 a_6 a_5.
\end{align}

\section{Analysis of the stealth Schwarzschild solution}
\label{sec:stealthstab}

As mentioned above, there exists an exact asymptotically flat Schwarzschild solution when $\eta$ and $\Lambda_\text{b}$ vanish, with a non-trivial scalar profile (\ref{eq:stealthJohn})-(\ref{eq:Xstealth}). In this case, the parameter $q$ is no longer related to the coupling constants of the action and is in fact a free parameter. It should also be emphasized that the solution (\ref{eq:stealthJohn})-(\ref{eq:Xstealth}) is the unique static and spherically symmetric solution with a linearly time dependent scalar field and $\eta=\Lambda_\text{b}=0$. The procedure for determining the effective metric of scalar perturbations, described in Sec.~\ref{sec:effmetrics}, breaks down for this background. It is not possible to carry on after Eq.~(\ref{eq:H0H1Lag}), and to express the fields $H_0$ and $H_1$ in terms of $\pi_\tx{s}$. The reason is that for the stealth Schwarzschild solution $H_0$ and $H_1$ cannot be simultaneously expressed in terms of the master variable $\pi_s$ from the Lagrangian introduced in (\ref{eq:piH0H1Lag}). Therefore, one needs to find another way of extracting the scalar mode from the second-order Lagrangian (\ref{eq:H012Lag}) which now reads
\be
\begin{split}
\label{eq:H012Lagbis}
\mc{L}^{(2)}_\tx{s} &= c_1 H_0 \dot{H}_2+c_2H_0'H_1+c_3H_0'H_2+c_4H_1\dot{H}_2+c_6H_2^2+c_7H_0H_2
\\
&\quad+c_8H_1H_2,
\end{split}
\ee
as $c_5=0$ for the relevant background. The equation of motion for $H_2$ following from (\ref{eq:H012Lag}) is algebraic in terms of $H_2$, so as before one can find $H_2$ in terms of $H_0$ and $H_1$,
\be
\label{H2}
H_2 = -\dfrac{1}{2c_6}(-c_1\dot{H}_0-c_4 \dot{H}_1+c_3 H_0'+c_7 H_0+c_8H_1).
\ee
Substituting (\ref{H2}) in (\ref{eq:H012Lagbis}) and rearranging terms, one can write (\ref{eq:H012Lagbis}) as
\be
\begin{split}
\label{eq:H01Lag}
\mc{L}^{(2)}_\tx{s} &= a_1(\dot{H}_0 + a_2 H_0' + a_3 \dot{H}_1 + a_5 H_0 + a_6 H_1)^2
\\
&\quad+ a_7 H_0^2+a_8 H_1^2+a_9 H_0 H_1,
\end{split}
\ee
where the coefficients $a_i$ are given above, in Sec.~\ref{sec:coeffmonop}. Let us now introduce new variables $x(t,r)$ and $y(t,r)$ grouping together the time and space derivatives:
\begin{equation}
\begin{split}
H_1 & = \frac{c_1}{c_4}(x-\frac{c_1}{c_3}y),
\\
H_0 &= \frac{c_1}{c_3}y.
\end{split}
\end{equation}
Indeed the Lagrangian (\ref{eq:H01Lag}) then takes the form
\begin{equation}
\label{Lmaster1}
\mc{L}^{(2)}_\tx{s} = \mathcal{P}^2 + \mathcal{A}y^2 + \mathcal{B}xy + \mathcal{C}x^2,
\end{equation}
where
\begin{equation}
\mathcal{P}= \dot{x}-y'+\tilde{a}_1 x + \tilde{a}_2 y,
\end{equation}
and
\begin{equation}
\begin{split}
\tilde{a}_1 &= \frac{2 c_2 c_6-c_3 c_8}{c_3 c_4},\\
\tilde{a}_2 &= \frac{c_4 c_1 c_3'-c_3 c_4 c_1'+c_8 c_1^2-c_4 c_7 c_1}{c_1 c_3 c_4},\\
\mathcal{A} &= \frac{c_1^2 \left[c_4 \left(c_2 c_1'+c_1 c_2'+2 c_4 c_5\right)-c_1 c_2 c_4'\right]}{2 c_3^2 c_4^2},\\
\mathcal{B} &= \frac{c_1^2 c_2 \left(c_1 c_8-c_4 c_7\right)}{c_3^2 c_4^2},\\
\mathcal{C} &= \frac{c_1^2 c_2 \left(c_2 c_6-c_3 c_8\right)}{c_3^2 c_4^2}.\\
\end{split}
\end{equation}
Variation of (\ref{Lmaster1}) with respect to $y$ yields the constraint
\begin{equation}
\label{constraintxy}
2\mathcal{P}' + 2\mathcal{A}y+\mathcal{B}x = 0.
\end{equation}
The above constraint (\ref{constraintxy}) contains $y''$, $y'$, $\dot{x}'$ and it may be seen as an equation which determines $y$ in terms of $x$ and its derivatives. To find $y$ from (\ref{constraintxy}), the use of nonlocal (in space) operators is in general required. Since we are only interested in the absence of ghost and gradient instabilities, however, we do not need to know the exact expression of $y$ in terms of $x$. This means that we focus on higher derivative terms, i.e., we neglect terms proportional to $x$ ($y$) with respect to those proportional to $\dot{x}$ or $x'$ ($y'$). With this approximation in mind, Eq.~(\ref{constraintxy}) becomes
\begin{equation}
\dot{x}' - y''=0,
\end{equation}
which after integration over $r$ and setting to zero the integration constant yields
\begin{equation}
\label{constraintxy2}
\dot{x}=y'.
\end{equation}
By the same token, Eq.~(\ref{constraintxy}) shows that the term $\mathcal{P}^2$ in (\ref{Lmaster1}) is of lower order in derivatives in comparison with the last three terms, because from (\ref{constraintxy}) one can see that $\mathcal{P}$ is of lower order compared to $x$ and $y$. As a consequence, to the leading order in derivatives, the Lagrangian (\ref{Lmaster1}) is
\begin{equation}
\label{Lmaster2}
\mc{L}_\tx{s;\:Kin}^{(2)} = \mathcal{A}y^2 + \mathcal{B}xy + \mathcal{C}x^2,
\end{equation}
where the subscript ``Kin'' stresses that only higher order terms (kinetic part) are left in the Lagrangian. Let us then introduce $\tilde{\pi}$ as
\begin{equation}
\label{xpi}
x = \tilde\pi'.
\end{equation}
From (\ref{constraintxy2}), one easily obtains
\begin{equation}
\label{ypi}
y = \dot{\tilde\pi},
\end{equation}
where the integration constant is set to zero. Finally, substituting (\ref{xpi}) and (\ref{ypi}) in (\ref{Lmaster2}), one can find the kinetic part of the Lagrangian for the scalar perturbations:
\begin{equation}
\label{Lmaster3}
\mc{L}_\tx{s;\:Kin}^{(2)} =-\frac12 \left(\tilde{\mathcal{S}}^{tt}\dot{\tilde\pi}^2 +2 \tilde{\mathcal{S}}^{tr}\dot{\tilde\pi}\tilde\pi' + \tilde{\mathcal{S}}^{rr}\tilde\pi'^2\right),
\end{equation}
where
\begin{equation}
\label{SABC}
\tilde{\mathcal{S}}^{tt} = - 2\mathcal{A}, \quad \tilde{\mathcal{S}}^{tr} = - \mathcal{B}, \quad \tilde{\mathcal{S}}^{rr} = - 2\mathcal{C}.
\end{equation}
One can obtain the same result for the de Sitter black hole by following the above method rather than (\ref{eq:piLag}).
Indeed, first of all, the hyperbolicity condition for (\ref{Lmaster3}) reads
\begin{equation}
\label{hyperbolicity}
\mc{D}= \tilde{\mc{S}}^{tt}\tilde{\mc{S}}^{rr}-(\tilde{\mc{S}}^{tr})^2<0.
\end{equation}
The explicit expression for $\mc{D}$ in terms of $c_i$ is given by
\begin{equation}
\label{Determ}
\mc{D} = - \frac{c_1^4 c_2 \left\{c_2 \left(c_4 c_7-c_1 c_8\right){}^2-2 \left(c_2 c_6-c_3 c_8\right) \left[c_4
\left(c_2 c_1'+c_1 c_2'+2 c_4 c_5\right)-c_1 c_2 c_4'\right]\right\}}{c_3^4 c_4^4}.
\end{equation}
One can also verify that
\begin{equation}
\mc{D}= -\frac{c_1^4 c_2}{c_3^4 c_4^4 c_6^2} \Delta,
\end{equation}
where $\Delta$ is defined in Eq.~(\ref{Ddef}). In terms of $\mc{D}$, the hyperbolicity condition found in (\ref{eq:hypscal}) reads
\begin{equation}
\frac{c_1^8}{16 c_6^4 \mc{D}}<0.
\end{equation}
As long as $\mc{D}<0$, i.e., the hyperbolicity condition (\ref{hyperbolicity}) is satisfied for $\tilde{\pi}$, the hyperbolicity condition is also satisfied for $\pi$. Moreover, for $\mc{D}<0$, the variables $\tilde{\pi}$ and $\pi$ and the kinetic matrices for $\tilde{\pi}$ and $\pi$ are related as
\begin{equation}
\label{Srelation}
\mathcal{S}^{ab} = -\frac{c_1^4}{4 c_6^2 \mc{D}}\, \tilde{\mathcal{S}}^{ab},\quad
\pi = \frac{2 c_6}{c_1^2} \sqrt{-\mc{D}}\,\tilde{\pi}.
\end{equation}
where indices $a$ and $b$ are either $t$ or $r$.

The advantage of the Lagrangian (\ref{Lmaster3}) obtained here is that it also allows to treat the case of the stealth Schwarzschild black hole, for which the method of Sec.~\ref{sec:effmetrics} fails. Indeed, for the stealth solution it turns out that $\Delta=\mc{D}=0$ (in other words, $H_0$ and $H_1$ are linearly dependent). However, the kinetic matrix $\tilde{\mathcal{S}}^{ab}$ remains finite, see (\ref{SABC}), while the kinetic matrix $\mathcal{S}^{ab}$ diverges, as can be seen from (\ref{Srelation}).

For the Lagrangian (\ref{Lmaster3}), the vanishing determinant of the kinetic matrix means that the equation of motion is parabolic (for all $r$). Per se, this fact does not necessarily mean that the perturbations are pathological on the considered background. For instance, in the case of the k-essence Lagrangian $\mathcal{L}_2= G_2(X)$, for solutions where $G_{2X}(X)=0$ with {\it timelike} $X$, the perturbations behave as dust, i.e., they are governed by a wave equation with $c_\tx{s}^2=0$. The determinant of the kinetic matrix in this case is also zero, since only the ($tt$) component of the kinetic matrix is non-vanishing. For the stealth solution, the kinetic matrix reads
\begin{equation}
\label{kinmatrs}
\tilde{\mathcal{S}}^{ab} \propto
\begin{bmatrix}
\dfrac{2m r}{(2m-r)^2} & \dfrac{\sqrt{2m r}}{2m-r} \\
\dfrac{\sqrt{2m r}}{2m-r} & 1
\end{bmatrix},
\end{equation}
Notice that, for $r\gg m$, all the terms of (\ref{kinmatrs}) apart from $\tilde{\mathcal{S}}^{rr}$ tend to zero. The global factor of Eq. (\ref{kinmatrs}) may have any sign, depending on the parameters of the model and the (arbitrary) value of $q$. When this global factor is negative, the dynamics of the perturbations indeed corresponds to dust (i.e., a vanishing velocity, similarly to the example of k-essence described above), but the infinitely thin cone of propagation tends towards the $r$ axis. This, together with the fact that the spin-2 cone has a ``usual'' behavior at $r\to \infty$, makes the stealth solution pathological. It corresponds to a limit of panels (i) and (j) of Fig.~\ref{fig:CausalCones} when the dashed (blue) cone is infinitely thin. On the other hand, for a positive global factor in Eq.~(\ref{kinmatrs}), the scalar dynamics corresponds to the limit of panels (c) and (d) of Fig.~\ref{fig:CausalCones} when the dashed (blue) cone totally opens, i.e., its sound velocity is infinite. In that case, the scalar field is no longer a propagating degree of freedom.

\clearpage

\phantomsection

\addcontentsline{toc}{chapter}{Bibliography}
\bibliographystyle{jhep}
\newpage
\thispagestyle{empty}
\cleardoublepage
\bibliography{biblio}

\newpage
\thispagestyle{empty}

\phantomsection
\addcontentsline{toc}{chapter}{Abstract}
\thispagestyle{empty}
\begin{textblock}{10}(2,.3)
\textblockcolour{white}
\logoED
\end{textblock}

\begin{textblock}{13.9}(1.3,1.6)
\textblockcolour{white}
\vfill
\setlength{\fboxsep}{5mm}
\noindent\fbox{\parbox{\linewidth-2\fboxrule-2\fboxsep}{
\large{\textbf{Titre :~}\PhDTitleFR}\\

\small{\textbf{Mots clefs :~} \PhDkeywordsFR 
\begin{multicols}{2}
\textbf{R\'esum\'e :~}\PhDsumFR
\end{multicols}}
}}

\vfill
\bigskip

\vfill
\setlength{\fboxsep}{5mm}
\noindent\fbox{\parbox{\linewidth-2\fboxrule-2\fboxsep}{
\large{\textbf{Title :~}\PhDTitleEN}\\

\small{\textbf{Keywords :~} \PhDkeywordsEN 
\begin{multicols}{2}
\textbf{Abstract :~}\PhDsumEN
\end{multicols}}
}}

\vfill
\end{textblock}

\begin{textblock}{13}(13,14.6)
\textblockcolour{white}
\includegraphics[height=2cm]{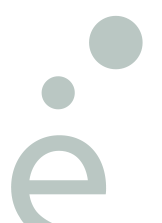}
\end{textblock}

\parindent=0pt 
\begin{textblock}{10}(2.2,14.8)
\textblockcolour{white}
\color{blue!20!red!45!black}{\footnotesize{\textbf{Universit\'e Paris-Saclay}\\Espace Technologique / Immeuble Discovery \\
Route de l'Orme aux Merisiers RD 128 / 91190 Saint-Aubin, France}}
\end{textblock}

\end{document}